\documentclass{aa}  
\usepackage{natbib}
\usepackage{amsmath}
\usepackage[utf8]{inputenc}
\usepackage{arydshln}

\usepackage{graphicx}
\graphicspath{{./}{figures/}}
\usepackage{txfonts}
\usepackage{hyperref}
\newcommand\kms{km$\,$s$^{-1}$}

\newcommand{\hi}{H\,{\sc i}}

\begin{document}

   \title{Disturbed, diffuse, or just  missing?} 
   \subtitle{A global study of the \hi \ content of Hickson compact groups}

   \author{M.~G.~Jones\inst{1,2}\fnmsep\thanks{jonesmg@arizona.edu} \and
          L.~Verdes-Montenegro\inst{1} \and
          J.~Moldon\inst{1} \and
          A.~Damas~Segovia\inst{1,3} \and
          S.~Borthakur\inst{4} \and
          S.~Luna\inst{1,5} \and
          M.~Yun\inst{6} \and
          A.~del~Olmo\inst{1} \and J.~Perea\inst{1} \and 
          J.~Cannon\inst{7} \and
          D.~Lopez Gutierrez\inst{7,8} \and
          M.~Cluver\inst{9,10} \and
          J.~Garrido\inst{1} \and S.~Sanchez\inst{1} 
          }
    \authorrunning{Jones et al.}
    \titlerunning{The \hi \ content of Hickson Compact Groups}

   \institute{Instituto de Astrof\'{i}sica de Andaluc\'{i}a (CSIC), Glorieta de la Astronom\'{i}a, 18008 Granada, Spain 
   \and
   Steward Observatory, University of Arizona, 933 North Cherry Avenue, Rm. N204, Tucson, AZ 85721-0065, USA
   \and
   Max-Planck-Institut f\"{u}r Radioastronomie, Auf dem H\"{u}gel 69, D-53121 Bonn, Germany
   \and
   School of Earth and Space Exploration, Arizona State University, 781 Terrace Mall, Tempe, AZ 85287, USA
   \and
   EGI Foundation, Science Park 140, 1098 XG Amsterdam, Netherlands
   \and
   Astronomy Department, University of Massachusetts, Amherst, MA 01003, USA
   \and
   Department of Physics \& Astronomy, Macalester College, 1600 Grand Avenue, Saint Paul, MN 55105, USA
   \and
   Physics Department, Harvard University, 17 Oxford Street, Cambridge, MA 02138, USA
   \and
   Centre for Astrophysics and Supercomputing, Swinburne University of Technology, John Street, Hawthorn 3122, Victoria, Australia
   \and
   Department of Physics and Astronomy, University of the Western Cape, Robert Sobukwe Road, Bellville, South Africa
             }

   \date{\today}

 
  \abstract
   {Hickson compact groups (HCGs) are dense configurations of four to ten galaxies, whose \hi \ morphology appears to follow an evolutionary sequence of three phases, with gas initially confined to galaxies, then significant amounts spread throughout the intra-group medium, and finally with almost no gas remaining in the galaxies themselves. It has also been suggested that several groups may harbour a diffuse \hi \ component that is resolved out by interferometric observations.}
   {The \hi \ deficiency of HCGs is expected to increase as the \hi \ morphological phase progresses along the evolutionary sequence. If this is the case, \hi \ deficiency would be a rough proxy for the age and evolutionary state of a HCG. We aim to test this hypothesis for the first time using a large sample of HCGs and to investigate the evidence for diffuse \hi \ in HCGs.}
   {We performed a uniform reduction of all publicly available VLA \hi \ observations (38 HCGs) with a purpose-built pipeline that also maximises the reproducibility of this study. The resulting \hi \ data cubes were then analysed with the latest software tools to perform a manual separation of emission features into those belonging to galaxies and those extending into the intra-group medium. We thereby classified the \hi \ morphological phase of each group as well as quantified their \hi \ deficiency compared to galaxies in isolation.}
   {We find little evidence that \hi \ deficiency can be used as a proxy for the evolutionary phase of a compact group in either of the first two phases, with the distribution of \hi \ deficiency being consistent in both. However, for the final phase, the distribution clearly shifts to high \hi \ deficiencies, with more than 90\% of the expected \hi \ content typically missing. Across all HCGs studied, we identify a few cases where there is strong evidence for a diffuse gas component in the intra-group medium, which might be detectable with improved observations. We also classify a new sub-phase where groups contain a lone \hi-bearing galaxy, but are otherwise devoid of gas. }
   {The new morphological phase we have identified is likely the result of an evolved, gas-poor group acquiring a new, gas-rich member. The large spread of \hi \ deficiencies in the first two morphological phases suggests that there is a broad range of initial \hi \ content in HCGs, which is perhaps influenced by large-scale environment, and that the timescale for morphological changes is, in general, considerably shorter than the timescale for the destruction or consumption of neutral gas in these systems.}


   \maketitle
%

\section{Introduction}

\citet{Hickson1982} catalogued 100 groups of four to ten members packed in extremely dense configurations, but located in relatively low density environments on a larger scale, which are commonly referred to as Hickson compact groups (HCGs). The high number density of HCGs is coupled with their relatively low velocity dispersions \citep{Hickson+1992}. Together, this make them ideal sites for strong, frequent galaxy--galaxy interactions. These interactions are visible through a myriad of neutral gas (\hi) tails, bridges, and clumps \citep[e.g.][]{Verdes-Montenegro+1997,Sulentic+2001,Verdes-Montenegro+2005,Serra+2013,Konstantopoulos+2013,Jones+2019}, as well as line emission from shocks \citep{Rich+2010,Vogt+2013,Cluver+2013,Duarte+2019}, apparent rapid morphological transformations \citep{Tzanavaris+2010,Plauchu-Frayn+2012,Alatalo+2015,Eigenthaler+2015,Zucker+2016,Lisenfeld+2017}, and in some cases the presence of a diffuse hot intra-group medium \citep[IGrM,][]{Belsole+2003,Desjardins+2013,OSullivan+2014b}.

Early single dish studied of HCGs revealed them to be deficient in \hi \ gas \citep{Williams+Rood1987,Huchtmeier1997}, while the first interferometric maps began to reveal the complex morphologies of HCGs in neutral gas \citep{Williams+1991,Williams+2002,Williams+vanGorkom+1995}. These findings led \citet{Verdes-Montenegro+2001} to pose the question: where is the neutral gas in HCGs? This seminal study of 72 HCGs observed with single dish telescopes confirmed and expanded on the initial findings that HCGs were \hi-deficient. A subset (16 HCGs) of the sample was followed up with Very Large Array (VLA) observations, allowing for the spatial distribution and kinematics of the gas to be studied in detail. 
Based on this imaging of 16 groups, \citet{Verdes-Montenegro+2001} proposed an evolutionary sequence for HCGs, with their phase assigned based on their \hi \ morphology. In Phase 1 of the sequence, \hi \ emission is predominantly confined to the discs of the galaxies in the group. In Phase 2 significant tidal features become evident and a significant fraction of the observed \hi \ is seen outside of galaxies. Finally, in Phase 3 almost all the gas has been removed from the galaxies or is simply undetected.\footnote{The sequence of \citet{Verdes-Montenegro+2001} originally included a Phase 3b for galaxies evolving within a common \hi \ cloud. However, there are no longer any convincing example of this proposed phenomenon \citep[e.g.][]{Verdes-Montenegro+2002}, with the possible exception of HCG~49, and we do no consider this a valid separate phase.} 

If this morphological sequence tracks the evolution of a HCG, then it would be expected that the \hi \ deficiency of the group would also generally increase with increasing morphological phase as gas (both atomic and molecular) is consumed or ionised and galaxies transform to earlier morphological types \citep{Cluver+2013,Lisenfeld+2017}. However, the validity of this proxy has never been verified for a larger sample of HCGs beyond the original sample used to formulate the proposed sequence. We note that although \hi \ content is merely one aspect of these extremely complex systems, a viable proxy of evolutionary state based solely on total \hi \ content, even if only approximate, would be a valuable tool for categorising compact groups identified in future surveys regardless of whether or not they are resolved in \hi.

\citet{Borthakur+2010} expanded on the work of \citet{Verdes-Montenegro+2001} by following up on 22 HCGs with deep \hi \ observations with the Green Bank Telescope (GBT). These observations recovered more flux than the previous VLA observations, somewhat reducing the estimated \hi \ deficiency of certain groups. However, more important was the distribution of this additional gas, which in some cases suggested the existence of a diffuse \hi \ component, spread over up to 1000~\kms \ and not recovered by the interferometer. \citet{Borthakur+2015} performed GBT \hi \ mapping for four HCGs with the richest IGrM content, revealing that the additional gas mostly traces the higher column density features mapped with the VLA, suggesting that its origin is also largely tidal.

In this work we return to the evolutionary sequence proposed by \citet{Verdes-Montenegro+2001} in order to test the long-standing assumption that the \hi \ evolutionary phase of HCGs is a proxy for their \hi \ deficiency, and vice versa. To do this we have compiled all publicly available VLA \hi \ observations of HCGs, 38 groups in total. These were reduced in as uniform a manner as possible, allowing revised \hi \ deficiency measurements of 38 groups, and the \hi \ morphological classification of 32. With this dataset we address the question of where the gas in HCGs has gone and discuss the pathways along which these groups likely evolve.

In the following section we outline how these data were compiled and reduced. In Section \ref{sec:sep_features} we discuss how \hi \ features were manually separated in each group and present moment maps and spectra of all groups. In Section \ref{sec:results} we present our results, and comparisons to previous works. Finally, in Section \ref{sec:discuss} we discuss our revised \hi \ morphological classification scheme, before presenting our conclusions in Section \ref{sec:conclusions}. In addition to scientific analysis of these data, the entire process from acquisition of the raw data, through their reduction and analysis, has been performed with the utmost attention paid to reproducibility and open science. Our efforts in this direction are discussed in detail in Appendix \ref{app:repo}.

\section{Data compilation and reduction}
\label{sec:data_reduction}

\begin{table*}
\centering
\caption{HCG \hi \ sample}
\label{tab:HCGs}
\begin{tabular}{ccccccccccc}
\hline \hline
HCG & $N_\mathrm{mem}$ & RA & Dec & $cz_\odot$ & Dist. & VLA & $\Delta v_\mathrm{chan}$ & $\sigma_\mathrm{rms}$ & Beam Size & $N_\mathrm{HI}\; (4\sigma)$ \\
 & & $\mathrm{deg}$ & $\mathrm{deg}$ & $\mathrm{km\,s^{-1}}$ & $\mathrm{Mpc}$ & Config. & $\mathrm{km\,s^{-1}}$ & $\mathrm{mJy\,beam^{-1}}$ & arcsec & $\mathrm{cm^{-2}}$ \\ \hline
2 & 3 & 7.87517 & 8.43126 & 4326 & 50 & D & 10.3 & 0.74 & $69.3 \times 51.4$ & $1.9 \times 10^{19}$ \\
7 & 4 & 9.84960 & 0.87818 & 4224 & 50 & C,D & 20.6 & 0.3 & $35.0 \times 28.0$ & $4.0 \times 10^{19}$ \\
10 & 4 & 21.53076 & 34.69095 & 4761 & 50 & DnC & 20.6 & 0.39 & $61.6 \times 49.7$ & $1.7 \times 10^{19}$ \\
15 & 6 & 31.91244 & 2.13827 & 7042 & 96 & DnC & 20.6 & 0.48 & $60.7 \times 47.5$ & $2.2 \times 10^{19}$ \\
16 & 5 & 32.38881 & -10.16298 & 3977 & 49 & C,D & 20.6 & 0.41 & $38.9 \times 31.7$ & $4.3 \times 10^{19}$ \\
19 & 3 & 40.68807 & -12.41181 & 4253 & 53 & C,D & 10.3 & 0.47 & $41.0 \times 27.0$ & $3.9 \times 10^{19}$ \\
22 & 4 & 45.88027 & -15.67570 & 2626 & 37 & CnB,D & 20.6 & 0.56 & $50.9 \times 36.8$ & $3.9 \times 10^{19}$ \\
23 & 5 & 46.77701 & -9.58547 & 4921 & 65 & C & 20.6 & 0.54 & $25.3 \times 20.0$ & $13.9 \times 10^{19}$ \\
25 & 4 & 50.18220 & -1.05192 & 6343 & 82 & DnC & 20.6 & 0.5 & $65.4 \times 56.6$ & $1.7 \times 10^{19}$ \\
26 & 7 & 50.47584 & -13.64586 & 9618 & 130 & C & 10.3 & 0.63 & $25.5 \times 16.8$ & $13.5 \times 10^{19}$ \\
30 & 4 & 69.11918 & -2.83239 & 4645 & 61 & DnC & 20.6 & 0.51 & $57.9 \times 44.8$ & $2.6 \times 10^{19}$ \\
31 & 5 & 75.40372 & -4.25671 & 4068 & 53 & CnB & 10.3 & 0.65 & $14.6 \times 12.1$ & $33.6 \times 10^{19}$ \\
33 & 4 & 77.69969 & 18.03465 & 7795 & 107 & C & 20.6 & 0.82 & $17.5 \times 15.4$ & $39.4 \times 10^{19}$ \\
37 & 5 & 138.39859 & 30.01417 & 6741 & 97 & DnC & 20.6 & 0.74 & $49.5 \times 47.4$ & $4.1 \times 10^{19}$ \\
38 & 3 & 141.91194 & 12.28081 & 8652 & 123 & D & 4.9 & 1.24 & $85.6 \times 57.6$ & $1.6 \times 10^{19}$ \\
40 & 5 & 144.72717 & -4.85196 & 6628 & 99 & DnC & 20.6 & 0.33 & $58.8 \times 44.8$ & $1.6 \times 10^{19}$ \\
47 & 4 & 156.45175 & 13.73157 & 9508 & 136 & C,D & 4.9 & 0.49 & $31.0 \times 20.0$ & $5.0 \times 10^{19}$ \\
48 & 2 & 159.44020 & -27.08746 & 2352 & 34 & DnC & 20.6 & 0.78 & $54.3 \times 37.4$ & $5.0 \times 10^{19}$ \\
49 & 5 & 164.15204 & 67.17920 & 9939 & 139 & DnC & 10.3 & 0.4 & $47.4 \times 37.1$ & $2.1 \times 10^{19}$ \\
54 & 5$^\ast$ & 172.31364 & 20.57852 & 1412 & 27 & C & 10.3 & 0.46 & $20.8 \times 16.9$ & $11.9 \times 10^{19}$ \\
56 & 5 & 173.13297 & 52.94860 & 8110 & 116 & C,D & 4.9 & 0.49 & $23.5 \times 19.6$ & $6.8 \times 10^{19}$ \\
57 & 8 & 174.46054 & 21.98504 & 9032 & 135 & C & 4.9 & 0.62 & $23.7 \times 18.9$ & $8.8 \times 10^{19}$ \\
58 & 5 & 175.54908 & 10.31703 & 6138 & 85 & DnC,D & 20.6 & 0.25 & $65.8 \times 57.0$ & $0.9 \times 10^{19}$ \\
59 & 4 & 177.10671 & 12.72623 & 4008 & 60 & C,D & 4.9 & 0.63 & $24.0 \times 16.0$ & $10.4 \times 10^{19}$ \\
61 & 3 & 183.09972 & 29.17781 & 3956 & 60 & C & 4.9 & 0.35 & $26.2 \times 18.8$ & $4.6 \times 10^{19}$ \\
62 & 4 & 193.28399 & -9.22410 & 4239 & 60 & C & 4.9 & 0.62 & $24.8 \times 14.9$ & $10.8 \times 10^{19}$ \\
68 & 5 & 208.42036 & 40.32794 & 2401 & 35 & D & 10.3 & 0.68 & $58.3 \times 54.8$ & $2.0 \times 10^{19}$ \\
71 & 5 & 212.76906 & 25.48492 & 9199 & 131 & C,D & 4.9 & 0.44 & $26.0 \times 20.0$ & $5.4 \times 10^{19}$ \\
79 & 4 & 239.80365 & 20.75163 & 4369 & 59 & C & 10.3 & 0.5 & $20.8 \times 17.4$ & $12.5 \times 10^{19}$ \\
88 & 4 & 313.09503 & -5.75790 & 6032 & 74 & C & 20.6 & 0.3 & $22.6 \times 17.2$ & $10.1 \times 10^{19}$ \\
90 & 4 & 330.52343 & -31.96680 & 2635 & 33 & DnC & 41.2 & 0.55 & $49.1 \times 39.9$ & $5.1 \times 10^{19}$ \\
91 & 4 & 332.30172 & -27.77593 & 7195 & 92 & DnC & 20.6 & 0.66 & $51.3 \times 47.0$ & $3.5 \times 10^{19}$ \\
92 & 4 & 339.00215 & 33.96577 & 6614 & 87 & B,C,D & 20.6 & 0.06 & $15.0 \times 15.0$ & $3.7 \times 10^{19}$ \\
93 & 4 & 348.85099 & 18.98311 & 5136 & 64 & DnC & 20.6 & 0.4 & $59.7 \times 53.3$ & $1.6 \times 10^{19}$ \\
95 & 4 & 349.88240 & 9.49185 & 11615 & 153 & CnB & 20.6 & 0.32 & $21.4 \times 19.3$ & $10.1 \times 10^{19}$ \\
96 & 4 & 351.99291 & 8.77408 & 8725 & 116 & C & 20.6 & 0.21 & $26.9 \times 18.3$ & $5.4 \times 10^{19}$ \\
97 & 5 & 356.86224 & -2.30542 & 6579 & 85 & DnC & 20.6 & 0.44 & $63.2 \times 49.5$ & $1.8 \times 10^{19}$ \\
100 & 5 & 0.33654 & 13.13256 & 5461 & 67 & DnC & 20.6 & 0.47 & $58.7 \times 50.5$ & $2.1 \times 10^{19}$ \\
\hline
\end{tabular}
\tablefoot{Columns: (1) HCG ID number, (2) number of group members considered in this work, (3 \& 4) group coordinates (J2000), (5) heliocentric radial velocity (re-calculated as the mean velocity of the group members), (6) group distance calculated via the Cosmicflows-3 model \citep{Tully+2016,Kourkchi+2020}, which have uncertainties of $\sim$3~Mpc, (7) VLA configurations the target group was observed with, (8) channel width, (9) rms noise, (10) synthesised beam size, (11) 4$\sigma$ column density sensitivity (over 20~\kms). \\
$^\ast$ The five `members' of HCG~54 are now thought to be clumps all associated with a single merging system.
}
\end{table*}

The interferometric data used in this work were compiled using the (J)VLA\footnote{We note that in this work we endeavour to use `VLA' to refer to historical data from the Very Large Array and `JVLA' for data from the upgraded Karl G. Jansky Very Large Array.} data archive and were originally observed as part of the projects AB651, AG645, AK580, AM559, AR251, AV206, AV221, AV227, AV230, AV275, AV285, AW234, AW272, AW351, AW500, AW568, AW601, AY155, AY160, AY86, AZ125, MYUN, and 13A-387. Together these projects observed 38 of the original 100 HCGs (HCG 2, 7, 10, 15, 16, 19, 22, 23, 25, 26, 30, 31, 33, 37, 38, 40, 47, 48, 49, 54, 56, 57, 58, 59, 61, 62, 68, 71, 79, 88, 90, 91, 92, 93, 95, 96, 97, 100) from \citet{Hickson1982} in the \hi \ line. The full list of target groups and a summary of their VLA observations are shown in Table \ref{tab:HCGs}. 

The key goals of our methodology here is to create a data reduction process that is as uniform as possible across all these heterogeneous data sets and to make it as reproducible as possible (see Appendix \ref{app:repo}). To this end we developed a pipeline for the reduction of historical VLA \hi \ data (but can also be used for JVLA data), which makes up the majority of the data for HCGs. In the following paragraphs we provide an overview of this pipeline. The code can be found on \texttt{github}\footnote{\url{https://github.com/AMIGA-IAA/hcg_hi_pipeline}}.

\subsection{Pipeline and reduction overview}

The majority of the VLA data was observed with channel resolutions of approximately 49 or 98 kHz, corresponding to velocity resolutions of about 10 and 20~\kms, respectively. The JVLA project (13A-387) used a channel resolution of 7.8~kHz ($\sim$1.6~\kms), however, we chose to average these data over three channels to expedite the reduction process and to make them more directly equivalent to the VLA data, without degrading their velocity resolution beyond $\sim$5~\kms.

Our data reduction pipeline is based in \texttt{CASA} \citep[Common Astronomy Software Applications, v5.4,][]{CASA} and \texttt{Python} and is designed specifically to reduce \hi \ spectral line data, particularly for historical VLA data sets. It provides a consistent and uniform data calibration, but allows manually tuned parameters to be adapted when needed. These are input as a parameters file with separate parameters for each section of the reduction process. The pipeline is designed so that each step can be run individually or the full pipeline run end-to-end (for a given set of observations). There is also an interactive mode where the pipeline will pause and query the user for any missing parameters rather than failing. 
The pipeline has been encoded using \texttt{CGAT-core} \citep{cgatcore} workflow management system, which allows defining the dependencies between the pipeline steps by using \texttt{Python} decorators. If previous steps or parameters are altered then the pipeline will know that it must re-run these steps before subsequent steps can be executed. The pipeline also performs automatic logging, recording every command that \texttt{CASA} executes, flagging errors automatically and halting the workflow when it is required. Subsequent steps cannot be executed without the successful completion of the prerequisite steps in the pipeline.

The reduction pipeline begins by reading in the data and applying any manual flags that the user has defined in a separate file. It then proceeds with standard flagging procedures, which were typically to remove antennae with more than 5.0 m of shadowing, and to remove the first 5~s of each scan (although in general these value can be modified in the pipeline parameters file). The \texttt{CASA} algorithm \texttt{tfcrop} is then typically run on all target fields, for example science targets and all calibrators. Again this can be, and was in a few cases, disabled in the parameters file. The pipeline then queries the user to identify the purpose of each target field (i.e. the science target, and the flux, bandpass and phase calibrators), which are not stored in the metadata of historical VLA observations. The pipeline can then proceed with standard calibration steps to calibrate the gains, phases, and absolute flux scale. 

After the first round of calibration the \texttt{rflag} algorithm is run to remove any remaining radio frequency interference (RFI). Although caution is advised when using this algorithm on spectral line data, typically for \hi \ observations the line is sufficiently weak that it is not apparent in the visibilities data, and this is not a concern. However, this step was disabled for all the JVLA data as the RFI required careful manual removal in many of these observations, as well as for a selection of the VLA data. If \texttt{rflag} is run then the pipeline repeats a second round of calibration, if not then it proceeds to split the data and perform continuum subtraction. When the science targets are split from the rest of the data the pipeline automatically identifies overlapping spectral windows and merges them together (this option can be disabled).

Before proceeding with the continuum subtraction the pipeline queries the user to define a line-free range of channels. Single dish spectra in the literature, in particular \citet{Borthakur+2010}, were used whenever available to identify genuinely emission free channels for each group. In some cases this required a 0th order fit to the continuum because line-free channels only existed on one side of the band as the bandwidths of the historical VLA data were narrow. However, for the JVLA data the bandwidth was sufficiently broad that an identical range of channels could be used in most cases without the risk of impinging on the groups' \hi \ line emission. Once the line-free channel range was defined for each group the \texttt{uvcontsub} algorithm was used to subtract the continuum. In most cases a 1st order fit was used, however, in several of the VLA data sets only a 0th order fit can be made as very few of the channels are free from line emission, or the observation is made up of two separate spectral windows that only overlap where there is line emission. Using a 0th order fit often leads to minor continuum artefacts in the final cubes, which then must be manually excluded when generating the source masks and moment maps.

The final step of the \texttt{CASA} pipeline is to image data and generate the final cubes. The pipeline images the data from each project and target separately using the \texttt{tclean} task in \texttt{CASA}. In most cases automatic masking and multi-scale clean were used. However, in a few cases the sidelobes of the synthesised beam were not sufficiently suppressed for multi-scale clean to be used, while in a few other cases using a simple primary beam mask gave better results than automated masking. Typically the data were cleaned down to a threshold of 2.5$\sigma$ within the mask. In most cases the Brigg's robust parameter was set to 2.0 in order the maximise the recovery of extended emission, at the expense of some angular resolution. This choice of weighting is likely non-optimal for many other science cases, but the robust parameter can be changed in the pipeline parameters files and the reduction repeated with a single command. In cases where a single target was observed in multiple projects a secondary script was used to combine the observations before following the same approach for the final imaging. The pipeline also generates simple moment zero maps using a simple threshold. However, these are only intended for quality control purposes and are not used in our analysis. There is also a ``cleanup'' function which can automatically remove unwanted files after execution in order to save disk space.

All the parameters files to re-execute or modify the steps of the pipeline on another machine, starting with the raw data from the VLA archive, are publicly available through \texttt{Zenodo}\footnote{\url{https://doi.org/10.5281/zenodo.6366659}}. In addition, the final image cubes and moment zero maps are stored there.

\subsection{Source masking}

All source masking was performed using the \texttt{SoFiA} package \citep{SoFiA,Serra+2015}. All \texttt{SoFiA} parameters files for each group are provided in the repository associated with this paper, which can be referred to for the exact details of the masking of any individual group. However, in general we used the smooth and clip algorithm within \texttt{SoFiA} with Gaussian smoothing kernels of approximately 0.5, 1.0, and 2.0 times the beam diameter. This was combined with boxcar smoothing in the spectral direction, but the number of channels smoothed over depended on the original resolution of the data. The historical VLA data with a resolution of $\sim$10~\kms \ was smoothed over zero, two, and three channels, whereas the data with $\sim$20~\kms \ resolution was only smoothed over zero and two channels. The newer, much higher spectral resolution, JVLA data were smoothed over zero, four, and six channels (roughly corresponding to 5.0, 20, and 30~\kms, after the initial three-channel averaging already performed). When constructing the masks typically a signal-to-noise ratio (S/N) threshold of 4.0 was used and a source reliability of over 95\% was enforced. Sources within half a beam width and within two or three channels (six channels for the JVLA data) were merged and then the masks dilated to recover as much extended flux as possible. In some cases it was necessary to manually exclude certain regions of the cube to avoid residual continuum artefacts being falsely identified as sources. 

The rationale behind this approach to source masking was not to separate out different features at this stage, indeed it is preferable to blend them into single SoFiA sources to recover as much extended flux as possible. The separation of features, described in Section \ref{sec:sep_features}, was performed after the mask generation using \texttt{SlicerAstro} \citep{SlicerAstro}.

\subsection{Distance estimates and uncertainties}

Distances to all groups were calculated from their radial velocities and the flow model\footnote{\url{http://edd.ifa.hawaii.edu/CF3calculator/}} of the Cosmicflows-3 project \citep{Graziani+2019,Kourkchi+2020}. Although \citet{Kourkchi+2020} do not provide uncertainty estimates for as part of the distance calculator, an approximate estimate can be made by assuming a typical peculiar velocity (not associated with large scale motions) of around 200~\kms. For a Hubble constant of 70~$\mathrm{km\,s^{-1}\,Mpc^{-1}}$, this equates to a minimum distance uncertainty of $\sim$3~Mpc. The distance estimates given in Table \ref{tab:HCGs} should be considered accurate to approximately this level.

Generally the velocity of each group member was taken from \citet{Hickson+1992} and the median velocity of the groups re-calculated after new members were added and false members removed. However, upon inspection of the \hi \ emission in HCGs it became apparent that the published redshifts in \citet{Hickson+1992} were erroneous for a number of galaxies. In these cases we adopted the redshift value given in the NASA/IPAC Extragalactic Database (NED). The galaxies that we identified with erroneous redshifts were: HCGs 19c, 22c, 48a, 48b, 57h, 71c, 91a, 95d, and 100d. After updating these redshifts we re-calculated the median redshift of each group and updated our distance estimate accordingly.

\section{Separation of features}
\label{sec:sep_features}

For each group with sources detected in the VLA observations we attempted to separate features into three broad classes: those associated with and likely bound to a group member galaxy, extended features likely not bound to any galaxy, and finally galaxies and features outside the core group. The separation of features associated with galaxies from extended features is usually not a straightforward process in HCGs, and consideration of the full 3-dimensional spatial and kinematic information is often required. For example, HCG~92 contains a large tidal feature that overlaps (in projection) with multiple member galaxies, but is in fact not associated with any of them \citep{Williams+2002}. 

Each \hi \ cube was displayed with the 2D and 3D visualisation tools in \texttt{SlicerAstro} and the \texttt{SoFiA} source mask was manually segmented into separate features. In cases with well resolved galaxies (spatially and in velocity) detected at high S/N, this process is quite robust as regions of emission that strongly deviate from the kinematic structure of each galaxy can be readily identified and excised. However, in many cases sources are marginally resolved, blended, and/or detected at low S/N. Wherever possible we attempt to perform this separation, but we note that in these cases the process can become quite subjective. However, as the purpose of this separation is to be able to broadly classify HCGs based on their \hi \ morphology, a certain level of uncertainty and subjectivity can be tolerated.

There is also the issue that some HCGs originally had false members, which have since been excluded (making some groups triplets). Triplets are included in our analysis, but are distinguished from $N \geq 4$ groups wherever possible. Other groups have new members which were too low surface brightness (LSB) to be included in the original catalogue or were too separated from the other members (with no apparent connection in optical images), but are clearly connected in \hi. The goal of this work is not to re-define or refine the HCG catalogue, hence we take the following simple approach: potential new members which are separated from the core (original) groups by more than $\sim$100~kpc will not be included as new members, unless they have a definite connection in \hi, indicating that they have already begun interacting with the core group. We note that in some cases such galaxies may be within (or entering for the first time) the DM halo of the CG, however, in this work we are seeking to characterise the \hi \ morphology, content, and evolution of the core group members and therefore these peripheral members are tangential to our goals unless they have already begun interactions with the core group. Within the core of each group, galaxy--galaxy projected separations (for nearest neighbours) are almost always less than 100~kpc. Hence this represents a reasonable cutoff separation for considering new members.

A galaxy must contain \hi \ for it to be detected in our observations, which clearly leads to some form of bias, as gas-poor galaxies in a similar configuration may not be considered members where gas-rich ones would be, in the absence of a uniform, deep, and highly complete redshift survey covering all HCGs, the introduction of such biases on some level is unavoidable. We have thus favoured clarity and simplicity over attempting to redefine these compact groups, for which frequently there would be insufficient data. 

In the following subsections we show a contour map of \hi \ emission (moment 0) overlaid on an optical image (from DECaLS, SDSS, or POSS) for each group. We favour DECaLS \citep{Dey+2019} images wherever they exist, as these are the deepest of the three, followed by SDSS \citep{York+2000} and POSS \citep{Reid+1991}. We use the $r$-band (R-band for POSS) images obtained directly from the public image cutout services of each survey. In cases where we were able to separate features, the group member galaxies are shown in blue contours, extended features are shown in green, and non-members are shown in red. For cases where it was not possible to separate features, the moment zero map is shown with multi-coloured contours, starting at 4$\sigma$ (black) and each subsequent contour being twice the previous one. Dashed black contours indicate -4$\sigma$.

In addition, a spectrum is shown for each group. This spectrum includes all emission within the \texttt{SoFiA} source mask, including non-members and features. Note that in some cases this includes emission well beyond the region shown in the optical images of each group. Where it exists a GBT spectrum \citep{Borthakur+2010} is overlaid on the VLA spectrum, along with a weighted version of the VLA spectrum that accounts for the response of the GBT beam (and any difference in the pointing centres). We model the GBT beam as a circular 2D Gaussian with a HPBW of 9.1\arcmin. The primary beam-corrected VLA data are weighted (in the image plane) with a Gaussian with a width such that its convolution with the VLA synthesised beam (assumed to be circular, with a width equal to the average of the major and minor axes), gives a Gaussian of HPBW 9.1\arcmin. This thereby approximately accounts for the beam response of the GBT observations, which in cases where there is emission near the edge (or beyond) the GBT primary beam can be essential to making a fair comparison. Finally, the moment one (velocity field) contour maps overlaid on the optical images of each group (with detected \hi \ emission) are shown in Appendix \ref{app:mom1}.

\subsection{HCG~2}

\begin{figure}[h]
    \centering
    \includegraphics[width=\columnwidth]{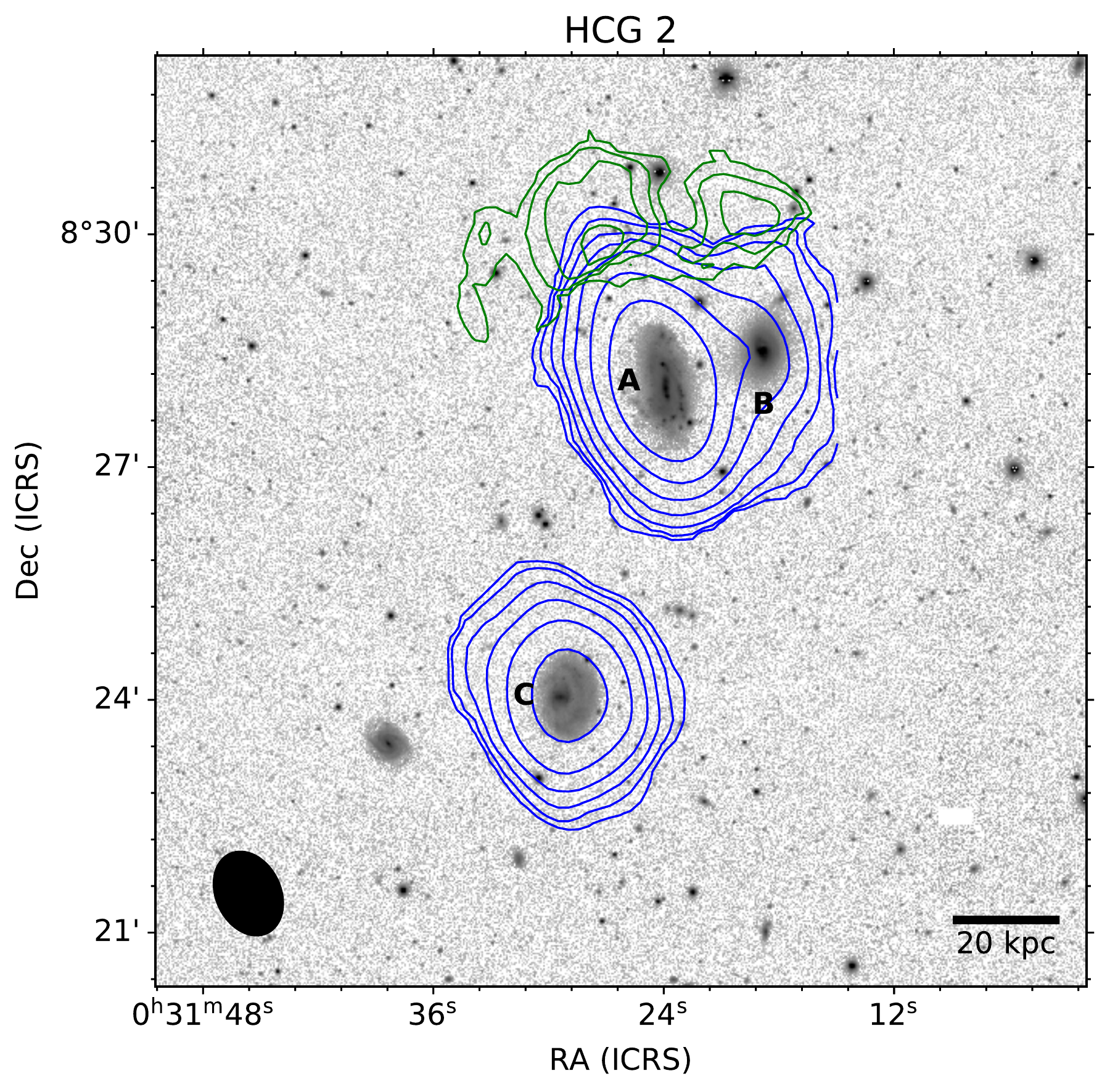}
    \caption{Integrated \hi \ emission (moment 0) contours overlaid on a DECaLS $r$-band image. HCG members are shown with blue contours, extended features in green, and non-member galaxies in red (none in this panel). The VLA synthesised beam is shown in the bottom left as a solid black ellipse and a scale bar indicating 20~kpc is in the lower right. The contours start at 4$\sigma$ (over 20~\kms) and each subsequent contour is double the one before it. The minimum contour level is listed for each group in Table \ref{tab:HCGs}. }
    \label{fig:HCG2_split_overlay}
\end{figure}

\begin{figure}[h]
    \centering
    \includegraphics[width=\columnwidth]{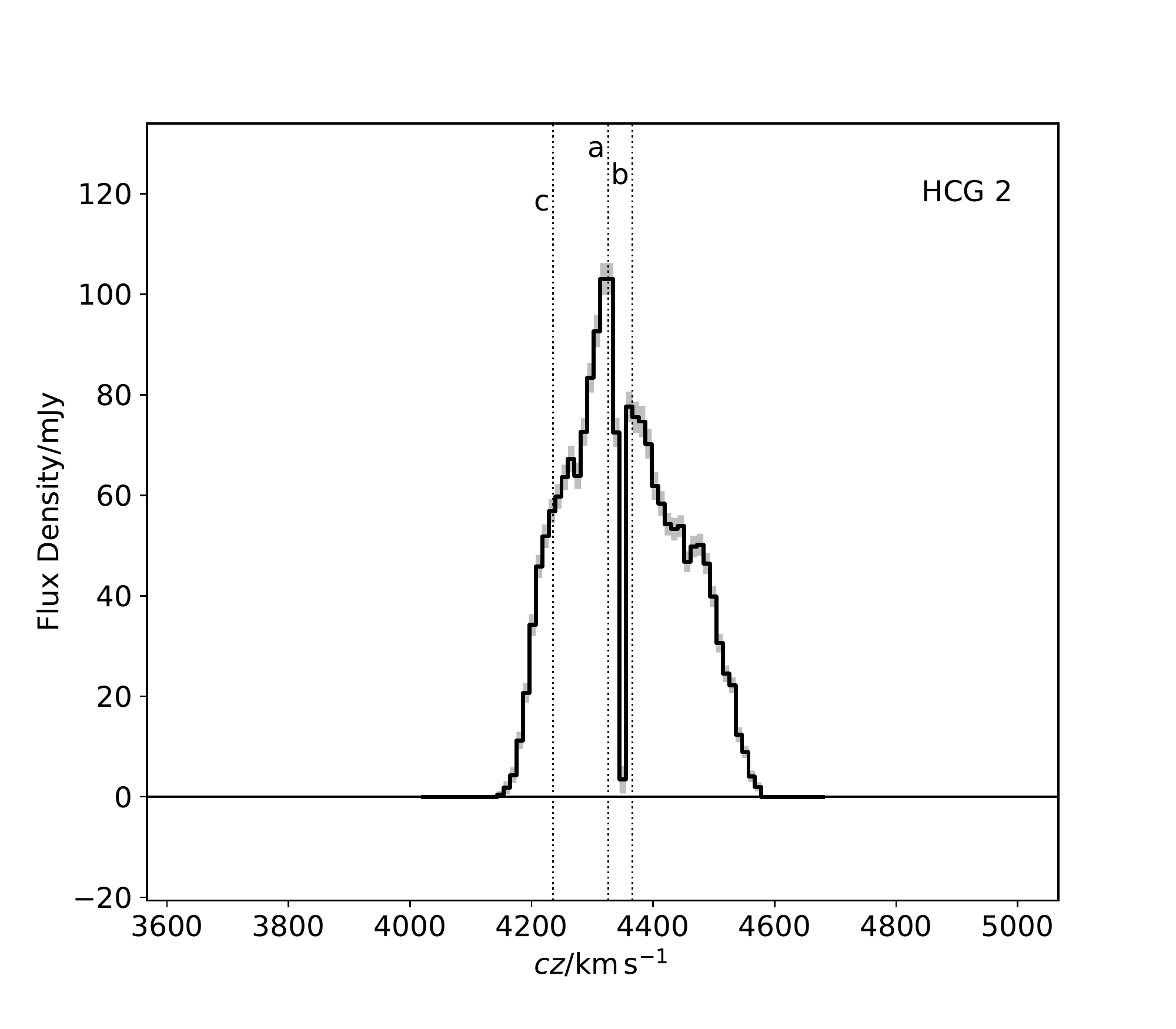}
    \caption{VLA spectrum of all detected \hi \ emission (solid black line) within the primary beam. The light grey shading around the black line indicates the uncertainty based on the rms noise and the number of pixels included in the source mask for a given channel. The vertical dotted lines indicate the velocities of group member galaxies. Regions of the spectrum where all values are exactly zero indicate spectral ranges where there are no pixels in the source mask.}
    \label{fig:HCG2_spec}
\end{figure}

HCG~2 is a triplet (HCG~2d is a background object) of late-type galaxies at approximately 4300~\kms \ (Figure \ref{fig:HCG2_split_overlay}). The global moment map and velocity field of the group clearly shows that HCG~2a and c are strongly detected. HCG~2c shows minimal signs of disturbance other than the misalignment of its iso-velocity contours with its minor axis, suggesting a probable warp. HCG~2a on the other hand is blended with emission that appears to be from HCG~2b, as it is co-spatial with the optical source and has a separate velocity structure (though is scarcely larger than a single beam) that occurs at the expected redshift. We therefore attribute this emission to HCG~2b as best as possible. There is a small, faint extension to the north of HCG~2a which we designate as an extended feature as it does not conform to the velocity structure of either HCG~2a or b. Finally, it should also be noted that one, central channel of the HCG~2 was almost entirely flagged, which will slightly influence the flux measurements of all sources. This channel is clearly visible in the integrated spectrum (Figure \ref{fig:HCG2_spec}).

\subsection{HCG~7}

\begin{figure}[h]
    \centering
    \includegraphics[width=\columnwidth]{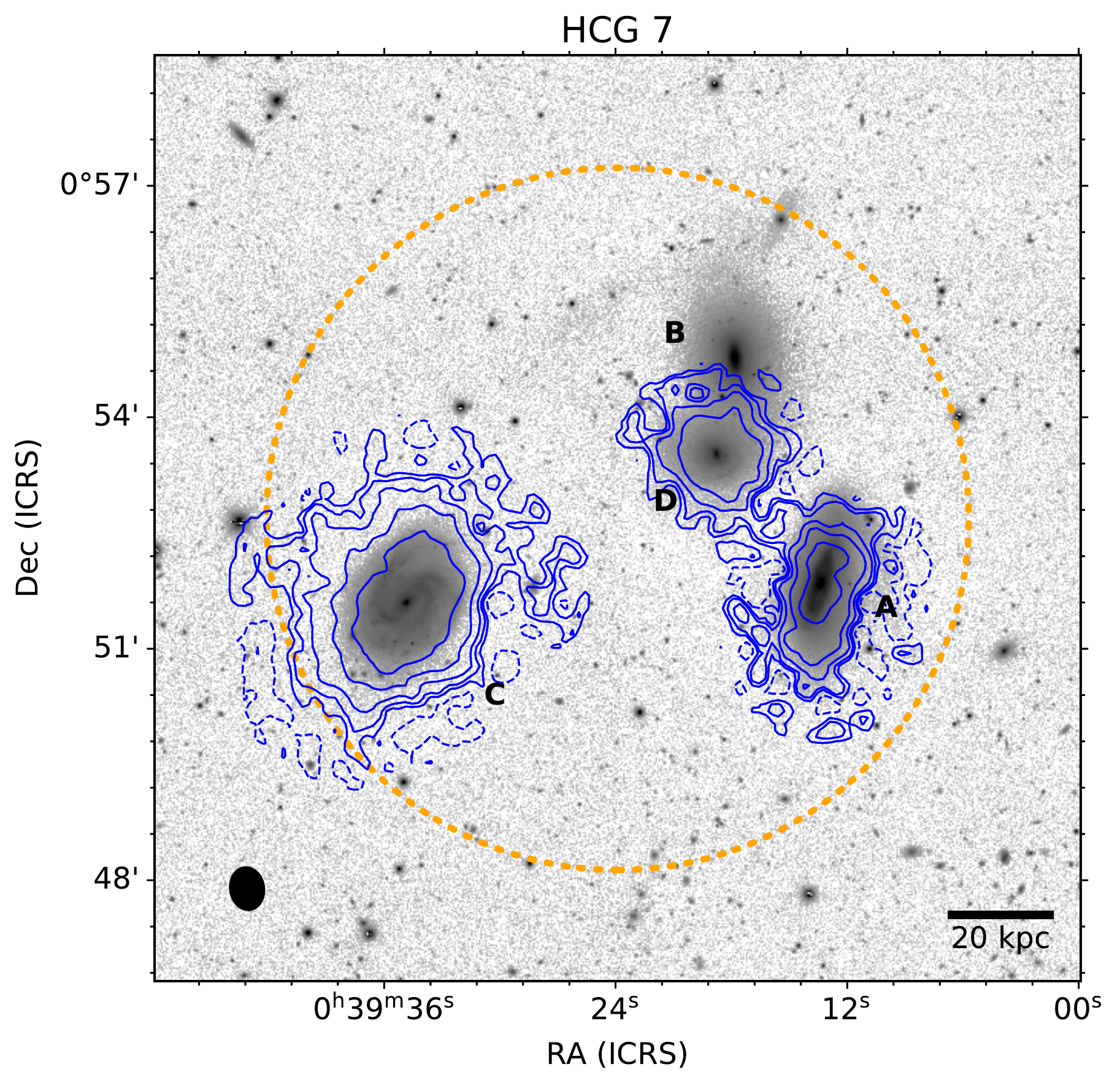}
    \caption{As in Figure \ref{fig:HCG2_split_overlay}. The orange dashed circle indicates the primary beam of the GBT observation of this group.}
    \label{fig:HCG7_split_overlay}
\end{figure}

\begin{figure}[h]
    \centering
    \includegraphics[width=\columnwidth]{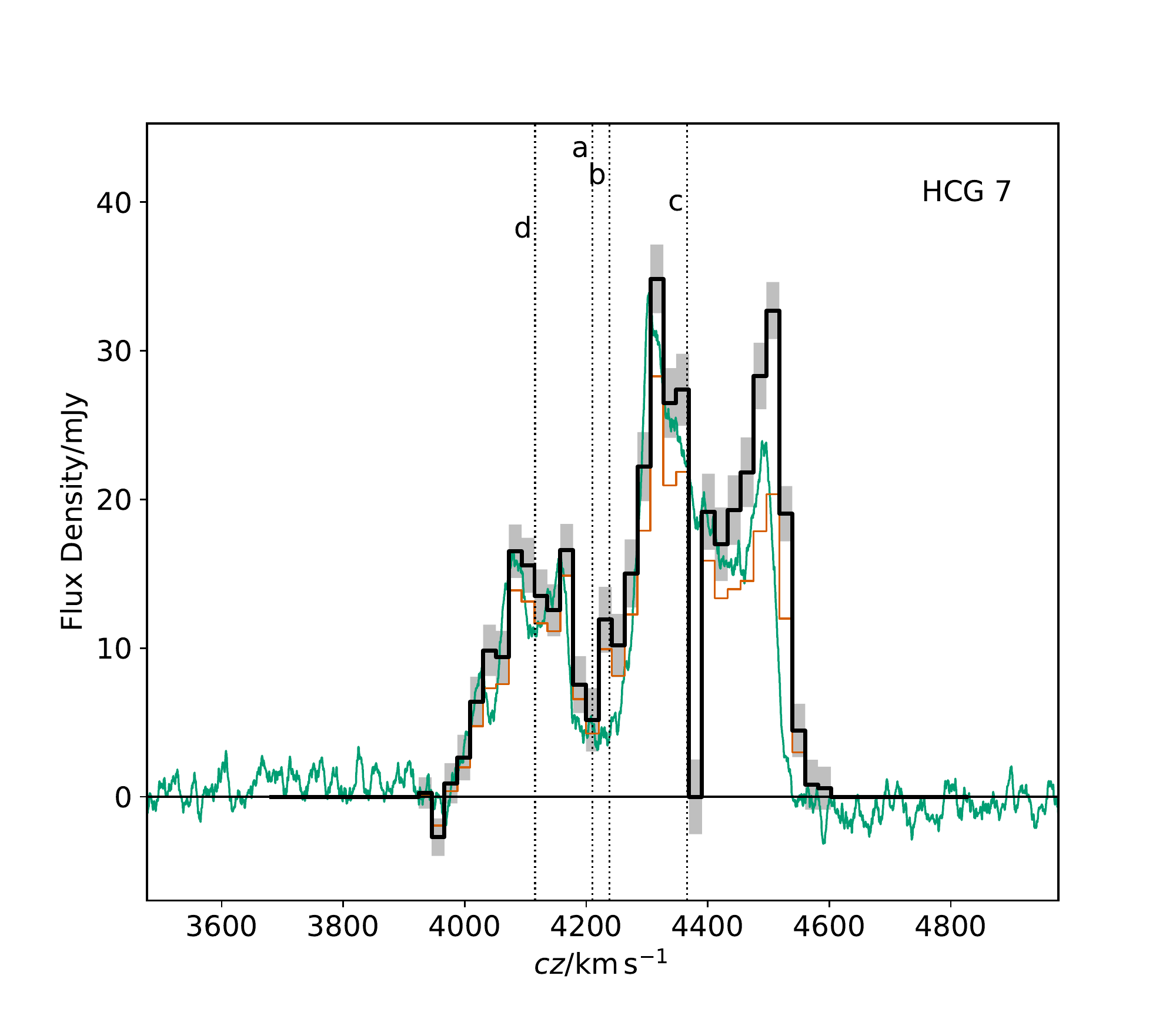}
    \caption{As in Figure \ref{fig:HCG2_spec}. The green (high velocity resolution) spectrum is from the GBT observation of this group \citep{Borthakur+2010}, and the orange spectrum is the VLA spectrum weighted to match the GBT primary beam response.}
    \label{fig:HCG7_spec}
\end{figure}

HCG~7 is a compact configuration of four (three late-type and one lenticular) galaxies at approximately 4200~\kms. The moment maps indicate that HCG~7a, c, and d are all strongly detected (Figure \ref{fig:HCG7_split_overlay}). 
HCG~7b, the lenticular, does not appear to be detected.
HCG~7c is a little separated from the other three group members and has a mostly regular velocity field. 
The \texttt{SoFiA} mask already separates the emission from HCG~7a and d, so not manual intervention was required. None of the galaxies in the group appear to be strongly disturbed in \hi \ (with caveat that HCG~7b is undetected), and as such no separate extended features were identified. One channel of the cube is entirely flagged, as is apparent in the integrated spectrum (Figure \ref{fig:HCG7_spec}).

\subsection{HCG~10}

\begin{figure}[h]
    \centering
    \includegraphics[width=\columnwidth]{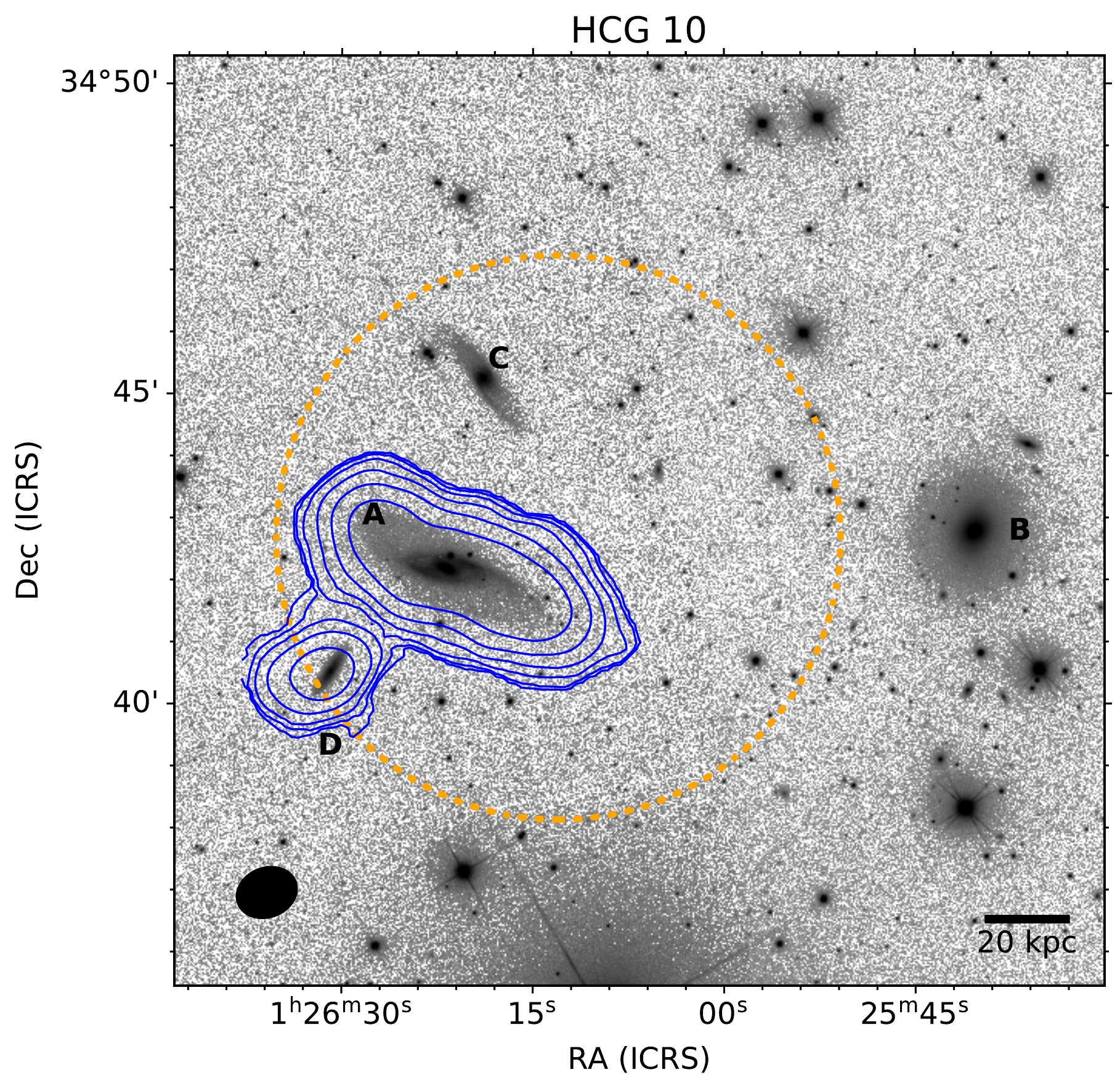}
    \caption{As in Figure \ref{fig:HCG7_split_overlay}.}
    \label{fig:HCG10_split_overlay}
\end{figure}

\begin{figure}[h]
    \centering
    \includegraphics[width=\columnwidth]{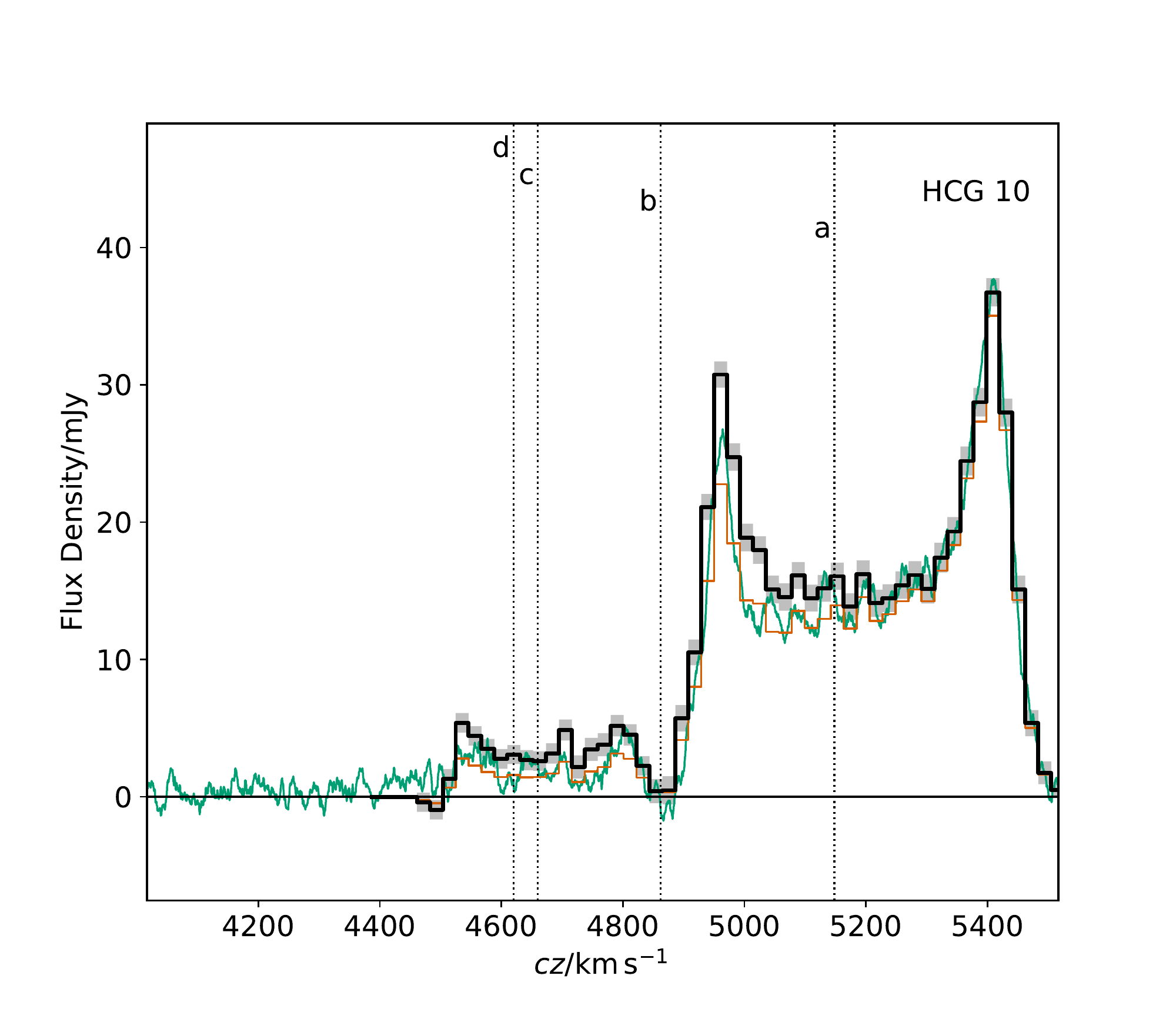}
    \caption{As in Figure \ref{fig:HCG7_spec}.}
    \label{fig:HCG10_spec}
\end{figure}

HCG~10 is a quartet made up of three tightly packed spirals and one elliptical slightly further afield, all within the range 4600-5200~\kms. Both HCG~10a and d are strongly detected, but no others. In this case the mask generated by \texttt{SoFiA} was not split further at all. It had already identified the two galaxies separately and there is little sign of extended features (Figures \ref{fig:HCG10_split_overlay} \& \ref{fig:HCG10_spec}), though in the \hi \ cube the western side of HCG~10a appears to be somewhat kinematically disturbed.

\subsection{HCG~15}

\begin{figure}[h]
    \centering
    \includegraphics[width=\columnwidth]{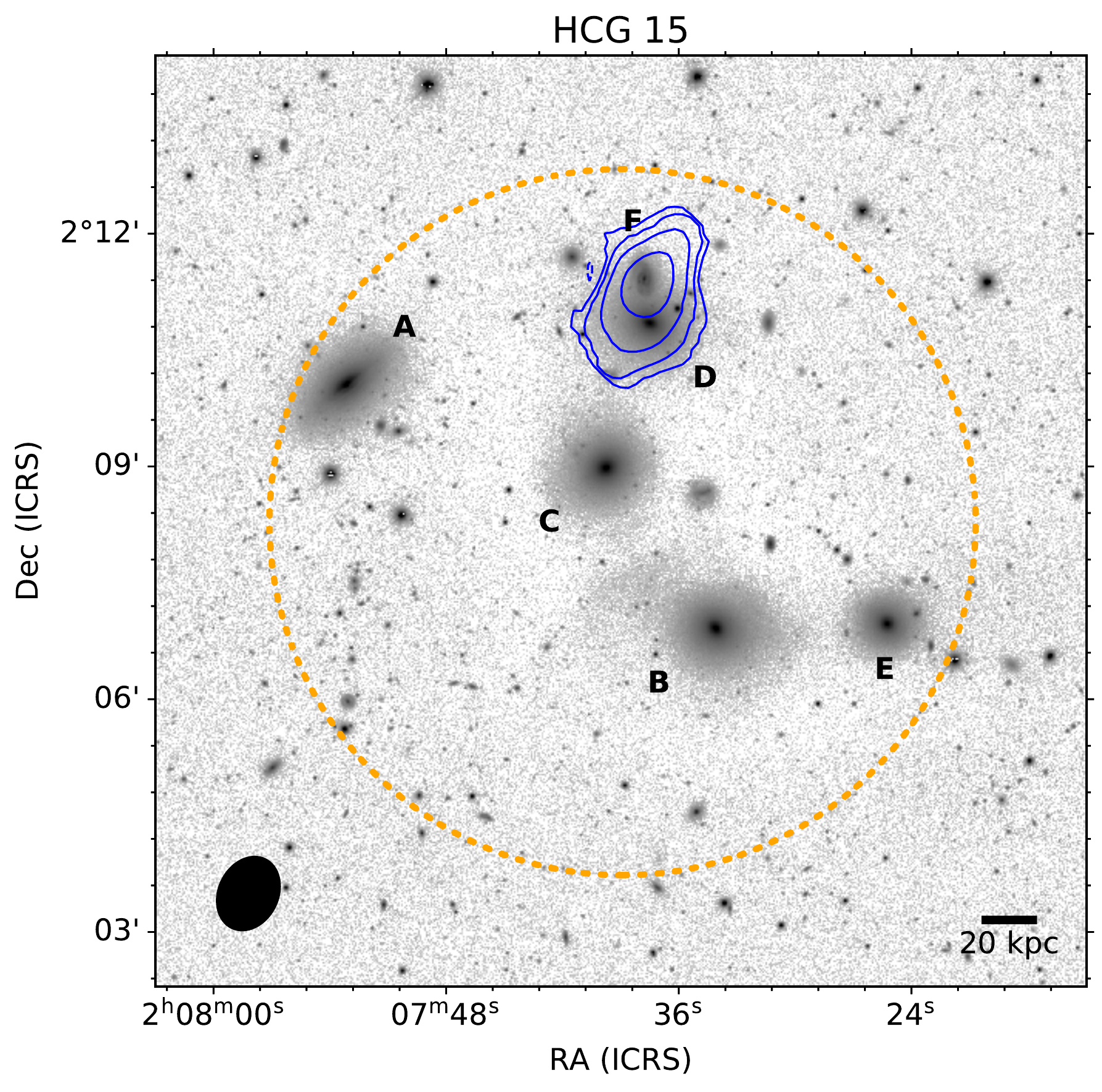}
    \caption{As in Figure \ref{fig:HCG7_split_overlay}.}
    \label{fig:HCG15_split_overlay}
\end{figure}

\begin{figure}[h]
    \centering
    \includegraphics[width=\columnwidth]{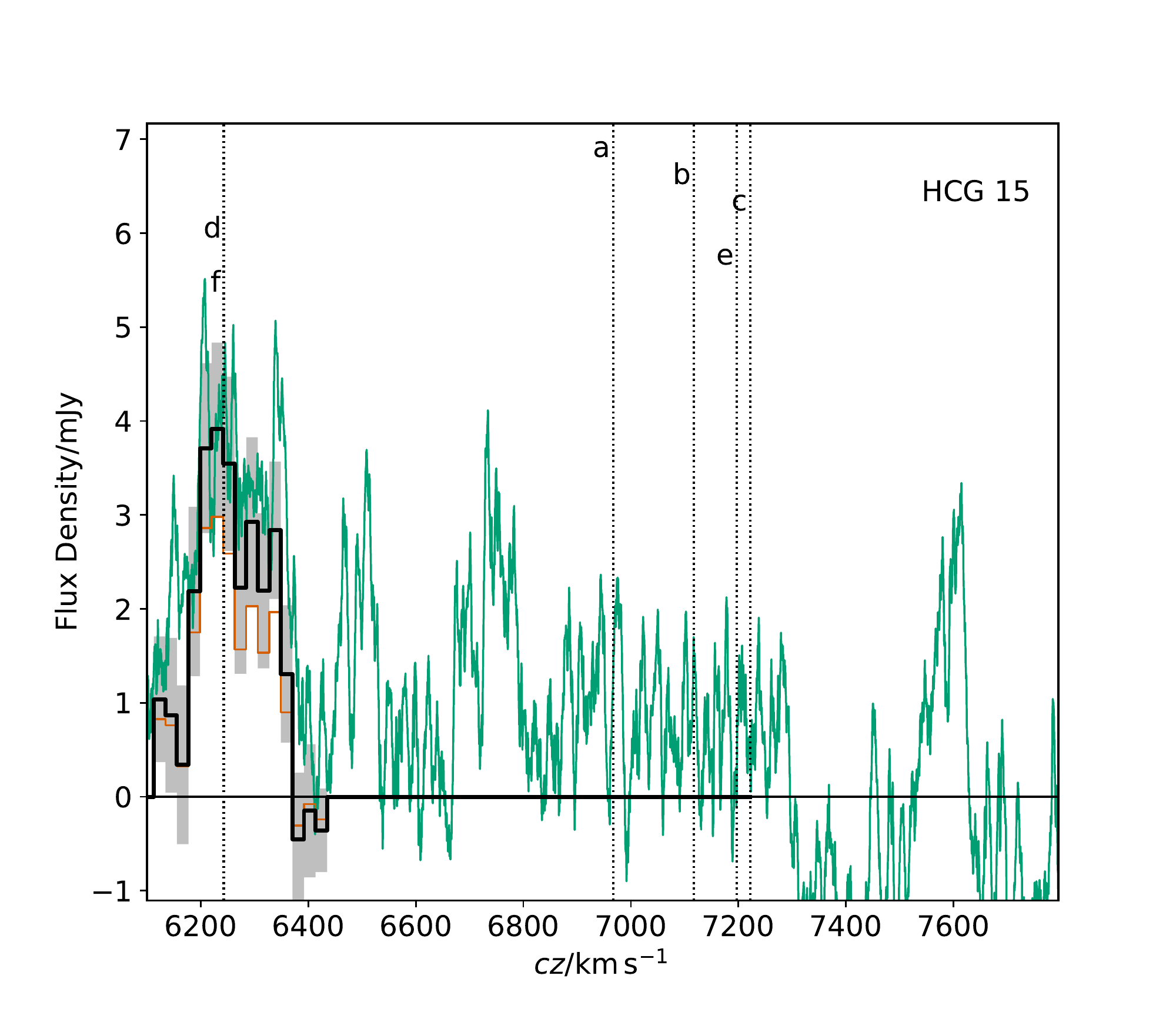}
    \caption{As in Figure \ref{fig:HCG7_spec}.}
    \label{fig:HCG15_spec}
\end{figure}

HCG~15 is a collection of six galaxies in a compact, inverted `T' formation. Most members are at $\sim$7000~\kms, but HCG~15d and f are at $\sim$6200~\kms. Only HCG~15f is detected in \hi, on the very edge of the group. At the (poor) resolution of the data no clear disturbance or extended features could be identified and thus no manual separate was performed (Figures \ref{fig:HCG15_split_overlay} \& \ref{fig:HCG15_spec}).

\subsection{HCG 16}

\begin{figure}[h]
    \centering
    \includegraphics[width=\columnwidth]{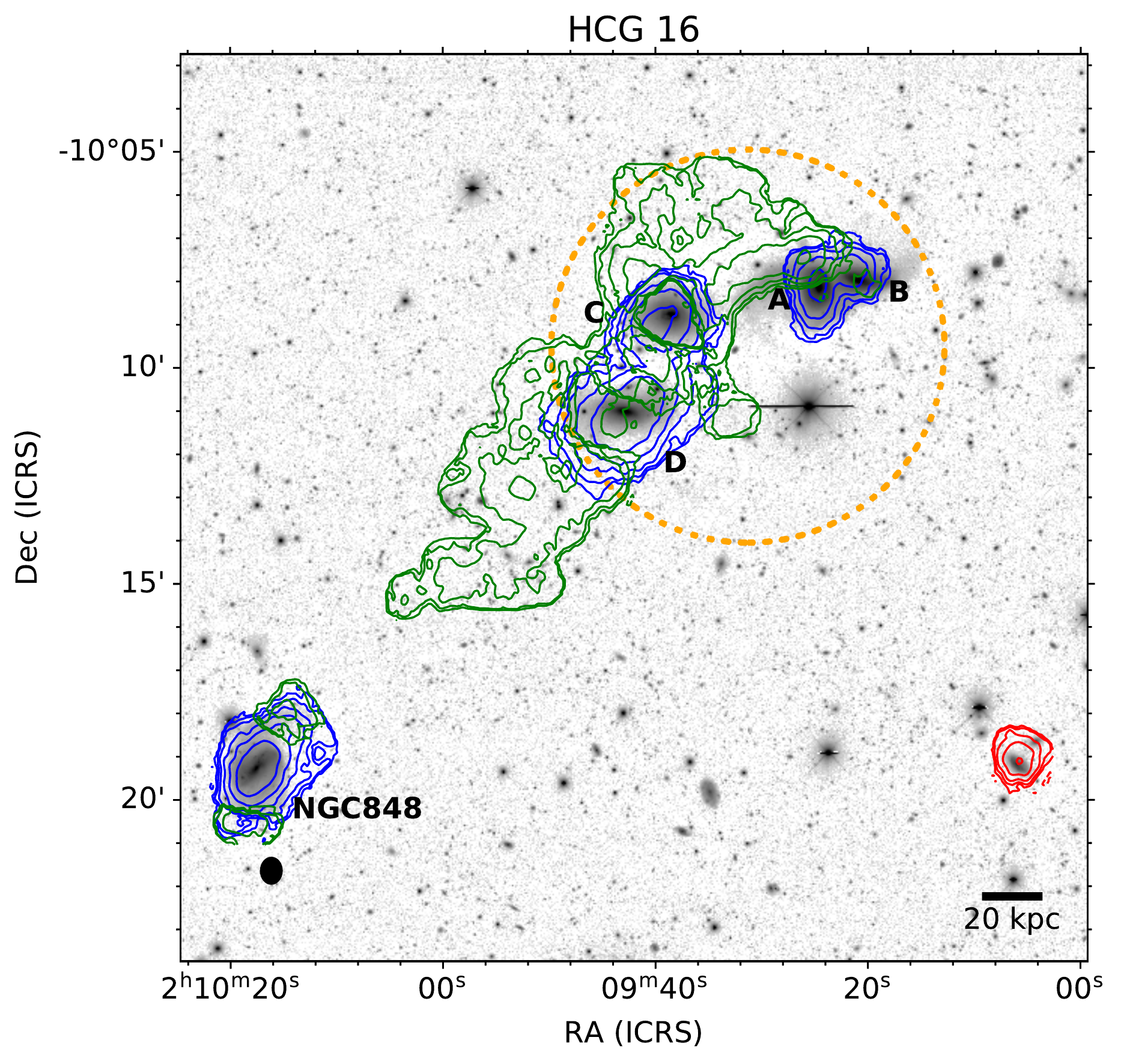}
    \caption{As in Figure \ref{fig:HCG7_split_overlay}.}
    \label{fig:HCG16_split_overlay}
\end{figure}

\begin{figure}[h]
    \centering
    \includegraphics[width=\columnwidth]{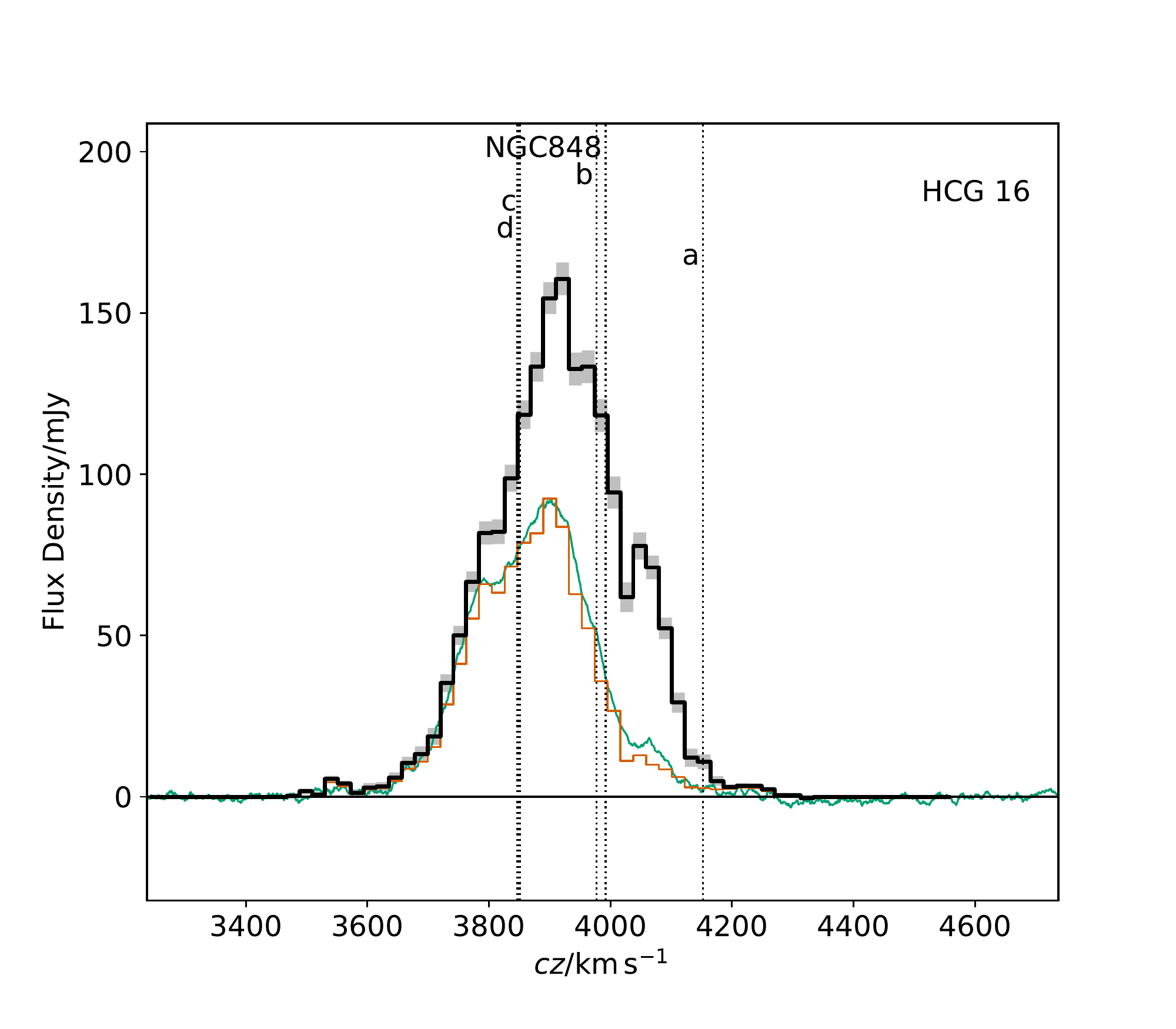}
    \caption{As in Figure \ref{fig:HCG7_spec}.}
    \label{fig:HCG16_spec}
\end{figure}

HCG~16 is a compact, linear group of five galaxies at $\sim$4000~\kms. \citet{Jones+2019} performed a detailed study of HCG~16 based on the same archival VLA data. Here we attempt an equivalent separation of features to that study, however, we note that the automated approach to reducing the data that is used here results in slightly more D-array data being flagged relative to C-array than in \citet{Jones+2019}, which in turn leads to a slightly lower column density sensitivity and slightly finer spatial resolution. In addition, in this work the \texttt{SoFiA} source masks are clipped at 4$\sigma$ as opposed to 3.5$\sigma$ in \citet{Jones+2019}.

All the core members of HCG~16 (a, b, c, and d) are detected as well as NGC~848 and PGC~8210. The former has clearly previously interacted with the core group and is at the head of a $\sim$160 kpc long tidal tail (Figure \ref{fig:HCG16_split_overlay}), while the latter is a dwarf galaxy that has likely never entered the core group before (we do not consider it a group member). The emission from HCG~16a and b are almost blended together, but have been separated as best as possible, there is a high column density \hi \ bridge between HCG~16c and d, as well as numerous tidal features and clumps throughout the group. The integrated \hi \ spectrum of the group for one continuous (in velocity) feature (Figure \ref{fig:HCG16_spec}).

\subsection{HCG~19}

\begin{figure}[h]
    \centering
    \includegraphics[width=\columnwidth]{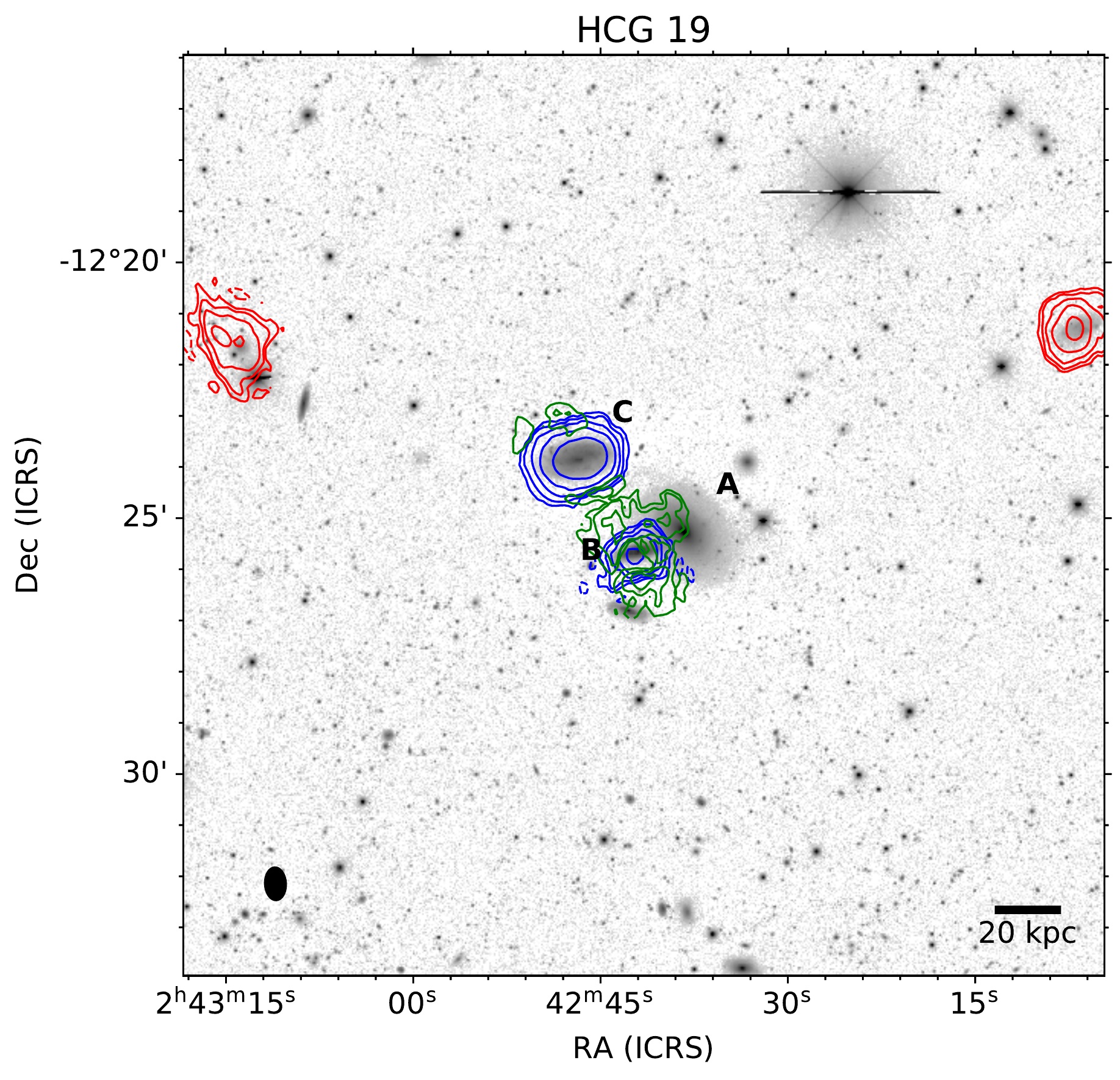}
    \caption{As in Figure \ref{fig:HCG2_split_overlay}.}
    \label{fig:HCG19_split_overlay}
\end{figure}

\begin{figure}[h]
    \centering
    \includegraphics[width=\columnwidth]{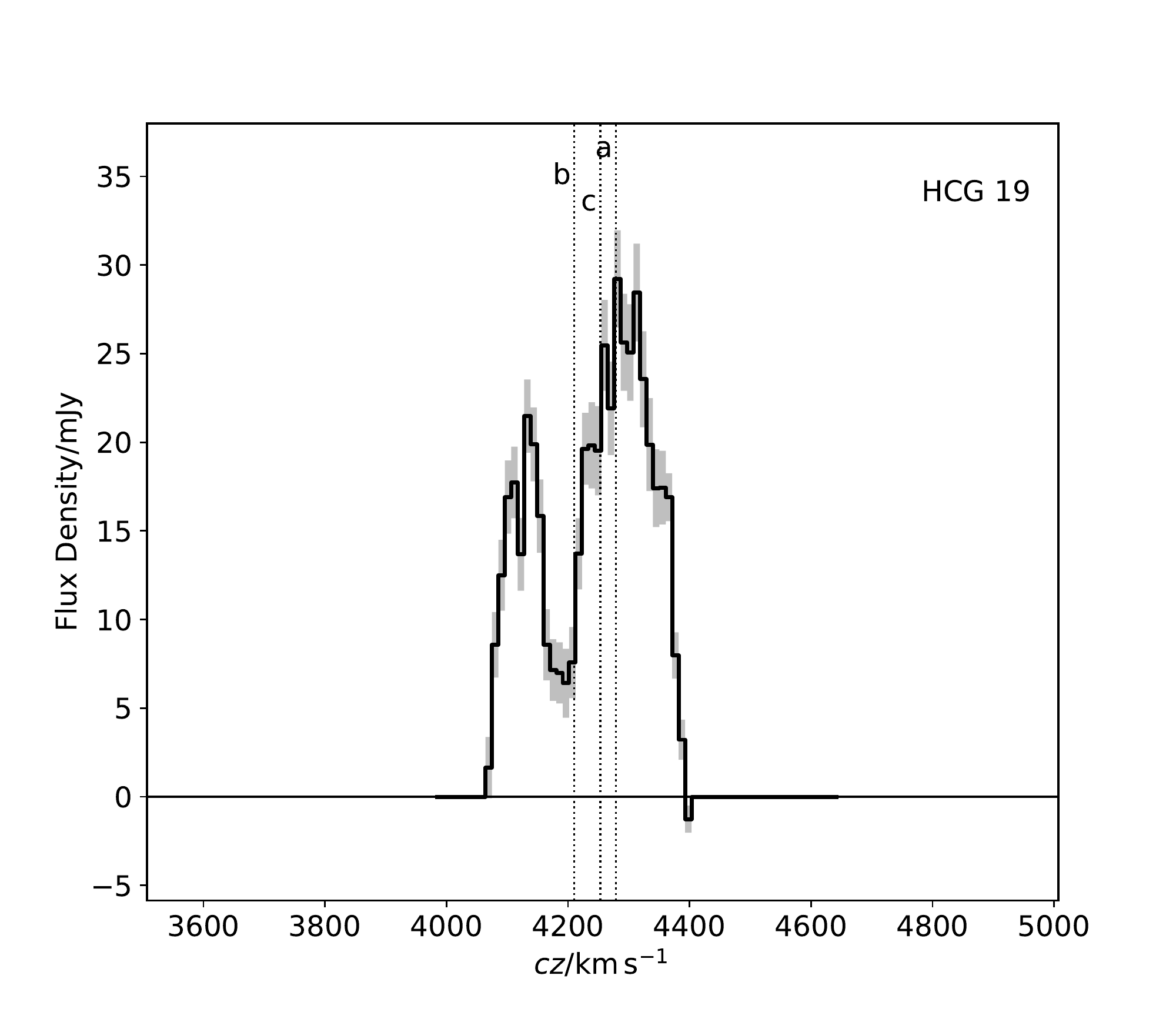}
    \caption{As in Figure \ref{fig:HCG2_spec}.}
    \label{fig:HCG19_spec}
\end{figure}

HCG~19 is a triplet (HCG~19d is a background object) of two late-type and one early-type galaxies at 4200~\kms. The \texttt{SoFiA} mask for HCG~19 was generated in a slightly different manner to most other groups owing to a troublesome continuum source artefact to the west of the group. To prevent the artefact from being included in the mask, positivity of sources was enforced, which in turn prevented the use of \texttt{SoFiA}'s reliability estimation, and therefore required the threshold level to be raised to 5.5$\sigma$ (from 4$\sigma$) to eliminate spurious sources.

In the core of the group HCG~19b and c are clearly detected, while HCG~19a is not detected. Further afield there are also strong detections of WISEA~J024313.79-122138.9 and WISEA~J024206.39-122118.3, which are low surface brightness (LSB) galaxies likely falling towards the group for the first time, but are still separated from the core group by $\sim$100 kpc. We therefore do not include them in our measurement of \hi \ content. For brevity we refer to these as HCG~19W1 and HCG~19W2.

With the exception of HCG~19c, the core galaxies do not appear to be strongly disturbed in \hi. However, there are numerous incipient tidal features on the outskirts of the galaxies, including a bridge between HCG~19 a and c. Although there is not much flux in these features and they are difficult to trace due to the poor noise properties of this cube, we have attempted to separate them from the emission of the galactic discs (Figures \ref{fig:HCG19_split_overlay} \& \ref{fig:HCG19_spec}).

\subsection{HCG~22}

\begin{figure}[h]
    \centering
    \includegraphics[width=\columnwidth]{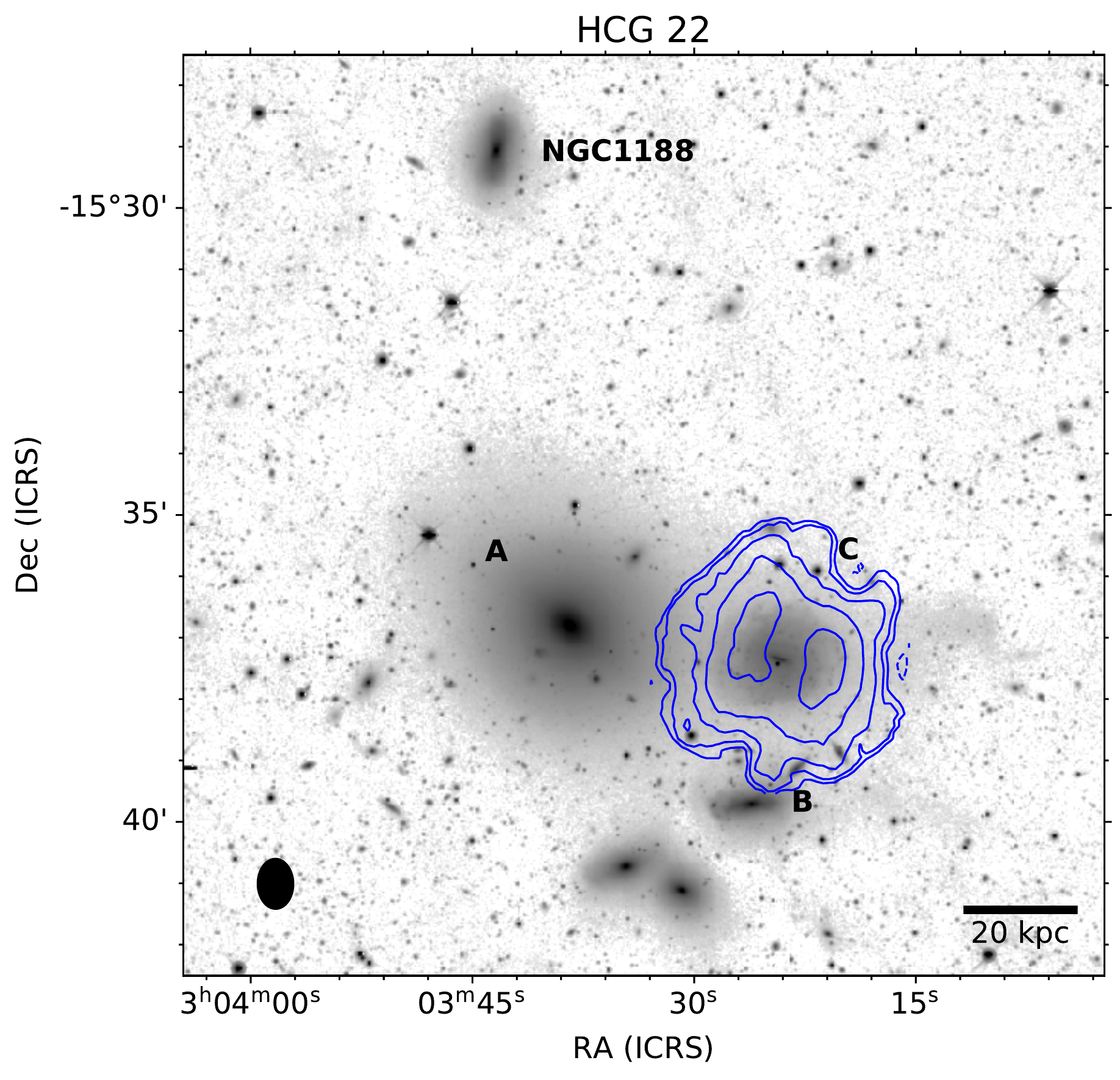}
    \caption{As in Figure \ref{fig:HCG2_split_overlay}.}
    \label{fig:HCG22_split_overlay}
\end{figure}

\begin{figure}[h]
    \centering
    \includegraphics[width=\columnwidth]{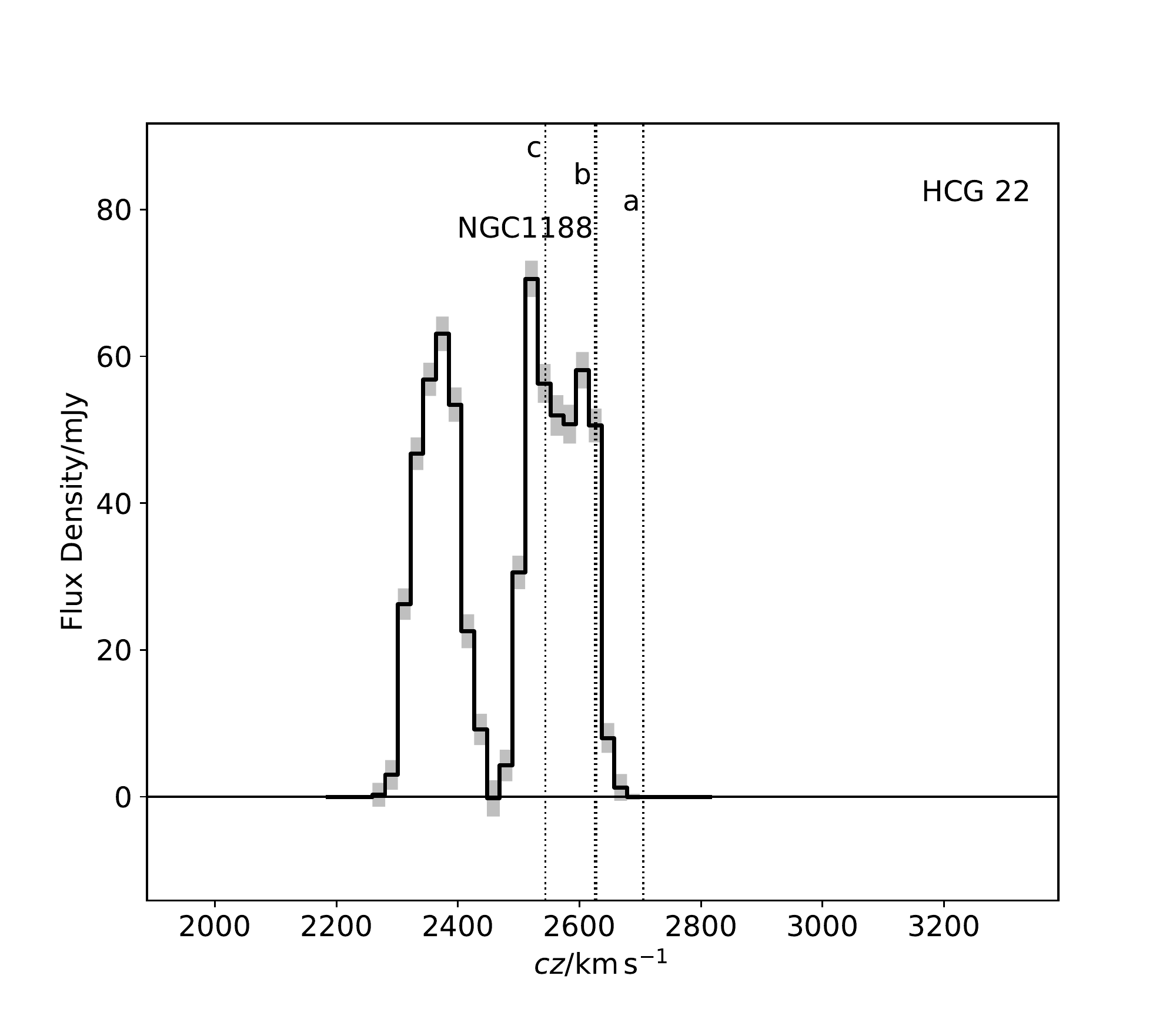}
    \caption{As in Figure \ref{fig:HCG2_spec}.}
    \label{fig:HCG22_spec}
\end{figure}

HCG~22 is an `S' shaped configuration of four galaxies (two early-types and two late-types) at $\sim$2600~\kms. Two original members of the group (HCG~22d and e) were subsequently shown to be background galaxies. The archival observations of HCG~22 form a very broad (in RA) mosaic of the group, however, only two galaxies, HCG~22c and NGC~1231 are detected within this field (Figures \ref{fig:HCG22_split_overlay} \& \ref{fig:HCG22_spec}). The latter is separated from the core group by over 20\arcmin \ and we do not consider it as a group member at present. HCG~22c shows a clean velocity field and we do not see evidence for tidal features in the cube. We also note that the tabulated redshift for HCG~22c from \citep{Hickson+1992} appears to be erroneous (Figure \ref{fig:HCG22_spec}). NGC~1231 shows some minor disturbances in the outskirts of its disc which we separate from its disc emission (even though it will not be included in our analysis of the \hi \ content of the group). This is likely a result of interactions with its neighbour, NGC~1209. We also note that we consider NGC~1188 as a member of HCG~22, as it is at the same redshift and close to the edge of the group. It is not detected in \hi.

\subsection{HCG~23}

\begin{figure}[h]
    \centering
    \includegraphics[width=\columnwidth]{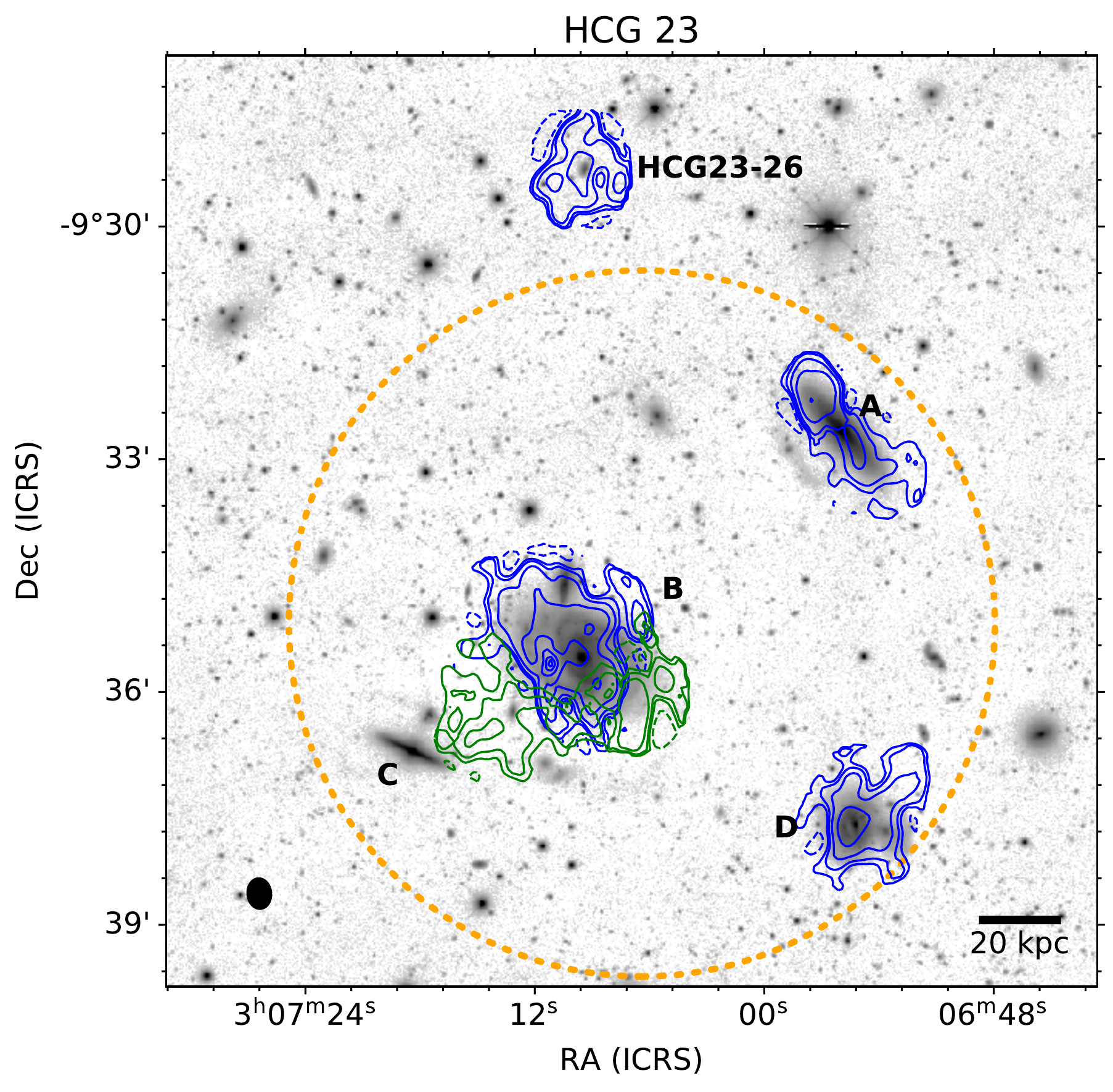}
    \caption{As in Figure \ref{fig:HCG7_split_overlay}.}
    \label{fig:HCG23_split_overlay}
\end{figure}

\begin{figure}[h]
    \centering
    \includegraphics[width=\columnwidth]{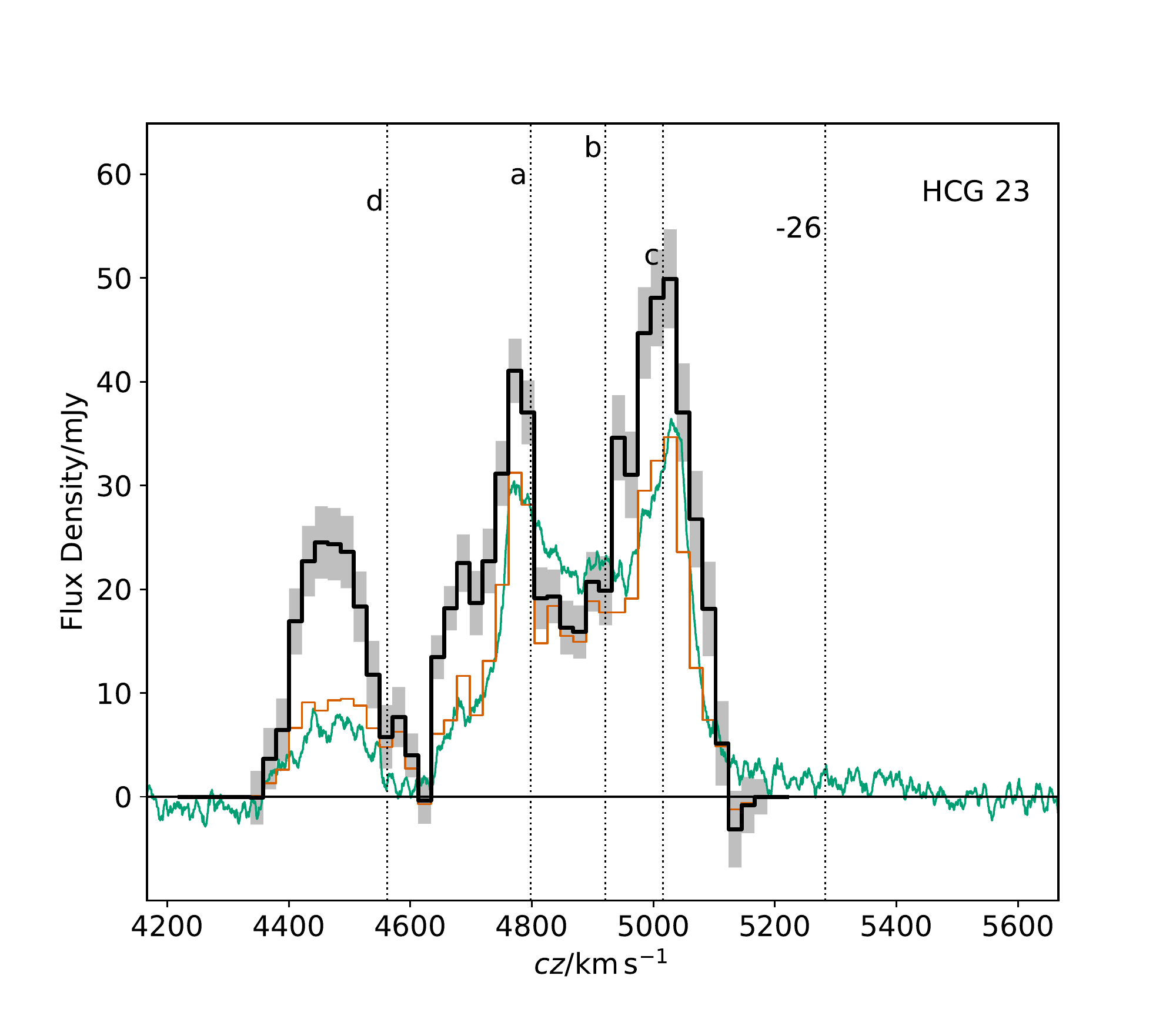}
    \caption{As in Figure \ref{fig:HCG7_spec}.}
    \label{fig:HCG23_spec}
\end{figure}

HCG~23 is a quintet of four late-type and one lenticular galaxies in a `T' configuration in the range 4500-5300~\kms. The narrow bands available for the historical VLA and the original observing strategy for this group, with the overlap of the two spectral windows centred at the redshift of the group's peak \hi \ emission, led to imperfect continuum subtraction and poorly behaved noise for this group. This complicates the interpretation of faint features in this cube.

In the core group HCG~23a, b, and d are detected, HCG~23c is undetected. Outside the core group there are three further detections: HCG~23-26 \citep{deCarvalho+1997}, PGC~987787, and PGC~011654. We consider the first of these as a group member, but the latter two are separated from the core group by well over 100~kpc and show no signs of past interaction with the group. 

HCG~23b has two small tidal tails emanating from it. The first extends SE towards HCG~23c, and the second towards the SW (Figures \ref{fig:HCG23_split_overlay} \& \ref{fig:HCG23_spec}). The remaining galaxies don't show any clear tidal features above the noise level. We also note that the emission from the almost edge-on HCG~23a is split into two separate sources by \texttt{SoFiA}, which we combine into one.

\subsection{HCG~25}

\begin{figure}[h]
    \centering
    \includegraphics[width=\columnwidth]{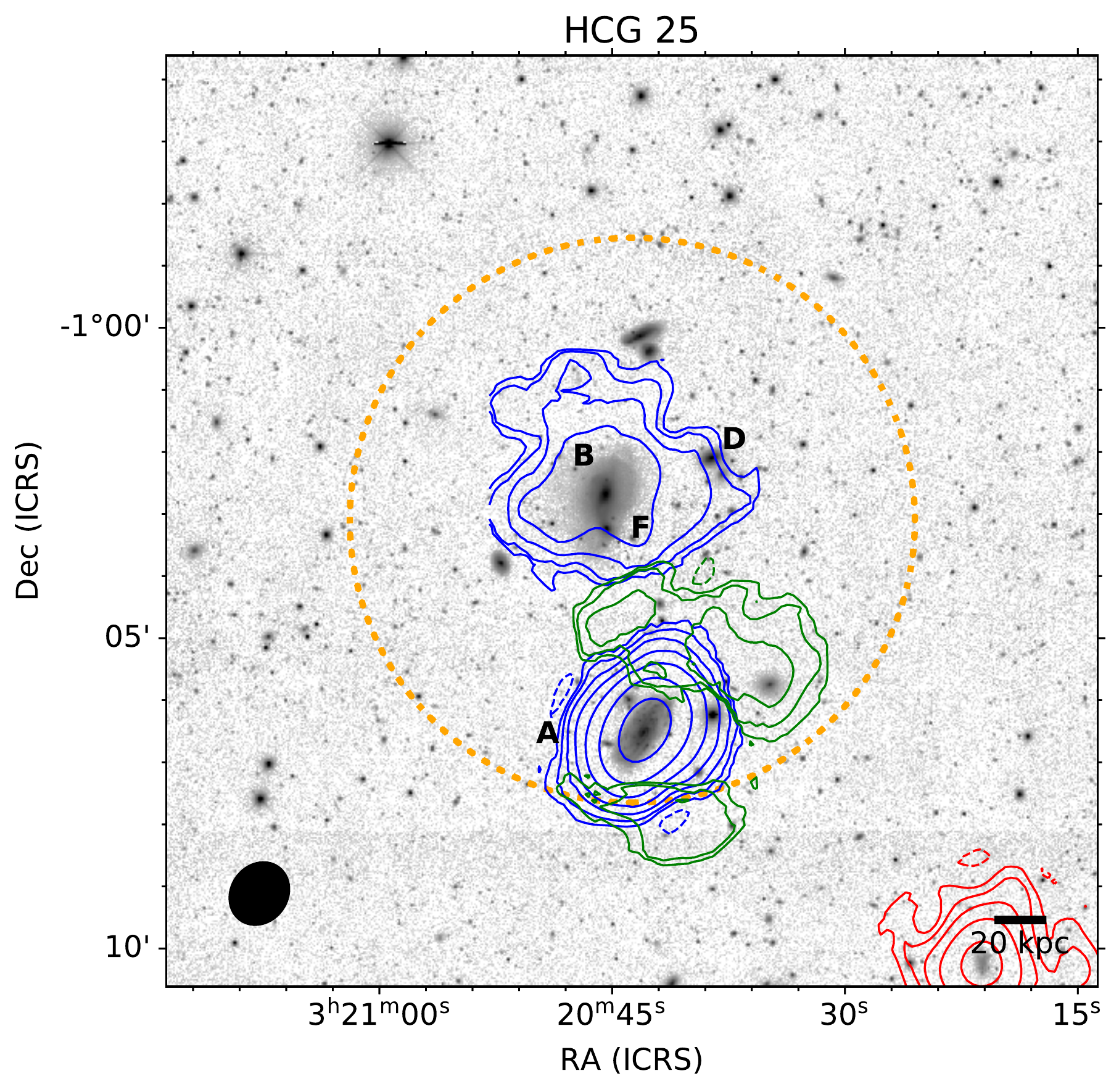}
    \caption{As in Figure \ref{fig:HCG7_split_overlay}.}
    \label{fig:HCG25_split_overlay}
\end{figure}

\begin{figure}[h]
    \centering
    \includegraphics[width=\columnwidth]{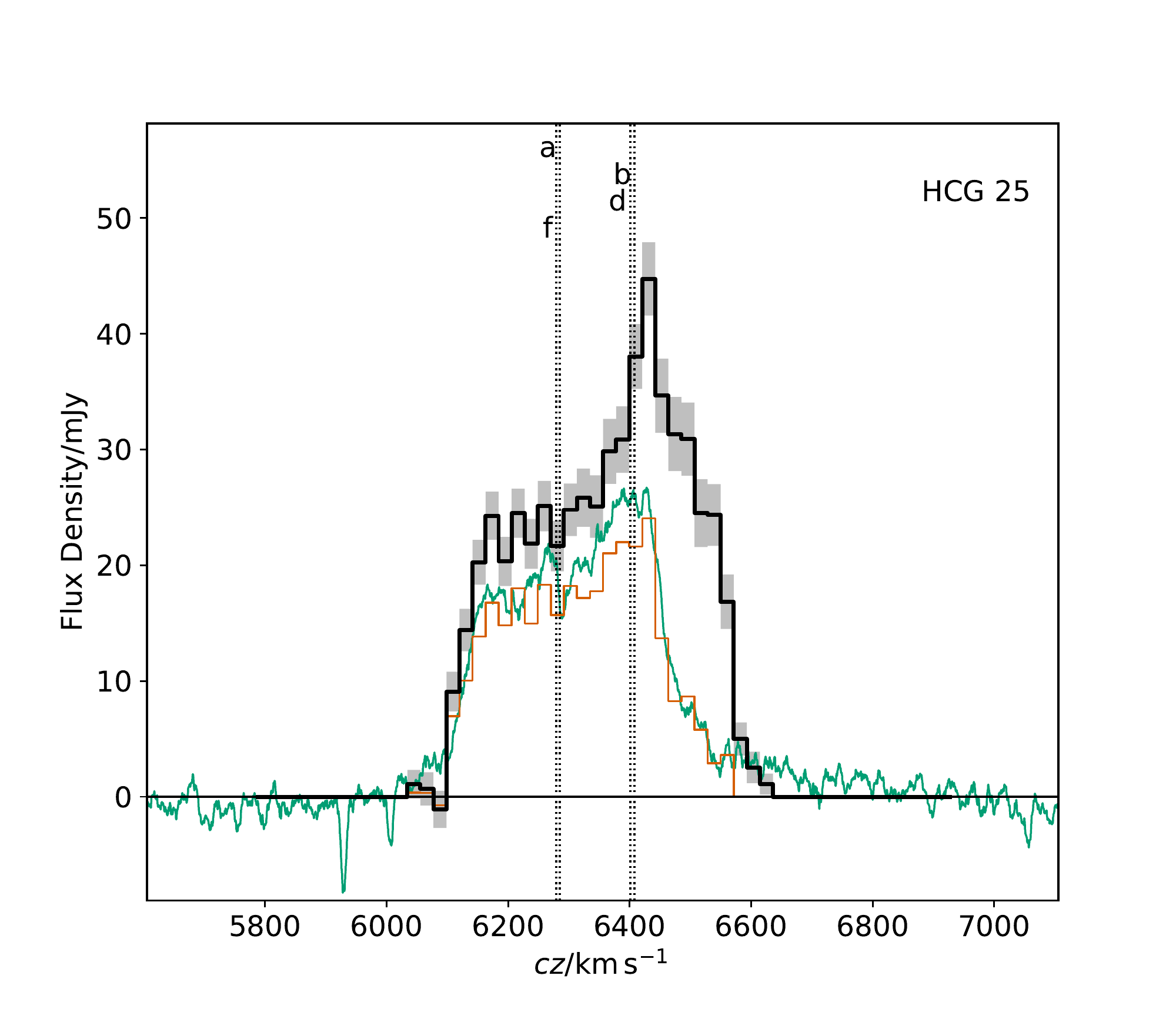}
    \caption{As in Figure \ref{fig:HCG7_spec}.}
    \label{fig:HCG25_spec}
\end{figure}

HCG~25 is a quartet of two late-type and two lenticular galaxies at $\sim$6300~\kms (HCG~25c, e and g are all background objects). The moment zero map of this group shows four clear detections (Figures \ref{fig:HCG25_split_overlay} \& \ref{fig:HCG25_spec}), HCG~25a and b in the core group, PGC~135673 slightly to the north, and the LSB galaxy 2SLAQ~J032021.10-011013.6 to the SW, subsequently classified as an ultra-diffuse galaxy by \citet{Roman+2017}. For brevity we refer to the latter as HCG~25S1. Both of these detections of the periphery of the group are separated from the core group by well over 100 kpc and we do not consider them as current members.

HCG~25a and b are connected by a faint \hi \ bridge, which we attempt to separate from the emission of the two galaxies themselves. However, this is complicated by the fact that it is at the limit of the spatial resolution and that, although HCG~25b is a strongly detected source overall, it is low S/N in individual channels. Thus the separation is largely based on the form of the emission from HCG~25a and the optical locations of the galaxies. HCG~25a also has an apparent incipient tidal tails connected to it on the NW and SW sides, which we also separate from the disc emission.

\subsection{HCG~26}

\begin{figure}[h]
    \centering
    \includegraphics[width=\columnwidth]{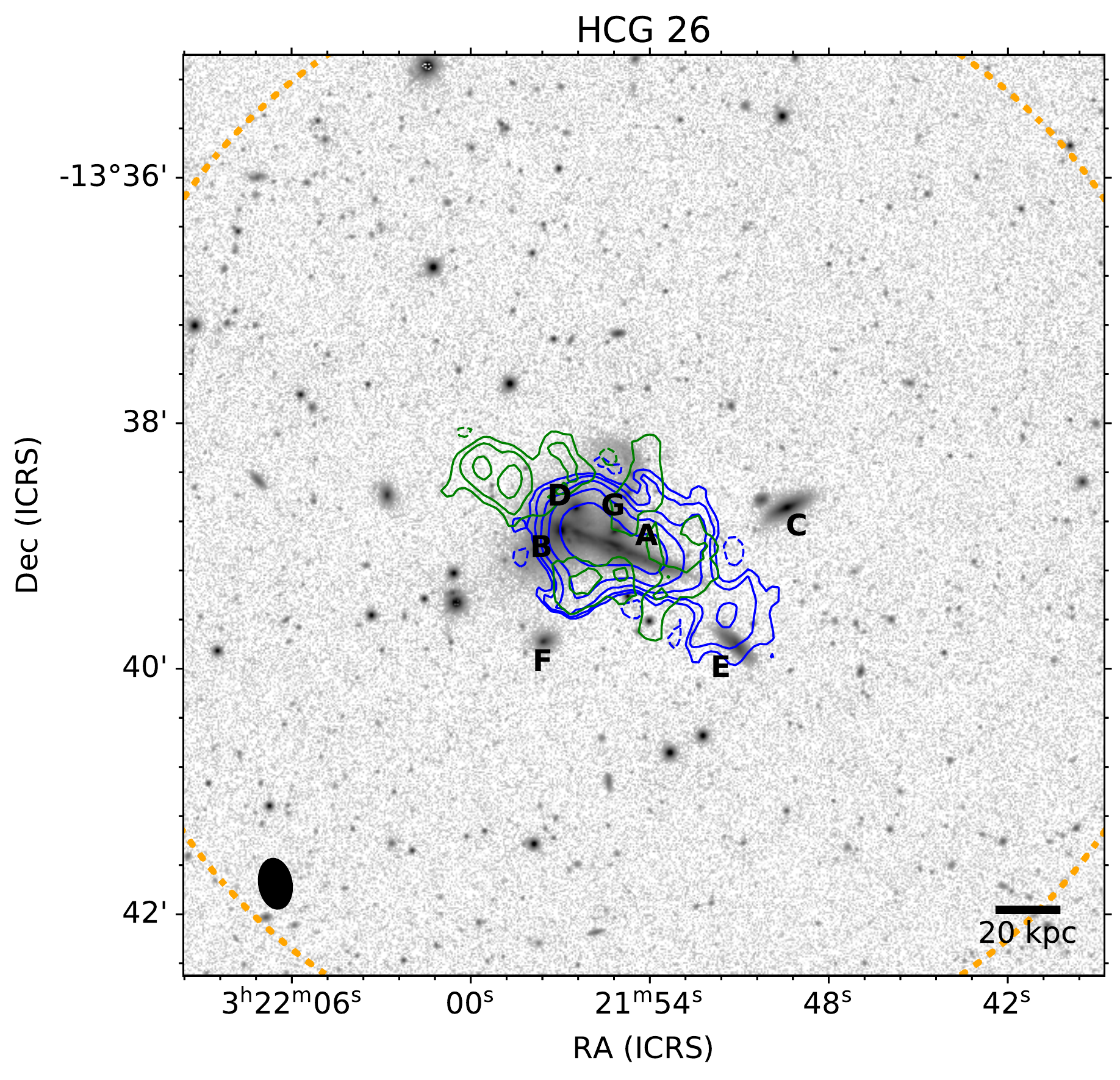}
    \caption{As in Figure \ref{fig:HCG7_split_overlay}.}
    \label{fig:HCG26_split_overlay}
\end{figure}

\begin{figure}[h]
    \centering
    \includegraphics[width=\columnwidth]{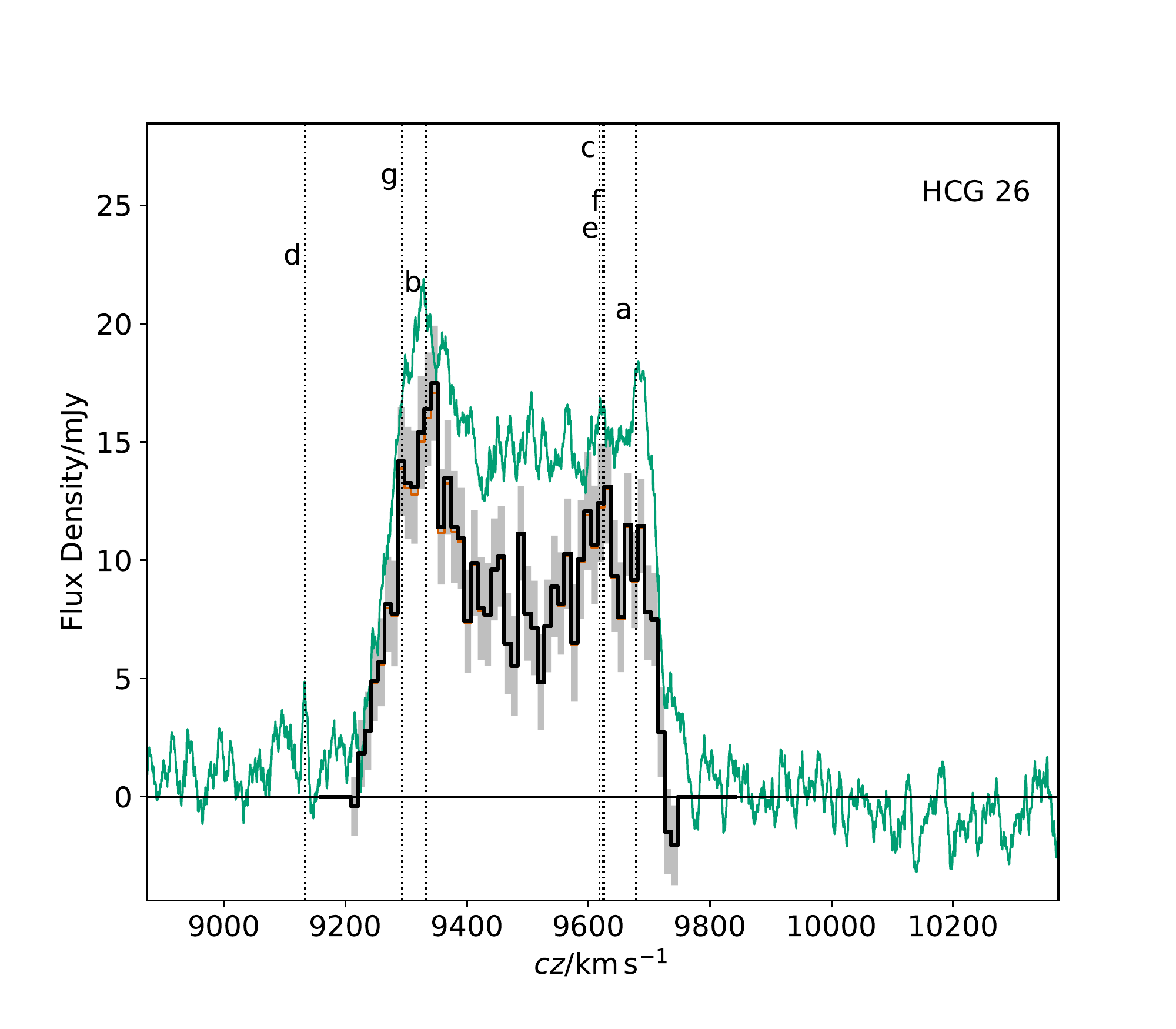}
    \caption{As in Figure \ref{fig:HCG7_spec}.}
    \label{fig:HCG26_spec}
\end{figure}

HCG~26 is a septet of galaxies between 9100-9700~\kms, dominated by the edge-on late-type HCG~26a (Figure \ref{fig:HCG26_split_overlay}). The vast majority of the \hi \ emission detected in HCG~26 belongs to HCG~26a (although there appears to an offset between the catalogued redshift and the \hi \ emission, Figure \ref{fig:HCG26_spec}). HCG~26e is also faintly detected as well as some tidal features. The velocity structure of HCG~26a dominates almost the entire band and appears to be mostly regular, however, there are two clear tidal features, one extending towards HCG~26e and another extending in the opposite directions towards the NE. There are also numerous apparent minor clumps of \hi \ emission that do not follow the velocity structure of HCG~26a. These were excised wherever possible, but the limited resolution and S/N of the features presented a challenge and some fainter features have likely not been successfully separated. HCG~26e itself appears significantly perturbed, not displaying a clear velocity structure, but given the resolution and low S/N of the detection it is difficult to draw more detailed conclusions. There also appears to be a $\sim$30\% offset between the GBT and VLA spectra of this group. Given the similar forms of the spectral profiles, this is most likely the result of differing absolute calibrations or continuum subtraction. However, the root cause could not be definitely identified.

\subsection{HCG~30}

\begin{figure}[h]
    \centering
    \includegraphics[width=\columnwidth]{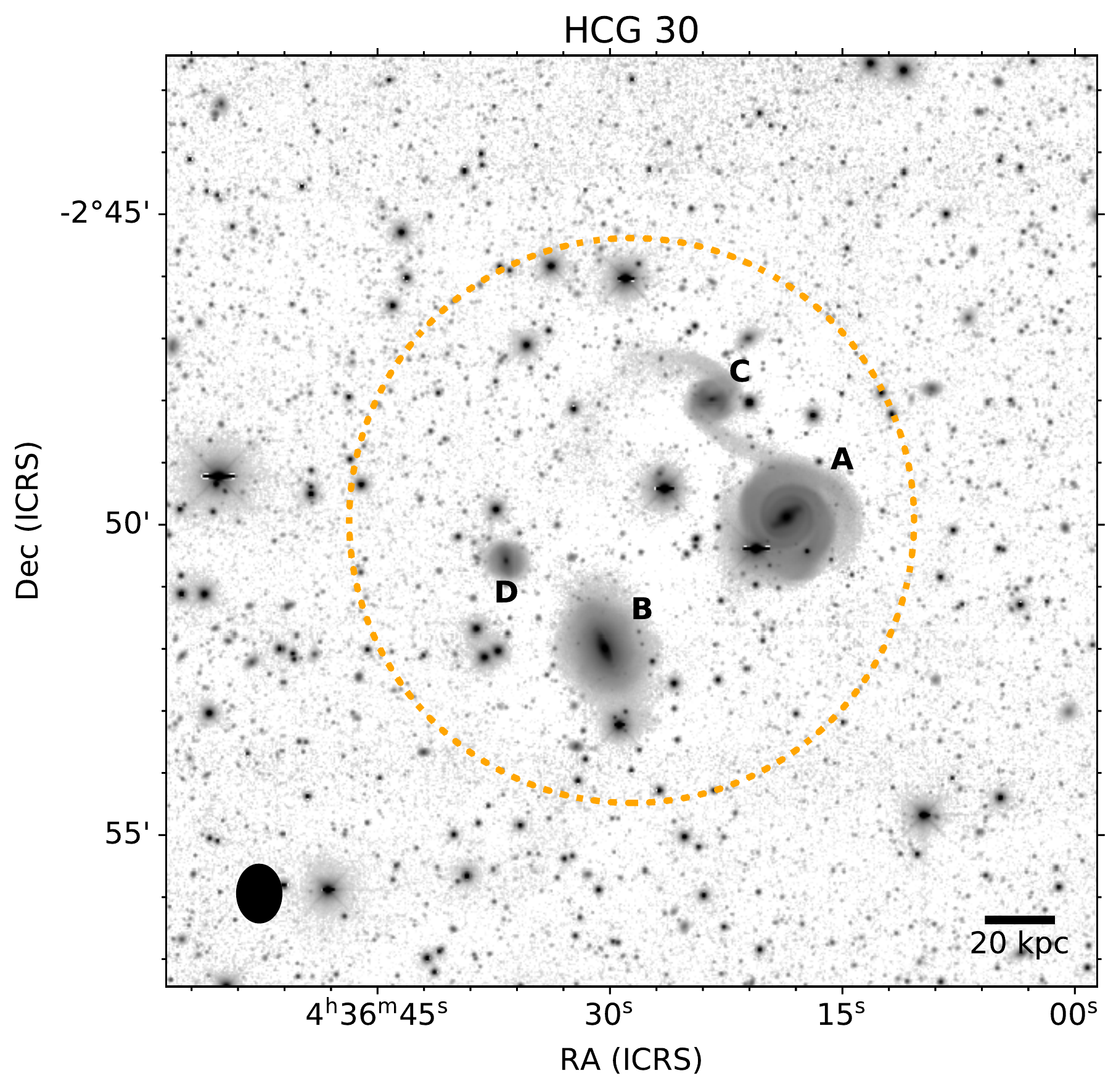}
    \caption{As in Figure \ref{fig:HCG7_split_overlay}.}
    \label{fig:HCG30_split_overlay}
\end{figure}

\begin{figure}[h]
    \centering
    \includegraphics[width=\columnwidth]{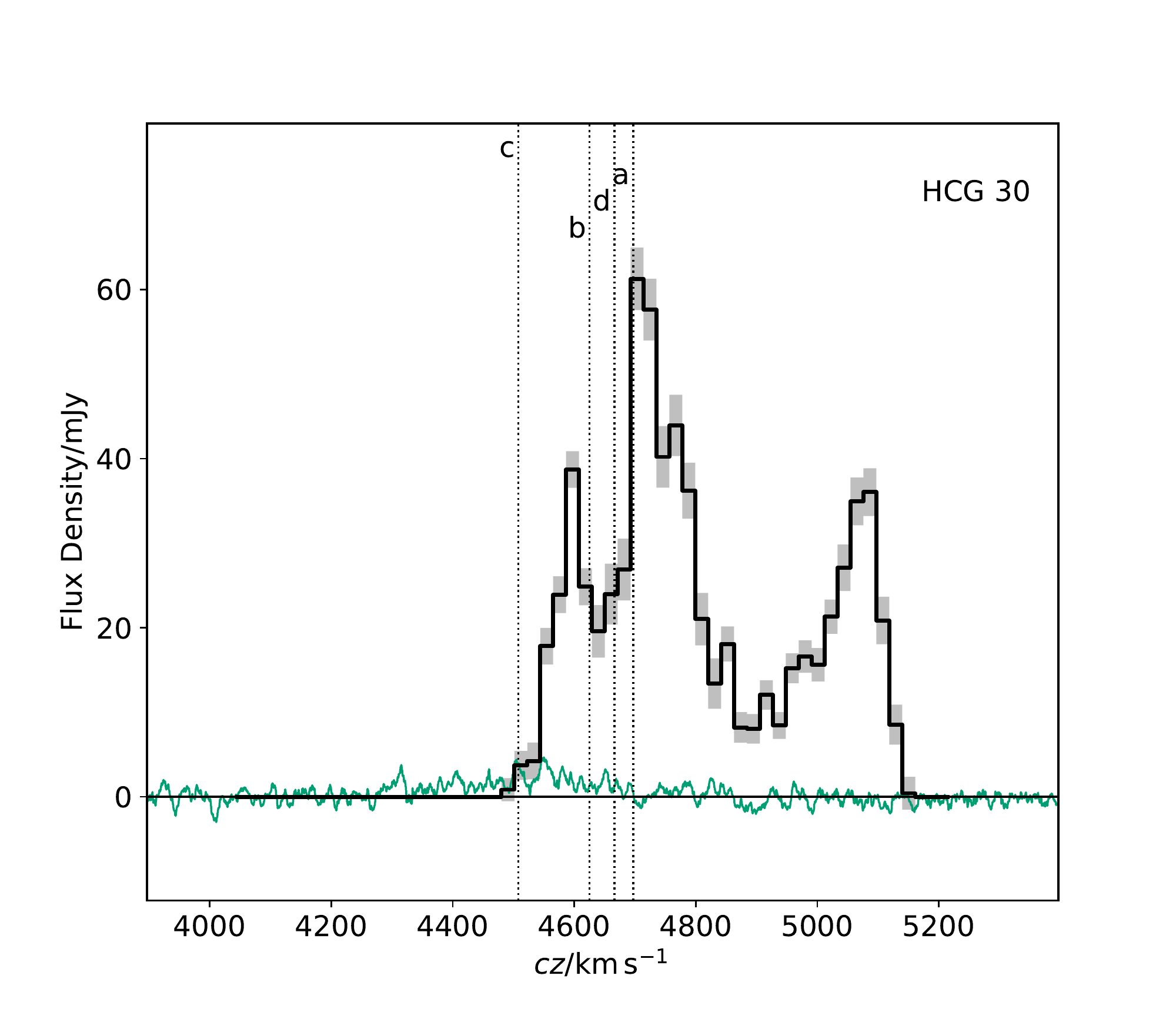}
    \caption{As in Figure \ref{fig:HCG7_spec}. Note that the emission in the VLA spectrum here is all emission beyond the core group, but still within the primary beam of the VLA. Therefore, it is not shown in Figure \ref{fig:HCG30_split_overlay}.}
    \label{fig:HCG30_spec}
\end{figure}

HCG~30 is a quartet of three late-type and one lenticular galaxies at $\sim$4600~\kms. None of the core members of HCG~30 are detected in the \hi \ cube (Figures \ref{fig:HCG30_split_overlay} \& \ref{fig:HCG30_spec}). However, three galaxies, 6dFGS~gJ043650.5-030237 (which we refer to as HCG~30dF1), NGC~1618, and NGC~1622, are strongly detected close to the edge of the VLA primary beam. None of these three detections appear particularly disturbed in \hi \ (at the resolution of the data) and it is unlikely that they have ever interacted with the core group. Thus we do not consider them in our \hi \ analysis of the group. NGC~1618 and NGC~1622 are both split into two separate sources by \texttt{SoFiA}, which we manually combine, but other than this no modification or separation of features was performed. The emission from these three detections accounts for the entirety of the flux in the VLA spectrum (Figure \ref{fig:HCG30_spec}) and when the data are weighted to compare to the GBT spectrum there is no flux remaining.

In addition it should be noted that three channels (4364--4406 \kms) in this data set were entirely flagged and a number of (primarily negative) low-level artefacts remain in the cube. Thus, positivity was enforced when creating the \texttt{SoFiA} mask and the threshold raised from 4$\sigma$ to 5$\sigma$ to increase reliability.

\subsection{HCG~31}

\begin{figure}[h]
    \centering
    \includegraphics[width=\columnwidth]{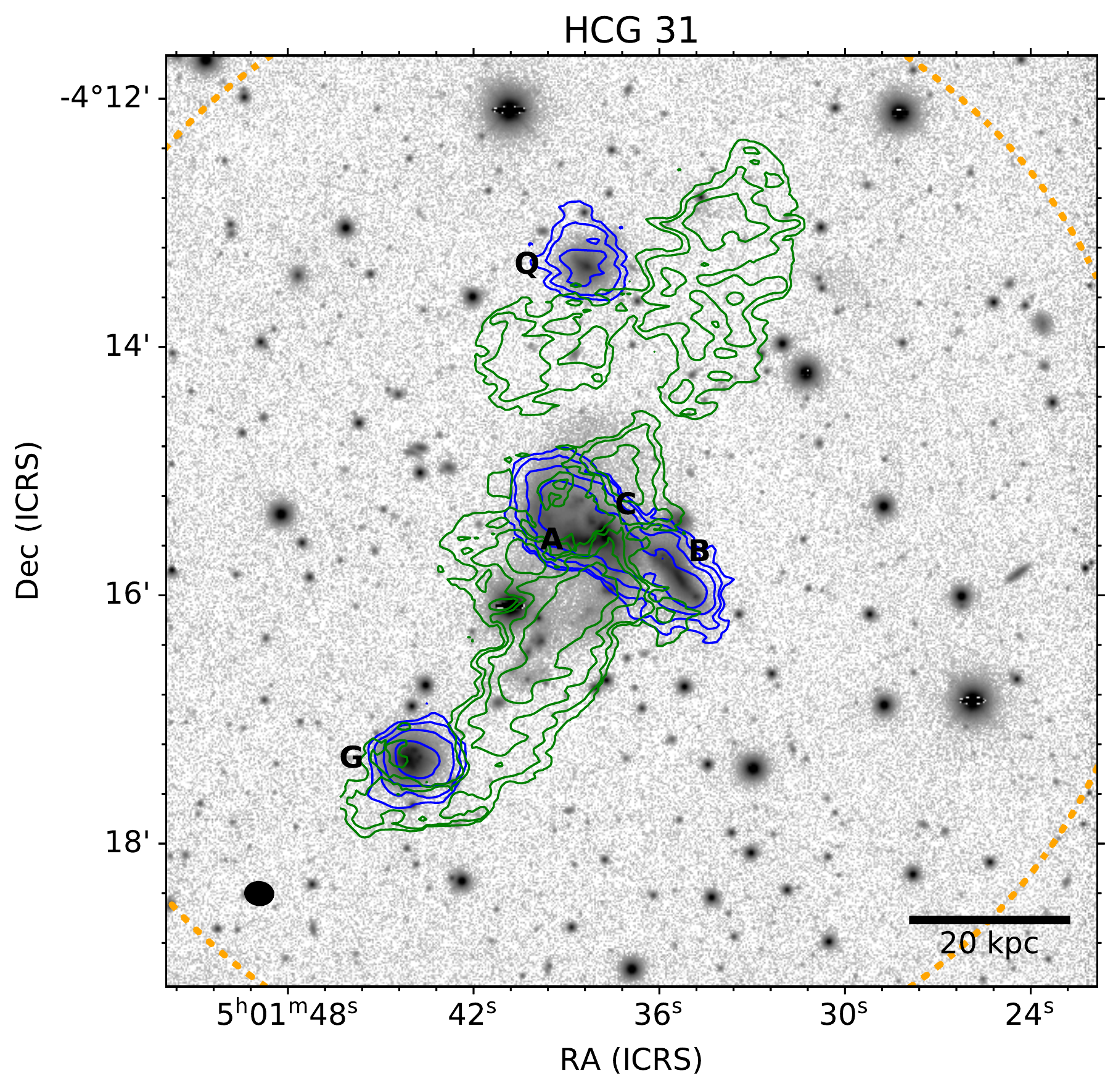}
    \caption{As in Figure \ref{fig:HCG7_split_overlay}.}
    \label{fig:HCG31_split_overlay}
\end{figure}

\begin{figure}[h]
    \centering
    \includegraphics[width=\columnwidth]{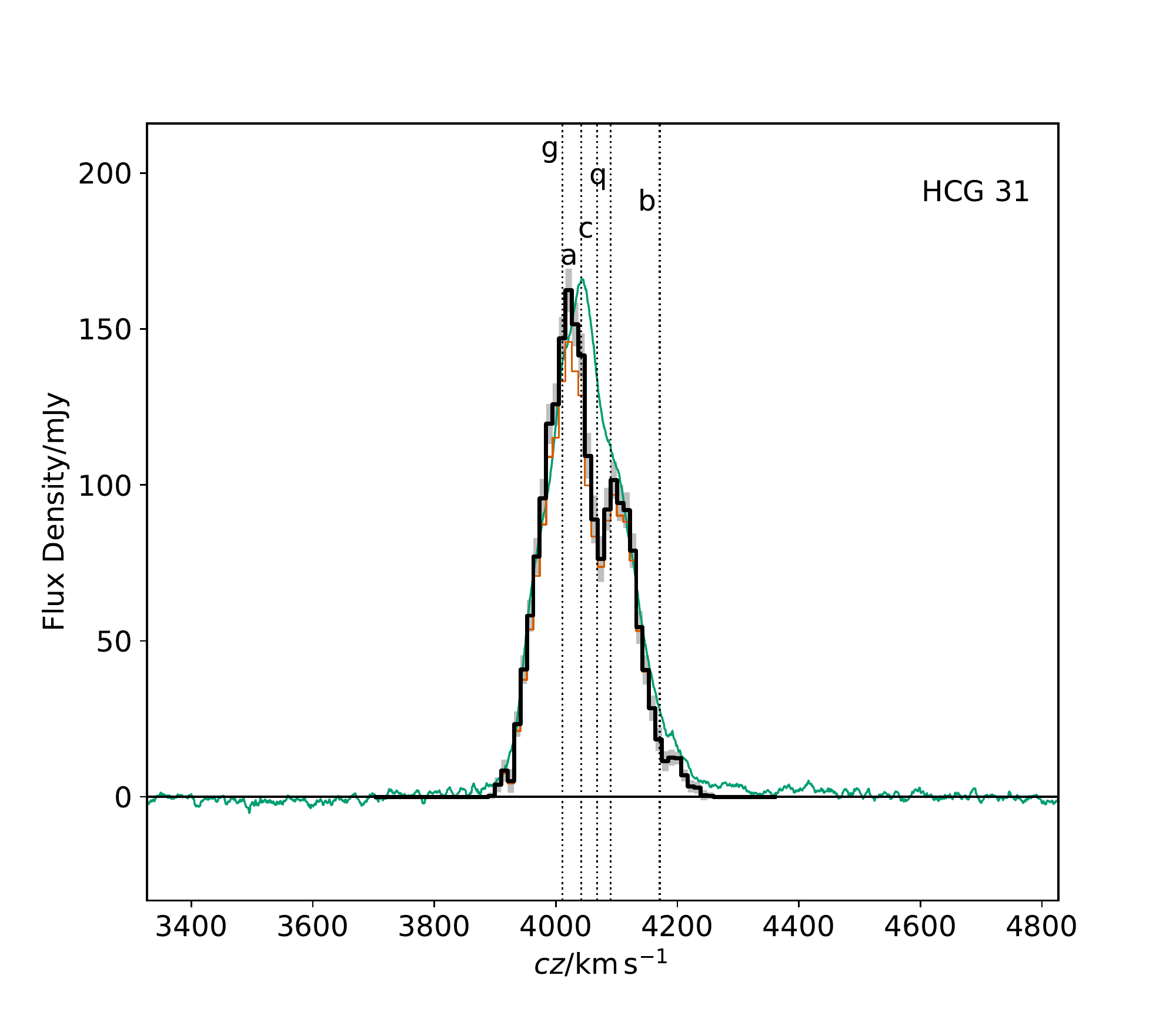}
    \caption{As in Figure \ref{fig:HCG7_spec}.}
    \label{fig:HCG31_spec}
\end{figure}

HCG~31 is quintet of late-type galaxies at $\sim$4100~\kms, three of which overlap with each other on the plane of the sky and are likely in the process of merging. 
The \hi \ emission in HCG~31 is an extremely complicated mixture of dwarf galaxies and tidal features (Figure \ref{fig:HCG31_split_overlay}), studied in detail by \citet{Verdes-Montenegro+2005} based on an independent reduction of the same VLA data. HCG~31a, b, c, g (IC~399), and q (WISEA~J050138.33-041321.2) are all detected in \hi. Most of these galaxies are highly disturbed and are embedded in an almost continuous set of tidal features which spans the entire group. HCG~31e and f are also detected in \hi, but are candidate tidal dwarf galaxies embedded in an \hi \ tail \citep[e.g.][]{Mendes+2006}, and we therefore consider them as tidal features rather than normal galaxies. \citet{Verdes-Montenegro+2005} also noted the \hi \ detection of PGC~3080767 to the south of the group, however, this is separated from the rest of the group by over 200~kpc and we do not consider it a member. There are also some blue clumps visible in the DECaLS image within the extended \hi \ features around HCG~31q (to the north of the core group), which may be other cases of in situ star formation. We also note that the absolute flux scale resulting from our calibration appears considerably higher than that of \citet{Verdes-Montenegro+2005}, which greatly improves how well the integrated \hi \ profile of the group matches to GBT observations shown Figure \ref{fig:HCG31_spec} \citep[c.f.][]{Borthakur+2010}.

Reliable separation of galaxies and tidal material in this group is extraordinarily challenging, in particular the galaxies HCG~31a and c, which are not only blended with tidal material, but also each other. The reported \hi \ masses for these two galaxies and, to a lesser extent, the other galaxies in the group are undoubtedly quite uncertain (for example, the approaching side of HCG~31b is deeply embedded in the complex of emission at the centre of the group). Despite this difficulty, what is clear, and of most importance for this work, is that a substantial fraction of the total \hi \ emission of this group originates in the IGrM.

\subsection{HCG~33}

\begin{figure}[h]
    \centering
    \includegraphics[width=\columnwidth]{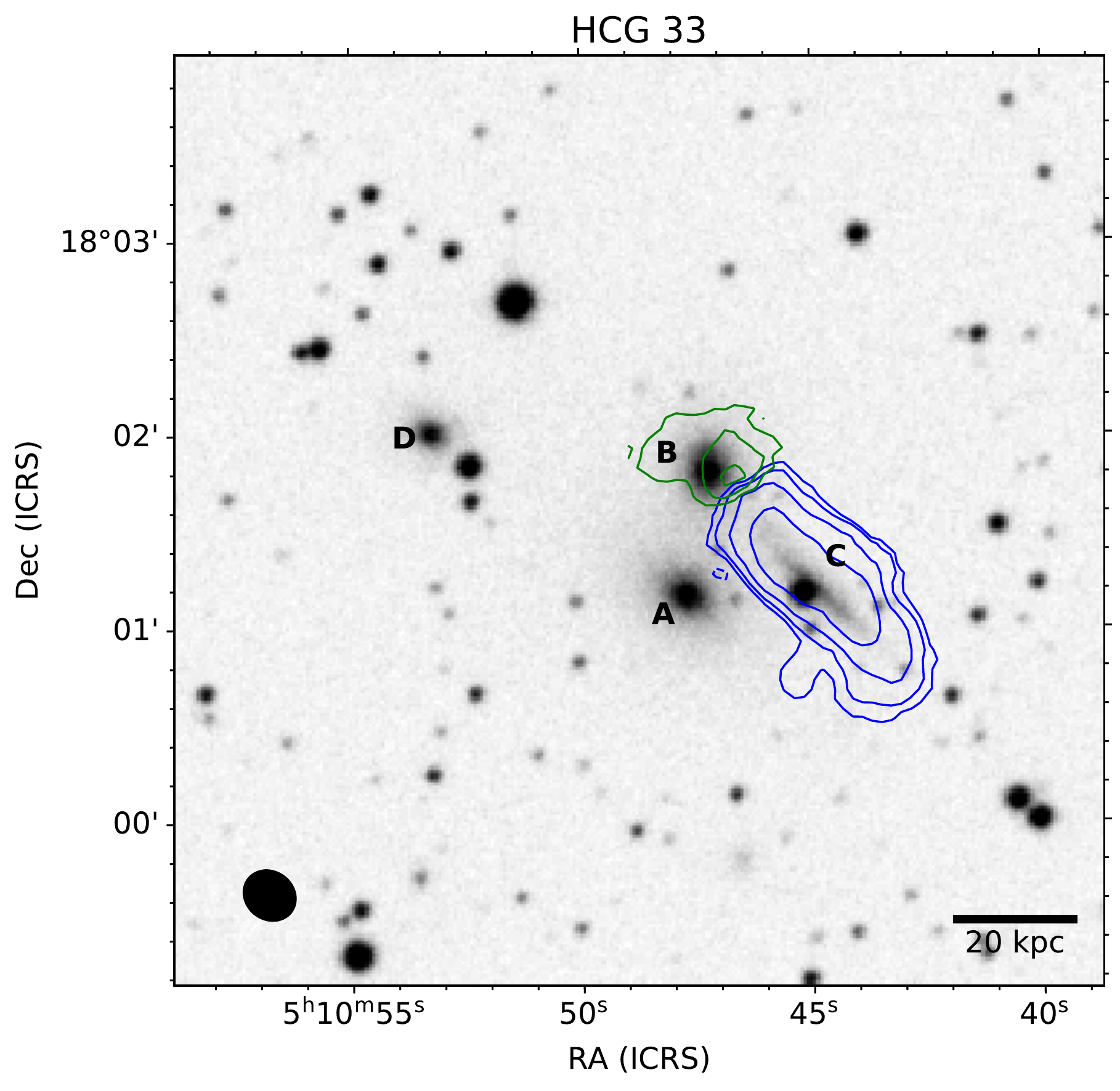}
    \caption{As in Figure \ref{fig:HCG2_split_overlay}, except the background image is POSS $R$-band.}
    \label{fig:HCG33_split_overlay}
\end{figure}

\begin{figure}[h]
    \centering
    \includegraphics[width=\columnwidth]{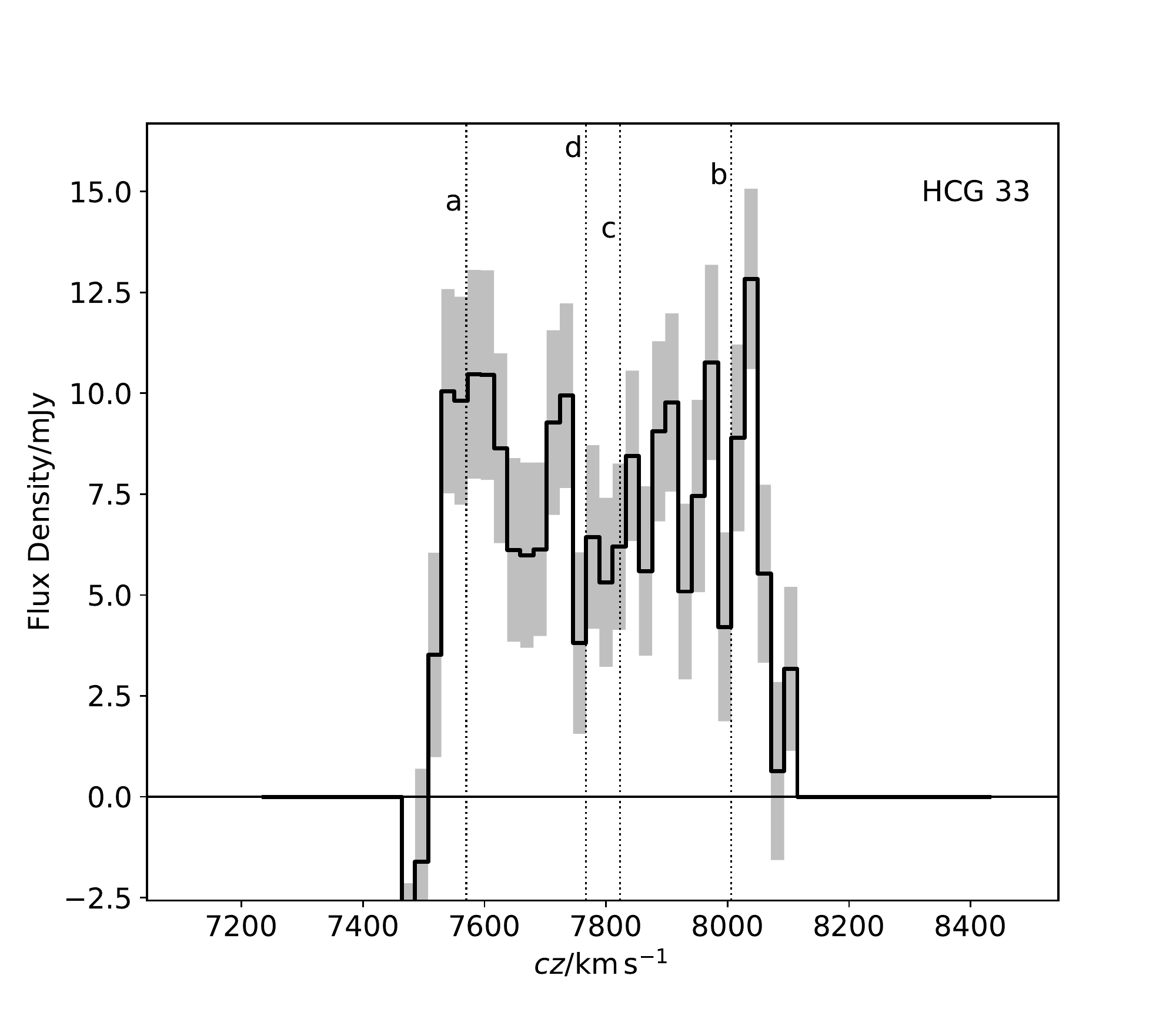}
    \caption{As in Figure \ref{fig:HCG2_spec}.}
    \label{fig:HCG33_spec}
\end{figure}

HCG~33 quartet of three early-type galaxies and a late-type galaxy in the redshift range 7500-8000~\kms.
Only HCG~33c is detected in \hi \ with the VLA (Figures \ref{fig:HCG33_split_overlay} \& \ref{fig:HCG33_spec}). The velocity structure of HCG~33c spans a large fraction of the observational bandwidth and is mostly regular except for a clear incipient tidal feature extending to the NE, which we separate from the galaxy's emission.

\subsection{HCG~37}

\begin{figure}[h]
    \centering
    \includegraphics[width=\columnwidth]{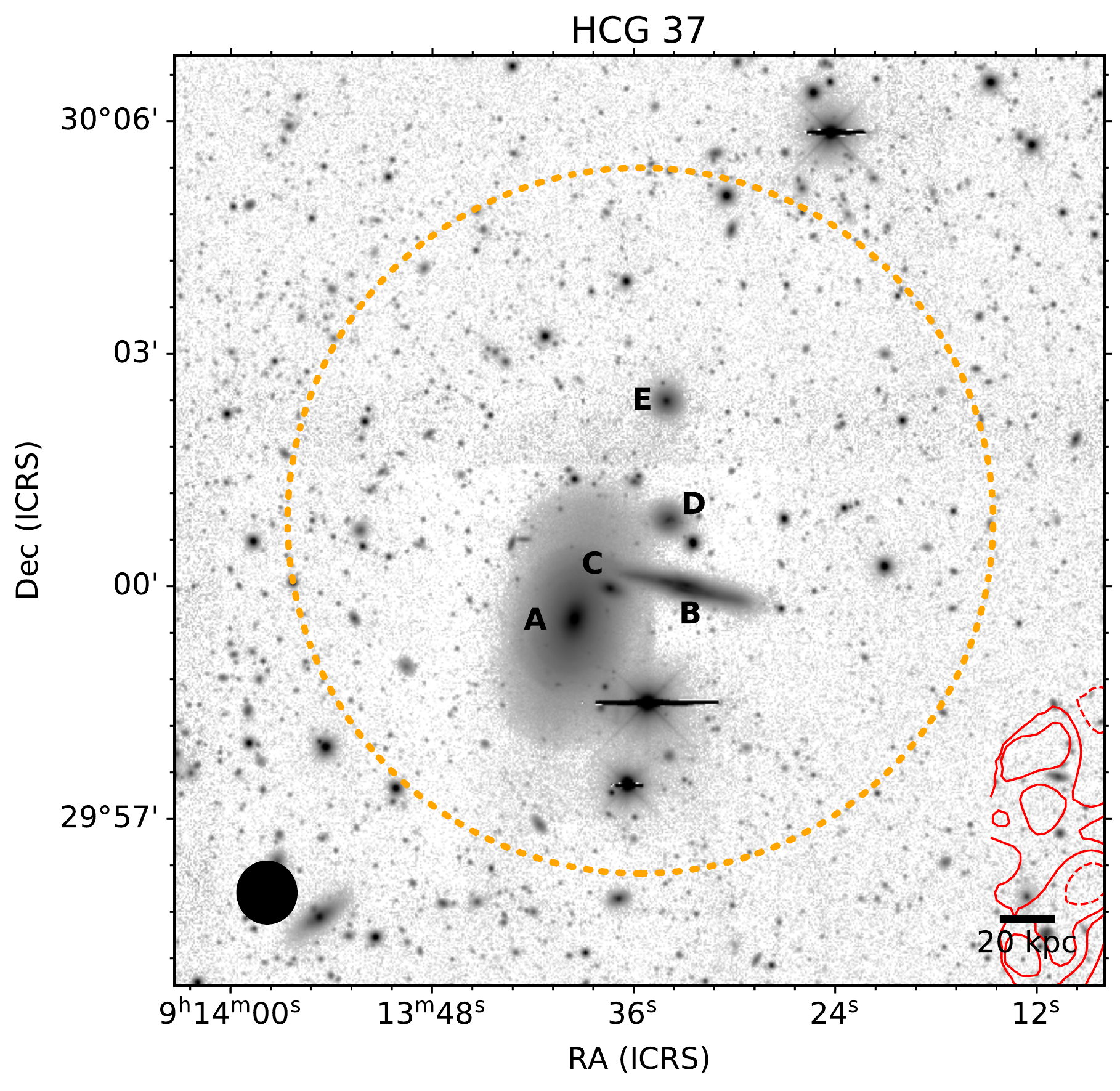}
    \caption{As in Figure \ref{fig:HCG7_split_overlay}.}
    \label{fig:HCG37_split_overlay}
\end{figure}

\begin{figure}[h]
    \centering
    \includegraphics[width=\columnwidth]{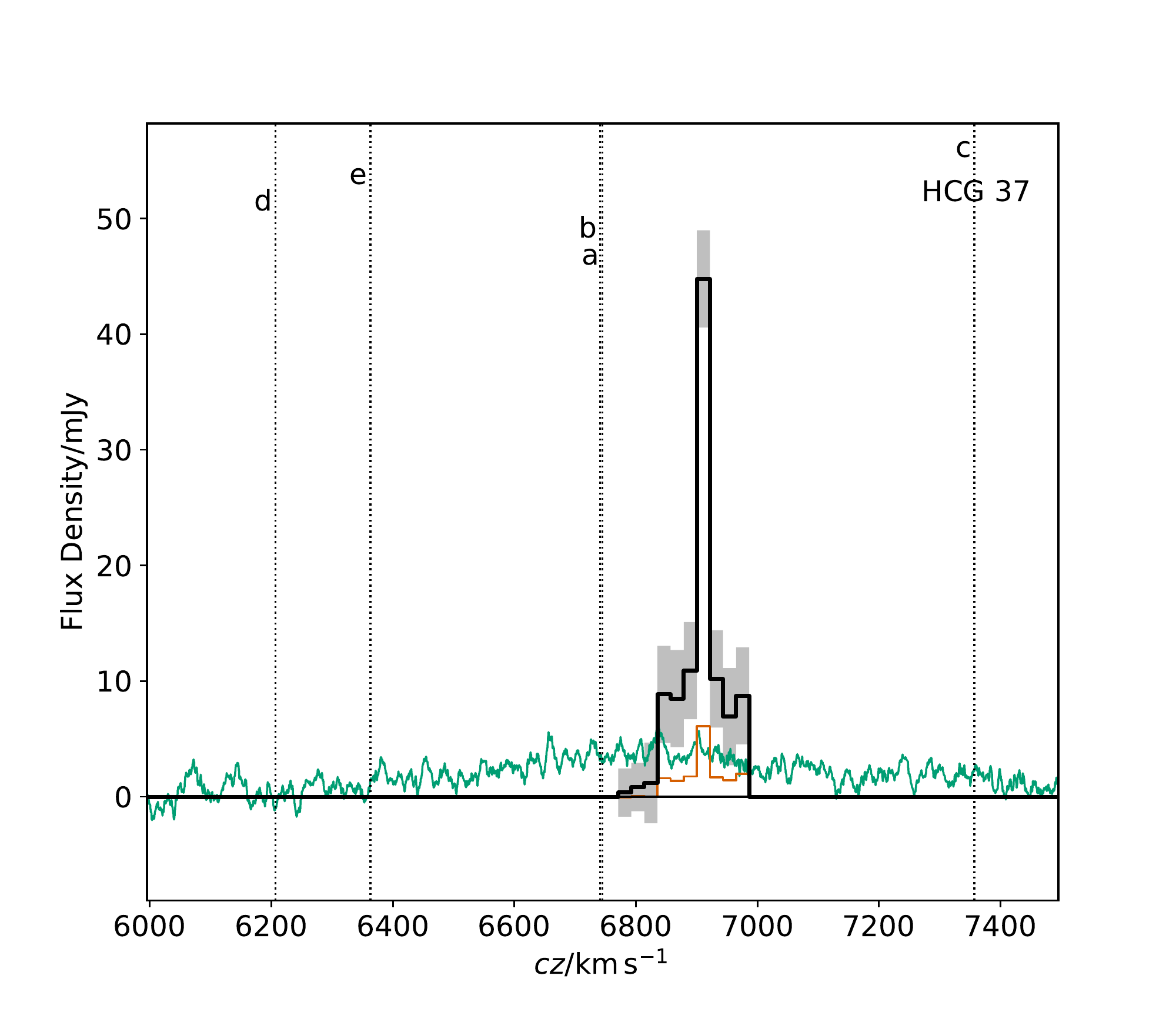}
    \caption{As in Figure \ref{fig:HCG7_spec}.}
    \label{fig:HCG37_spec}
\end{figure}

HCG~37 is a very compact quintet containing two early-type, one lenticular, and two late-type galaxies in the redshift range 6300-7400~\kms.
None of the core galaxies of HCG~37 are detection in \hi \ with the VLA (Figures \ref{fig:HCG37_split_overlay} \& \ref{fig:HCG37_spec}). The only feature above \texttt{SoFiA}'s threshold is an amorphous blob to the SW of the group. This feature may be spurious, however, its peak coincides with an uncatalogued LSB dwarf galaxy (at 09h13m05.95s +29d55m14.2s), which we refer to as HCG~37LSB1. We separated this feature into the densest clump that appears to be associated with an optical counterpart and the more diffuse emission. However, if associated with the LSB counterpart then this source is separated from the core group by well over 100~kpc and we do not consider it part of the compact group at present.

\subsection{HCG~38}

\begin{figure}[h]
    \centering
    \includegraphics[width=\columnwidth]{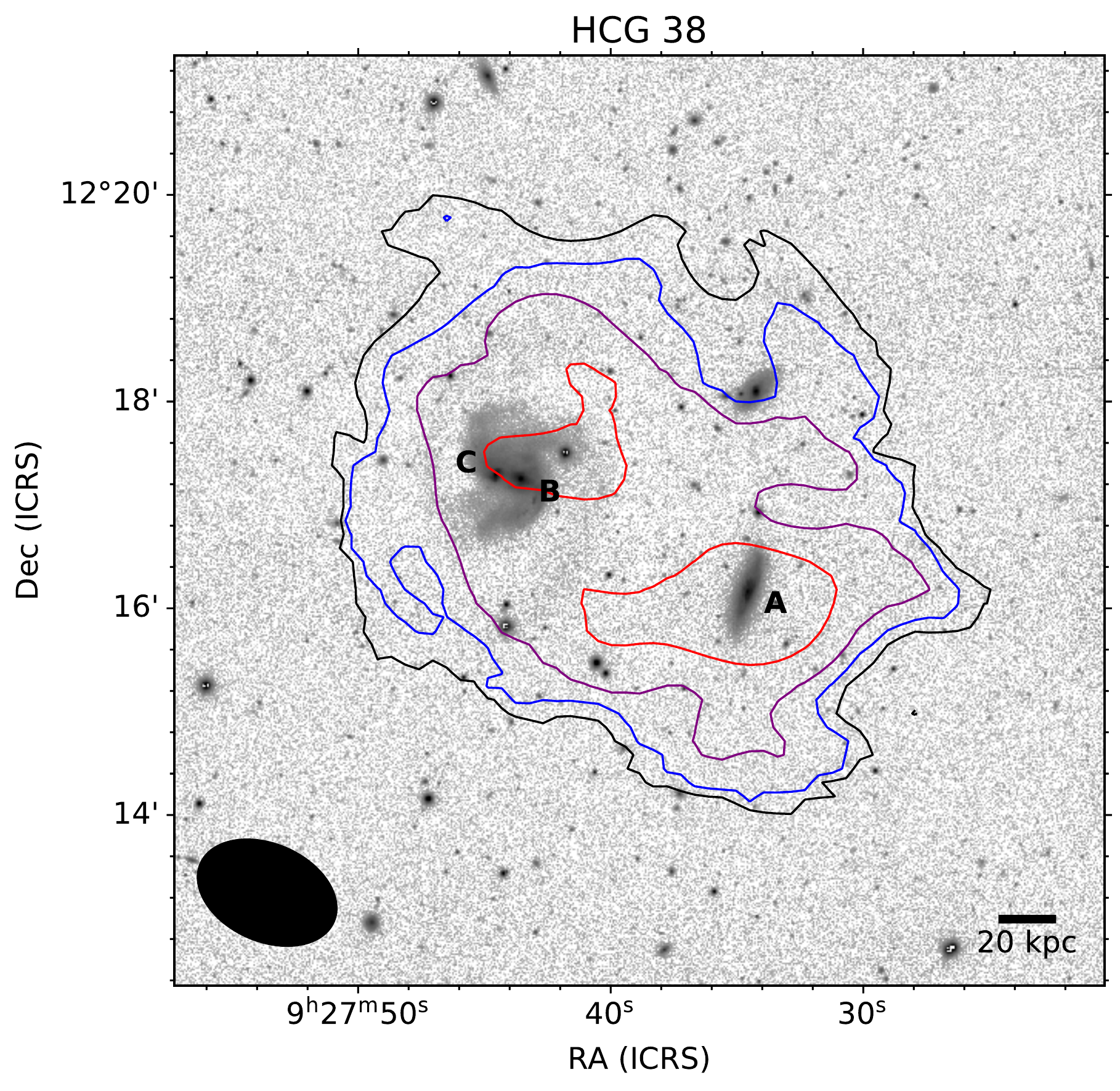}
    \caption{Integrated \hi \ emission (moment 0) contours overlaid on a DECaLS r-band image. Contours begin at 2$\sigma$ (solid black lines) and each subsequent contour is a factor of two greater. The dashed black contours indicate -2$\sigma$. The VLA synthesised beam is shown as a black ellipse in the bottom left and a 20~kpc scale bar is in the bottom right.}
    \label{fig:HCG38_overlay}
\end{figure}

\begin{figure}[h]
    \centering
    \includegraphics[width=\columnwidth]{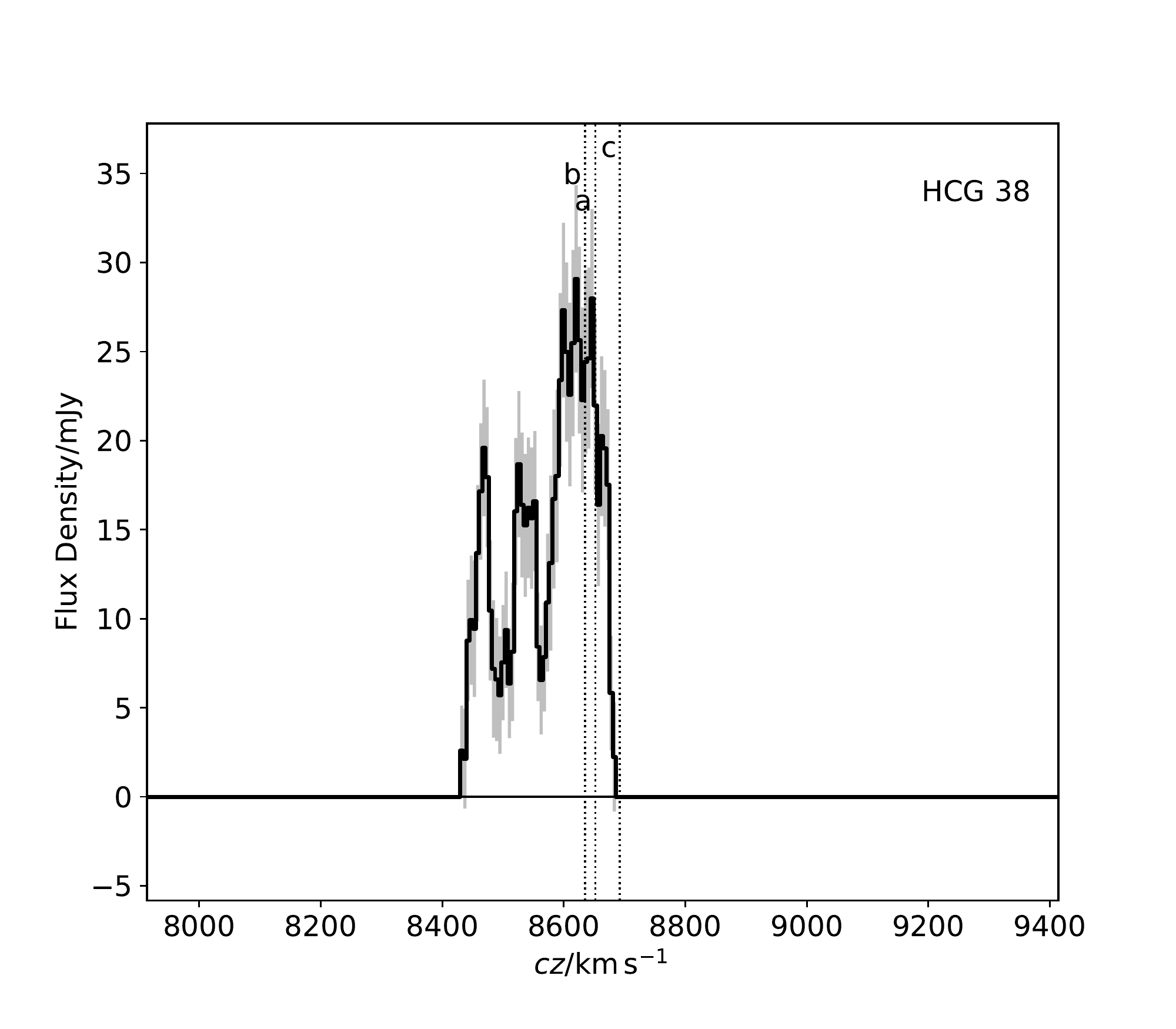}
    \caption{As in Figure \ref{fig:HCG2_spec}.}
    \label{fig:HCG38_spec}
\end{figure}

The single short observation of this group was hampered by RFI resulting in poor $uv$ coverage, a poorly behaved synthesised beam, and a noisy cube. We caution the reader that results for this group are less reliable than for most others.

This group is apparently a triplet of late type galaxies at 8700~\kms (the redshift of HCG~38d places it in the distant background), where two of the members appear to be merging HCG~38b and c. In our \hi \ cube we detect emission in the vicinity of all of these galaxies, as is apparent in the moment map (Figure \ref{fig:HCG38_overlay}). However, given the poor angular resolution and the noisy nature of the cube, emission is extremely challenging to confidently identify in individual channels and we thus do not attempt separation of features in this group. Will consider it only in terms of its global properties in the following analyses (Figure \ref{fig:HCG38_spec}).

\subsection{HCG~40}

\begin{figure}[h]
    \centering
    \includegraphics[width=\columnwidth]{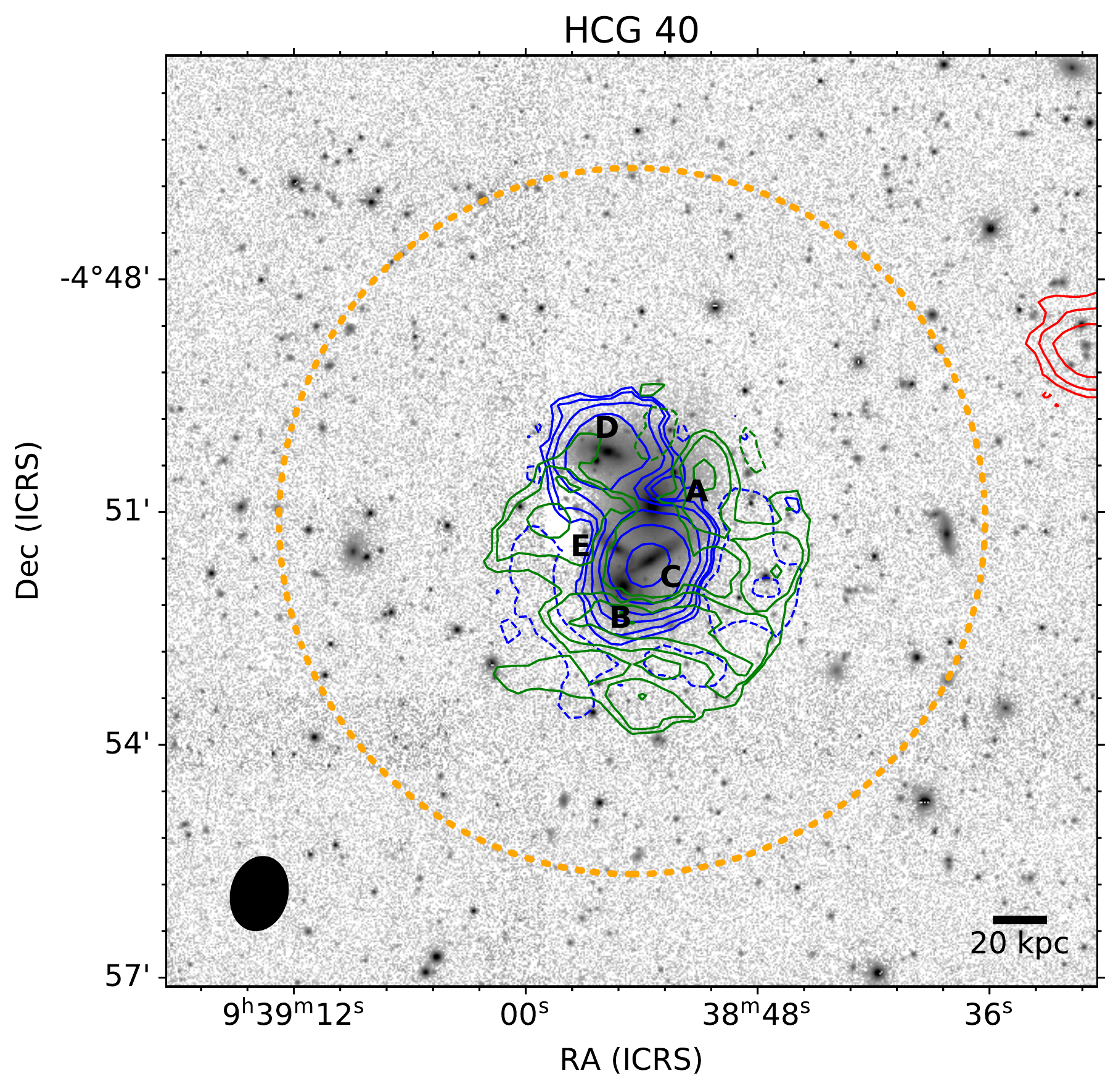}
    \caption{As in Figure \ref{fig:HCG7_split_overlay}.}
    \label{fig:HCG40_split_overlay}
\end{figure}

\begin{figure}[h]
    \centering
    \includegraphics[width=\columnwidth]{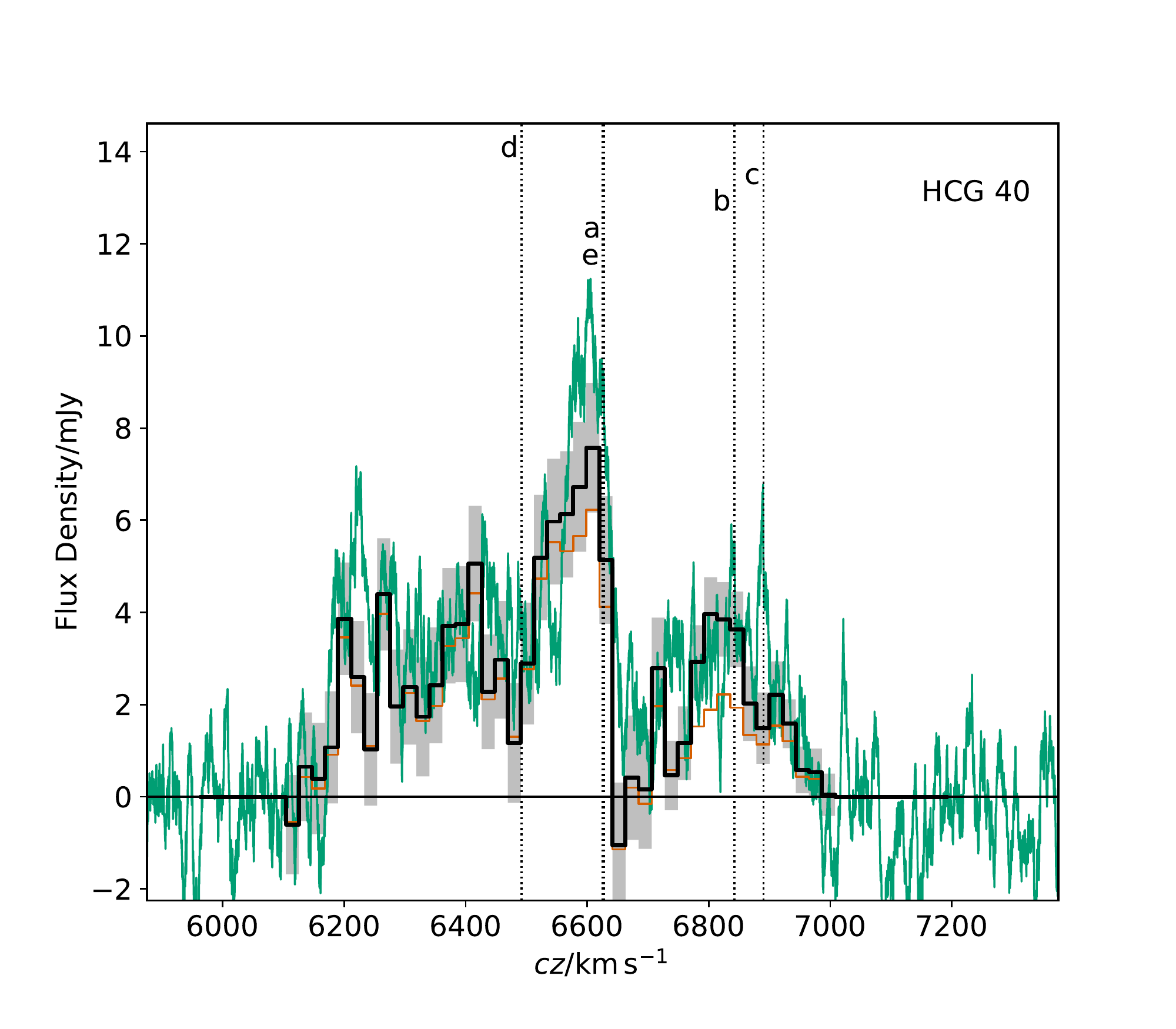}
    \caption{As in Figure \ref{fig:HCG7_spec}.}
    \label{fig:HCG40_spec}
\end{figure}

HCG~40 is an extremely compact configuration of five galaxies (three late-type and two early-type) in the redshift range 6400-6900~\kms, all of which overlap on the plane of the sky with their nearest neighbour. In the \hi \ cube the only well-defined velocity structures (in the vicinity of the core group) align with the major axes of HCG~40c and d. We therefore attribute most of the \hi \ to these two galaxies. Although some emission is classified as tidal features as it clearly does not conform to a disc-like structure, this is mostly low significance emission with minimal flux. To the NW of the group another \hi \ source is detected, probably associated with GALEXASC~J093831.05-044853.6. For brevity we will refer to this as HCG~40GLX1. This object is separated from the core group by $\sim$150 kpc and we do not consider it as a member.

\subsection{HCG~47}

\begin{figure}[h]
    \centering
    \includegraphics[width=\columnwidth]{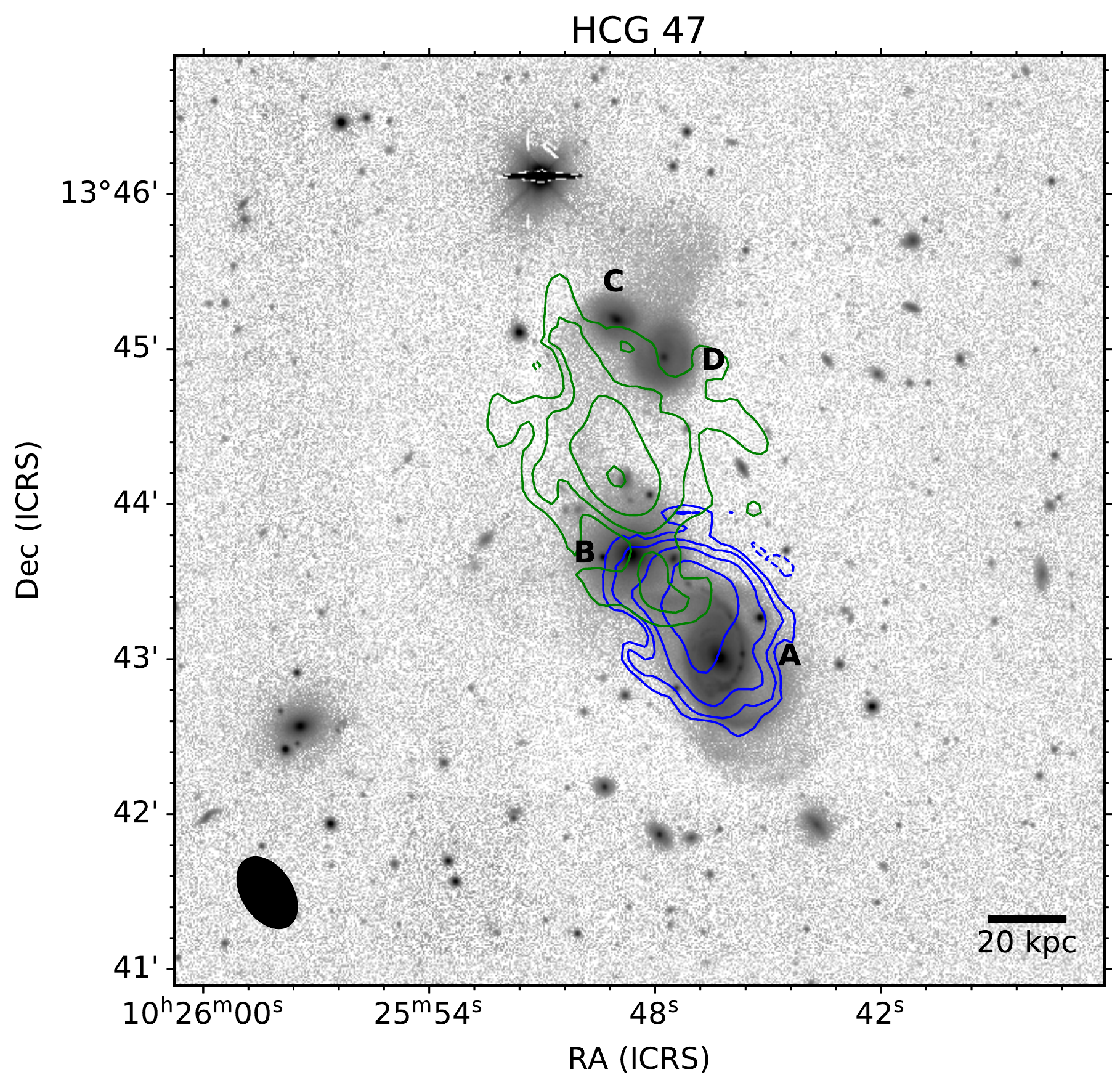}
    \caption{As in Figure \ref{fig:HCG2_split_overlay}.}
    \label{fig:HCG47_split_overlay}
\end{figure}

\begin{figure}[h]
    \centering
    \includegraphics[width=\columnwidth]{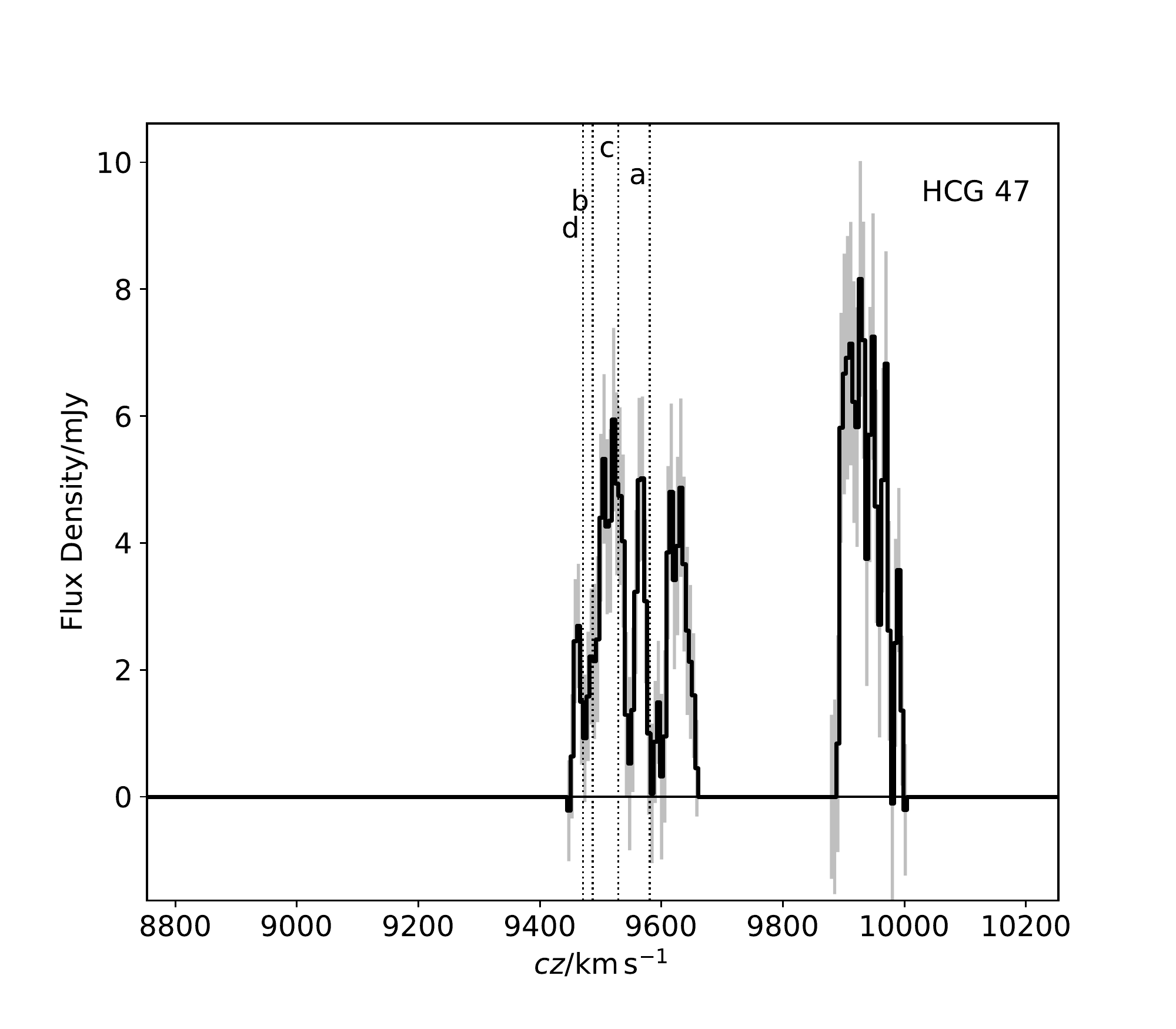}
    \caption{As in Figure \ref{fig:HCG2_spec}.}
    \label{fig:HCG47_spec}
\end{figure}

HCG~47 has the appearance of a pair of pairs, with the two larger galaxies, HCG~47a and b, in the south and the two smaller galaxies, HCG~47c and d, to the north, all of which are in the redshift range 9400-9600~\kms. \hi \ emission is detected in HCG~47a and b and towards HCG~47c and d (Figures \ref{fig:HCG47_split_overlay} \& \ref{fig:HCG47_spec}). However, much of this is a relatively low S/N, an issue which is exacerbated by the fact that the synthesised beam has significant side lobes due to the short observations and poor resulting $uv$ coverage. In addition, five channels ($\Delta v \approx 24$ \kms) near the centre of the cube were entirely, or almost entirely, flagged due to RFI.

The \texttt{SoFiA} mask contains two separate regions of \hi \ emission. One, upon inspection of the channel maps, appears to mostly coincide with HCG~47a and b, while the other lies in between the two pairs. Owing to the difficulties with the data quality, we simply denote the first region as being from HCG~47a and b combined (although likely the majority is from HCG~47a), while the latter region is classified as extended emission which does not coincide with a galaxy. The reader is cautioned that the results for this group a quite uncertain.

\subsection{HCG~48}

\begin{figure}[h]
    \centering
    \includegraphics[width=\columnwidth]{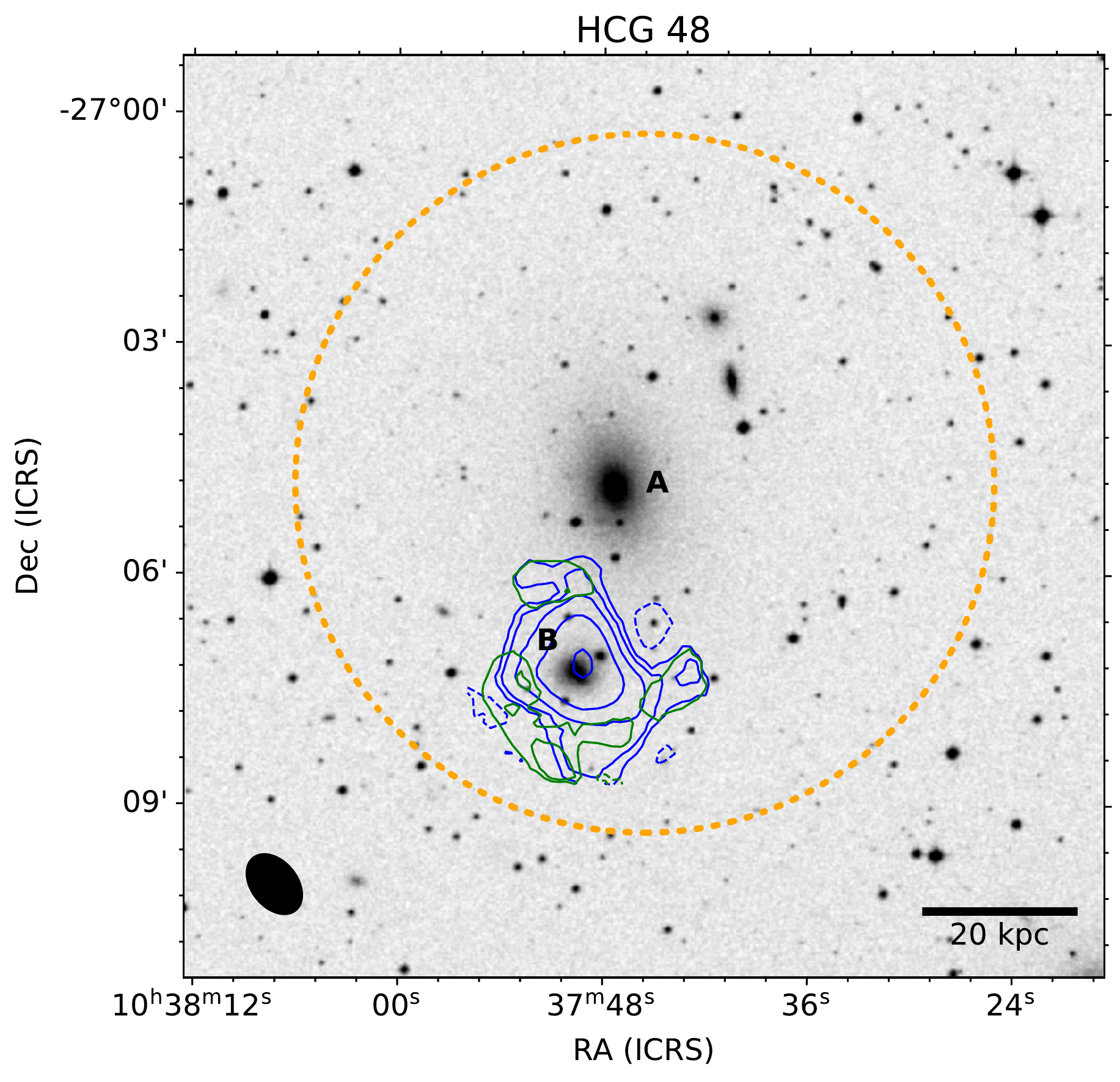}
    \caption{As in Figure \ref{fig:HCG7_split_overlay}.}
    \label{fig:HCG48_split_overlay}
\end{figure}

\begin{figure}[h]
    \centering
    \includegraphics[width=\columnwidth]{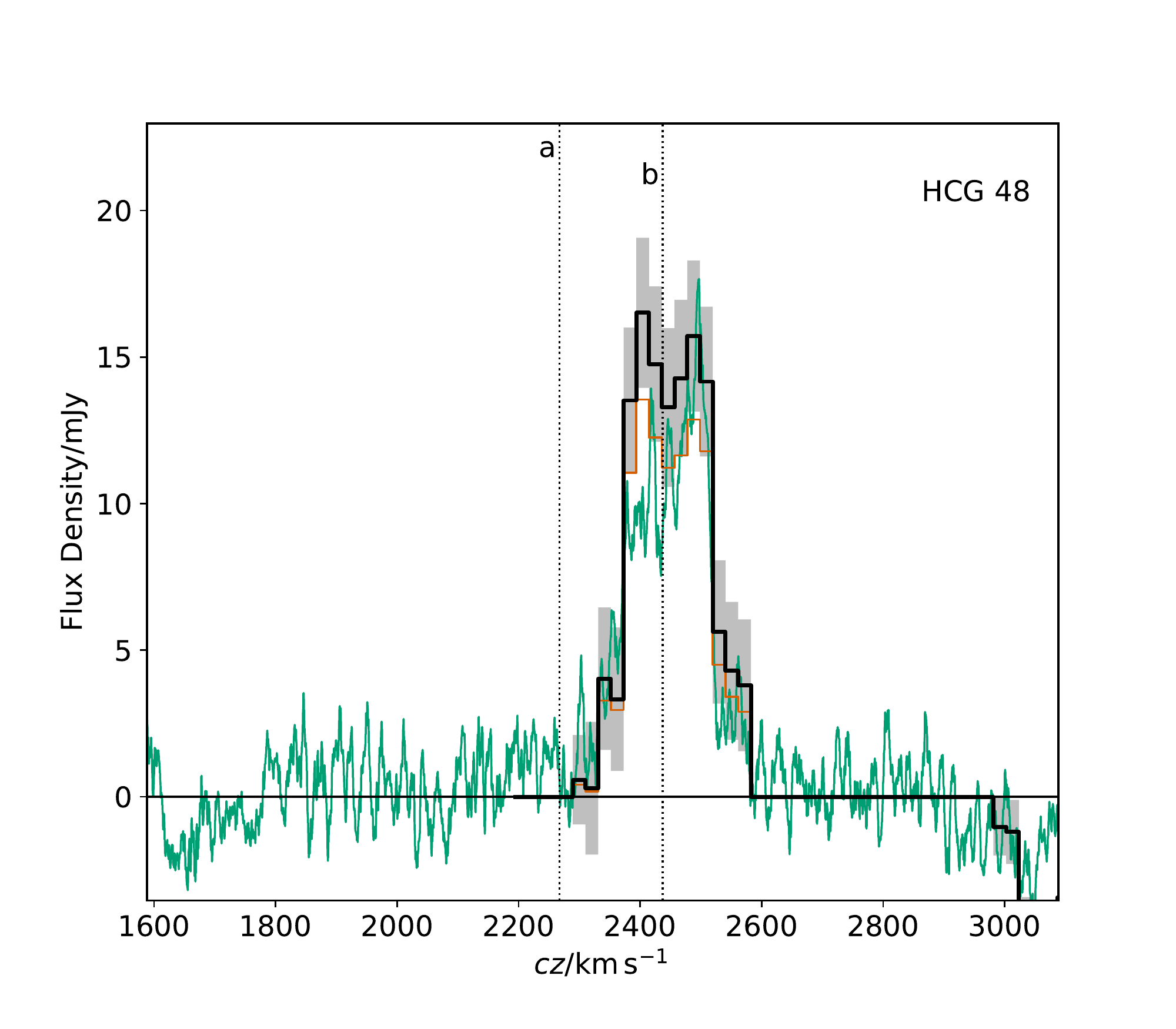}
    \caption{As in Figure \ref{fig:HCG7_spec}.}
    \label{fig:HCG48_spec}
\end{figure}

HCG~48 is likely a false group and is really just a pair of galaxies (HCG~48a, a large elliptical, and HCG~48b, a smaller late-type galaxy). The other two original members of this group were revealed to be a background pair once redshifts were obtained (we also note that there are several conflicting redshift measurements for these galaxies in the literature). In the VLA data only the \hi \ emission from HCG~48b is detected, as well as PGC~762452, approximately 165~kpc to the NW (Figures \ref{fig:HCG48_split_overlay} \& \ref{fig:HCG48_spec}). The \hi \ cube has mostly well-behaved noise, but there is a continuum artefact NW of the group, as well as some structured noise. There are a few minor, low significance \hi \ features connected to HCG~48b which we assign as tidal. In summary, as this HCG is unlikely to be a genuine group we do not consider it in the remainder of our analysis.

\subsection{HCG~49}

\begin{figure}[h]
    \centering
    \includegraphics[width=\columnwidth]{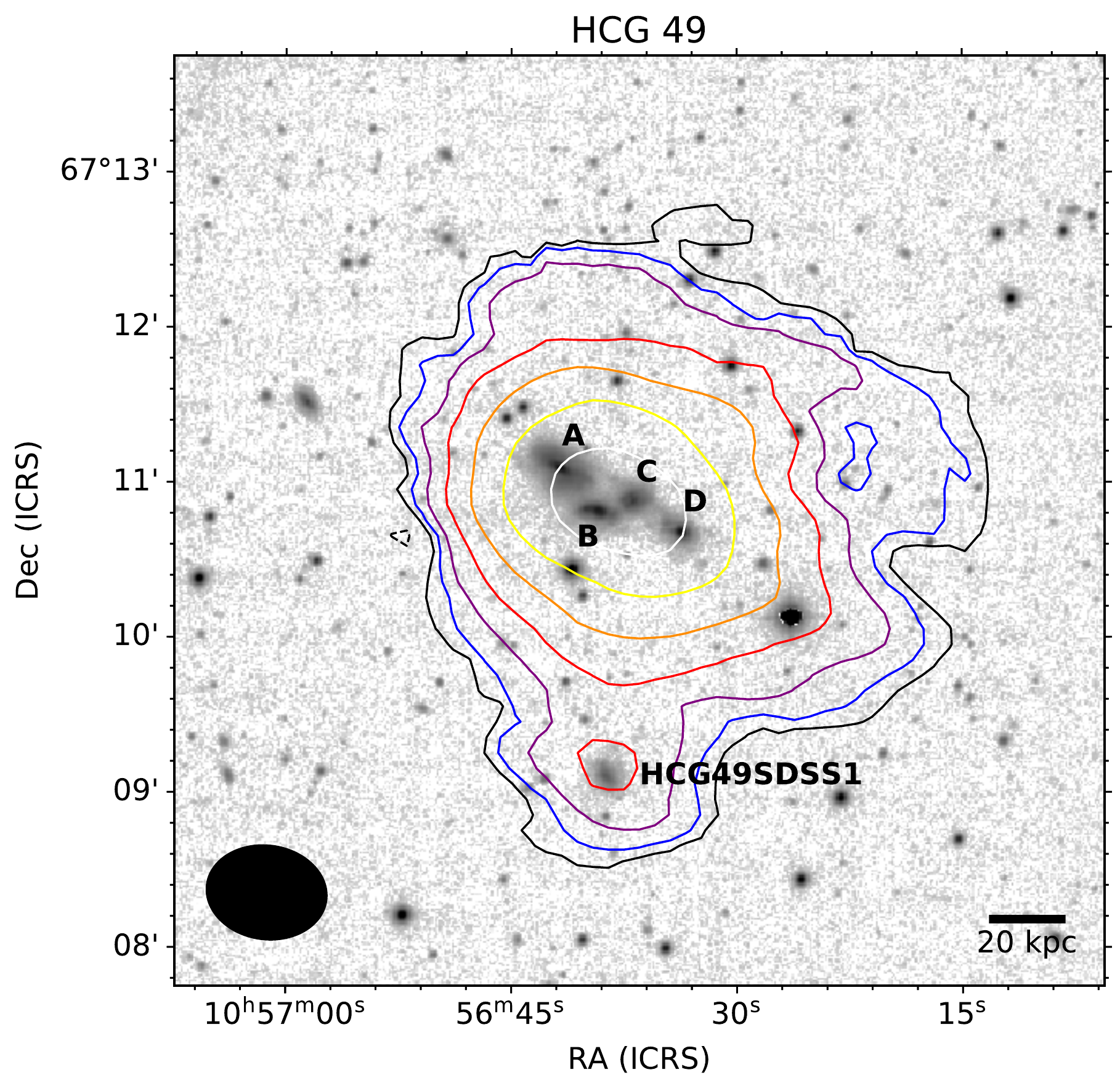}
    \caption{As in Figure \ref{fig:HCG38_overlay}.}
    \label{fig:HCG49_overlay}
\end{figure}

\begin{figure}[h]
    \centering
    \includegraphics[width=\columnwidth]{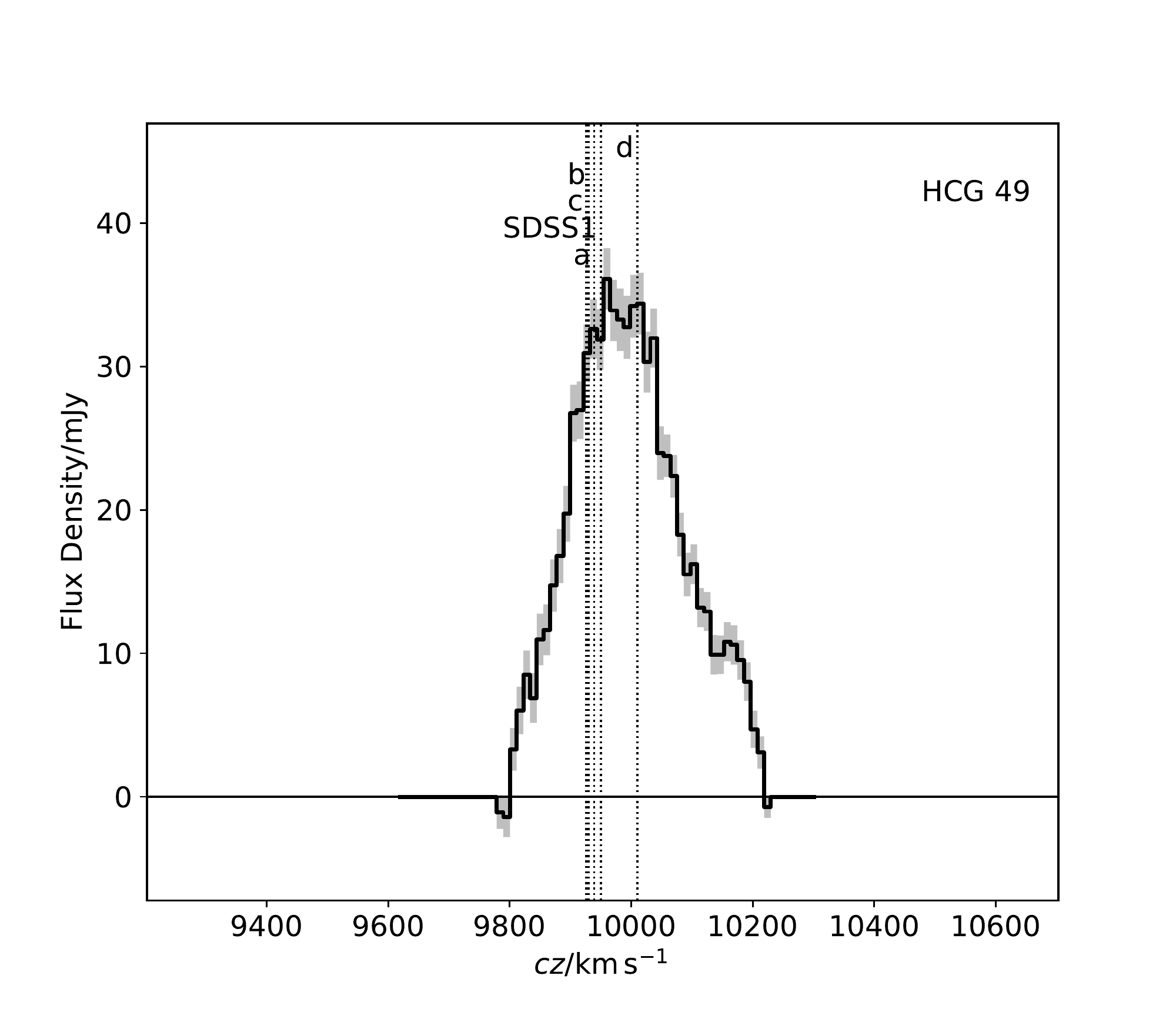}
    \caption{As in Figure \ref{fig:HCG2_spec}.}
    \label{fig:HCG49_spec}
\end{figure}

HCG~49 is an extremely compact configuration of three late-type galaxies and one early-type, with an additional late-type slightly to the south. All have redshifts of $\sim$10000~\kms. Owing to the distance of this group $\sim$140~Mpc, the physical resolution of the VLA DnC imaging is extremely poor. There is no possibility of reliably separating emission from the four core group members as this appears almost a one continuous cloud in the \hi \ cube (Figures \ref{fig:HCG49_overlay} \& \ref{fig:HCG49_spec}). Therefore, we approximately separate the emission of SDSS~J105638.63+670906.0 (HCG~40SDSS1, the galaxy slightly to the south), which does appear to be interacting with the core group (though resolution is a limiting factor), and SDSS~J105454.97+670919.4 (HCG~40SDSS2), detected about~350 kpc to the west (not considered a member). However, for the remainder of this work this group will only be considered in terms of its global properties.

\subsection{HCG~54}

\begin{figure}[h]
    \centering
    \includegraphics[width=\columnwidth]{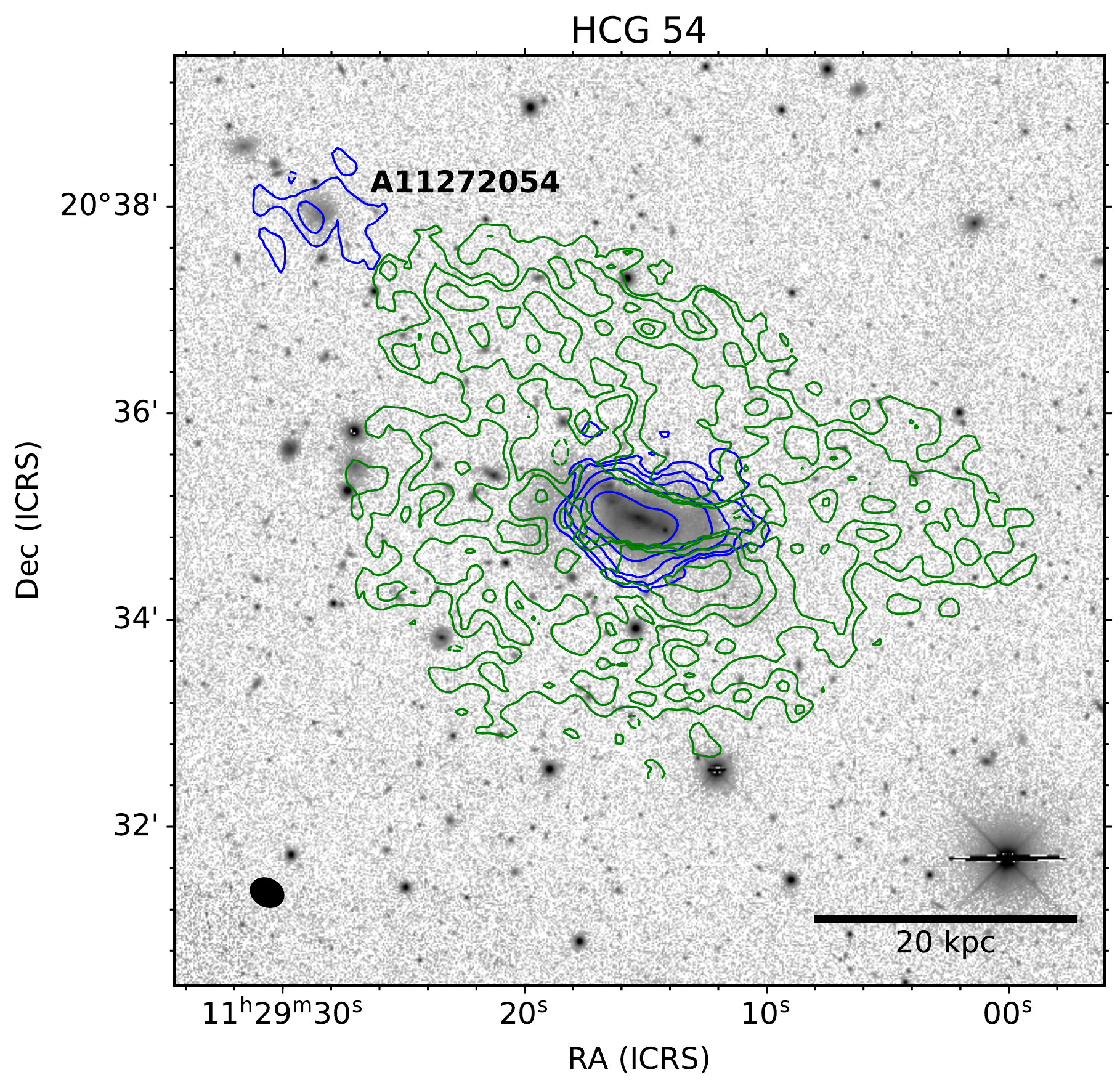}
    \caption{As in Figure \ref{fig:HCG2_split_overlay}.}
    \label{fig:HCG54_split_overlay}
\end{figure}

\begin{figure}[h]
    \centering
    \includegraphics[width=\columnwidth]{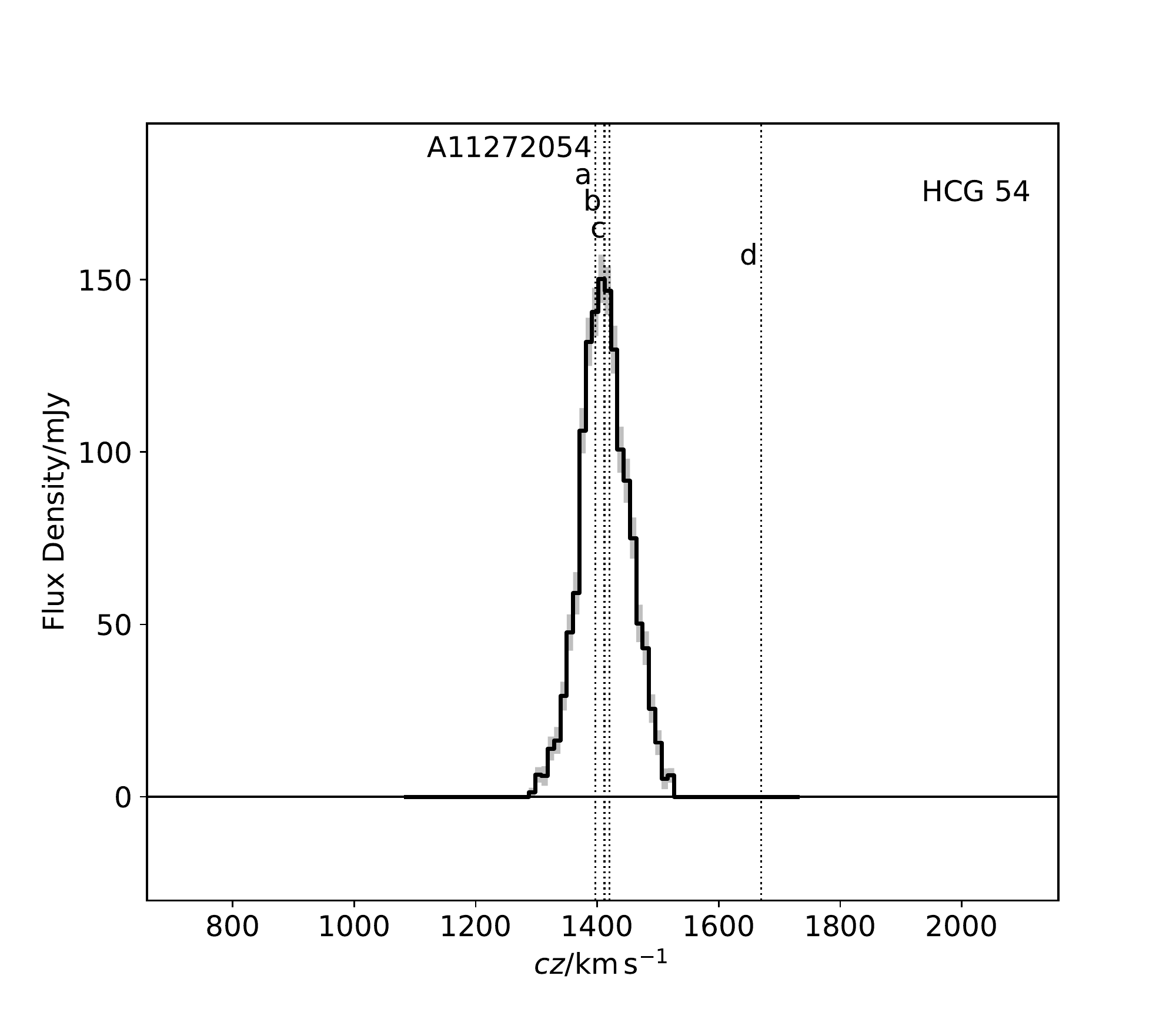}
    \caption{As in Figure \ref{fig:HCG2_spec}.}
    \label{fig:HCG54_spec}
\end{figure}

HCG~54 is a tight, linear arrangement of what were originally thought to be late-type dwarf galaxies \citep{Hickson1982}, but are likely all regions associated with a merger event of two or more galaxies \citep{Verdes-Montenegro+2002}. The \hi \ emission mapped with the VLA spans all the bright, star-forming components of this merger and includes a significant tidal tail which wraps around to the north of the group \citep[noted previously by][]{Verdes-Montenegro+2002}. The moment zero map of the group has a mottled appearance due to significant side lobes to the synthesised beam (Figures \ref{fig:HCG54_split_overlay} \& \ref{fig:HCG54_spec}).

We do not attempt to separate individual features in the main region of \hi \ emission, which forms a contiguous structure from one side of the merger to the other. However, we do separate the clear tidal extended features around the group. In particular, there is a main tail that wraps around the merger and ends in the galaxy A1127+2054. As this HCG mainly consists of a single merger in cannot be considered as a group in the same manner as the other groups and we therefore do not consider it in our analysis.

\subsection{HCG~56}

\begin{figure}[h]
    \centering
    \includegraphics[width=\columnwidth]{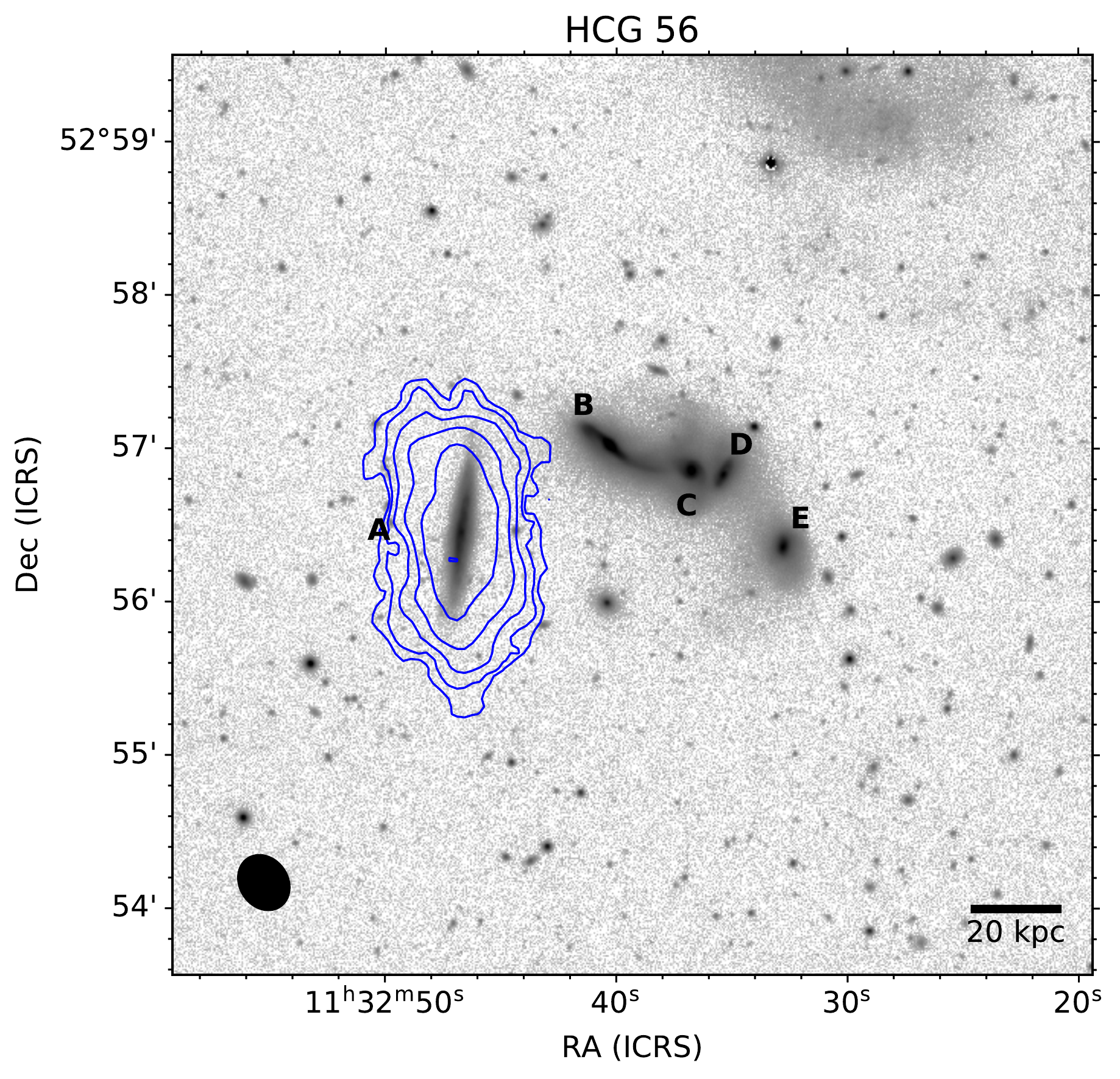}
    \caption{As in Figure \ref{fig:HCG2_split_overlay}. Note that the extend feature in the NW corner of this image is due to a large foreground galaxy just outside the frame.}
    \label{fig:HCG56_split_overlay}
\end{figure}

\begin{figure}[h]
    \centering
    \includegraphics[width=\columnwidth]{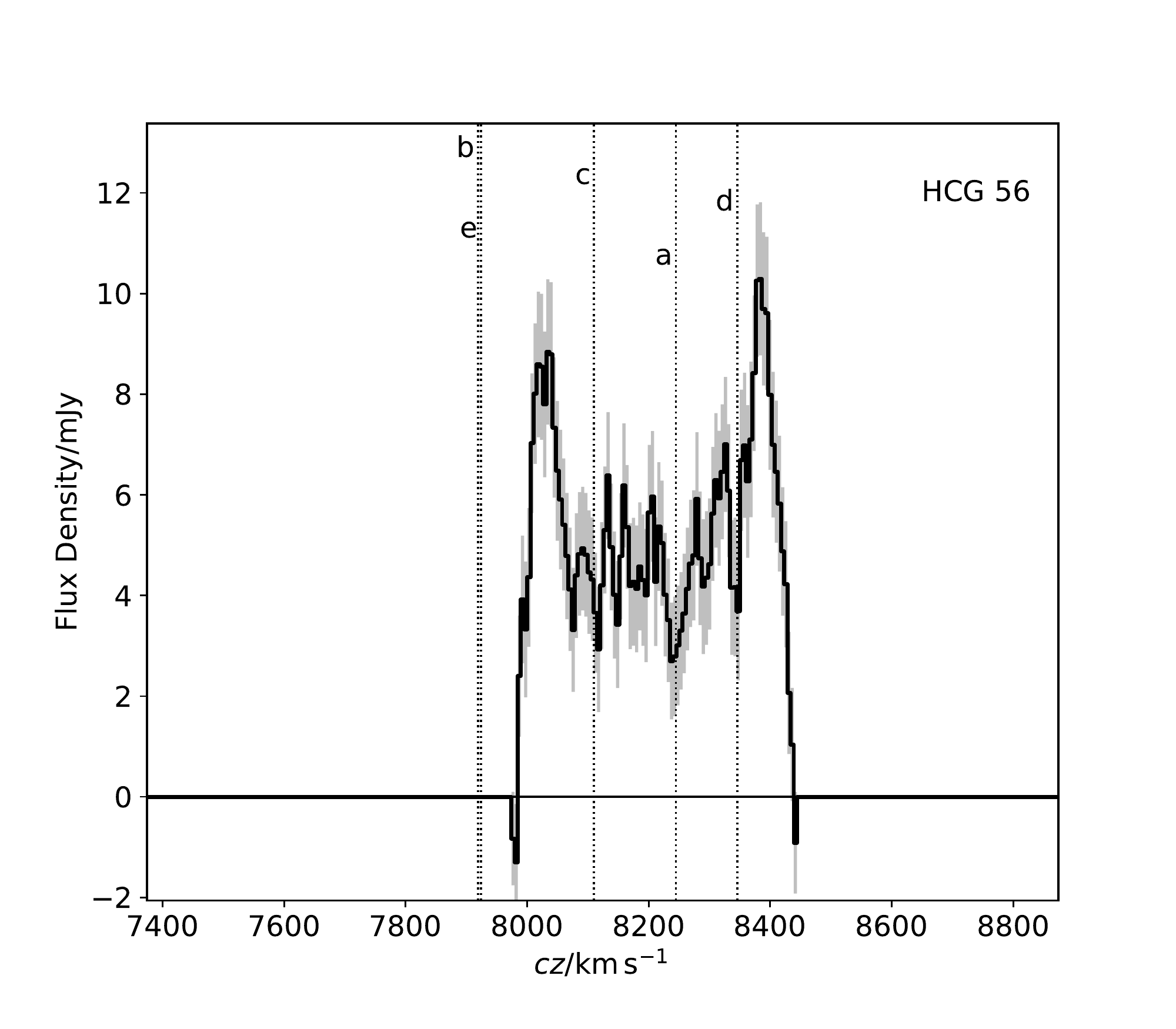}
    \caption{As in Figure \ref{fig:HCG2_spec}.}
    \label{fig:HCG56_spec}
\end{figure}

HCG~56 is a compact configuration of four lenticular and one late-type galaxies in the redshift range 7900-8400~\kms. Only HCG~56a is detected in \hi, which is an extremely high S/N detection with a clear, regular velocity structure (Figures \ref{fig:HCG56_split_overlay} \& \ref{fig:HCG56_spec}). No separation of features is required as this is the only source in the \texttt{SoFiA} mask and there are no clear signs of tidal tails. However, there are some low significance features which did not reach the threshold required to be included within the \texttt{SoFiA} mask, deeper mapping would be required to verify these features and reveal if they are genuine \hi \ associated with the remaining galaxies. 

\subsection{HCG~57}

\begin{figure}[h]
    \centering
    \includegraphics[width=\columnwidth]{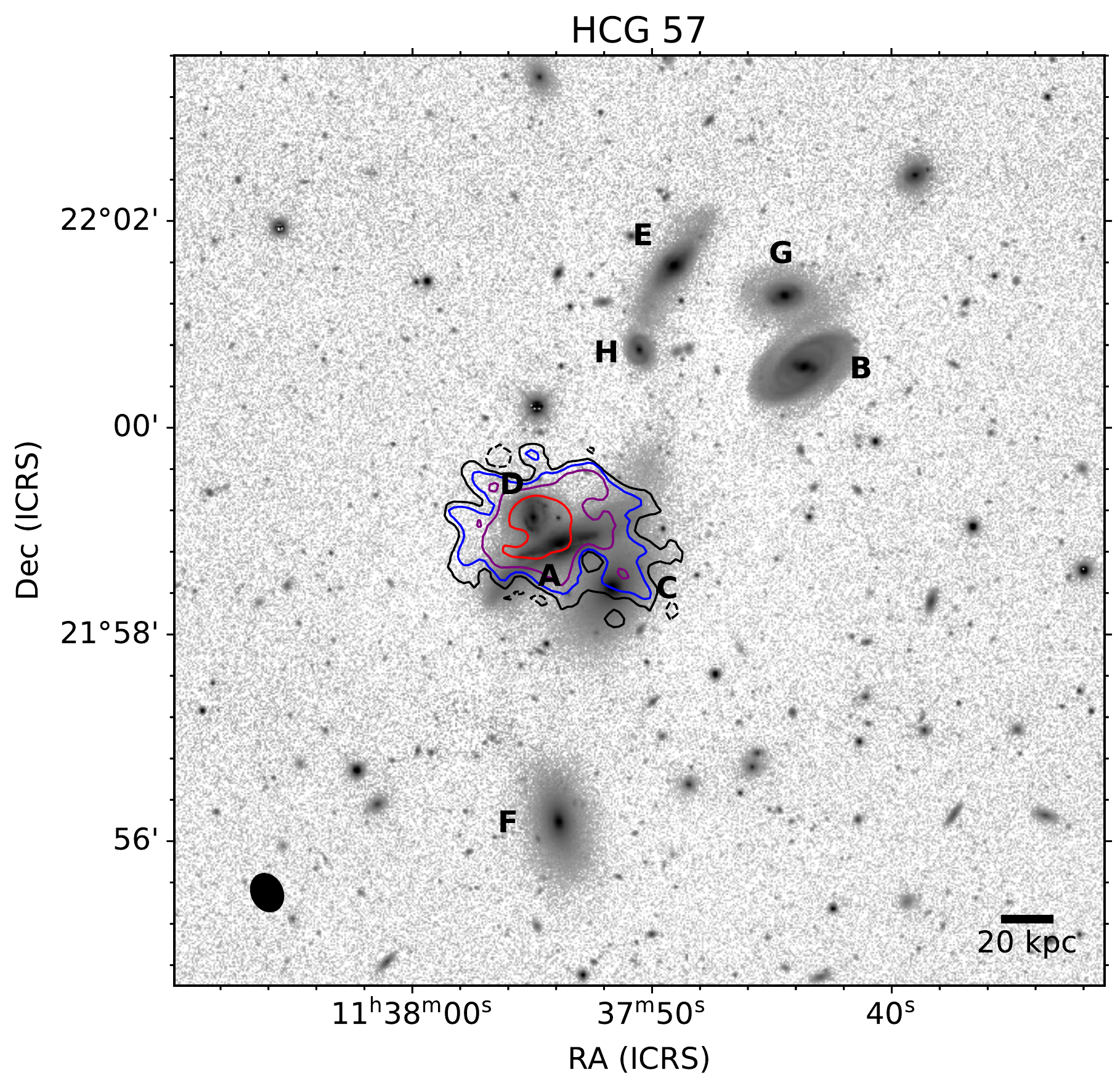}
    \caption{As in Figure \ref{fig:HCG38_overlay}.}
    \label{fig:HCG57_overlay}
\end{figure}

\begin{figure}[h]
    \centering
    \includegraphics[width=\columnwidth]{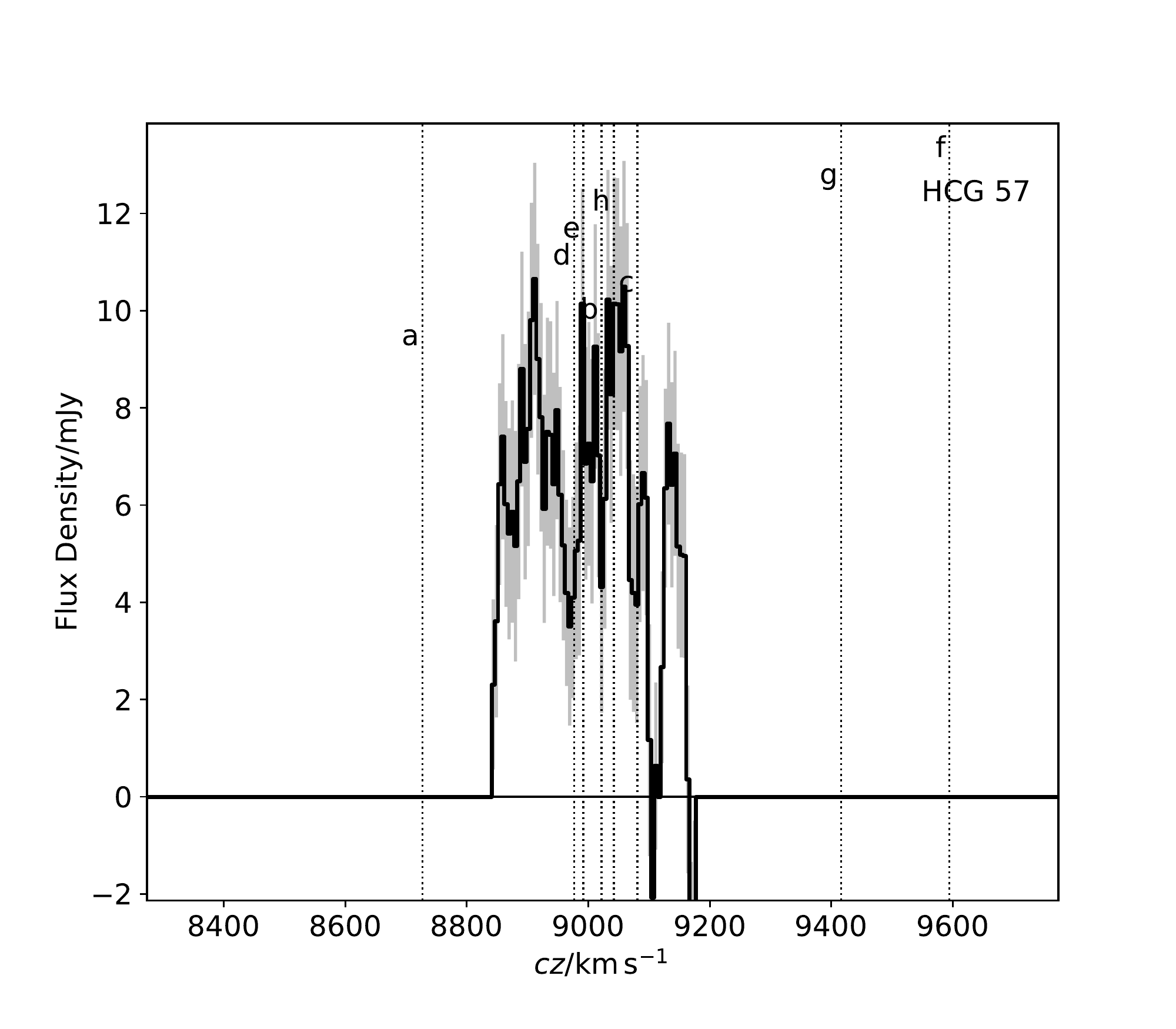}
    \caption{As in Figure \ref{fig:HCG2_spec}.}
    \label{fig:HCG57_spec}
\end{figure}

HCG~57 is a group of eight, mostly late-type, galaxies in the redshift range 8700-9600~\kms. The three galaxies at the centre of the group (HCG~57a, c, and d) appear to be detected in the VLA observations (Figures \ref{fig:HCG57_overlay} \& \ref{fig:HCG57_spec}). However, the data quality is not optimal, with a relatively poorly behaved synthesised beam and some residual structure in the noise that becomes visible when summing several channels. Given these limitation we do not attempt to separate this emission into distinct galaxies and tidal features, and will consider this group only in terms of its global \hi \ properties.

In addition, we detect faint \hi \ emission approximately 400~kpc to the SW of the group. This appears to coincide with an uncatalogued LSB galaxy. However, we note that given the low data quality this detection may be spurious and the apparent optical counterpart coincidental. 

\subsection{HCG~58}

\begin{figure}[h]
    \centering
    \includegraphics[width=\columnwidth]{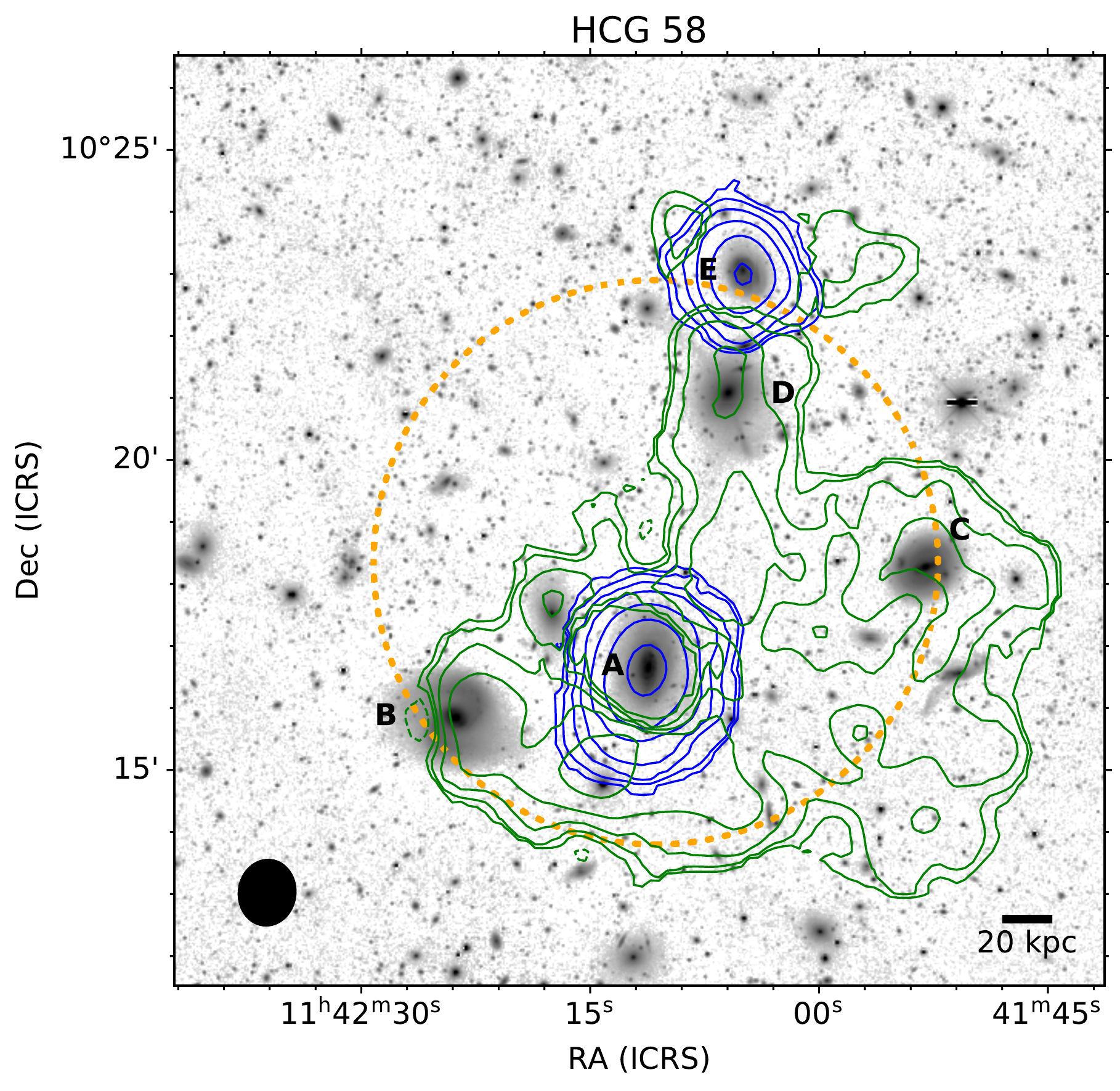}
    \caption{As in Figure \ref{fig:HCG7_split_overlay}.}
    \label{fig:HCG58_split_overlay}
\end{figure}

\begin{figure}[h]
    \centering
    \includegraphics[width=\columnwidth]{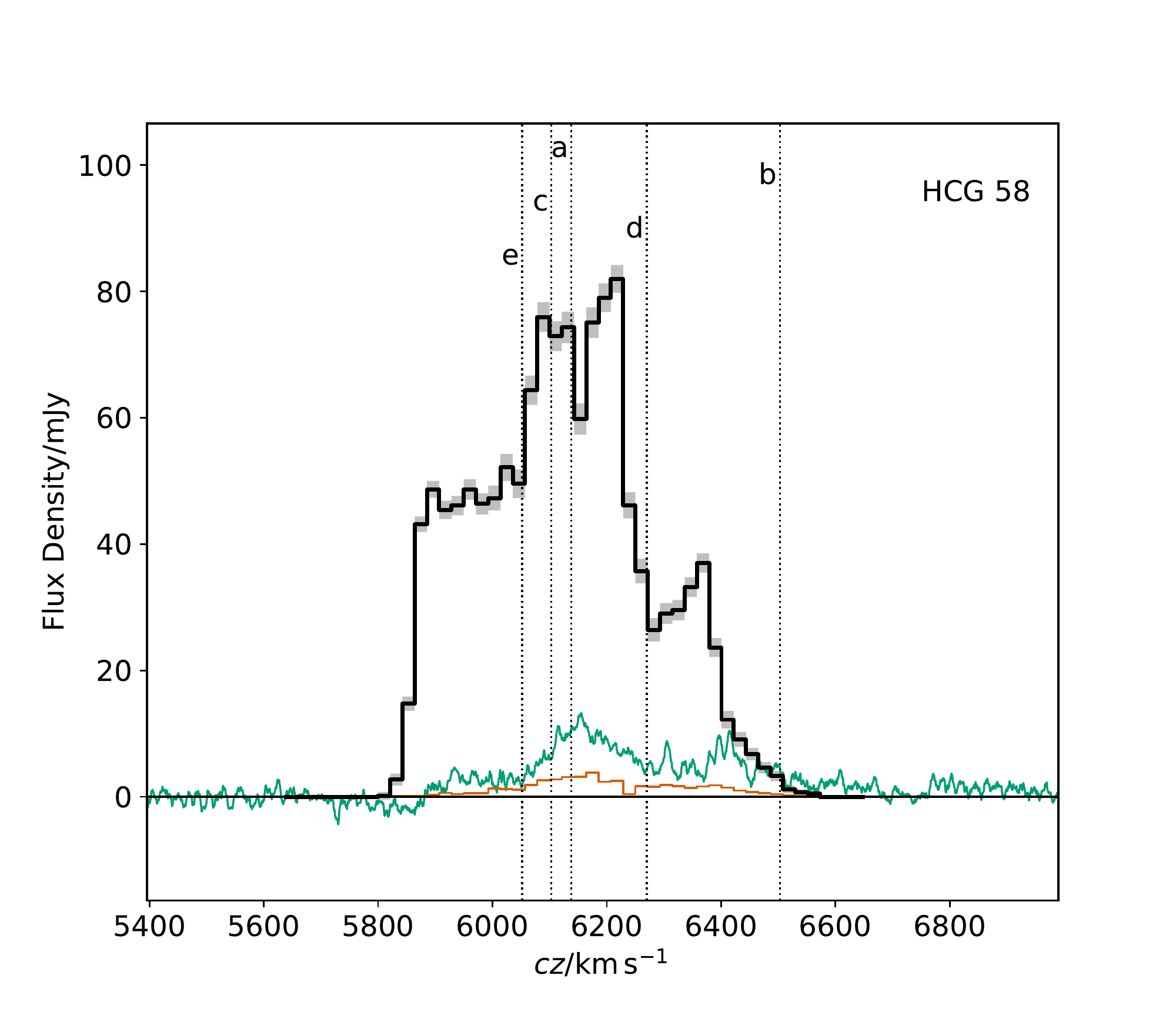}
    \caption{As in Figure \ref{fig:HCG7_spec}.}
    \label{fig:HCG58_spec}
\end{figure}

HCG~58 is a configuration of three late-type galaxies, one lenticular, and one early-type. \hi \ emission is detected in the VLA observations throughout the structure (Figures \ref{fig:HCG58_split_overlay} \& \ref{fig:HCG58_spec}). The core group is also surrounded by a number of other galaxies, separated from the core group by 200-500~kpc, several of which we also detect in \hi \ but do not consider as members of the core group. There is a continuum subtraction artefact to the south of the core group which was excluded when generating the \texttt{SoFiA} mask.

Although \hi \ is detected that is co-spatial with all the five group members, there are only clear overdensities associated with HCG~58a and e. The remaining galaxies appear so disturbed that they are virtually indistinguishable from the gas the fills the IGrM. We therefore separate only the high density gas associated with HCG~58a and e, and mark all remaining gas in the group as belonging to extended features.

The VLA spectrum weighted to the match the GBT beam response (orange line, Figure \ref{fig:HCG58_spec}) is down-weighted so much in comparison to the raw spectrum that it is significantly below the GBT spectrum. Given that HCG~58e is on the edge of the GBT primary beam (Figure \ref{fig:HCG58_split_overlay}) and much of the extended emission is beyond the beam, this comparison will be extremely sensitive to the approximate beam model we have assumed. In addition, any minor pointing offset in the GBT observation would also lead to significant differences in the comparison. Thus, although the GBT spectrum appears to contain considerably more flux, it is difficult to be certain that this indicates emission that was missed by the VLA.

\subsection{HCG~59}

\begin{figure}[h]
    \centering
    \includegraphics[width=\columnwidth]{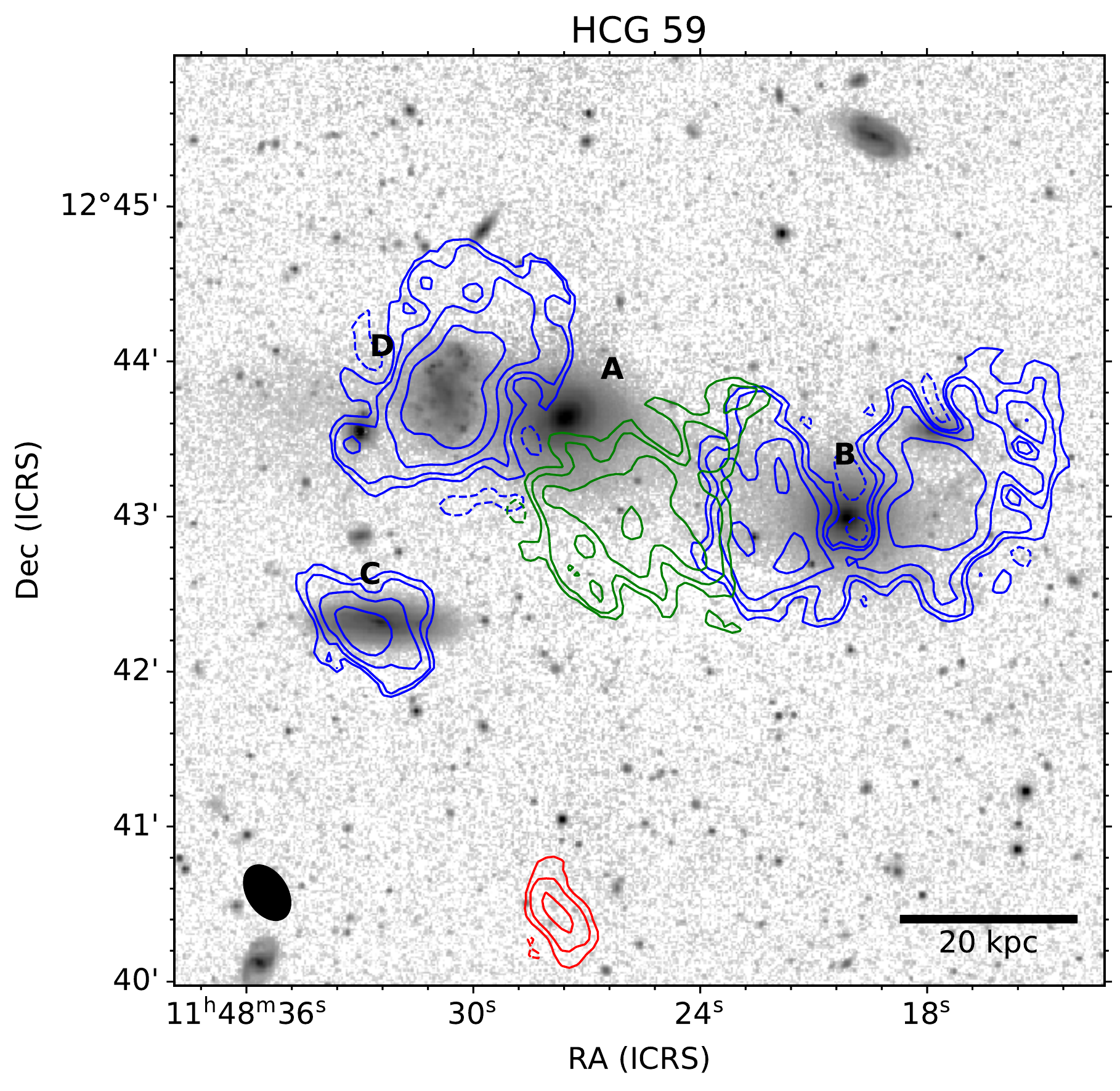}
    \caption{As in Figure \ref{fig:HCG2_split_overlay}.}
    \label{fig:HCG59_split_overlay}
\end{figure}

\begin{figure}[h]
    \centering
    \includegraphics[width=\columnwidth]{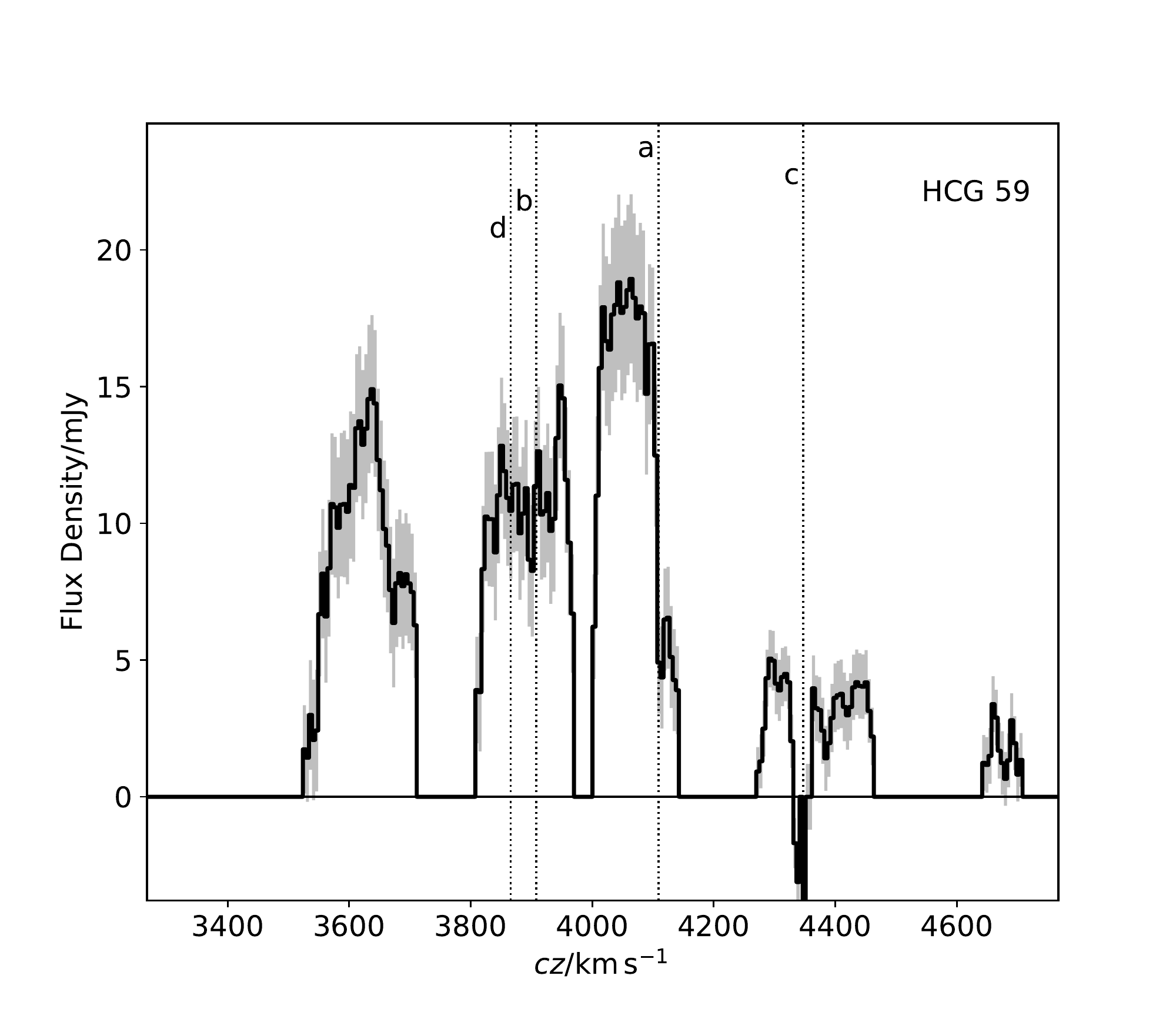}
    \caption{As in Figure \ref{fig:HCG2_spec}.}
    \label{fig:HCG59_spec}
\end{figure}

Unlike all other sources the \hi \ data for HCG 59 went through an additional step of self-calibration due to strong interference fringes originating from a bright, double continuum source near the edge of the primary beam. These additional steps greatly reduce the strength of artefacts in the \hi \ cube, however, there is still some residual structure in the noise and a high threshold (5$\sigma$) is used in \texttt{SoFiA} to attempt to eliminate them.

HCG~59 is a group of two spirals, one irregular and one elliptical galaxy arranged in an L-shape at a redshift of approximately 4000 \kms. All four appear to be detected in \hi \ with the VLA data, although the \hi \ emission of HCG~59c is interrupted by RFI (Figures \ref{fig:HCG59_split_overlay} \& \ref{fig:HCG59_spec}). Although the emission from the three of these galaxies appears to overlap in the moment map, but are separated by slight differences in their radial velocities. The spectrum of the group also implies that the catalogued redshift for HCG~59d is offset from its \hi \ velocity by over 200~\kms. A 5th \hi \ detection occurs just south of the core group. There does not appear to by any optical counterpart to this source (which we label HCG~59LSB1) and it is likely spurious.

The low quality of the data makes confident separation of features challenging and caution is advised when interpreting results of this group. However, while in the cases of HCG~59c and d the \hi \ emission appears to be well centred, for HCG~59a most of the \hi \ emission occurs between HCG~59a and b. The emission associated with HCG~59b also appears to have been broken into approaching and receding parts by \texttt{SoFiA}. We therefore attribute, as best as possible, emission associated with HCG~59b and label the remainer as an extended feature rather than gas bound to HCG~59a. This feature is likely an \hi \ bridge between the two galaxies.

\subsection{HCG~61}

\begin{figure}[h]
    \centering
    \includegraphics[width=\columnwidth]{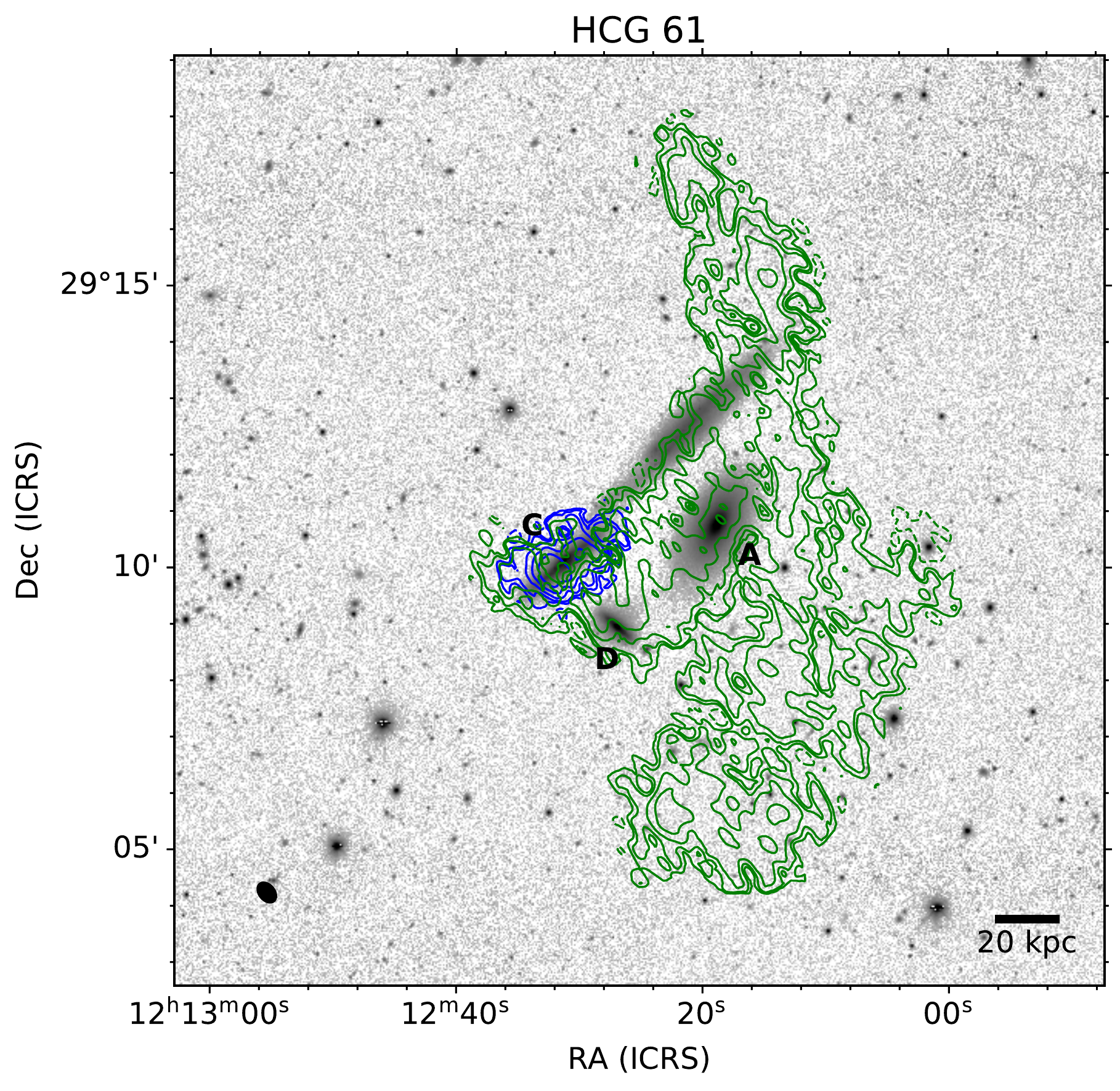}
    \caption{As in Figure \ref{fig:HCG2_split_overlay}.}
    \label{fig:HCG61_split_overlay}
\end{figure}

\begin{figure}[h]
    \centering
    \includegraphics[width=\columnwidth]{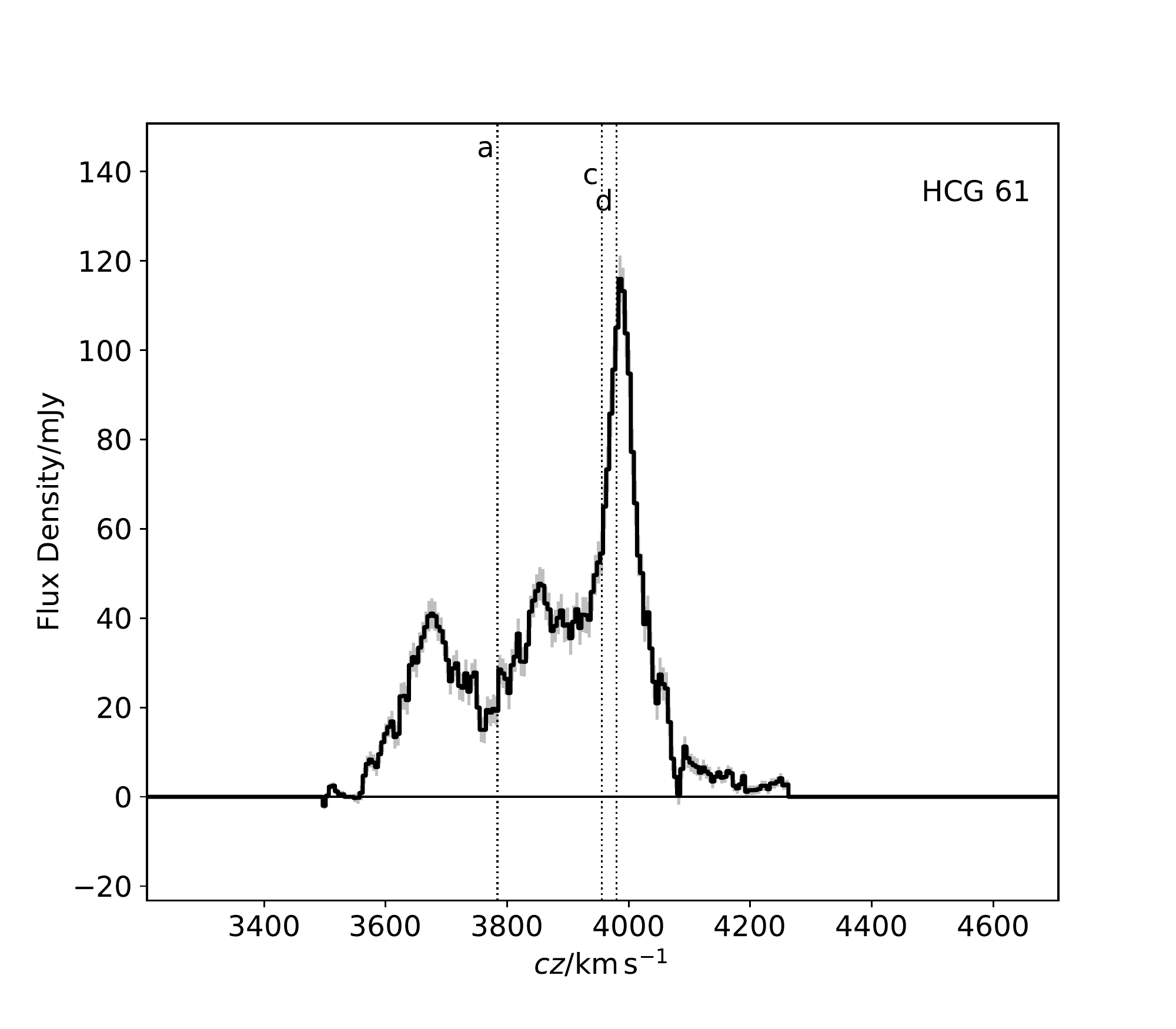}
    \caption{As in Figure \ref{fig:HCG2_spec}.}
    \label{fig:HCG61_spec}
\end{figure}

HCG~61 is a triplet made up of HCG~61a, c and d (HCG~61b is a foreground galaxy). HCG 61a and d are both classified as lenticular and HCG~61c is late-type, but edge-on. The \hi \ map from the VLA shows several enormous extended features reminiscent of HCG~92. The moment zero map appears mottled due to significant side lobes to the synthesised beam.

The most easily discernible galaxy in the \hi \ cube is HCG~61c as a faint, point-like emission region progresses along its edge-on disc when stepping through the channels of the cube in a relatively uniform manner. However, near its central velocity the signal almost entirely disappears and there are several apparent tidal features which we separate as best as possible. Aside from this gas, none of the remainder of the \hi \ emission is definitively associated with either of the other galaxies. Although some of this emission does coincide with HCG~61a and d, and there is some velocity gradient across each, both are embedded in the large extended features and it is unlikely that much of this gas is still bound to the galaxies. Therefore, we classify all remain emission in the group as tidal in nature (Figures \ref{fig:HCG61_split_overlay} \& \ref{fig:HCG61_spec}).

In addition to the \hi \ in the group we also detect PGC~4316478 about 220~kpc to the SE of the group. This source is at the very edge of the VLA primary beam and its flux is likely unreliable. We do not consider it a member of the group.
 
\subsection{HCG~62}

\begin{figure}[h]
    \centering
    \includegraphics[width=\columnwidth]{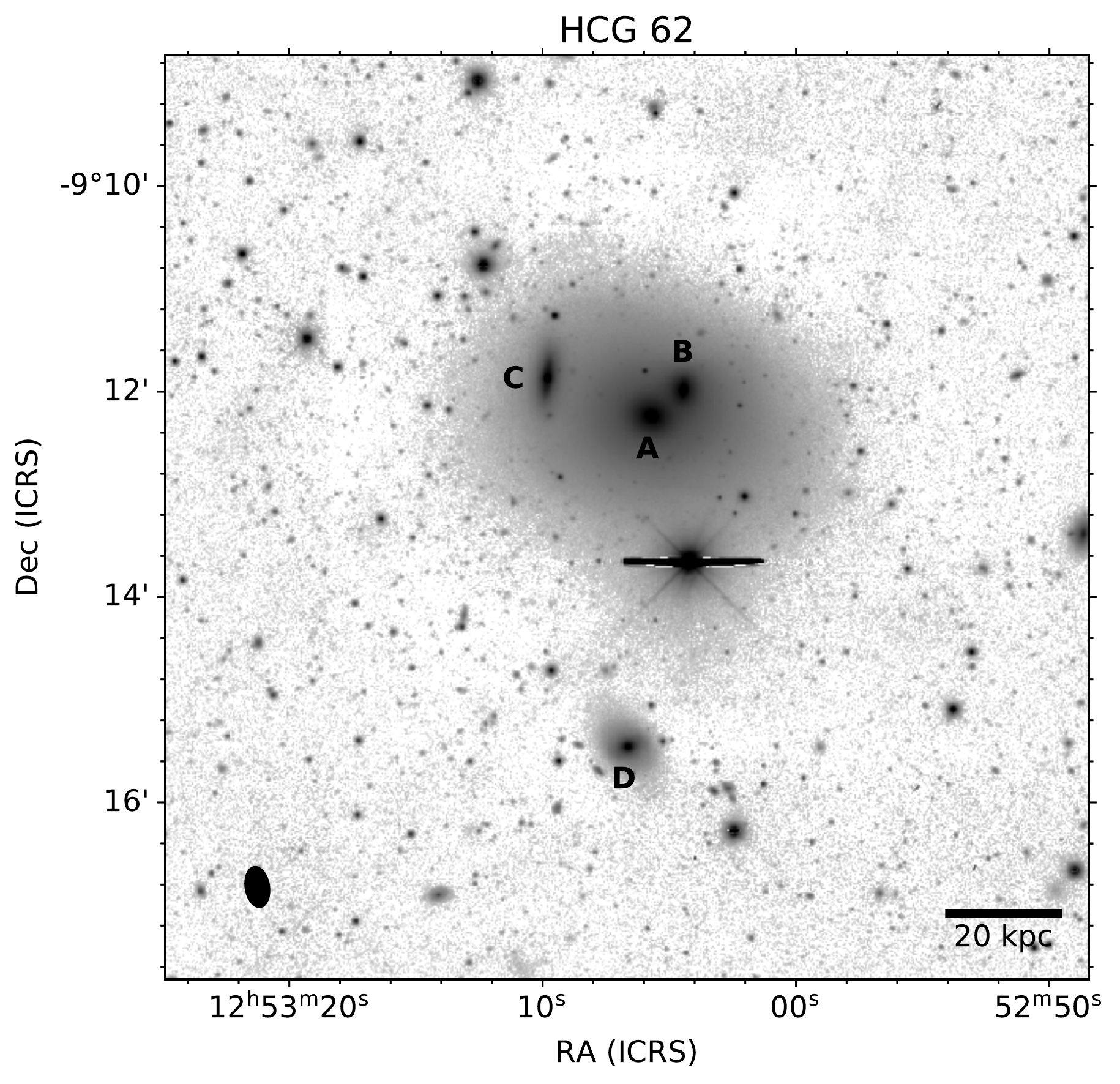}
    \caption{As in Figure \ref{fig:HCG2_split_overlay}, except no \hi \ emission is detected within the group.}
    \label{fig:HCG62_overlay}
\end{figure}

\begin{figure}[h]
    \centering
    \includegraphics[width=\columnwidth]{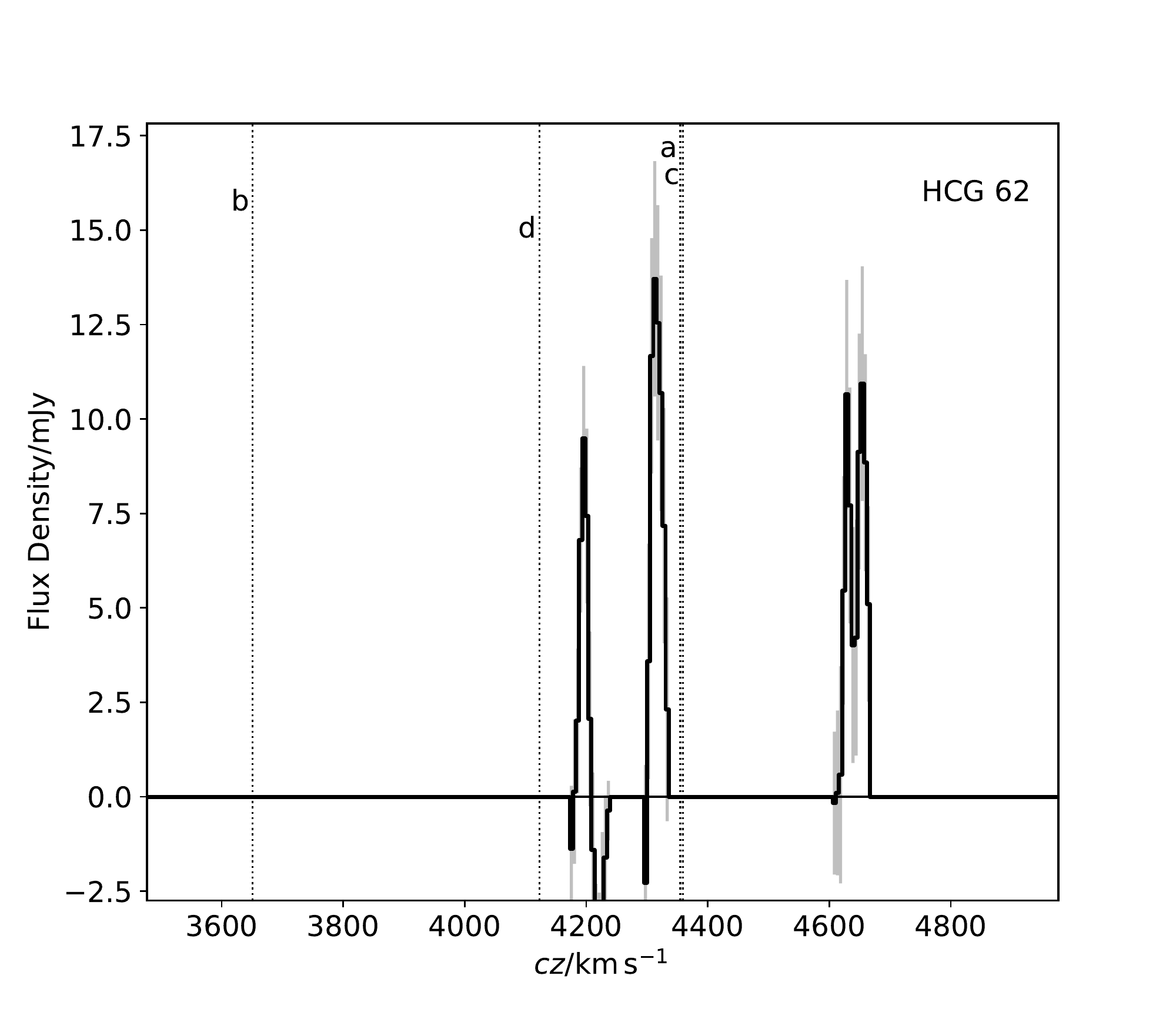}
    \caption{As in Figure \ref{fig:HCG2_spec}.}
    \label{fig:HCG62_spec}
\end{figure}

HCG~62 is a group of two lenticular and two elliptical galaxies in the range 3600-4400~\kms. None are detected in the JVLA observations of this group (Figures \ref{fig:HCG62_overlay} \& \ref{fig:HCG62_spec}). Three low significance, outlying clumps exist in the \hi \ spectral line cube. However, these have no apparent optical counterparts and are almost certainly spurious. Therefore, we consider this group as entirely undetected with the JVLA observations.

\subsection{HCG~68}

\begin{figure}[h]
    \centering
    \includegraphics[width=\columnwidth]{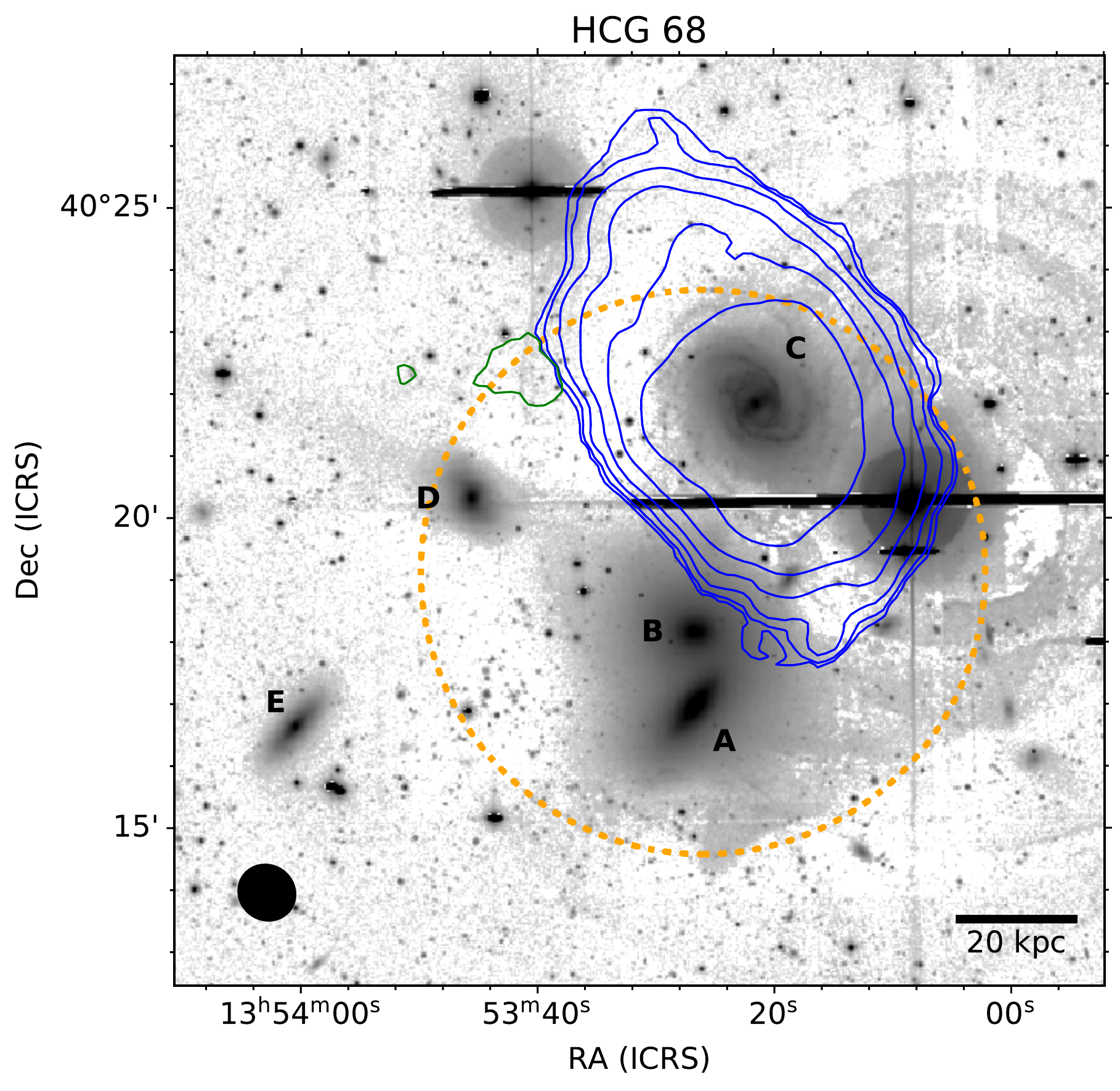}
    \caption{As in Figure \ref{fig:HCG7_split_overlay}.}
    \label{fig:HCG68_split_overlay}
\end{figure}

\begin{figure}[h]
    \centering
    \includegraphics[width=\columnwidth]{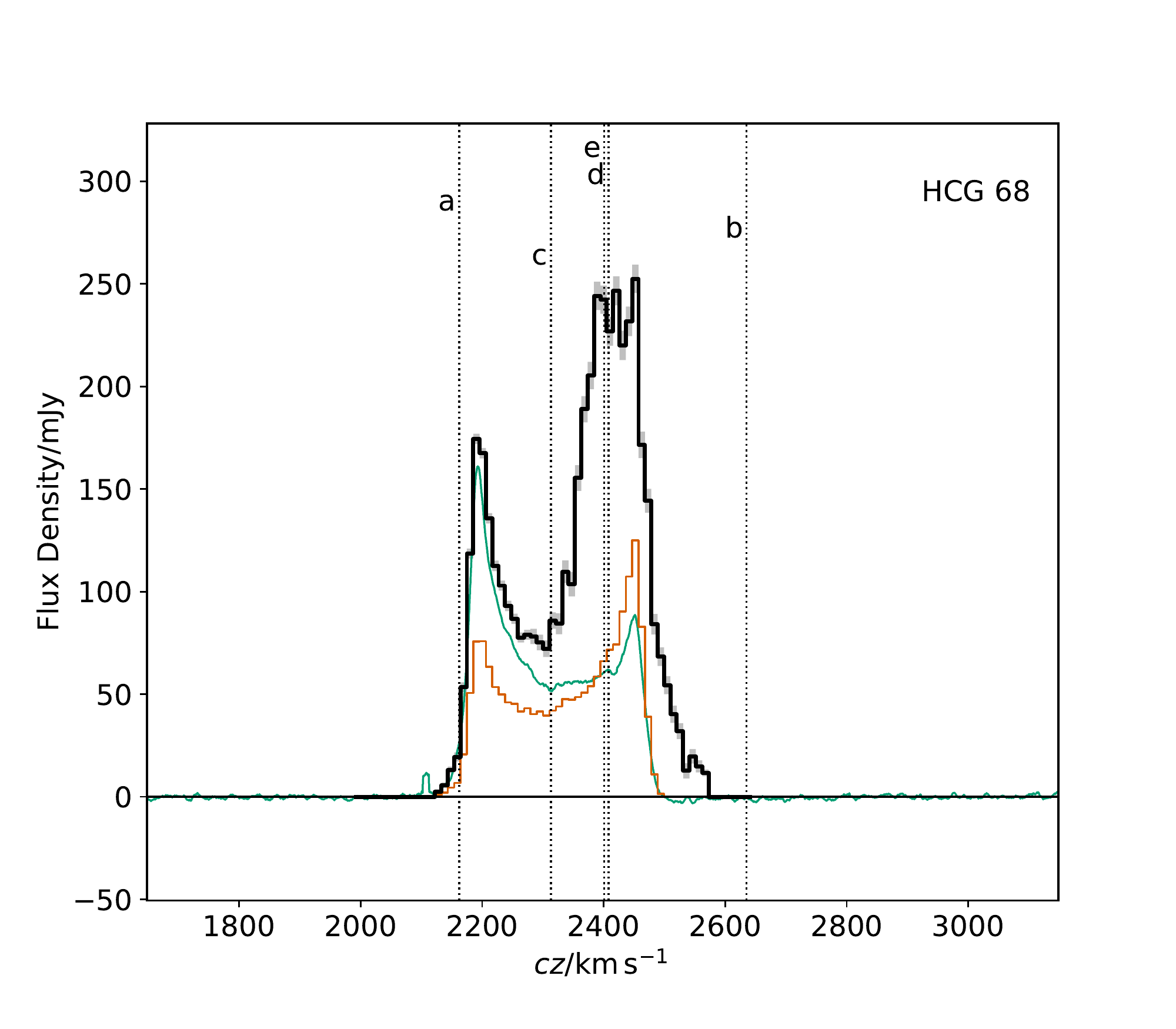}
    \caption{As in Figure \ref{fig:HCG7_spec}.}
    \label{fig:HCG68_spec}
\end{figure}

HCG~68 is a group of five galaxies, two lenticulars, two ellipticals, and one late-type, the only core group galaxy detected in \hi \ with the VLA observations (HCG~68c). The velocity structure of the \hi \ detected in HCG~68c is highly regular, with only a very minor, faint extension on the NW side which we separate as a tidal feature (but may simply be spurious), and a possible slight warp in its outer disc (Figures \ref{fig:HCG68_split_overlay} \& \ref{fig:HCG68_spec}).

In addition to HCG~68c, UGC~8841 is clearly detected in \hi \ about 150~kpc to the SW of the core group, and NGC~5371 is also detected a similar projected distance to the west of the group. The receding side of NGC~5371 is not visible as this is truncated by the edge of the bandpass of the VLA observation. We do not consider either of these galaxies as members of the core group.

The comparison of the VLA and GBT sprecta in Figure \ref{fig:HCG68_spec} shows a strange phenomenon where at low velocities the GBT spectrum (green line) contains more flux, but at high velocities the VLA weighted spectrum does (orange line). As with HCG~58 this is likely the result of the main source of \hi \ emission lying near the edge of the GBT beam. Even a small offset in pointing or a slight bias in the beam weighting we have used could result in more emission for the receding or approaching side of HCG~68c being included.

\subsection{HCG~71}

\begin{figure}[h]
    \centering
    \includegraphics[width=\columnwidth]{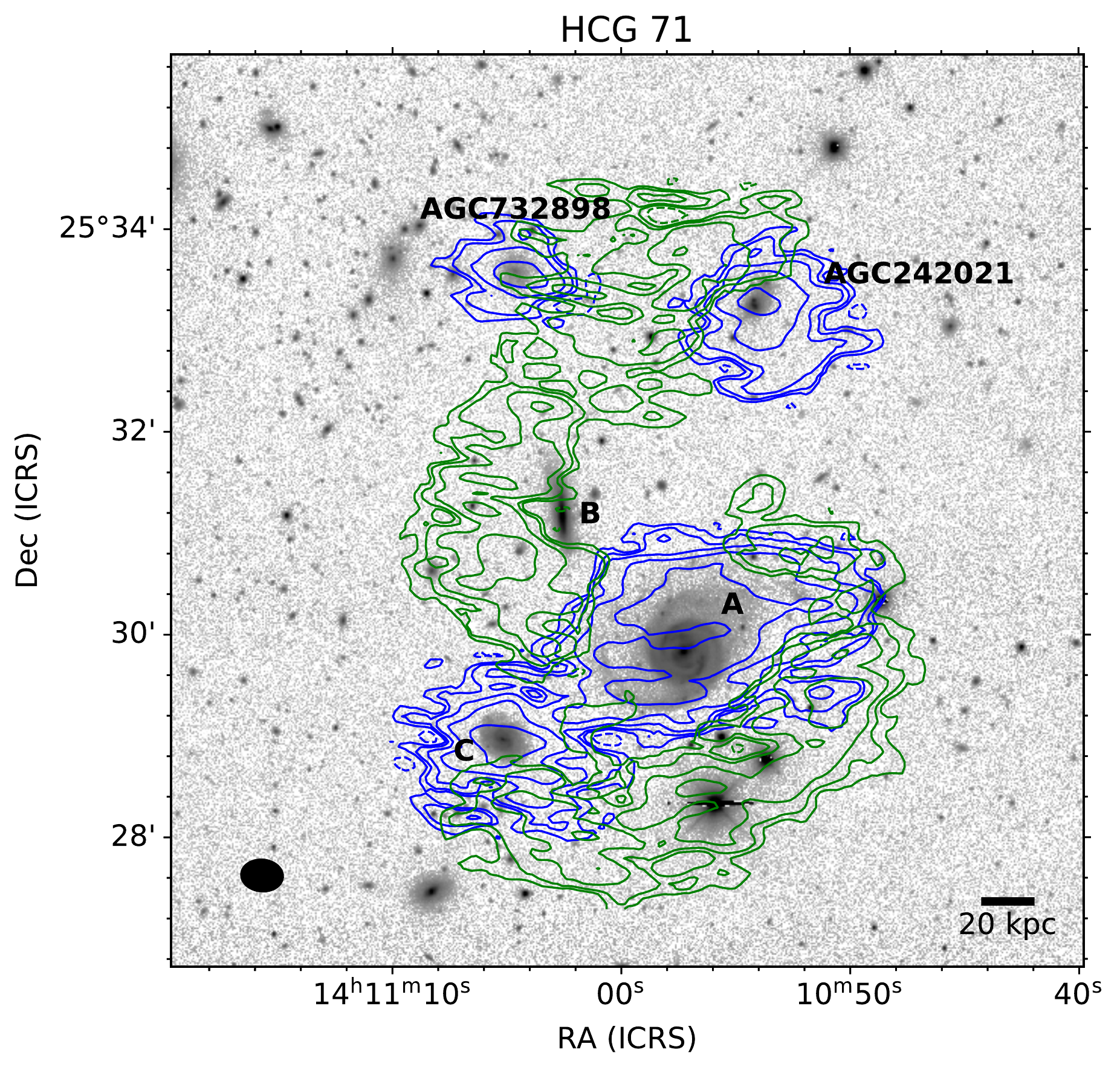}
    \caption{As in Figure \ref{fig:HCG2_split_overlay}.}
    \label{fig:HCG71_split_overlay}
\end{figure}

\begin{figure}[h]
    \centering
    \includegraphics[width=\columnwidth]{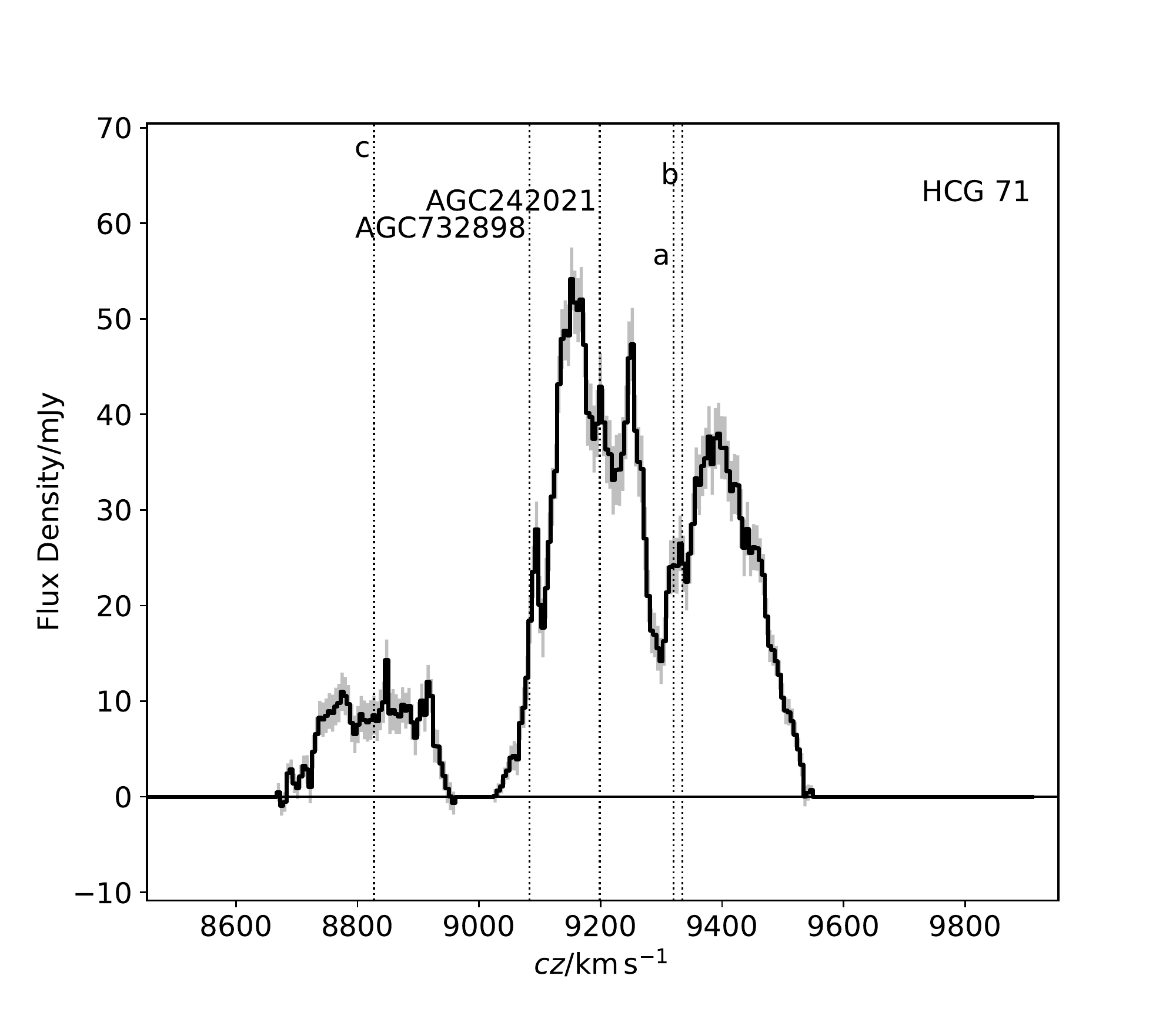}
    \caption{As in Figure \ref{fig:HCG2_spec}.}
    \label{fig:HCG71_spec}
\end{figure}

HCG~71 is a triangular configuration made up of the three late-type galaxies HCG~71a, b, and c (HCG~71d is a distant background galaxy) in the range 8800-9400~\kms. The \hi \ moment map is dominated by emission associated with HCG~71a, a near face-on spiral (Figures \ref{fig:HCG71_split_overlay} \& \ref{fig:HCG71_spec}). HCG~71c is also detected, but HCG 71b is not. Thes emission from HCG~71a stretches out to form a large loop to the north that connects to two LSB galaxies, AGC~732898 and AGC~242021, both of which were detected in \hi \ in the ALFALFA survey (though blended with other emission in the group) and which we consider group members. The moment zero map of the group has a slightly mottled appearance due to significant side lobes and elongation of the synthesised beam, and possible slight residual continuum artefacts.

HCG~71c is slightly separated from the rest of the group in velocity and thus it is straightforward to distinguish its emission. It appears to have two minor (low S/N) extensions to the south and west, which we excise in separate features. HCG~71a has two large tails emanating from the east and west. The latter stretches down to the SE and in the moment map apparently connects to HCG~71c. However, this is a projection effect as in velocity space it stretches away from, not towards, HCG~71c. There is a dense clump of \hi \ emission in the other tail that is at the same velocity as HCG~71b, but it does not quite overlap on the plane of the sky and therefore does not appear to be associated (although perhaps some of this gas originated from HCG~71b). Again, in projection, this tails appears to connect to AGC~732898 and AGC~242021, but veers away in velocity. Although only detected in a small number of channels AGC~732898 appears to have a regular structure and no extended features were identified around it. AGC~242021, on the other hand, appears highly disturbed in \hi. We separate the emission into galaxy and extended features, but note that this is quite uncertain for this object.

\subsection{HCG~79}

\begin{figure}[h]
    \centering
    \includegraphics[width=\columnwidth]{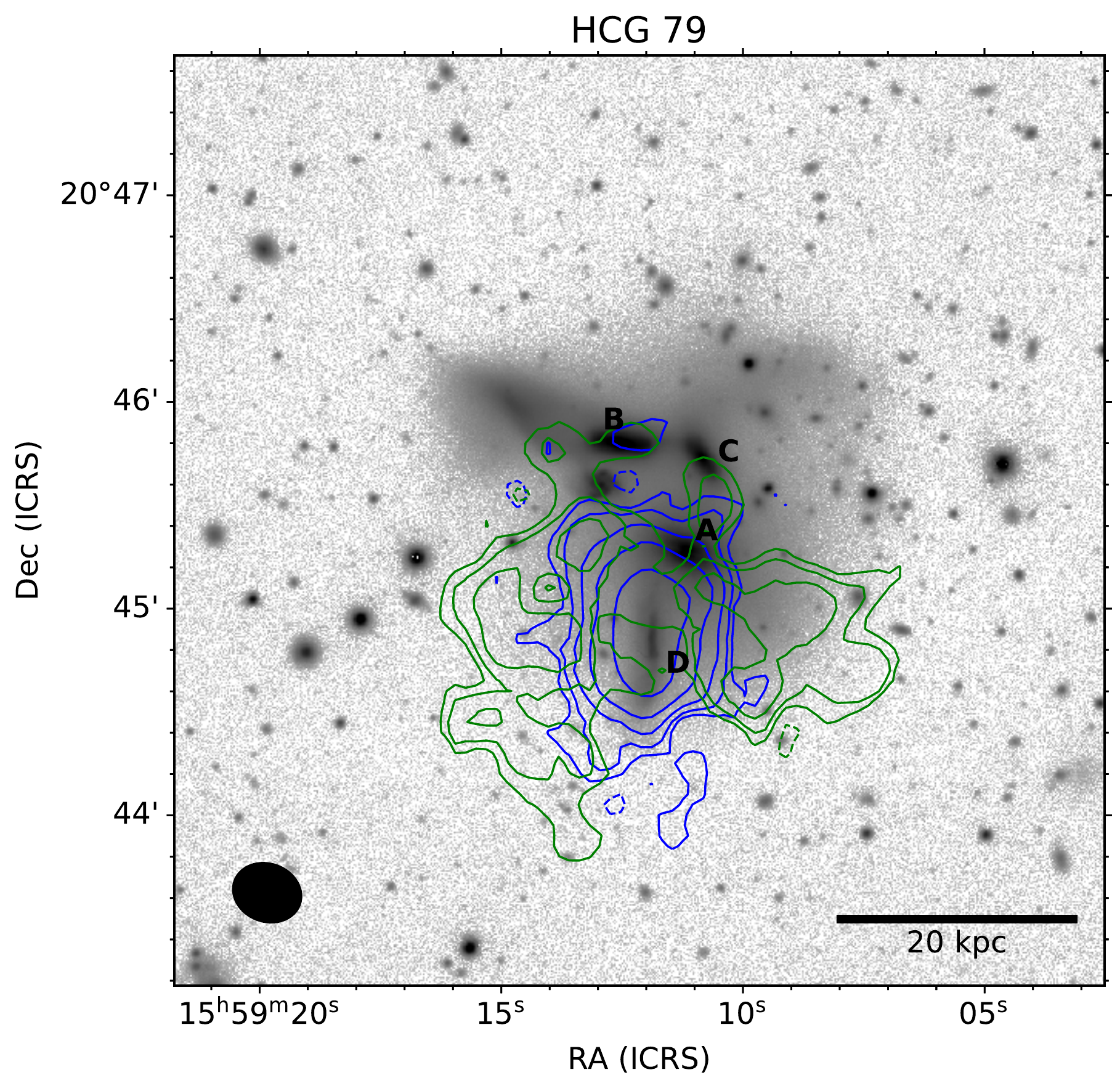}
    \caption{As in Figure \ref{fig:HCG2_split_overlay}.}
    \label{fig:HCG79_split_overlay}
\end{figure}

\begin{figure}[h]
    \centering
    \includegraphics[width=\columnwidth]{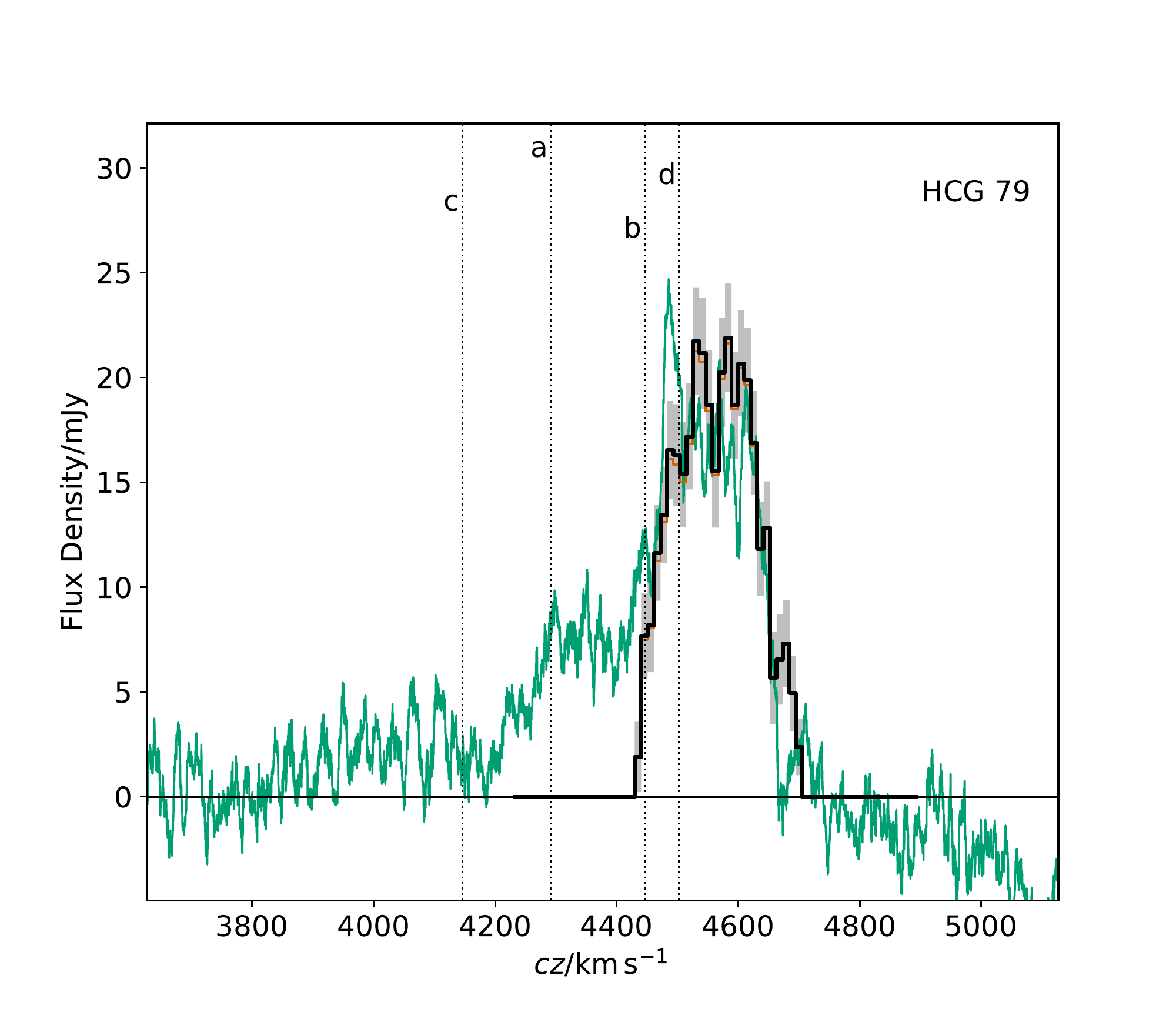}
    \caption{As in Figure \ref{fig:HCG2_spec}.}
    \label{fig:HCG79_spec}
\end{figure}

HCG~79 is a group of two lenticular, one elliptical, and one late-type galaxy, all of which overlap on the plane of the sky (a 5th galaxy, HCG~79e, is a background interloper) and have redshifts in the range 4100-4500~\kms. Most of the \hi \ emission detected with the VLA is centred on HCG~79d, the sole late-type galaxy in the group (Figures \ref{fig:HCG79_split_overlay} \& \ref{fig:HCG79_spec}). There are numerous faint features on the edge of HCG~79d that appear disturbed and are separated out. However, it should be noted that these are faint features that are subject to variations in the noise.

\subsection{HCG~88}

\begin{figure}[h]
    \centering
    \includegraphics[width=\columnwidth]{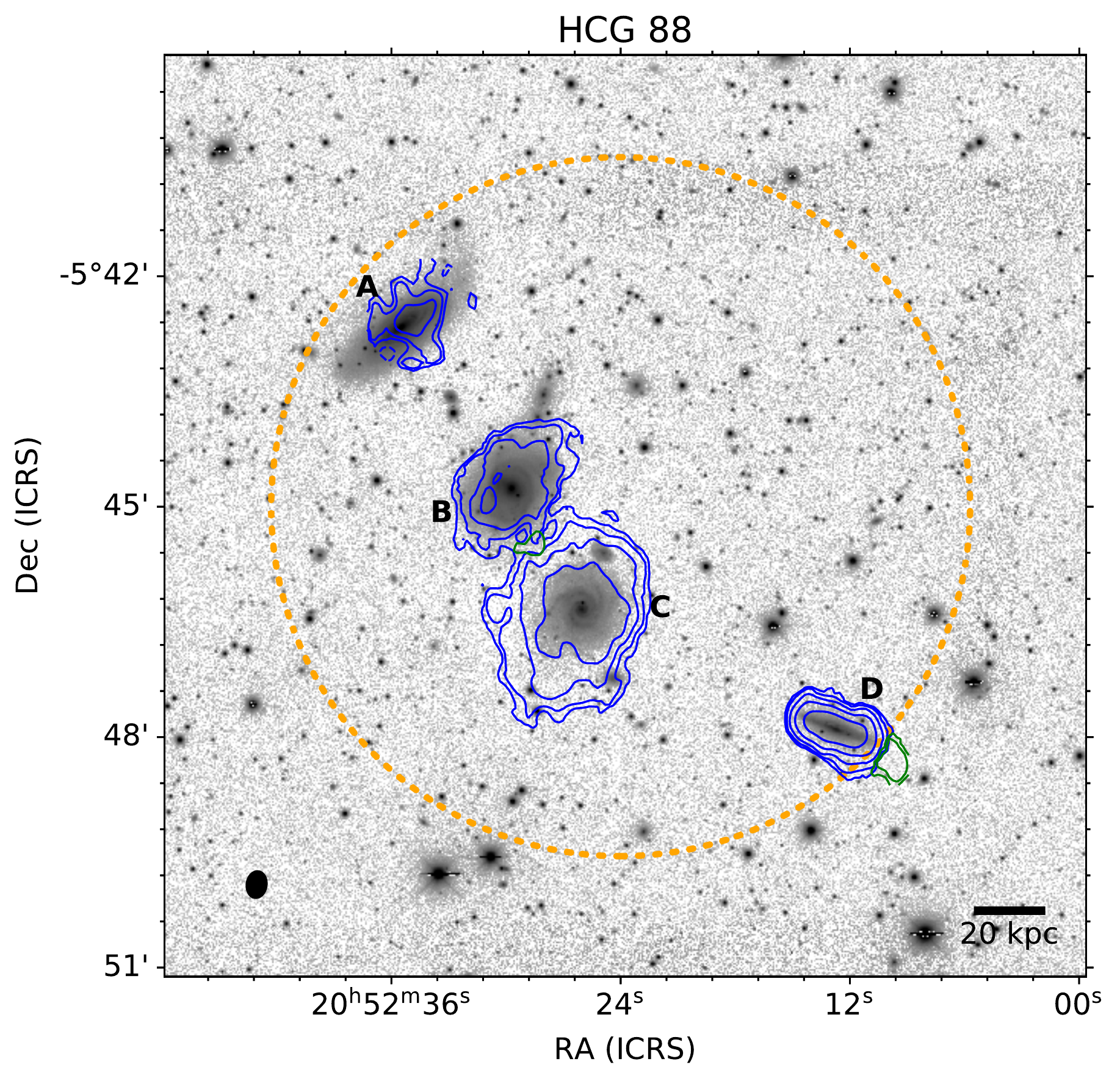}
    \caption{As in Figure \ref{fig:HCG7_split_overlay}.}
    \label{fig:HCG88_split_overlay}
\end{figure}

\begin{figure}[h]
    \centering
    \includegraphics[width=\columnwidth]{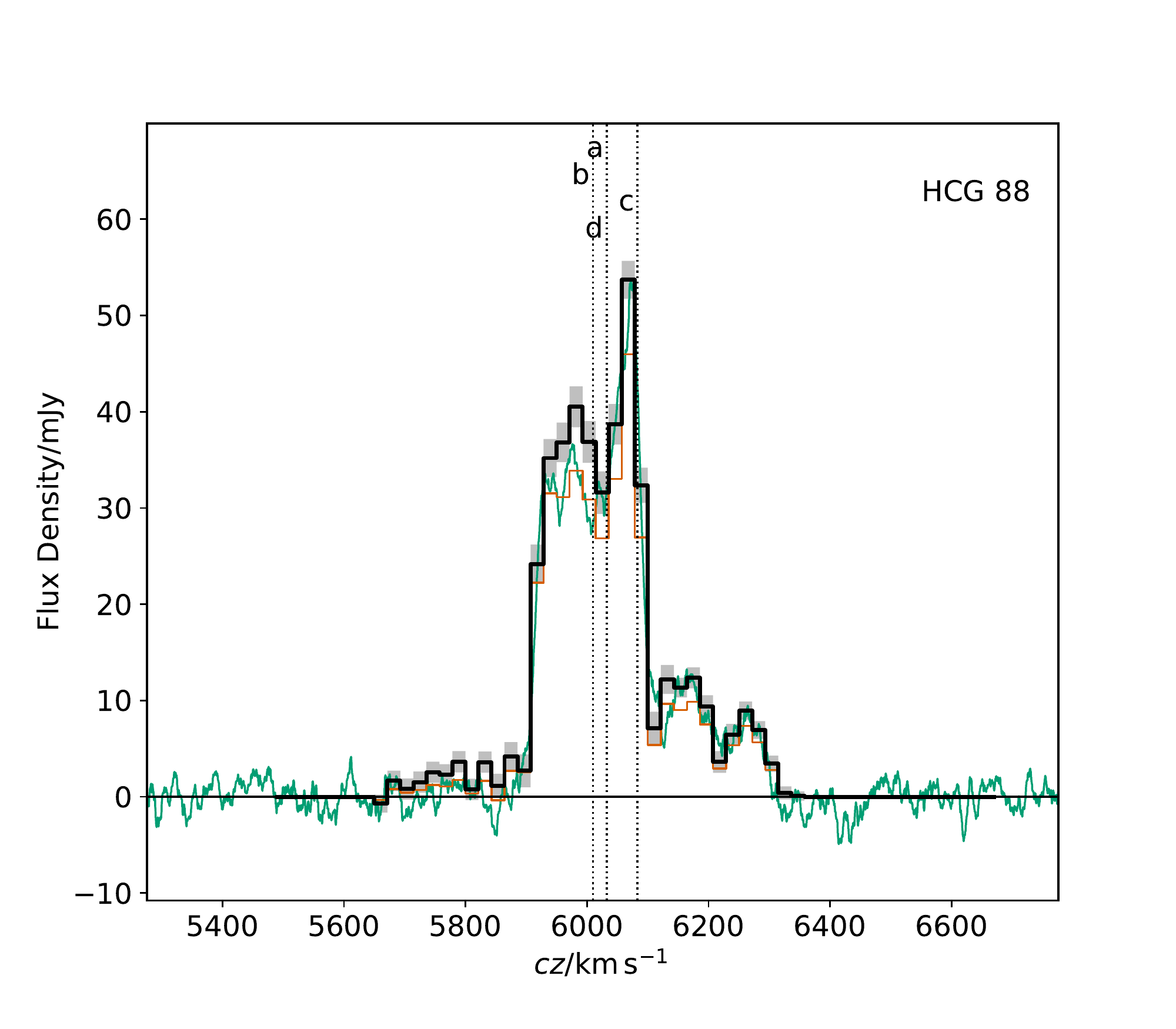}
    \caption{As in Figure \ref{fig:HCG2_spec}.}
    \label{fig:HCG88_spec}
\end{figure}

HCG~88 is a linear arrangement of four late-type galaxies all with redshifts of $\sim$6000~\kms. All four are detected in \hi \ with the VLA, and have appear mostly undisturbed, except for HCG 88a which appears to have a truncated \hi \ disc (Figures \ref{fig:HCG88_split_overlay} \& \ref{fig:HCG88_spec}). HCG~88b and c have overlapping \hi \ distributions. The regular velocity structure of the two galaxies implies that this \hi \ is still mostly bound to their respective discs, and therefore we separate the two galaxies as best as possible and assign very little emission to a bridge feature. HCG~88d also appears to have an incipient tidal tail on its SE side.

\subsection{HCG~90}

\begin{figure}[h]
    \centering
    \includegraphics[width=\columnwidth]{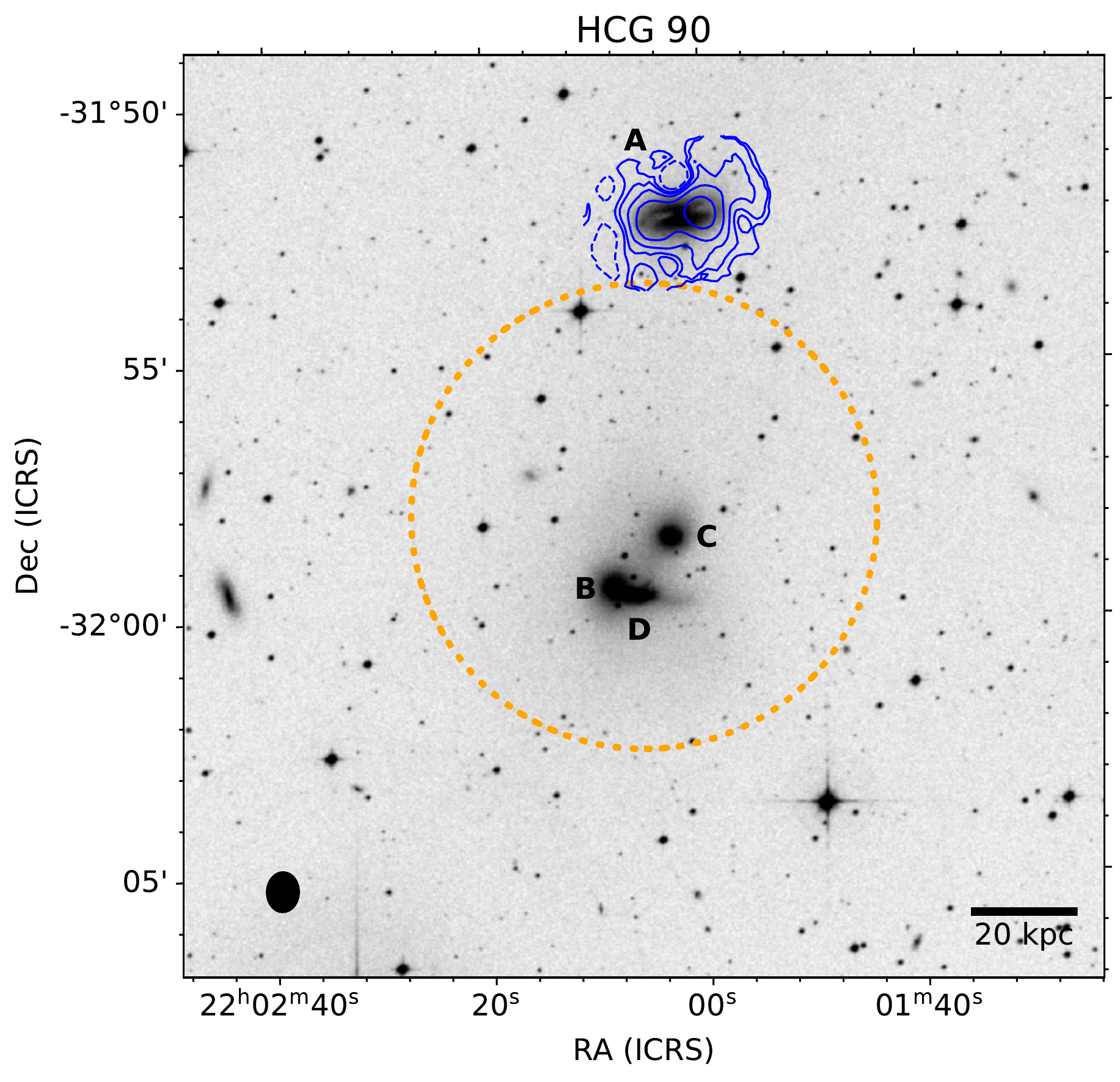}
    \caption{As in Figure \ref{fig:HCG7_split_overlay}, except the background image is POSS $R$-band.}
    \label{fig:HCG90_split_overlay}
\end{figure}

\begin{figure}[h]
    \centering
    \includegraphics[width=\columnwidth]{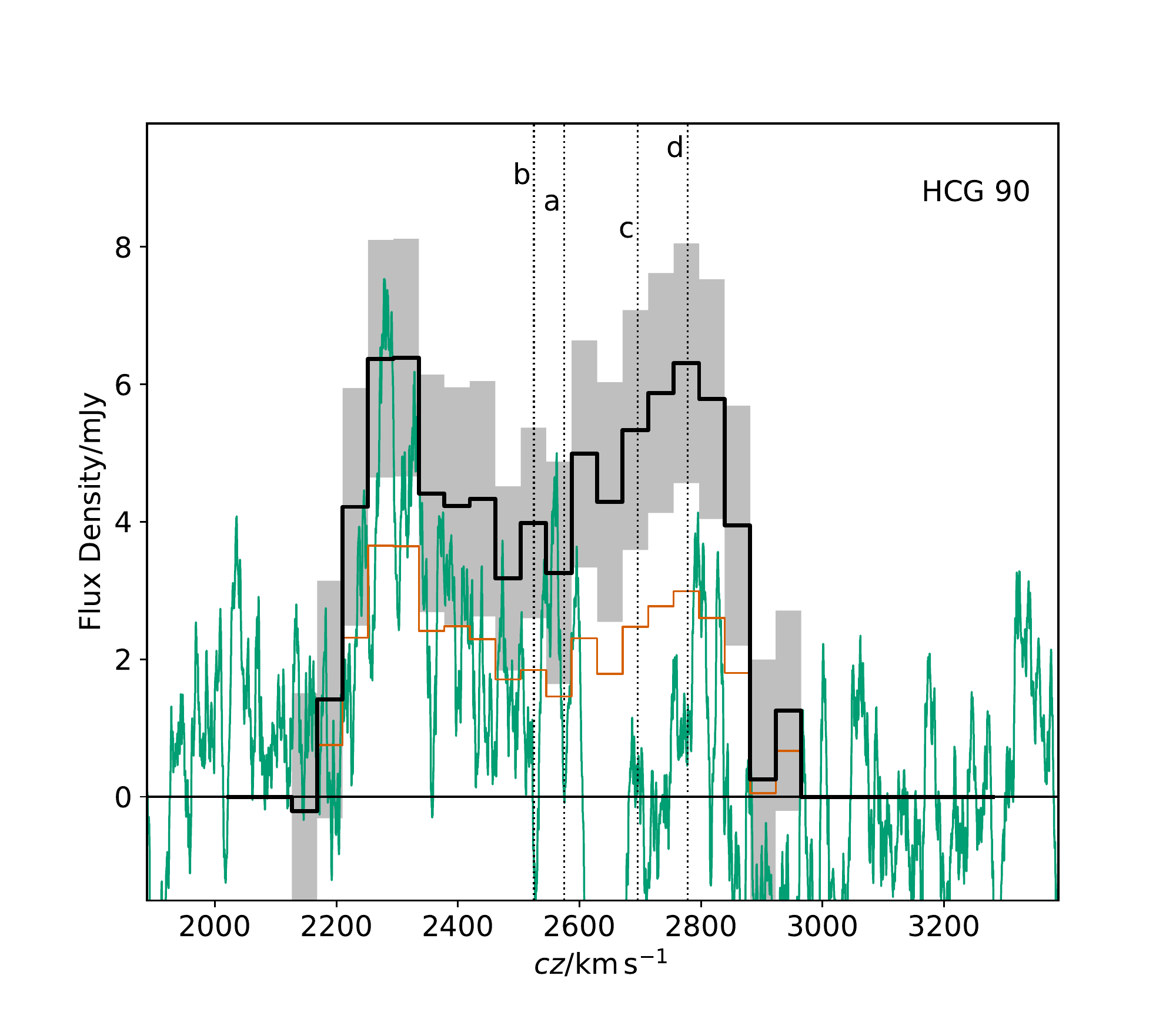}
    \caption{As in Figure \ref{fig:HCG7_split_overlay}.}
    \label{fig:HCG90_spec}
\end{figure}

HCG~90 is an extremely compact configuration of three early-type galaxies and a fourth late-type galaxy about 60~kpc to the north. The VLA observations of the group used a very broad channel width (195 kHz) in order to span the full velocity range of the group and beyond. However, the broad channels cause \texttt{SoFiA}'s reliability verification algorithm to fail, as too few sources (spurious plus real) are detected. Therefore, we raised the masking threshold to 5$\sigma$ in order to eliminate any spurious detections. 

We note that HCG~90 appears to be embedded at the centre of a much larger structure containing many tens of galaxies. There are several dwarf galaxies in the vicinity of the core group, some of which could be members \citep{deCarvalho+1997}, but which we had chosen to disregard, and instead we focus on the core group, as we have done with other HCGs. With the exception of PGC~198500, which is marginally detected (but excluded due to the higher threshold of the \texttt{SoFiA} mask), these additional members are all undetected in \hi \ and would minimally impact the estimated \hi \ deficiency of the group, owing to their low mass relative to the core members. 

\texttt{SoFiA} splits the emission from HCG~90a (the only core member detected) into two halves, which we combine (Figures \ref{fig:HCG90_split_overlay} \& \ref{fig:HCG90_spec}). No other modification is made to the mask.

\subsection{HCG~91}

\begin{figure}[h]
    \centering
    \includegraphics[width=\columnwidth]{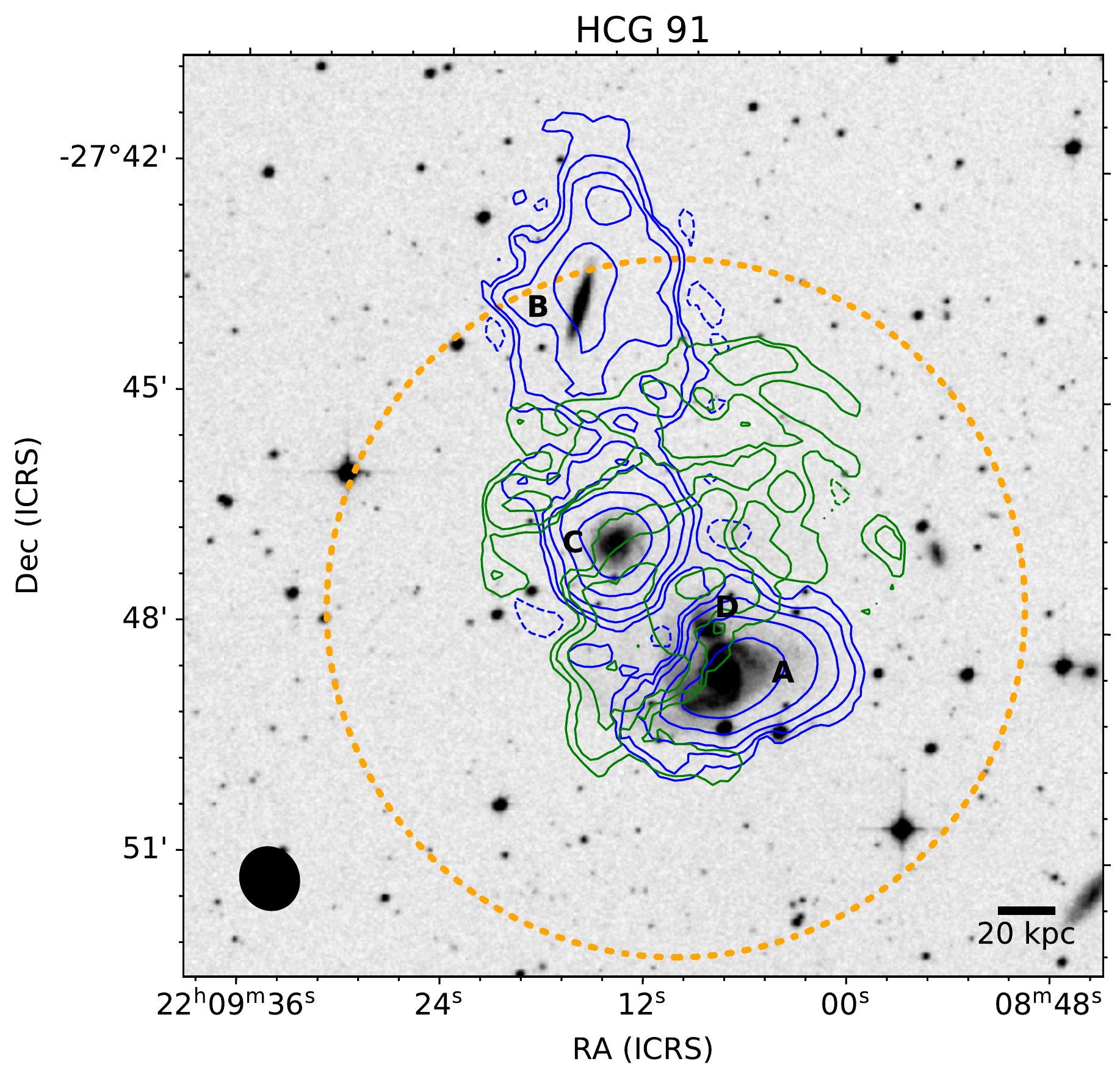}
    \caption{As in Figure \ref{fig:HCG7_split_overlay}, except the background image is POSS $R$-band.}
    \label{fig:HCG91_split_overlay}
\end{figure}

\begin{figure}[h]
    \centering
    \includegraphics[width=\columnwidth]{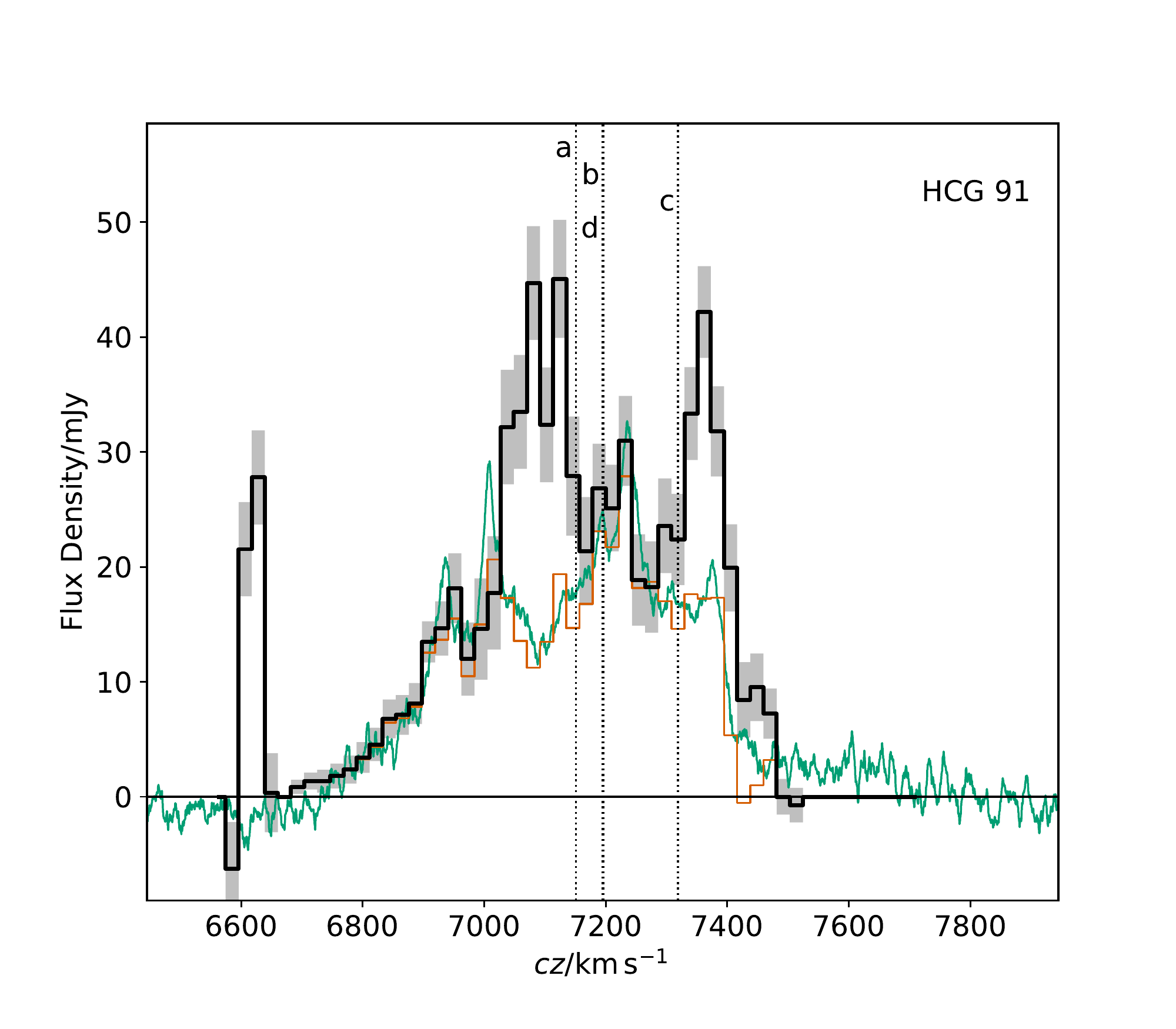}
    \caption{As in Figure \ref{fig:HCG7_spec}.}
    \label{fig:HCG91_spec}
\end{figure}

HCG~91 consists for four disc galaxies (in the redshift range 7100-7300~\kms) in an almost linear N-S configuration. The \hi \ distribution of this group has been studied in detail before by \citet{Vogt+2015}. Our analysis is based on a re-reduction of the same data. HCG~91a, b and c and all detected in \hi, but HCG~91d is not (Figures \ref{fig:HCG91_split_overlay} \& \ref{fig:HCG91_spec}). The main tidal feature in the group is a tail originating on the eastern side of HCG~91a and wrapping around to its north side. HCG~91c has a bright and narrow \hi \ distribution, whereas HCG~91b is quite broad (in velocity) reducing its S/N in each channel. There also appears to be a low S/N extended feature between HCG~91b and c, which we separate out as best as possible. Finally, there are three additional detections in the \hi \ separated from the core group by over 300~kpc. These are PGC~68187, PGC~68161, and a likely spurious detection to the SE for which we could not identify and optical counterpart. 

\subsection{HCG~92}

\begin{figure}[h]
    \centering
    \includegraphics[width=\columnwidth]{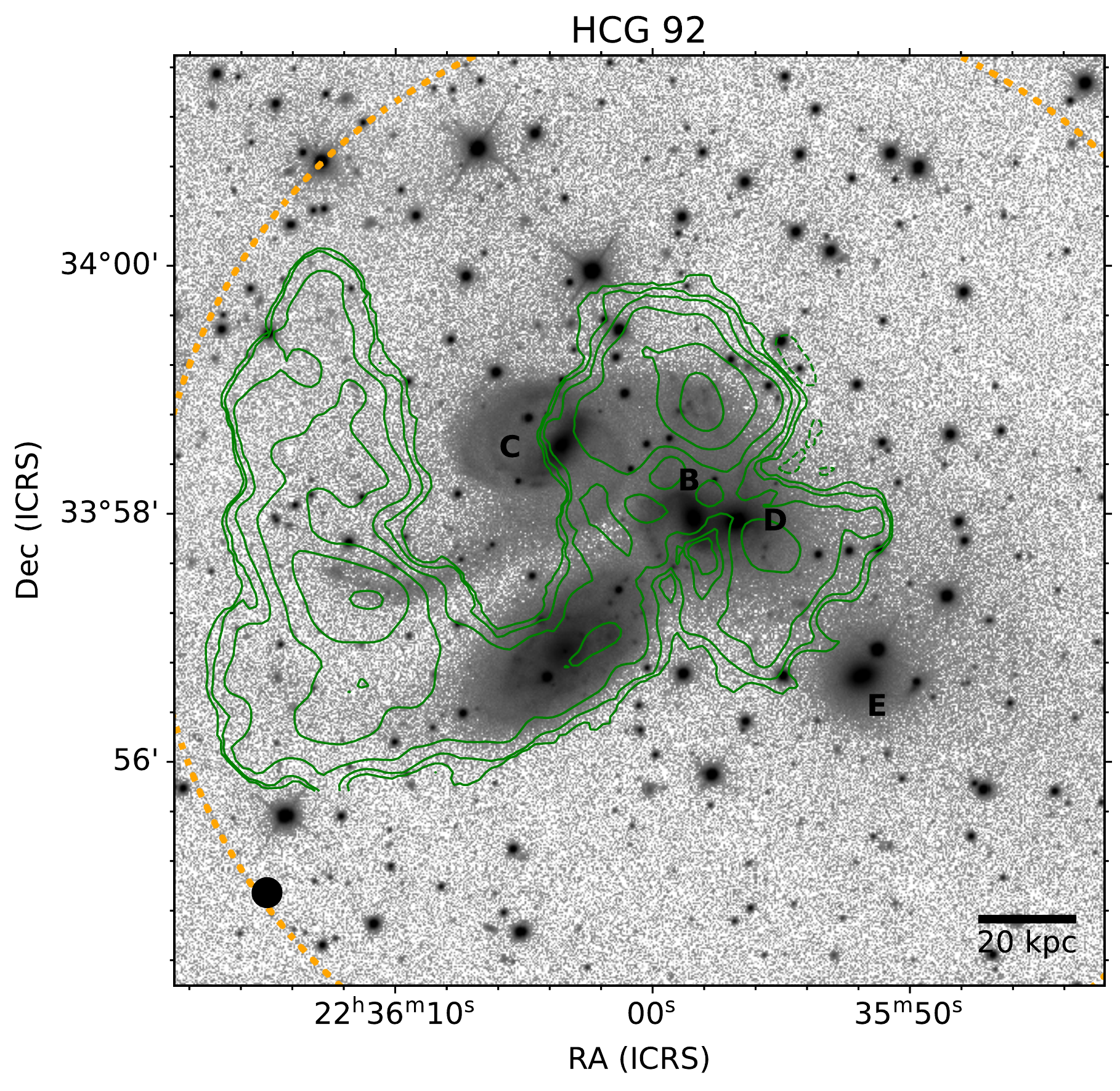}
    \caption{As in Figure \ref{fig:HCG7_split_overlay}.}
    \label{fig:HCG92_split_overlay}
\end{figure}

\begin{figure}[h]
    \centering
    \includegraphics[width=\columnwidth]{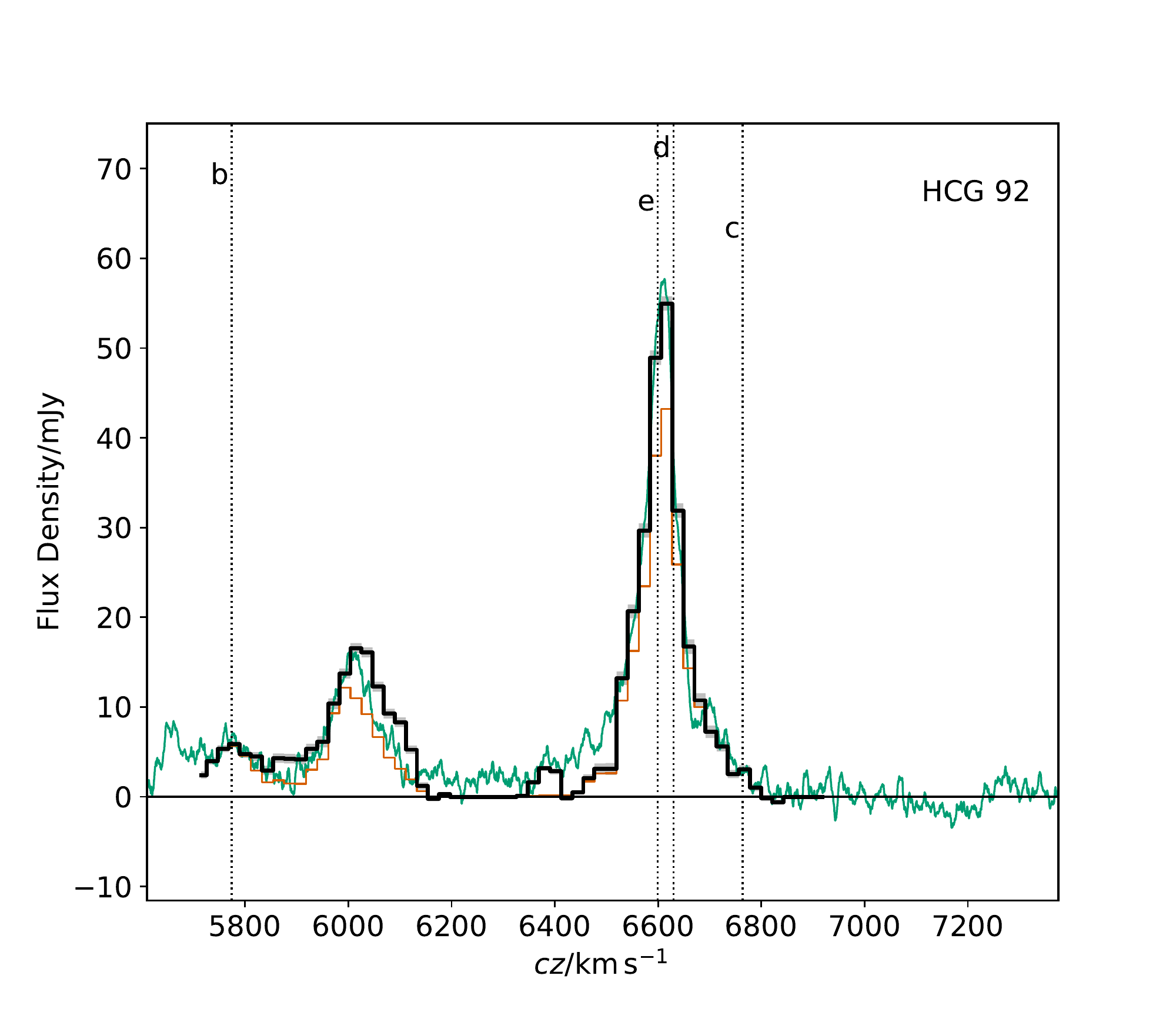}
    \caption{As in Figure \ref{fig:HCG7_spec}.}
    \label{fig:HCG92_spec}
\end{figure}

HCG~92 (or Stephan's Quintet) is an extremely compact configuration of galaxies that exhibits shocks and tidal tails in both neutral and ionised gas. It has been the target of over a hundred papers in IR, optical, UV, and X-ray imaging \citep[e.g.][]{Sulentic+1995,Sulentic+2001,Gallagher+2001,OSullivan+2009,Duc+2018}, radio continuum \citep[e.g.][]{vanderHulst+1981,Blazej+2020}, and line emission \citep[e.g.][]{Yun+1997,Williams+2002,Lisenfeld+2002,Appleton+2006,Guillard+2012,Konstantopoulos+2014,Duarte+2019}. 

The core group is made up of four members, as HCG~92a is a foreground interloper. Although the morphology and kinematics of the \hi \ gas in the group are extremely complex \citep{Williams+2002}, for our purposes it is quite straightforward. As discussed in detail by \citet{Williams+2002}, none of the \hi \ emission is located in the galaxies themselves. The majority is in an enormous L-shaped tail that extends to the east of the group, then there are three separate clouds on the western side of the group (two of which project on top of each other), but these do no coincide with the galaxies (Figures \ref{fig:HCG92_split_overlay} \& \ref{fig:HCG92_spec}).

To the east of the group we also detect AGC~320272 in \hi, however, this dwarf galaxy is $\sim$175~kpc from the core group and appears undisturbed in \hi. Therefore, we do not consider it as a member. In addition, we detect AGC~321245 about 250~kpc north of the core group, and finally there is a likely spurious detection in the \hi \ cube a similar distance to the NW. We also note that a few central channels of the cube were effectively lost due to inadequate continuum subtraction where the two sub-bands of the observation (just) overlap, however, this does not appear to overlap with any of the aforementioned \hi \ features.

\subsection{HCG~93}

\begin{figure}[h]
    \centering
    \includegraphics[width=\columnwidth]{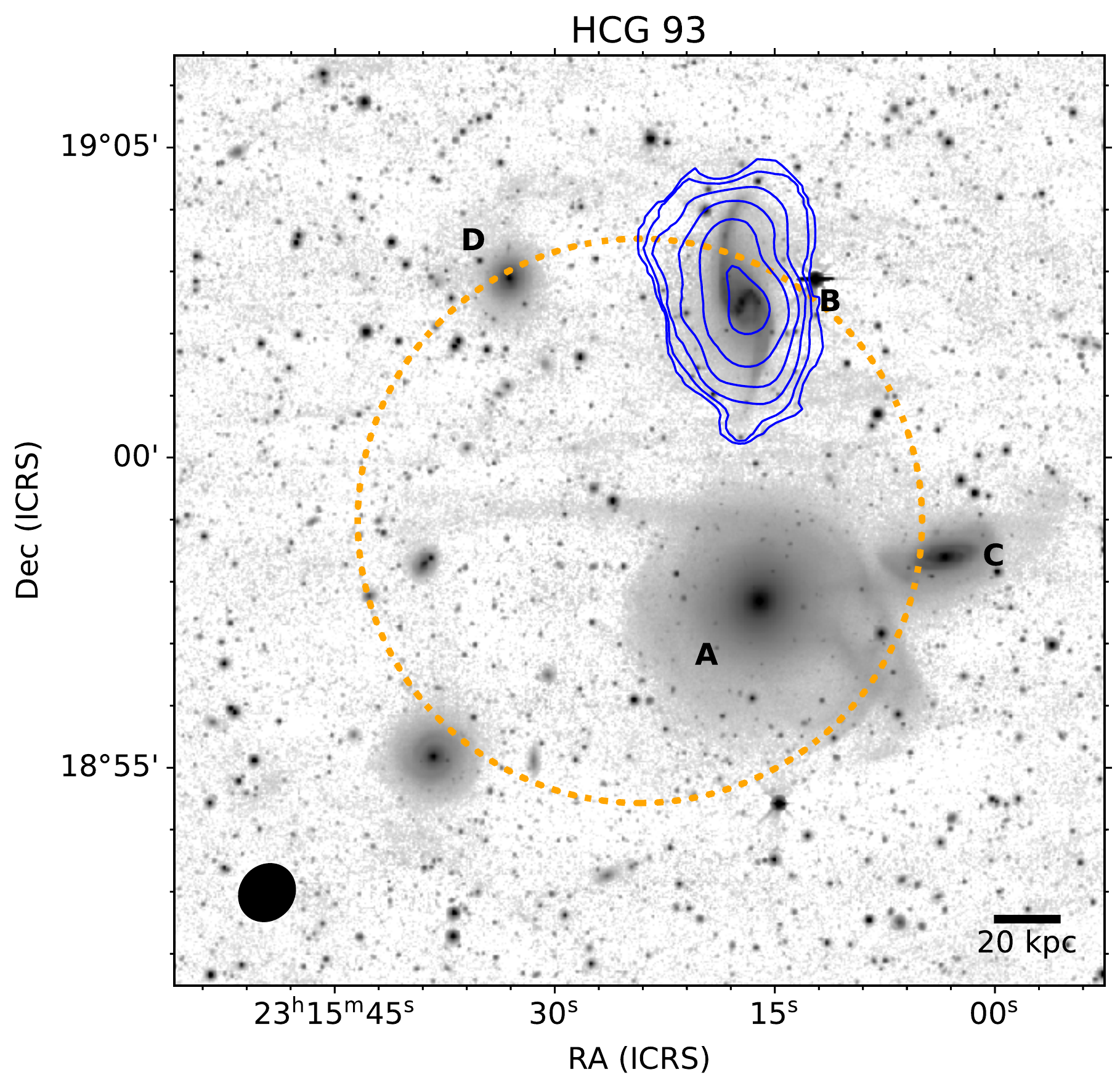}
    \caption{As in Figure \ref{fig:HCG7_split_overlay}.}
    \label{fig:HCG93_split_overlay}
\end{figure}

\begin{figure}[h]
    \centering
    \includegraphics[width=\columnwidth]{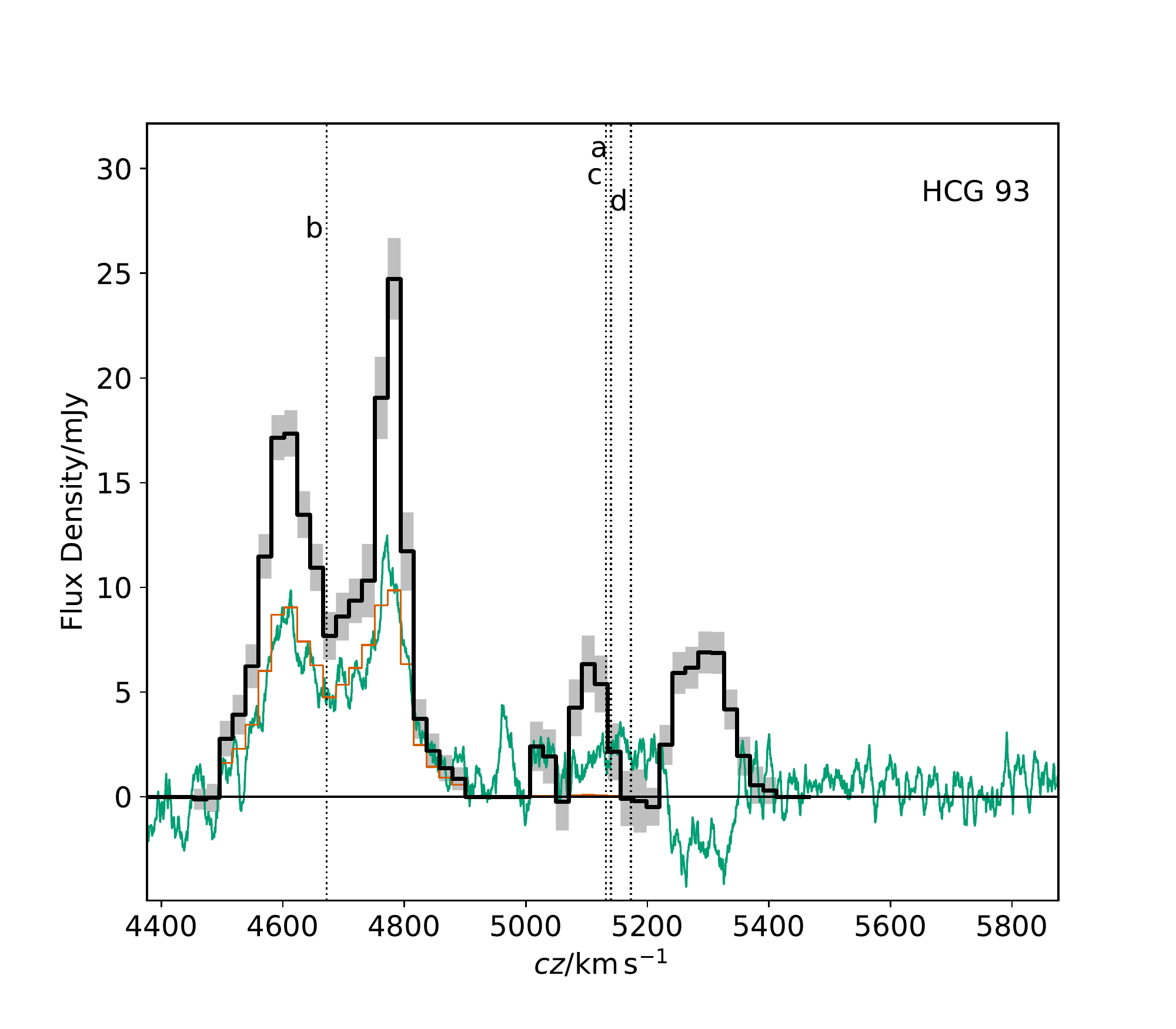}
    \caption{As in Figure \ref{fig:HCG7_spec}.}
    \label{fig:HCG93_spec}
\end{figure}

HCG~93 (projected adjacent to HCG~94, separated by only $\sim$30\arcmin, but by over 6000~\kms) is made up of three late-type and one elliptical galaxy. In the core group only HCG~93b, a loosely wound spiral, is the only galaxy detected in the VLA \hi \ map. In addition, we also detect three uncatalogued dwarf galaxies in the vicinity of the group, however, these are all separated from the core group by $\sim$200~kpc.

The \hi \ of HCG~93b appears quite regular and we do not separate off any tidal features; quite surprising given its optical appearance (Figures \ref{fig:HCG93_split_overlay} \& \ref{fig:HCG93_spec}). This may in part be the result of the fairly low resolution of the VLA data, however, there is certainly no highly extended \hi \ emission as is typical in gas-bearing galaxies experiencing strong tidal forces.

The first two of the three outlying dwarf galaxies (which we refer to as HCG~93LSB1--3) have clear optical counterparts in the DECaLS images. The third may be spurious as it is the lowest S/N of the three and has no apparent optical counterpart (although this may be hidden by its proximity to a bright star).

\subsection{HCG~95}

\begin{figure}[h]
    \centering
    \includegraphics[width=\columnwidth]{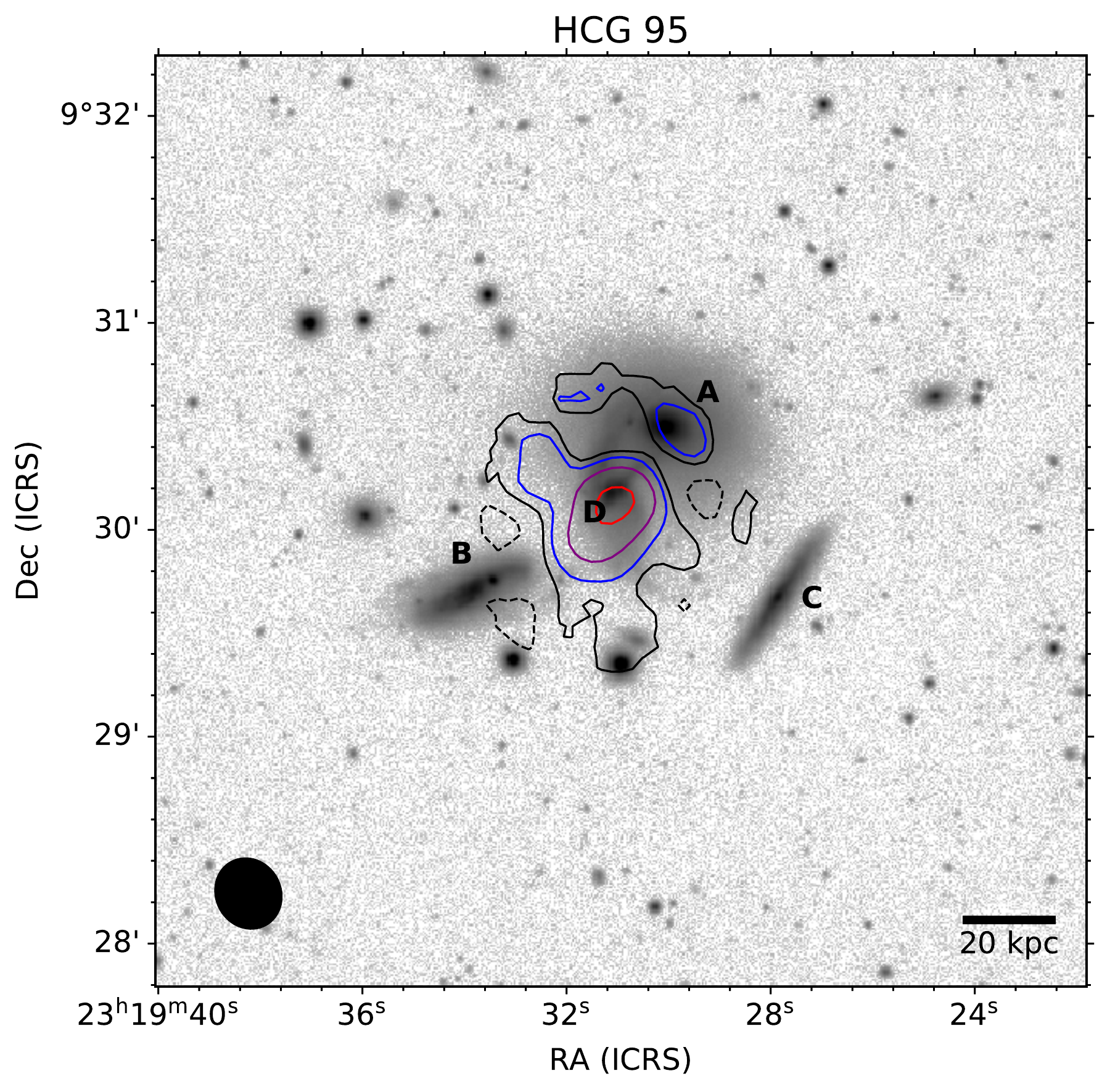}
    \caption{As in Figure \ref{fig:HCG38_overlay}.}
    \label{fig:HCG95_overlay}
\end{figure}

\begin{figure}[h]
    \centering
    \includegraphics[width=\columnwidth]{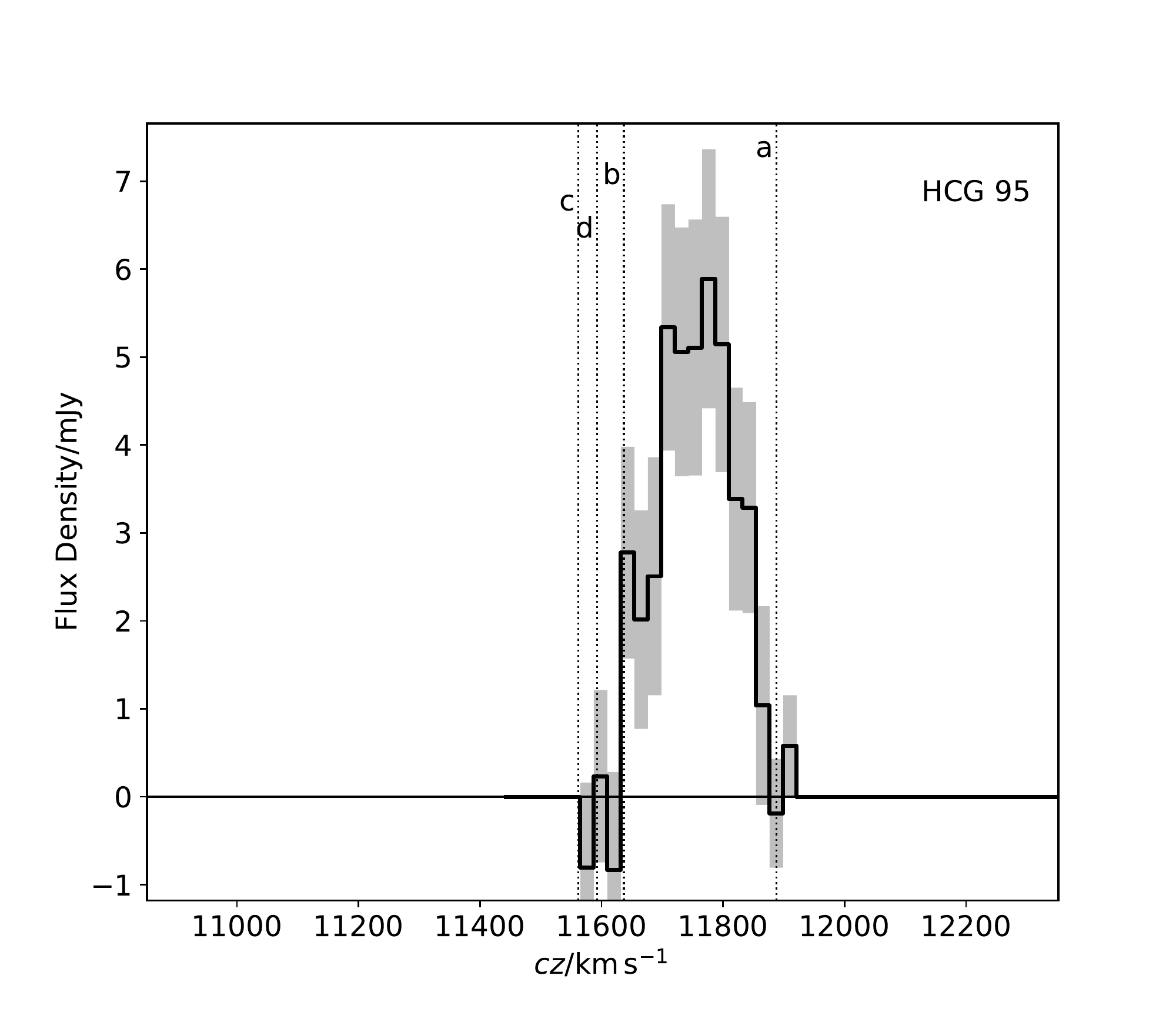}
    \caption{As in Figure \ref{fig:HCG2_spec}.}
    \label{fig:HCG95_spec}
\end{figure}

HCG~95 is a group of three late-type and one elliptical galaxies in the redshift range 11500-11900~\kms. Owing to its large distance (160~Mpc) the physical resolution of the VLA data are quite poor (even in CnB configuration). We therefore do not attempt to separate galaxies and extended features in this group. However, we note the detected \hi \ emission is mostly co-spatial with HCG~95d, though there does appear to be an extension towards HCG~95a (Figures \ref{fig:HCG95_overlay} \& \ref{fig:HCG95_spec}). 

Beyond 125~kpc outside of the core group we detect HCG~95f and HCG~95e in \hi. We note that these were not originally classified as HCG members by \citet{Hickson1982} and \citet{Hickson+1992}, but they were added after deep imaging searching for LSB galaxies \citep{Shi+2017}. For consistency with our approach to the other groups we do not consider these as members of the group. We also note that there are slight continuum subtraction artefacts remaining in this \hi \ cube.

\subsection{HCG~96}

\begin{figure}[h]
    \centering
    \includegraphics[width=\columnwidth]{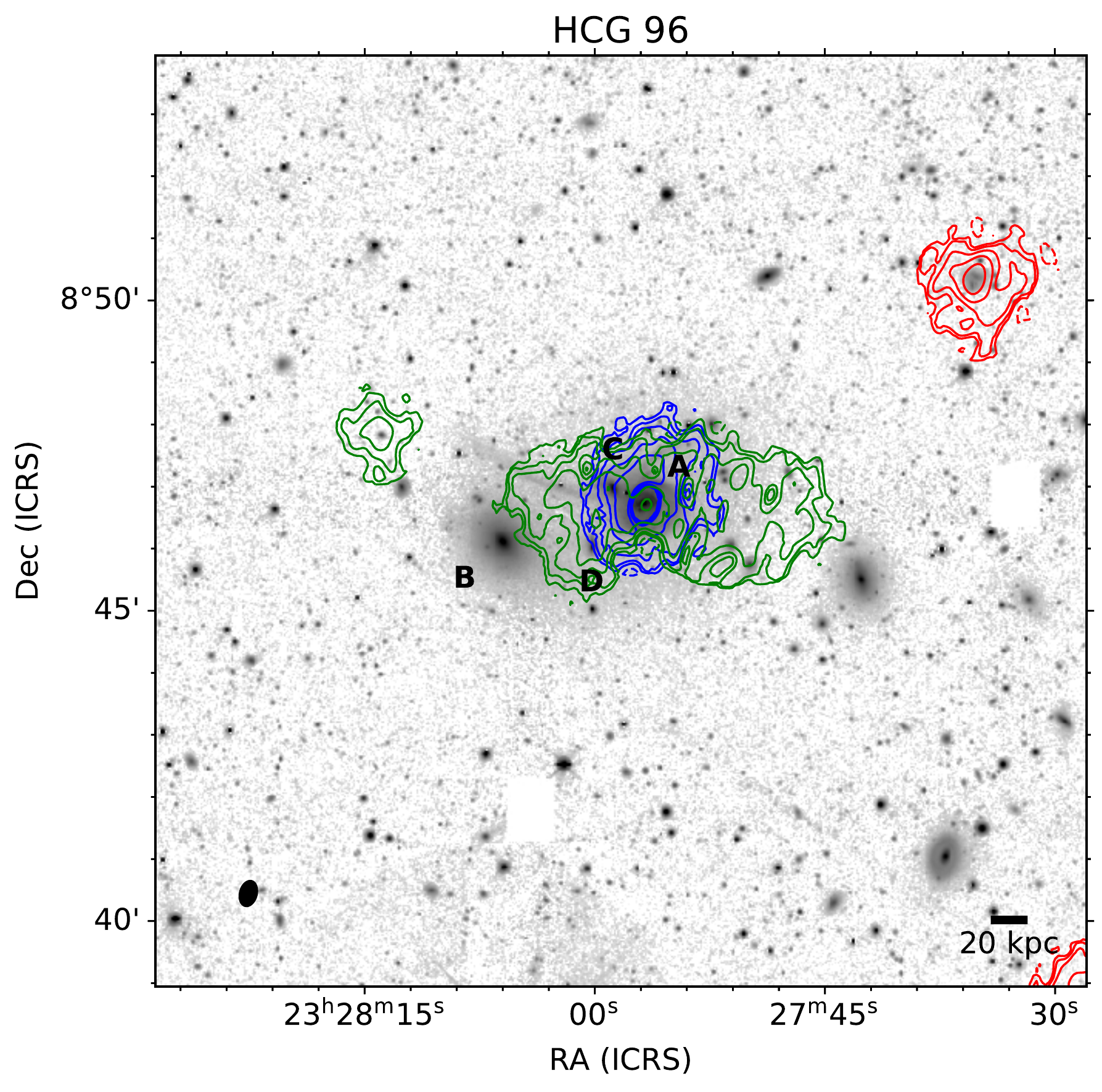}
    \caption{As in Figure \ref{fig:HCG2_split_overlay}.}
    \label{fig:HCG96_split_overlay}
\end{figure}

\begin{figure}[h]
    \centering
    \includegraphics[width=\columnwidth]{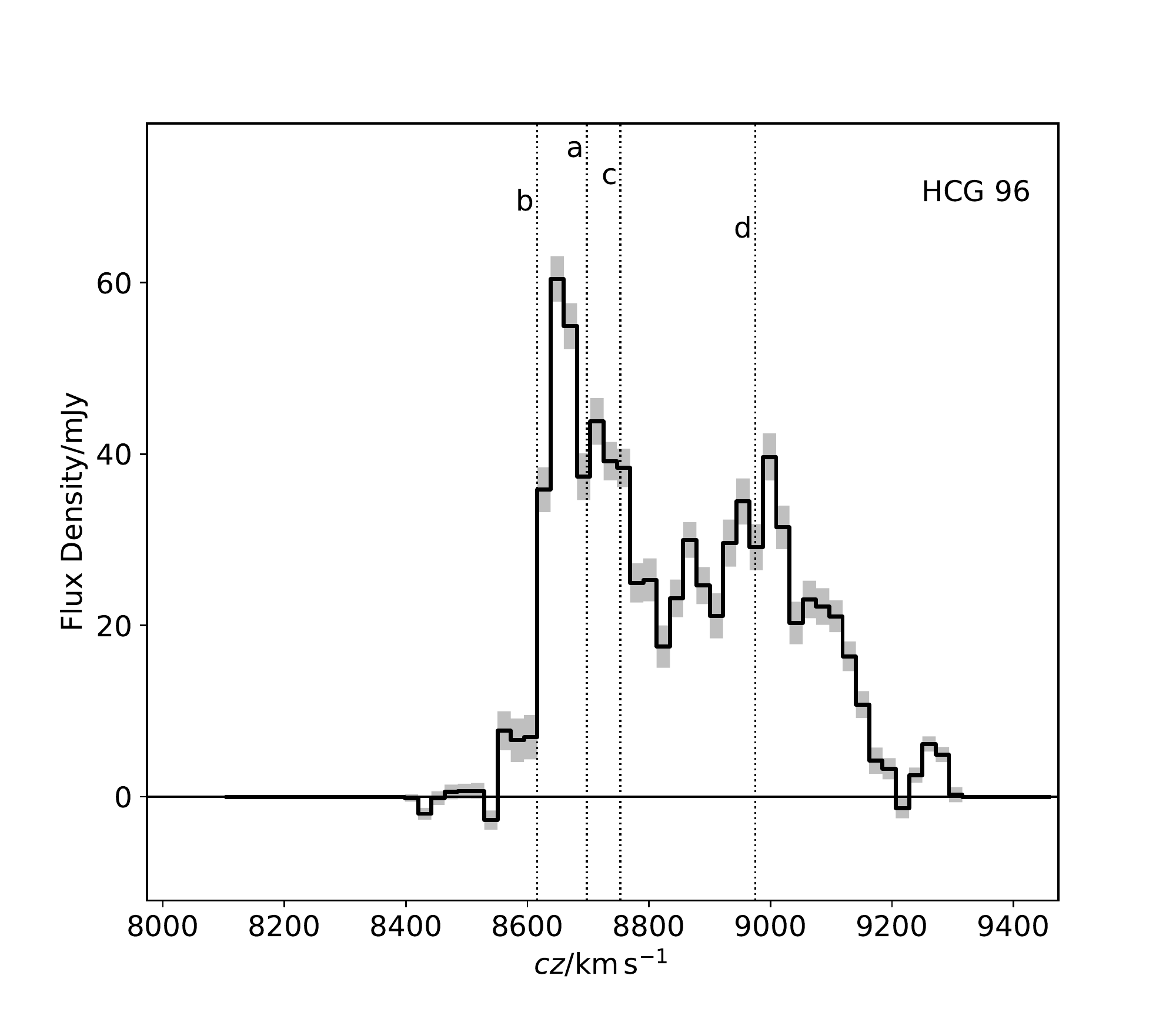}
    \caption{As in Figure \ref{fig:HCG2_spec}.}
    \label{fig:HCG96_spec}
\end{figure}

HCG~96 is a group of three late-type and one elliptical in the redshift range 8600-9000~\kms, two of which (HCG~96a and c) may be in the process of merging. There is a large extended region of \hi \ emission in the core group that appears to encompass HCG~96a, c, and d (Figures \ref{fig:HCG96_split_overlay} \& \ref{fig:HCG96_spec}). HCG~96b is undetected. There are also several additional \hi \ detections around the core ground. The closest of these is an uncatalogued compact, blue, irregular galaxy GALEXASC~J232813.78+084750.1 (which we refer to as HCG~96cl1). This object could plausibly be interacting with core group, or could even be a TDG forming from the ongoing major interactions. We therefore consider it as part of the group, but designate it as a tidal feature. The remaining detections outside of the group are GALEXASC~J232735.09+085018.2, GALEXASC~J232727.99+083853.7, GALEXASC~J232714.32+084125.5, PGC~71560, and three likely spurious detections near the edge of the primary beam with no apparent optical counterparts.

Although the extended emission in the core group covers all of HCG~96a, c, and d in projection, all of this emission appears to be associated with HCG~96a. HCG~96d is sufficiently separated in velocity that it would not fall within the most complex region of emission, and we can confidently conclude that it is undetected. In the case of HCG~96c there are clumps in the extended emission which might correspond to the galaxy, however, even if they do, they do not significantly stand out from the surrounding emission and are thus unlikely to be bound to the galaxy. Hence, all the emission in the core group that is not clearly consistent with the rotation of HCG~96a is classified as extended features. The \hi \ in the core group is also complicated by the presence of a strong \hi \ absorption feature in front of the nucleus of HCG~96b.

\subsection{HCG~97}

\begin{figure}[h]
    \centering
    \includegraphics[width=\columnwidth]{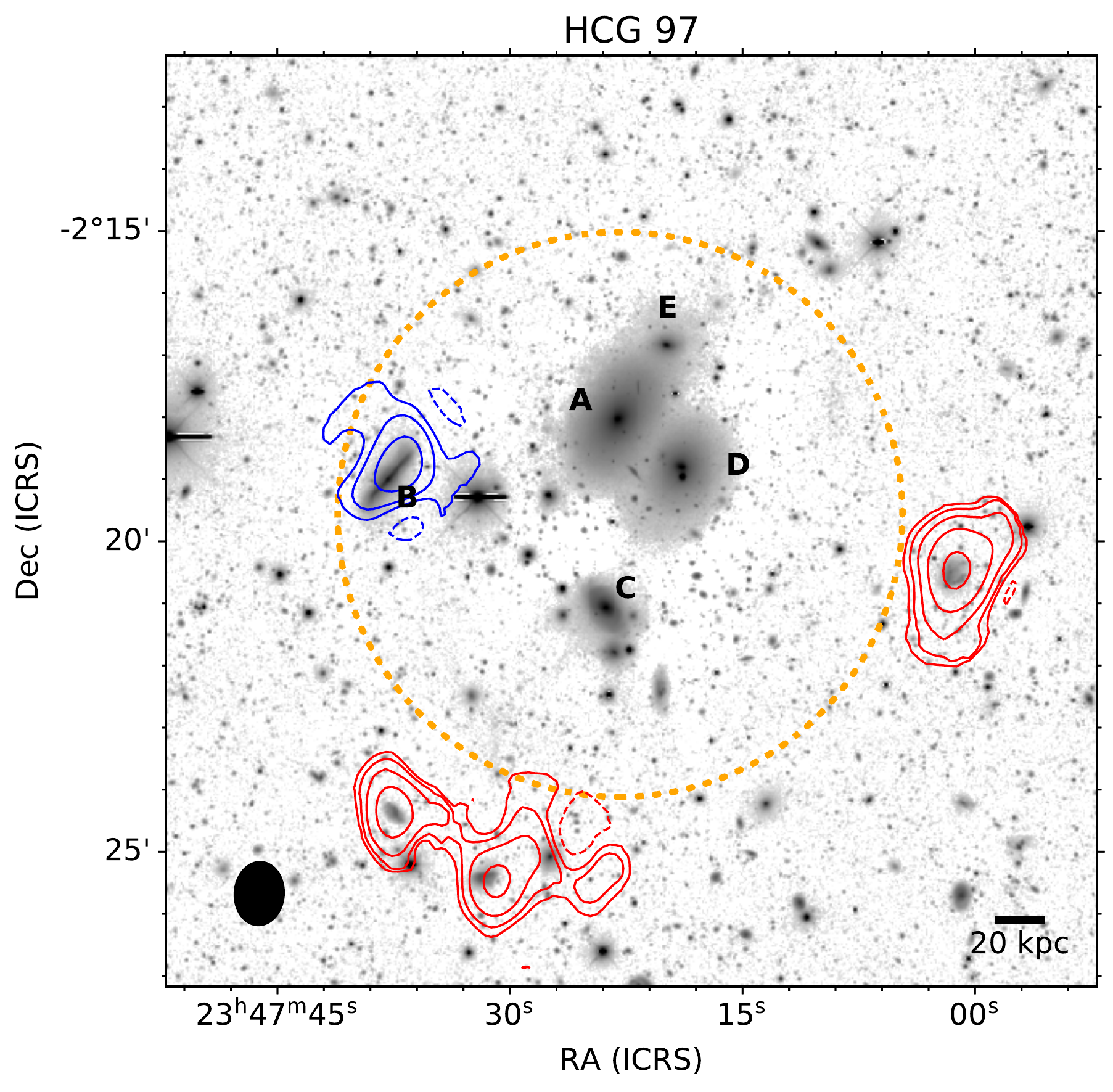}
    \caption{As in Figure \ref{fig:HCG7_split_overlay}.}
    \label{fig:HCG97_split_overlay}
\end{figure}

\begin{figure}[h]
    \centering
    \includegraphics[width=\columnwidth]{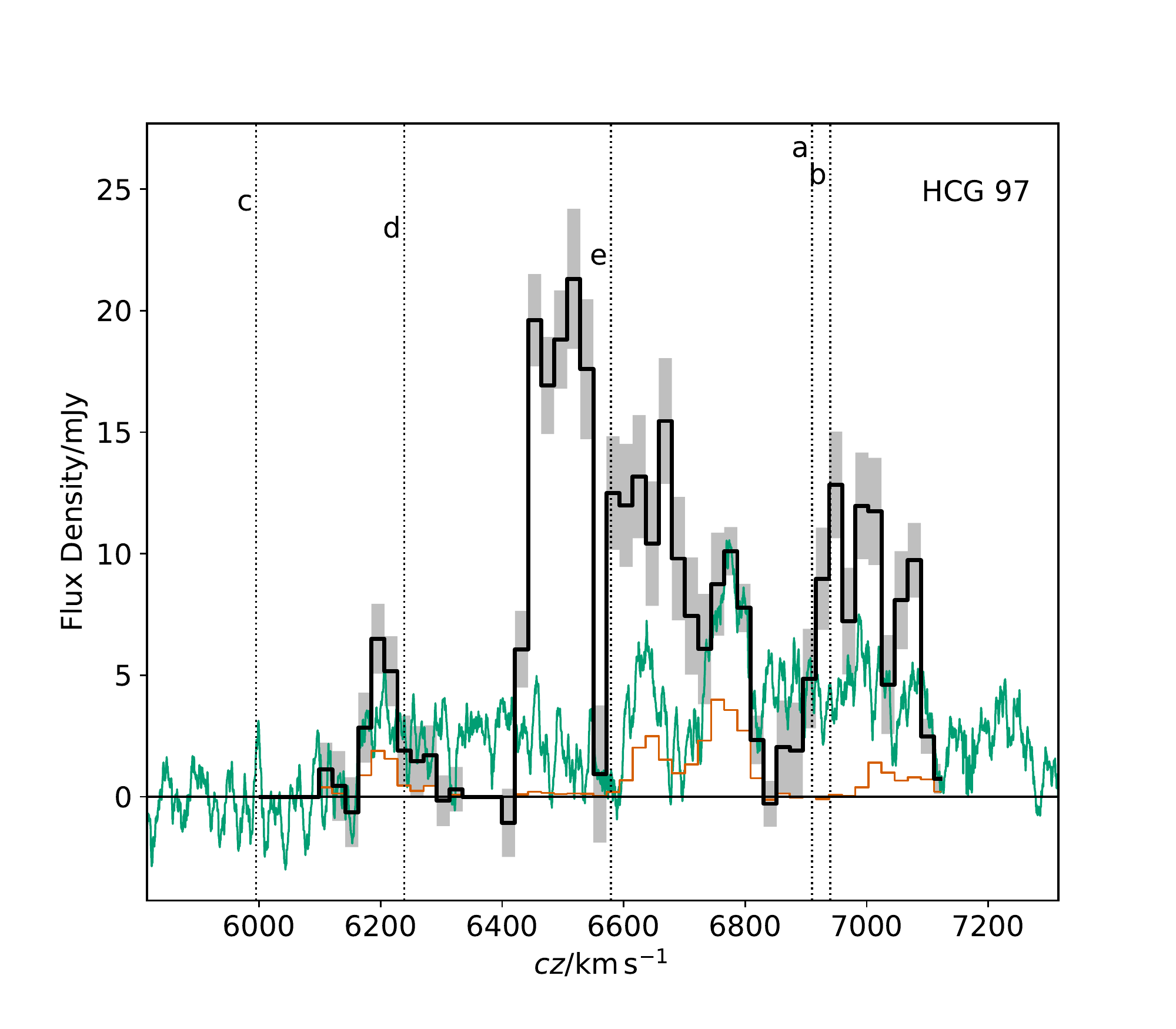}
    \caption{As in Figure \ref{fig:HCG7_spec}.}
    \label{fig:HCG97_spec}
\end{figure}

HCG~97 consists of five galaxies approximately in the redshift range 6000-7000~\kms, two ellipticals, two spirals, and one lenticular. Only one of these, HCG~97b, is detected in the VLA \hi \ observations (Figures \ref{fig:HCG97_split_overlay} \& \ref{fig:HCG97_spec}). However, a number of other objects are detected within the VLA primary beam (WISEA~J234701.65-022033.4, LEDA~1092439, PGC~72457, PGC~3080162, UM~177, PGC~1098512, and PGC~1092963, whose emission is truncated by the edge of the band), between about 100~kpc to 500~kpc away from the core group. The nearest of these lie on the edge of the beam of the GBT observation, and likely contributed some additional flux to the group's \hi \ mass measurement.

HCG~97b is an edge-on spiral and it appears that \hi \ is likely only detected on one side of the galaxy, with the emission from the receding side of the galaxy being too low S/N to be included in the source mask. This indicates that the galaxy is likely disturbed, but it is only marginally resolved in the VLA data.

\subsection{HCG~100}

\begin{figure}[h]
    \centering
    \includegraphics[width=\columnwidth]{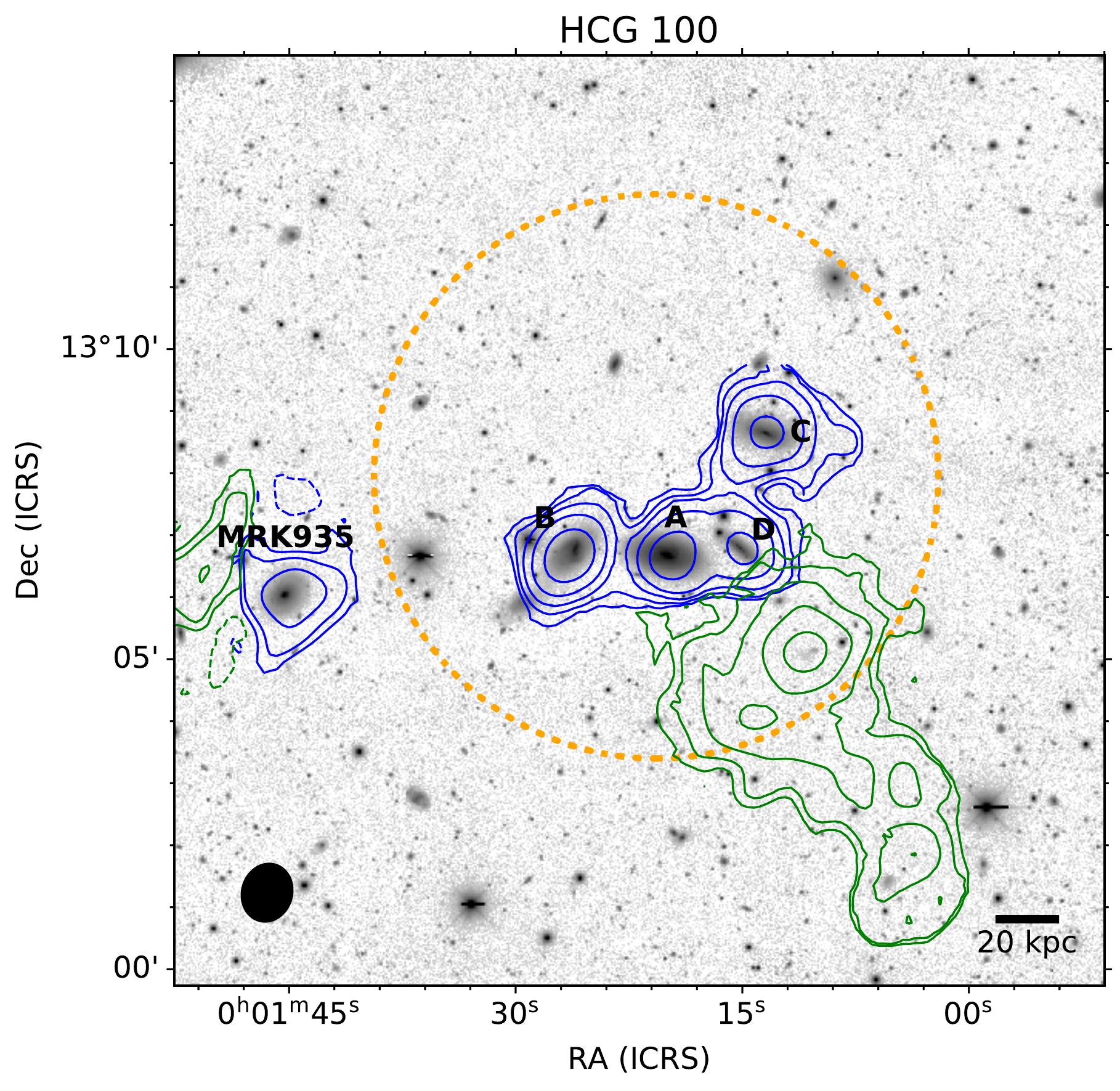}
    \caption{As in Figure \ref{fig:HCG7_split_overlay}.}
    \label{fig:HCG100_split_overlay}
\end{figure}

\begin{figure}[h]
    \centering
    \includegraphics[width=\columnwidth]{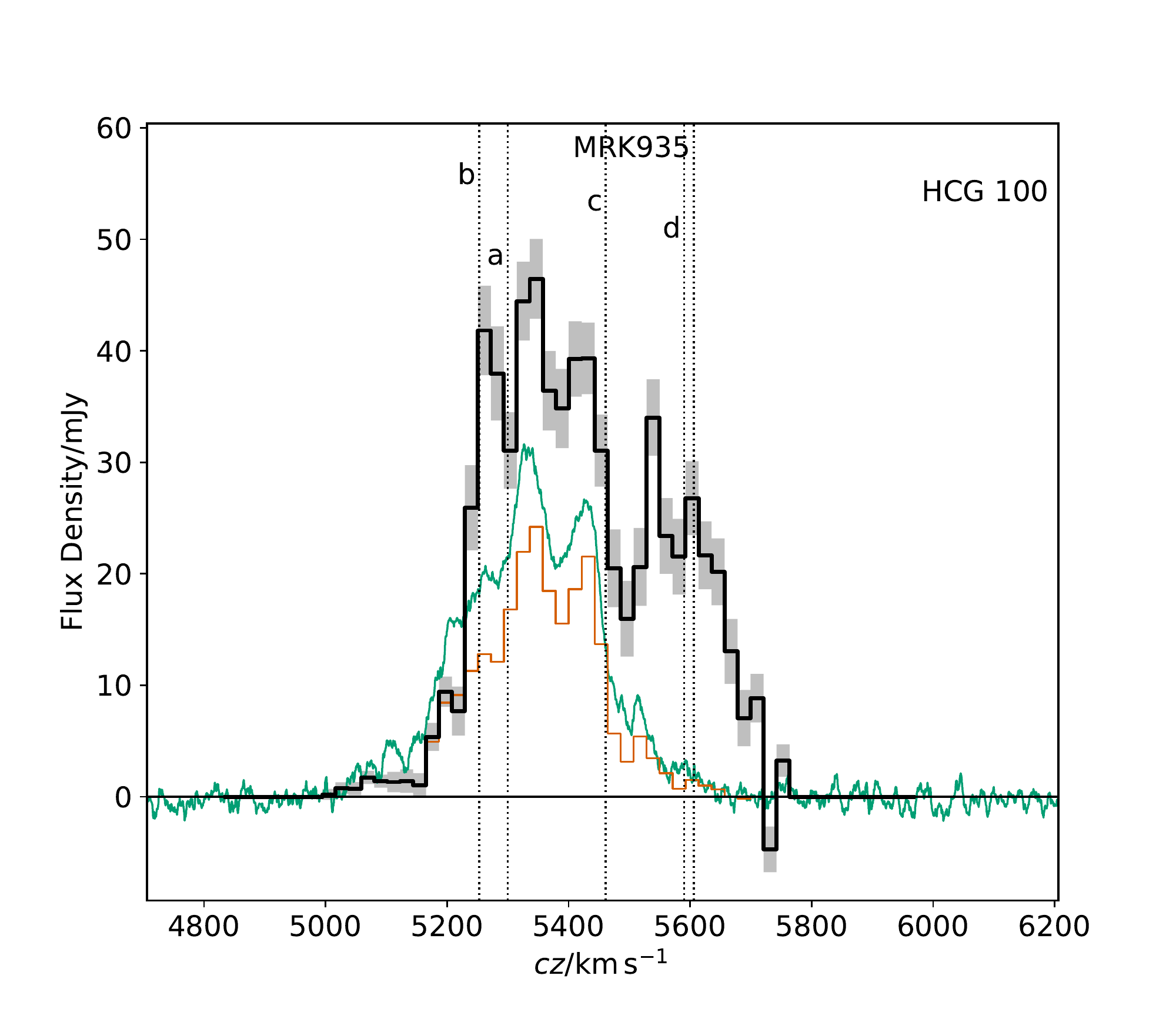}
    \caption{As in Figure \ref{fig:HCG7_spec}.}
    \label{fig:HCG100_spec}
\end{figure}

HCG~100 is made up of a chain of four late-type galaxies in the redshift range 5200-5600~\kms. All are detected in \hi \ in the VLA imaging, as well as an enormous extended feature, stretching over 100~kpc to the SW of the core group. At the tip of this tail there is a faint, blue optical counterpart, perhaps indicating the formation of a TDG \citep[e.g.][]{deMello+2012}. In addition, we detect MRK~935 about 85~kpc east of the core group.\footnote{We note that there are conflicting redshift measurements for MRK~935, however, the \hi \ detection is consistent with $cz_\odot = 5606$~\kms \ \citep{Petrosian+2007}.} The \hi \ in this galaxy appears heavily perturbed and it is likely interacting with the core group and we therefore consider it a member. Further to the SE we also detect \hi \ in NGC~7810 and AGC~105092.

We do not attempt to separate the four core galaxies from each other. Their optical discs almost overlap, and given that the are spread over only $\sim$400~\kms \ it is not possible to reliably separate them at the resolution of the VLA data. However, the majority of the emission in extended features appears to be in the SE tail, which is separated as a distinct feature. Although the detection of MRK~935 is quite low S/N we attempt to separate this into emission from the galaxy itself and extended emission. However, the faintness of this object (in \hi) mean that it has little bearing of the overall ratio of disc versus extended emission in the group.

\section{Results}
\label{sec:results}

In this section we present the global results for the \hi \ content of HCGs based on aggregating the results for individual groups from the previous section. First we compare the VLA measurements to those from the GBT, and then proceed calculate the \hi \ deficiency of each HCG.

\begin{figure}
    \centering
    \includegraphics[width=\columnwidth]{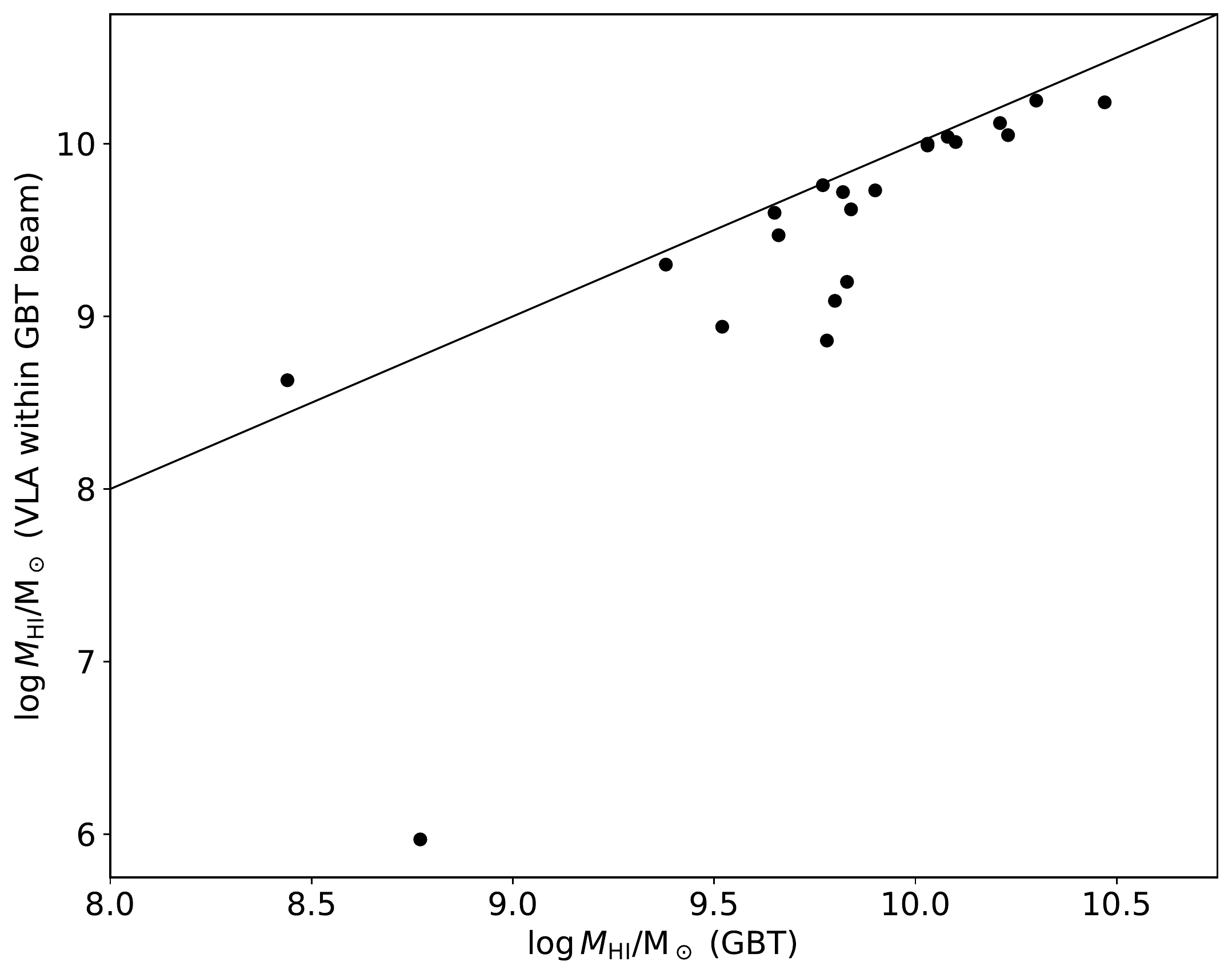}
    \caption{\hi \ mass of HCGs measured by the GBT (x-axis) and measured by the VLA with spatial weighting to match the falloff in sensitivity corresponding to the GBT beam (y-axis). The solid line indicates equality of the two. Note that the scale is not the same of both axes.}
    \label{fig:GBT_VLA_comp}
\end{figure}

\begin{figure}
    \centering
    \includegraphics[width=\columnwidth]{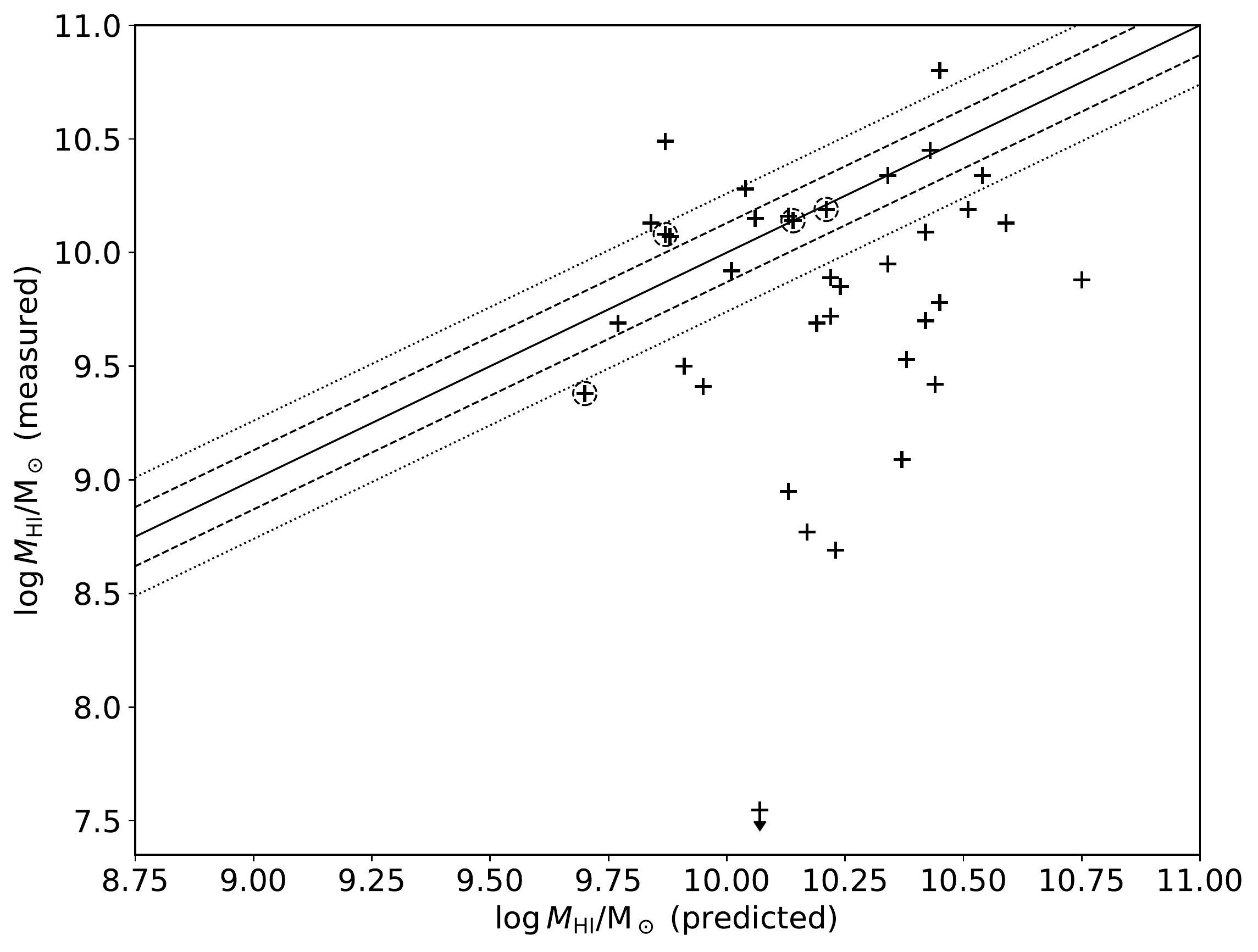}
    \caption{Measured versus predicted (with the \citet{Jones+2018} $L_{B}$--$M_\mathrm{HI}$ scaling relation) \hi \ masses of HCGs. All measured \hi \ masses are those from the VLA data, except for HCG~30 and 37, which use the GBT measurements. Points circled with black dashed lines are triplets. The solid black line indicates equality and the dashed and dotted lines indicate 1$\sigma$ and 2$\sigma$ scatters from the \citet{Jones+2018} relation (for a single galaxy). HCG~62 was undetected with the VLA (and has no GBT observation) and is marked as an upper limit.}
    \label{fig:HIdef}
\end{figure}

\begin{figure}
    \centering
    \includegraphics[width=\columnwidth]{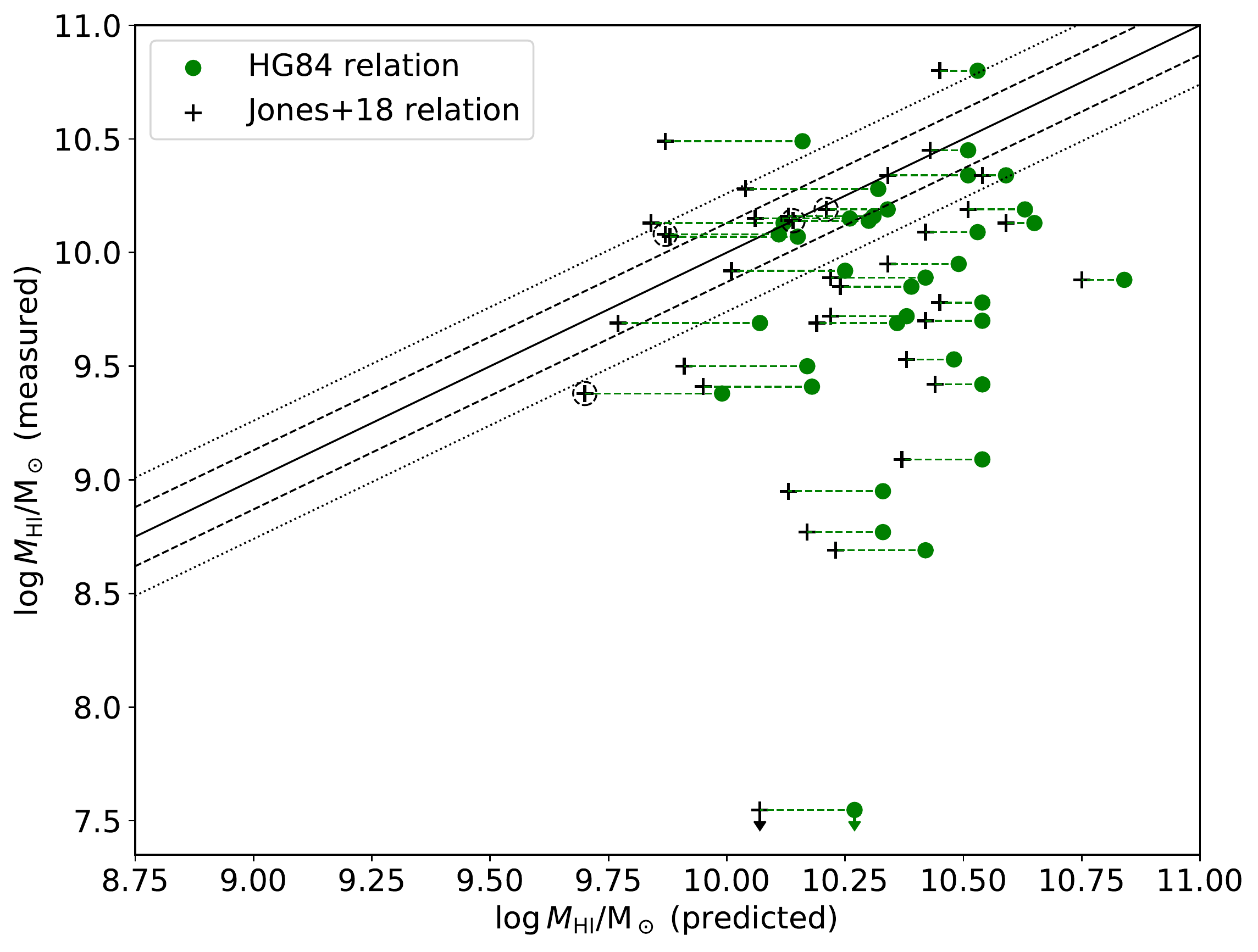}
    \caption{As for Figure \ref{fig:HIdef}, except now marked with green circles are the predicted \hi \ masses using the \citet{Haynes+Giovanelli1984} $L_{B}$--$M_\mathrm{HI}$ scaling relation. In general this relation predicts significantly higher \hi \ masses than those from the equivalent \citet{Jones+2018} relation.}
    \label{fig:HIdef_HG84}
\end{figure}

\begin{figure}
    \centering
    \includegraphics[width=\columnwidth]{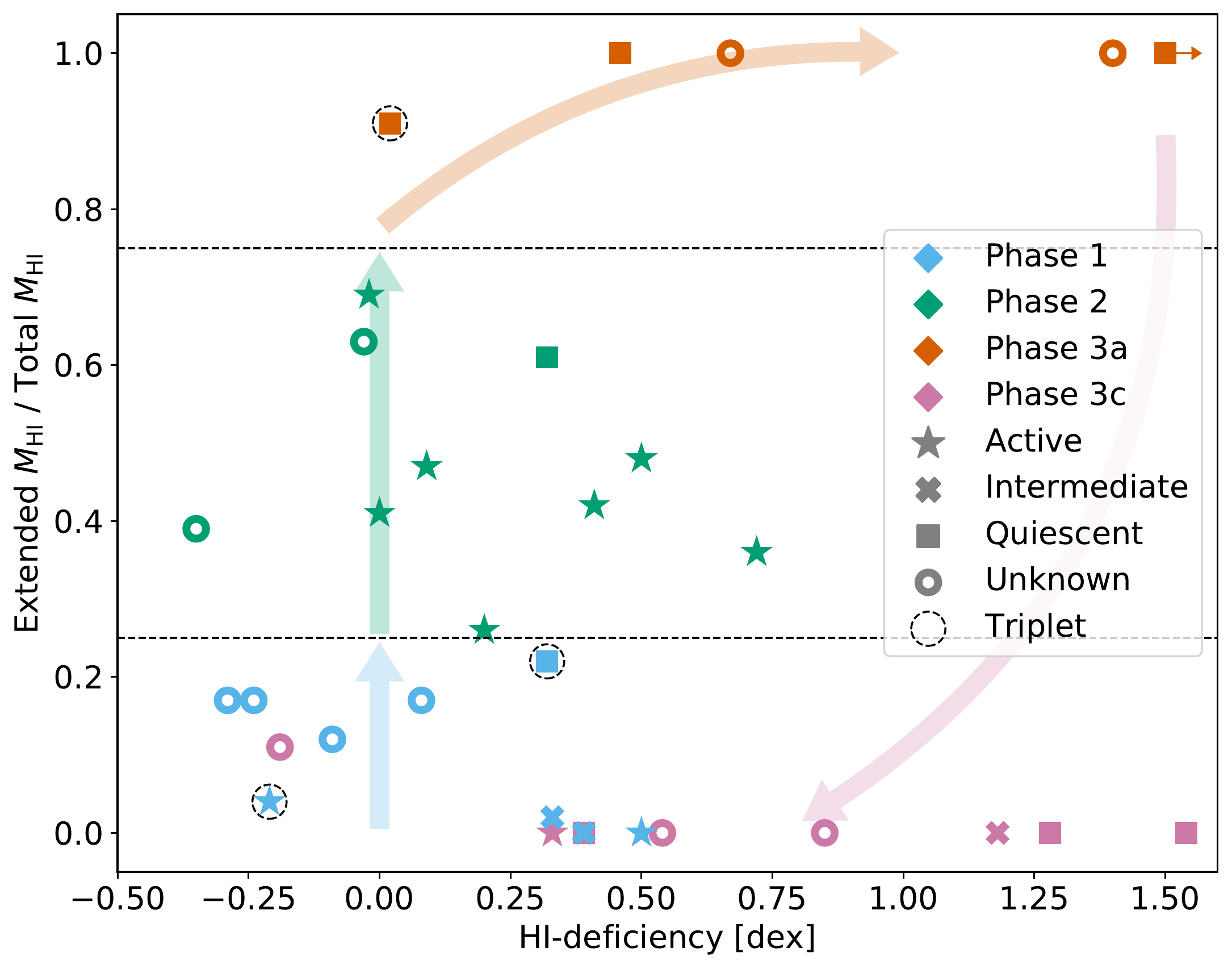}
    \caption{\hi \ deficiencies and extended \hi \ emission fraction for each (resolved) group in the sample. The colour of each point indicates its \hi \ morphological phase: blue for 1, green for 2,orangefor 3a, and pink for 3c. The marker shape indicates the IR classification \citep{Zucker+2016} of each group: stars for groups dominated by IR active galaxies, crosses for groups not dominated by either active or quiescent (or transition) galaxies, squares for groups dominated by quiescent (or transition) galaxies, and rings for groups with too many members without IR classifications. Markers enclosed in a dashed black circle correspond to triplets. The lone upper limit plotted in the top-right corner corresponds to HCG~62 that was entirely undetected in \hi \ with the VLA and has no GBT observation. The horizontal dashed lines indicate 25\% and 75\% extended emission. These thresholds entirely determine which groups are classified as Phase 2 or 3a. Phase 1 and 3c are distinguished from each other based on the number of galaxies detected in \hi \ (Section \ref{sec:classification}). The shaded arrows indicate our proposed evolutionary path of groups through this parameter space.}
    \label{fig:tidal_fraction}
\end{figure}

\begin{figure}
    \centering
    \includegraphics[width=\columnwidth]{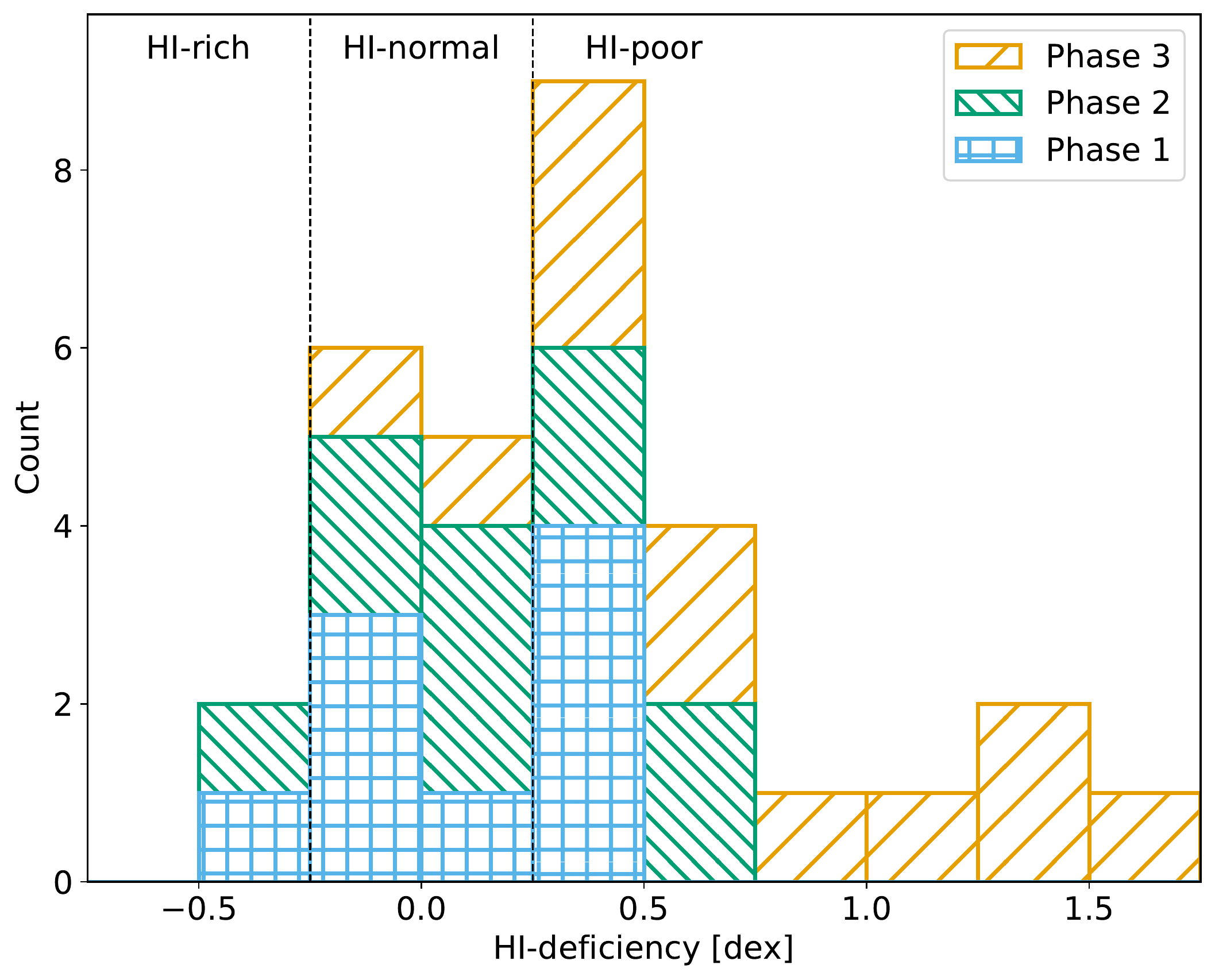}
    \caption{Histogram of \hi \ deficiencies of HCGs. The three phases are shown individually by differently hatched bars. The vertical black dashed lines indicate approximately the typical 1$\sigma$ uncertainty (away from zero) in the measure of \hi \ deficiency for an individual galaxy \citep{Jones+2018}. Note that HCG~62 is not included as there is only an upper lower limit on its \hi \ deficiency.}
    \label{fig:HIdefhist}
\end{figure}

\begin{figure}
    \centering
    \includegraphics[width=\columnwidth]{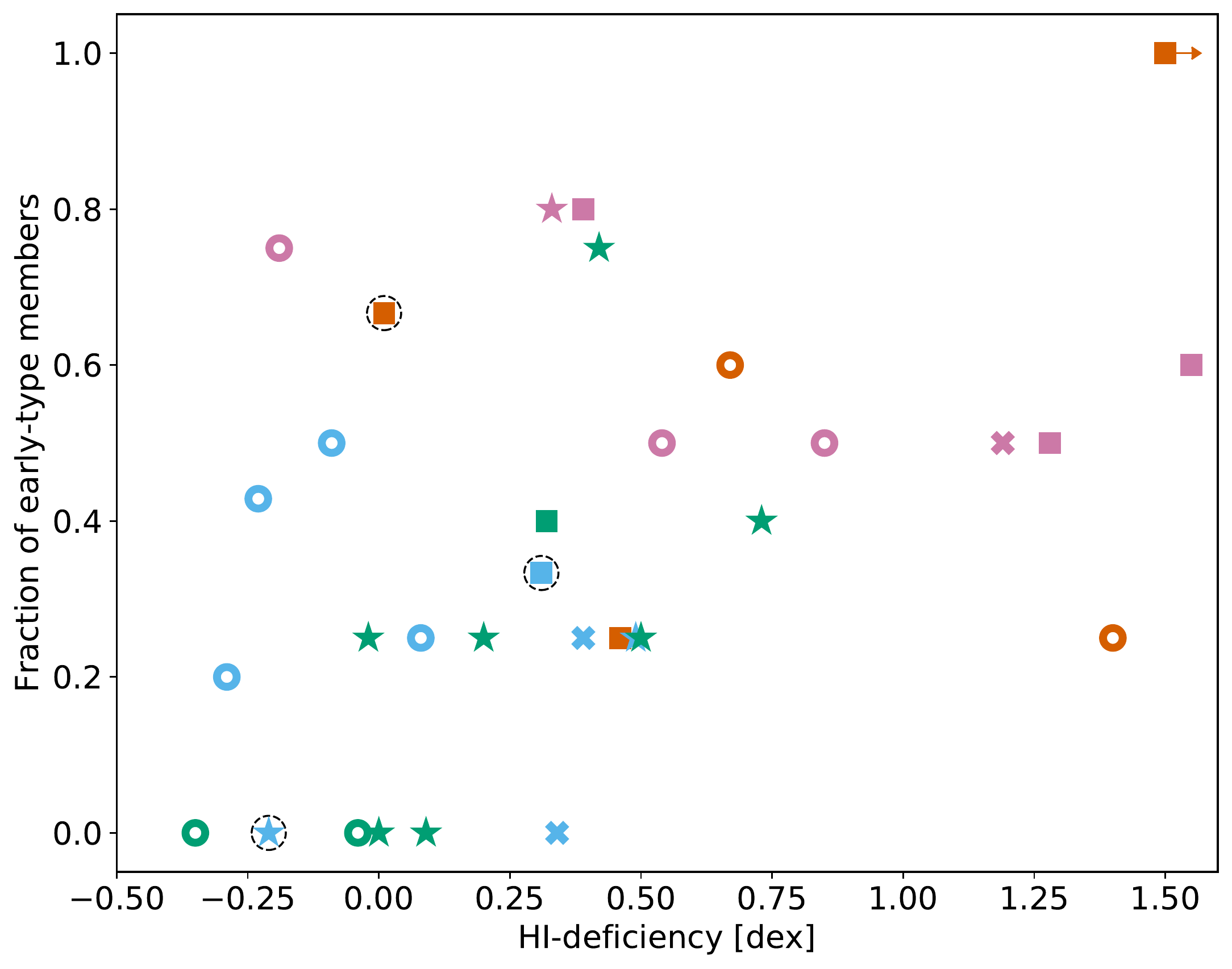}
    \caption{\hi \ deficiencies of HCGs versus the fraction of early-type members. In this case early-types are defined as lenticular or elliptical. The markers follow the same scheme as in Figure \ref{fig:tidal_fraction}.}
    \label{fig:ETG_fraction}
\end{figure}

\begin{table*}
\centering
\caption{\hi \ deficiency of HCGs}
\label{tab:HIdef}
\begin{tabular}{cccccccccc}
\hline \hline
HCG & $\log M_\mathrm{HI,VLA}$ & $\log M_\mathrm{HI,GBT}$ & $\log M_\mathrm{HI,pred}$ & \hi-def. & \hi-def. & $\log M_\mathrm{HI,gals}$ & $\log M_\mathrm{HI,exfs}$ & $f_\mathrm{exfs}$ & \hi \ Phase \\ 
 & $[\mathrm{M_\odot}]$ & $[\mathrm{M_\odot}]$ & $[\mathrm{M_\odot}]$ & (VLA) & (GBT) & $[\mathrm{M_\odot}]$ & $[\mathrm{M_\odot}]$ &  &  \\ \hline
2$\ddagger$ & $10.08\pm0.04$ &  & $9.87 \pm 0.12$ & $-0.21$ &  & 10.06 & 8.66 & 0.04 & 1 \\
7 & $9.73\pm0.05$ & 9.65 & $10.22 \pm 0.16$ & $0.49$ & $0.57$ & 9.73 &  & 0.0 & 1 \\
10 & $9.85\pm0.04$ & 9.77 & $10.24 \pm 0.12$ & $0.39$ & $0.47$ & 9.85 &  & 0.0 & 1 \\
15 & $9.09\pm0.07$ & 9.52 & $10.37 \pm 0.09$ & $1.28$ & $0.85$ & 9.09 &  & 0.0 & 3c \\
16 & $10.34\pm0.05$ & 10.08 & $10.34 \pm 0.09$ & $-0.0$ & $0.26$ & 10.11 & 9.95 & 0.41 & 2 \\
19$\ddagger$ & $9.39\pm0.05$ &  & $9.7 \pm 0.13$ & $0.31$ &  & 9.28 & 8.72 & 0.22 & 1 \\
22 & $9.41\pm0.04$ &  & $9.95 \pm 0.15$ & $0.54$ &  & 9.41 &  & 0.0 & 3c \\
23 & $10.13\pm0.05$ & 10.03 & $9.84 \pm 0.12$ & $-0.29$ & $-0.19$ & 10.05 & 9.37 & 0.17 & 1 \\
25 & $10.15\pm0.05$ & 10.1 & $10.06 \pm 0.13$ & $-0.09$ & $-0.04$ & 10.1 & 9.23 & 0.12 & 1 \\
26 & $10.27\pm0.05$ & 10.47 & $10.04 \pm 0.11$ & $-0.23$ & $-0.43$ & 10.19 & 9.49 & 0.17 & 1 \\
30 & $<7.78$ & 8.77 & $10.17 \pm 0.13$ & $>2.39$ & $1.4$ &  &  &  & 3a \\
31 & $10.17\pm0.05$ & 10.22 & $10.13 \pm 0.19$ & $-0.04$ & $-0.09$ & 9.74 & 9.96 & 0.63 & 2 \\
33 & $10.07\pm0.06$ &  & $9.88 \pm 0.12$ & $-0.19$ &  & 10.02 & 9.1 & 0.11 & 3c \\
37 & $<8.35$ & 9.78 & $10.45 \pm 0.14$ & $>2.1$ & $0.67$ &  &  &  & 3a \\
38$\ddagger$ & $10.14\pm0.05$ &  & $10.14 \pm 0.12$ & $-0.0$ &  &  &  &  &  \\
40 & $9.69\pm0.07$ & 9.84 & $10.42 \pm 0.12$ & $0.73$ & $0.58$ & 9.5 & 9.25 & 0.36 & 2 \\
47 & $9.69\pm0.05$ &  & $10.19 \pm 0.13$ & $0.5$ &  & 9.41 & 9.37 & 0.48 & 2 \\
48$\dagger$ & $8.85\pm0.06$ & 8.78 & $9.44 \pm 0.17$ & $0.59$ & $0.66$ &  &  &  &  \\
49 & $10.49\pm0.04$ &  & $9.87 \pm 0.11$ & $-0.62$ &  &  &  &  &  \\
54 & $9.25\pm0.05$ &  & $8.99 \pm 0.17$ & $-0.26$ &  &  &  &  &  \\
56 & $9.89\pm0.04$ &  & $10.22 \pm 0.11$ & $0.33$ &  & 9.89 &  & 0.0 & 3c \\
57 & $9.88\pm0.05$ &  & $10.75 \pm 0.09$ & $0.87$ &  &  &  &  &  \\
58 & $10.19\pm0.04$ & 9.83 & $10.51 \pm 0.1$ & $0.32$ & $0.68$ & 9.78 & 9.98 & 0.61 & 2 \\
59 & $9.69\pm0.05$ &  & $9.77 \pm 0.13$ & $0.08$ &  & 9.61 & 8.92 & 0.17 & 1 \\
61$\ddagger$ & $10.2\pm0.04$ &  & $10.21 \pm 0.13$ & $0.01$ &  & 9.13 & 10.16 & 0.91 & 3a \\
62 & $<7.54$ &  & $10.07 \pm 0.12$ & $>2.53$ &  &  &  &  & 3a \\
68 & $9.95\pm0.04$ & 9.83 & $10.34 \pm 0.11$ & $0.39$ & $0.51$ & 9.95 & 7.4 & 0.0 & 3c \\
71 & $10.8\pm0.05$ &  & $10.45 \pm 0.14$ & $-0.35$ &  & 10.59 & 10.39 & 0.39 & 2 \\
79 & $9.49\pm0.05$ & 9.66 & $9.91 \pm 0.12$ & $0.42$ & $0.25$ & 9.26 & 9.11 & 0.42 & 2 \\
88 & $10.08\pm0.05$ & 10.02 & $10.42 \pm 0.11$ & $0.34$ & $0.4$ & 10.08 & 8.4 & 0.02 & 1 \\
90 & $8.94\pm0.06$ & 8.43 & $10.13 \pm 0.1$ & $1.19$ & $1.7$ & 8.94 &  & 0.0 & 3c \\
91 & $10.34\pm0.06$ & 10.3 & $10.54 \pm 0.15$ & $0.2$ & $0.24$ & 10.21 & 9.76 & 0.26 & 2 \\
92 & $10.13\pm0.04$ & 10.23 & $10.59 \pm 0.11$ & $0.46$ & $0.36$ &  & 10.13 & 1.0 & 3a \\
93 & $9.53\pm0.04$ & 9.38 & $10.38 \pm 0.12$ & $0.85$ & $1.0$ & 9.53 &  & 0.0 & 3c \\
95 & $9.42\pm0.09$ &  & $10.44 \pm 0.12$ & $1.02$ &  &  &  &  &  \\
96 & $10.45\pm0.05$ &  & $10.43 \pm 0.14$ & $-0.02$ &  & 9.93 & 10.29 & 0.69 & 2 \\
97 & $8.68\pm0.1$ & 9.8 & $10.23 \pm 0.1$ & $1.55$ & $0.43$ & 8.68 &  & 0.0 & 3c \\
100 & $9.92\pm0.05$ & 9.89 & $10.01 \pm 0.11$ & $0.09$ & $0.12$ & 9.64 & 9.59 & 0.47 & 2 \\
\hline
\end{tabular}
\tablefoot{Columns: (1) HCG ID number, (2) total \hi \ mass from the VLA observations with uncertainties approximated as $\sigma_{M_\mathrm{HI}} = 235600 \left( \frac{D}{\mathrm{Mpc}} \right)^2\times 3\frac{\sigma_\mathrm{rms}}{\mathrm{Jy}} \times 100\;\mathrm{km\,s^{-1}}$, combined with an assumed 10\% uncertainty in absolute flux calibration (distance uncertainties are neglected), (3) total \hi \ mass from the GBT observations \citep{Borthakur+2010}, (4) predicted \hi \ mass based on B-band luminosity \citep{Jones+2018}, (5 \& 6) \hi \ deficiency of the entire group from the VLA and GBT mass measurements, respectively, (7) combined \hi \ mass of all member galaxies, (8) combined \hi \ mass of all extended features in the group, (9) fraction of total \hi \ mass in extended features, (10) \hi \ morphological classification. $\dagger$ Pair or false group. $\ddagger$ Triplet.
}
\end{table*}

\subsection{Discrepancies between single-dish and VLA fluxes}
\label{sec:fluxdiff}

As shown in Figure \ref{fig:GBT_VLA_comp} the vast majority of the VLA observations result in total fluxes (after weighting for the GBT beam response) that are just marginally below those observed with the GBT. This is expected, as in general an interferometer resolves out some extended emission and therefore does not recover the full flux seen with a single dish. However, there are a few points which require further investigation: a) the single point above the line of equality, b) the group of four points about 0.5~dex below the line, and c) HCG~30 which is about 2~dex below the line.

The first of these is formally impossible as an interferometer cannot detect more flux than a single dish, and would therefore normally point to a calibration, pointing, or continuum subtraction problem. This point corresponds to HCG~90. As can be seen in Figure \ref{fig:HCG90_split_overlay}, only HCG~90a is detected in the VLA map and this galaxy lies outside the HPBW of the GBT observation. Therefore, the comparison between the VLA and GBT fluxes will be extremely sensitive to exact nature of our (approximate) weighting for the GBT primary beam response. Furthermore, there is a strong \hi \ absorption feature at the centre of HCG~90a, which appears to impact the GBT spectrum more strongly than the VLA spectrum (Figure \ref{fig:HCG90_spec}). 

The second case, the four groups with about 0.5~dex lower VLA fluxes than GBT fluxes, also requires explanation. These groups are HCGs 15, 37, 58, and 97.
HCG~97 is another case similar to HCG~90, where much of the \hi \ emission occurs near or beyond the edge of the GBT primary beam, and therefore the comparison between the VLA and GBT spectra is likely to be unreliable (Figures \ref{fig:HCG97_split_overlay} \& \ref{fig:HCG97_spec}). However, the spectra do also show regions where emission is detected by the GBT, but not at all with the VLA. This may be a sign of diffuse \hi \ emission that was resolved out by the VLA.

The comparison of the VLA and GBT spectra (Figure \ref{fig:HCG15_spec}) for HCG~15 indicates that a considerable fraction of the \hi \ emission in the GBT spectrum is undetected with the VLA. We also note that the VLA spectrum presented here differs strongly from that presented by \citet{Borthakur+2010}, where even HCG~15f is not detected. The high rms noise of the VLA data relative to the signals in the GBT spectrum, might prevent them from being included in the source mask. Alternatively, these could be diffuse features that were resolved out by the VLA. We also note that the two highest peaks in the GBT spectrum (aside from HCG~15f) do not correspond to the redshifts of any of the group members, which could mean that they are additional LSB members, \hi-only features, or spurious in some way.

In the cases of HCG~30 and 37, there is no \hi \ emission detected in the group cores by the VLA (Figures \ref{fig:HCG30_spec} \& \ref{fig:HCG37_spec}), but in the GBT spectra there is broad ($\sim$1000~\kms \ wide), faint emission that might indicate the presence of a diffuse \hi \ component that the VLA is unable to detect \citep{Borthakur+2010}. These two groups are the most clear examples of this phenomena that we identified, with the other possibility being HCG~97. In all other cases (except HCG~15) the spectra from the VLA and the GBT are broadly consistent.

Two other groups warrant mention for their minor mismatches in their GBT and VLA spectra, HCG~26 and 79. The spectral profile of HCG~26 in the VLA data matches closely with that from the GBT data, however, it is consistently lower (Figure \ref{fig:HCG26_spec}). As the profile is so similar this likely indicates a slight calibration problem with one or both of these observations, rather than the presence of diffuse emission. For HCG~79 an entire emission feature appears to be absent from the VLA spectrum when compared to that of the GBT (Figure \ref{fig:HCG79_spec}). At present there is not a clear explanation for this mismatch. It seems unlikely that this feature would be too low S/N to be detected, meaning it could be a diffuse feature resolved out by the VLA, an artefact in the GBT data, or a problem with the continuum subtraction of the VLA data. However, we were unable to identify the root cause.

Aside from the few cases discussed above, the GBT flux is well recovered by the VLA observations. Therefore, as all our targets have VLA data, but not all have GBT data, we use the \hi \ deficiency of each group determined from the VLA data in most of our subsequent analysis. Switching to the GBT observations would make only marginal differences for all but a handful of groups, and in most of those cases would lead to a worse estimate (e.g. because of blended emission from non-members), not an improvement. The exception are HCG~30 and 37 for which we use the GBT measurements, as these are undetected with the VLA.

\subsection{\hi \ deficiency of HCGs}

\citet{Huchtmeier1997} and \citet{Verdes-Montenegro+2001} found that the vast majority of HCGs were highly \hi-deficient, motivating them to ask the question ``where is the neutral atomic gas in HCGs?" The distribution of observed versus predicted \hi \ masses is shown in Figure \ref{fig:HIdef}. We too find that many HCGs have significantly lower \hi \ masses than would be expected for non-interacting galaxies, while very few are significantly rich in \hi. We also see that triplets (circled with black dashed lines in Figure \ref{fig:HIdef}) appear to follow the same distribution as groups with more members. However, there are just four triplets in the sample.

We define \hi \ deficiency \citep[as in][]{Haynes+Giovanelli1984} to the be logarithmic decrement between the predicted and observed \hi \ mass, that is \hi-$\mathrm{def} = \log M_\mathrm{HI,pred} - \log M_\mathrm{HI,obs}$, where the predicted mass is found from the B-band luminosity and the scaling relation of \citet{Jones+2018}, measured based on isolated galaxies in the AMIGA \citep[Analysis of the interstellar Medium of Isolated GAlaxies;][]{Verdes-Montenegro+2005b} sample. The AMIGA sample offers a benchmark of \hi \ content of galaxies in the (near) absence of interactions. It is also an optically selected sample (as are HCGs) rather than an \hi-selected sample. Thus, the \citet{Jones+2018} relations offer a well-suited metric for HCGs, whereas other scaling relations in the literature \citep[e.g.][]{Toribio+2011,Brown+2017} would be less appropriate for this particular use case and could introduce a bias in the measures of \hi-deficiency.

We note, however, that no scaling relation will function perfectly as intended in a compact group environment. The myriad of ongoing interactions, starbursts, shocks etc. in these groups is bound to influence the B-band luminosities of the galaxies and complicate the use of this scaling relation. We chose to adopt B-band luminosity as the predictor for \hi \ mass as the alternative, optical diameter \citep{Jones+2018}, can be extremely uncertain in systems where tidal effects are significant. Although also affected by tidal interactions, B-band luminosity is usually dominated by the bulge and is therefore expected to be less severely impacted in most cases. Even so, we caution the reader that the uncertainty estimates on the predicted $M_\mathrm{HI}$ values in Tables \ref{tab:HIdef} \& \ref{tab:HIdef_mem} are based on the scatter of values for isolated galaxies and are likely underestimate the scale of the uncertainties for HCG galaxies.

The B-band magnitudes (and morphological types) of the member galaxies (Table \ref{tab:HIdef_mem}) for use with this scaling relation are taken from \citet{Hickson+1989}. In cases where we have identified new group members these values were taken from the HyperLeda database \citep{Makarov+2014}. To perform detailed photometry \citet{Hickson+1989} imaged HCG galaxies with the Canada-France-Hawaii telescope and overlapping galaxies were separated. For pairs the separation was performed by slicing the image at the nuclei of each galaxy and proportioning the light in between them based on the magnitude of the non-overlapping side of each galaxy. For high multiplicity blends the light in overlapping isophotes was divided amongst all the galaxies based on their magnitudes inside their non-overlapping isophotes. Finally, \citet{Hickson+1989} corrected these values for internal and Galactic extinction following \citet{RC2}.

For uniformity, we use the same relation for all galaxies with no consideration of morphological type. Although \citet{Jones+2018} did also fit separate relations for early, late, and very late types, these relations were considerably more uncertain than the main relation, and \citet{Jones+2018} largely advised against their use. We note, however, that this may lead to a slight bias towards higher \hi \ deficiencies for groups dominated by early-types. However, in practice these are mostly Phase 3 groups that mostly have high \hi \ deficiencies, regardless of the exact relation used. In addition, as these groups are likely more evolved than those dominated by late-types, and as we are interested in the comparison of their original and current \hi \ contents, it is uncertain what morphological types should be used for them in the event that a morphology-dependent relation were to be used, further supporting the choice to ignore morphological type when estimating \hi \ deficiency.

We find a similar result to \citet{Verdes-Montenegro+2001}, that the mean deficiency is 0.34~dex. We also note that the median deficiency is a fairer estimate of what is typical of the population, both because \hi \ deficiency is a non-linear quantity and because it allows for the inclusion of upper limits from undetected groups. This value is 0.33~dex, almost identical to the mean value. 

This close agreement between the average \hi \ deficiency that we find and that found by \citet{Verdes-Montenegro+2001}, 0.40~dex, disguises some fundamental differences between the results of that work and this work. The first is that the \citet{Jones+2018} $L_{B}$--$M_\mathrm{HI}$ scaling relation provides a less biased prediction of the \hi \ mass of a galaxy than the equivalent relation from \citet{Haynes+Giovanelli1984}. As explained in that work, this is primarily driven by the improved fitting methodology employed by \citet{Jones+2018} and also, to a lesser extent, because of the larger and more morphologically diverse sample used. This results in predicted \hi \ masses being significantly lower (on average) for the same groups compared to \citet{Haynes+Giovanelli1984}, as shown in Figure \ref{fig:HIdef_HG84}. However, even our predicted \hi \ masses based on the \citet{Haynes+Giovanelli1984} differ from those of \citet{Verdes-Montenegro+2001}, likely a result of changes in group membership and the use of a corrected form of the \citet{Haynes+Giovanelli1984} relation \citep[see Appendix C of ][]{Jones+2018}. In addition, our measurements of the \hi \ mass of HCGs (after correcting for the different distance estimates used) are typically about 0.1~dex larger than those of \citet{Verdes-Montenegro+2001}. This is likely due to the fact that the single-dish observations of many HCGs had primary beams that were either too small to include all the \hi \ flux from the group, or the flux from the outlying regions of a group was down weighted by the declining beam response. Serendipitously, these effects appear to have approximately cancelled each other out, resulting in a very similar average \hi \ deficiency being calculated.

\section{Discussion}
\label{sec:discuss}

In this section we discuss how groups are classified based on their \hi \ and IR morphology. In addition we assess whether the hypothesis that \hi \ deficiency should act as a proxy for morphological phase is supported by the observations that we have presented.

\subsection{\hi \ morphological classification scheme}
\label{sec:classification}

\citet{Verdes-Montenegro+2001} first proposed an evolutionary sequence for HCGs based on their \hi \ morphology. One of the goals of this work is to compare that morphological sequence to the \hi \ deficiencies of HCGs. With the uniform analysis of all available VLA observations of HCGs we have now greatly expanded the sample for which this comparison is possible, while the \hi \ deficiency has been revised for all these groups based on a more self-consistent definition \citep{Jones+2018}, as described above. However, before proceeding with this comparison, we first make slight adjustments to the \citet{Verdes-Montenegro+2001} scheme based on our expanded findings of the \hi \ morphology of our sample of HCGs. 

The original scheme split groups up into three main categories, phases 1, 2, and 3. Phase 1 groups were those where the majority (70\% or more) of the detected \hi \ was in features associated with the galaxies themselves. Phase 2 groups were those where a significant fraction (30-60\%) of the detected \hi \ was in extended features. Finally, Phase 3 were those groups that were either undetected in \hi \ or whose \hi \ was predominantly (60\% or more) in extended features, rather than in the galaxies themselves. The prototypical examples of each phase are HCGs 88, 16, and 92, in order of increasing evolutionary phase.

The first modification is very minor, and is to set the thresholds for being classified a Phase 2 or 3 as 25\% and 75\% of the detected \hi \ in extended features, respectively. These are slight a decrease and increase compared to \citep{Verdes-Montenegro+2001}, respectively, and a reflection of the apparent breaks between these classes in our data. In particular, the morphology of groups like HCG~31 and 58 (Figures \ref{fig:HCG31_split_overlay} \& \ref{fig:HCG58_split_overlay}) seem to fit best into Phase 2, as they have both significant extended \hi \ features, but also clear concentration remaining in the galactic discs. However, with the new analysis presented here and the original classification boundaries, these would have been classified as Phase 3. We note that the exact location of these thresholds is somewhat subjective, and there are cases where a particular HCG is essentially on the threshold and could be classified in one of two ways (e.g. HCG~19 or 91).

The second modification is more significant and concerns distinguishing groups which are Phase 1 from those that are Phase 3. \citet{Verdes-Montenegro+2001} split the Phase 3 classification into two subcategories, 3a and 3b. In that scheme, 3a represented groups with either very high fractions of \hi \ emission in extended features or no \hi \ detection at all. We keep this subcategory essentially unchanged (except the threshold mentioned above). Phase 3b was used to denote groups that appeared to be evolving within a common \hi \ envelope. 
However, improved observations and analyses \citep[e.g.][]{Verdes-Montenegro+2002} have demonstrated that no groups (with the possible exception of HCG~49) convincingly fit this scenario, and thus the classification is no longer required. In this work we add a new designation, Phase 3c, to indicate groups that would be classified as Phase 1, but where only a single galaxy is detected in \hi. There are a large number of groups where only one member of the group is detected in \hi, for example, HCGs 15, 22, 33, 48, 56, 68, 90, 93, 97. The detected galaxy is always a late-type, while the remaining group members are typically dominated by early-types. In at least some, if not all, of these cases it is likely that this lone detection is a recent addition to the group, and effectively leads to a rejuvenation of the \hi \ content of the group. Therefore, these groups are likely in a late stage of their evolution, but have recently gained a new member, hence the classification as Phase 3. 

We note that the possibility of new members is not limited to Phase 3 groups. For example, \citet{Jones+2019} found that HCG~16 has a recent addition to the group that is partly responsible for its extreme \hi \ morphology (Figure \ref{fig:HCG16_split_overlay}). However, in most Phase 1 and 2 groups it would be difficult to identify new members as these groups already typically host several members with normal \hi \ reservoirs. Indeed in Phase 1 it is not clear if the concept of a new member even has much meaning, as the lack of tidal features implies that the galaxies only recently entered into a compact configuration. In the case of Phase 3 groups, however, the incorporation of a new \hi-bearing member is striking as these are clear outliers both in terms of their \hi \ content and morphology.

As noted by \citet{Martinez-Badenes+2012} the abundance of lenticular galaxies increases with \hi \ phase, with Phase 3a groups being $\sim$30\% lenticulars, compared to $\sim$10\% in Phase 2 groups. As has been argued previously \citep{Sulentic+2001,Verdes-Montenegro+2001,Bekki+2011} this increase in abundance of lenticulars is likely the result of spirals being stripped and the increase in their abundance from Phase 2 to 3 is an indication that the latter phase is more evolved. To estimate the abundance of lenticulars in Phase 3c we deducted one member per group (i.e. removing the assumed recent addition) before calculation the fraction, which resulted in a value of $\sim$30\%, supporting the idea that these are also evolved groups, just with one new (gas-rich) member.

The diffuse light seen in Phase 3c groups also supports the idea that these are evolved groups. HCG~90 is an extreme case where over a third of its total light is in a diffuse form \citep{White+2003}. In HCG~22 the DECaLS images show clear shells and arcs around HCG~22b, signs of past interactions, and potentially even more extended (and extremely faint) features to the NW. HCG~93a also shows an extraordinary complex of diffuse features, and signs of similar (though less impressive) diffuse features are evident in all Phase 3c groups with DECaLS imaging. We are therefore confident that these are all evolved systems that have simply gained a new, gas-rich member.

In summary, our classification scheme is as follows:
\begin{itemize}
    \item Phase 1: $f_\mathrm{exfs} < 0.25$ and $N_\mathrm{det} > 1$
    \item Phase 2: $0.25 < f_\mathrm{exfs} < 0.75$
    \item Phase 3a: $f_\mathrm{exfs} > 0.75$ and/or $N_\mathrm{det} = 0$
    \item Phase 3c: $f_\mathrm{exfs} < 0.25$ and $N_\mathrm{det} = 1$
\end{itemize}
The fraction of \hi \ emission in extended features, $f_\mathrm{exfs}$, is the fraction of the total \hi \ emission in each group that was assigned as an extended feature (e.g. a gas tail or bridge) in Section \ref{sec:sep_features}. $N_\mathrm{det}$ is the number of galaxies in each group that were detected in resolved \hi \ imaging.

Figure \ref{fig:ETG_fraction} shows the early-type fraction of group members against the \hi \ deficiency of each group. Although there is no simple correspondence between the two, we see that almost all groups classified as either Phase 3a or 3c have 50\% or more of their members as early-type galaxies, and the reverse is true for Phases 1 and 2, even though no explicit reference to the morphological type of members is made in the classification scheme above. In general the Phase 1 and 2 HCGs are towards the lower-left corner of this figure (low \hi \ deficiencies and few early-type members). The \hi \ deficiencies of the Phase 3 HCGs cover a broad range (extending to much higher deficiency values than the Phase 1 and 2 groups), but these groups are confined almost exclusively to the top half of the plot. This figure thus supports the notion that the \hi \ evolutionary sequence above does in some way correspond to the temporal evolution of HCGs under the assumption that interactions in these groups drive morphological change \citep[e.g.][]{Cluver+2013,Zucker+2016,Lisenfeld+2017}.

The classification of each group is written in Table \ref{tab:HIdef} and plotted in Figure \ref{fig:tidal_fraction}. Figure \ref{fig:HIdefhist} show the histogram of the \hi \ deficiencies of all HCGs in our sample and split into the three phases. We see that the Phase 1 groups and the Phase 2 groups have a very similar distribution of \hi \ deficiencies, with the typical value being just marginally deficient. The Phase 3 groups extend from this range to \hi \ deficiencies of about 1.5~dex, and two groups are entirely undetected and have only limits on their \hi \ deficiencies. These findings are discussed further in Section \ref{sec:HIevo}.

\subsection{Comparison with IR activity}

In addition to highlighting the morphological phase and \hi \ deficiency of HCGs, Figure \ref{fig:tidal_fraction} also shows the IR classification of \citet{Zucker+2016} for each HCG. Star symbols indicate groups where more than 50\% of the member galaxies are classified as IR active, crosses indicate groups with no dominant classification,  squares indicate groups where more than 50\% of members are classified as quiescent or `canyon' (IR transition) galaxies, and rings indicate groups where too many galaxies are missing IR classifications in \citet{Zucker+2016} for them to be conclusively assigned to any of the three other categories.

Curiously there is not a strong correspondence between the \hi \ morphology and \hi \ deficiency, and the IR classifications. The clearest trend is that there are almost no groups dominated by quiescent galaxies in Phase 2. There is some correlation between galaxies being \hi-rich and those that are IR active (Table \ref{tab:HIdef_mem}), but there are also numerous cases of active galaxies that are undetected in \hi, as well as quiescent or canyon galaxies that have relatively normal \hi \ reservoirs. We note that these results are somewhat at odds with \citet{Walker+2016} who find a stronger correlation between the $g-r$ colours of galaxies in compact groups and the global \hi \ content of the groups. However, the finding that there are several galaxies that appear to strongly deviate from the expected correlation between activity and gas content, is in common with that work. We also note that by defining \hi \ content in terms of $M_\mathrm{HI}/M_\ast$, rather than \hi \ deficiency, \citet{Walker+2016} likely incurred a considerable bias, as lower stellar mass galaxies in the field are expected to be more \hi-rich (by that metric) as well as later morphological type \citep[e.g.][]{Huang+2012}.

We find that the majority of the IR active groups are actually those classified as Phase 2 by \hi \ morphology, not those in Phase 1 (although nearly half of the Phase 1 groups are unclassified). This may indicate that the ongoing interactions that lead to large amounts of \hi \ being spread throughout the IGrM more or less ensure that the participating galaxies will be actively forming stars. However, it should also be noted that some of the Phase 1 groups with the richest \hi \ contents are not classified as active, for example, HCG~23. Thus, provided a group is not strongly \hi-deficient, the connection between \hi \ and IR classification appears to be more closely related to morphology of the group than to the exact quantity of neutral gas that is available. 

There are some groups that are relatively \hi-deficient (e.g. HCGs 7, 40) yet are still IR active. Both HCG~7 and HCG~40 contain a mixture of quiescent and active galaxies (though active galaxies dominate). In all cases the quiescent galaxy(ies) are undetected in \hi, immediately raising the \hi-deficiency of the group. However, in some cases the active galaxies are also \hi-deficient themselves. Although we caution against over interpreting the \hi-deficiencies of individual galaxies (which are only expected, in ideal circumstances, to be accurate to 0.2~dex), some of these groups may be in the process of losing their \hi \ gas (potentially accelerated by interactions) but have not yet lost their molecular gas. Thus, interactions are still able to promote elevated levels of star formation activity. Figure 1 of \citet{Lisenfeld+2017} shows that IR activity and molecular gas richness are highly correlated in HCGs. Furthermore, \citet{Martinez-Badenes+2012} argue that the ongoing tidal interactions in HCGs might also enhance the efficiency of the conversion from \hi \ to H$_2$. All of the active galaxies in HCGs 7 and 40 have significant molecular gas reservoirs, typically $M_\mathrm{H_2}/M_\ast > 0.1$, which is the threshold where galaxies typically appear to transition from active to quiescent \citep{Lisenfeld+2017}. \hi \ is generally much more loosely bound than molecular gas and is therefore lost first, meaning that \hi-poor galaxies may still be quite H$_2$-rich and actively forming stars. However, this is presumably an unsustainable situation as they will rapidly deplete or disperse their molecular gas reservoirs.

Phase 3 groups are mostly classified as quiescent (or unclassified), although there is one example of an IR active Phase 3 group, HCG~56, which is a rather peculiar group containing mostly lenticular galaxies. This finding that Phase 3c groups are often quiescent reinforces our conclusion that most groups with a sole \hi \ detection are evolved groups (regardless of the value of their \hi \ deficiency).

\subsection{Evidence for a diffuse \hi \ component}

\citet{Borthakur+2010} compared the spectra of HCGs from GBT and VLA observations and raised the possibility that, in some groups, flux missed by the VLA observations could be in the form of a diffuse \hi \ component of the IGrM. In this work we have re-reduced all of the VLA observations which \citet{Borthakur+2010} compared to and have attempted to recover as much (extended) flux as possible, using both multi-scale clean (which was not widely used when the original reduction was performed) for imaging and \texttt{SoFiA} for masking. As discussed in Section \ref{sec:fluxdiff} the vast majority of our VLA spectra agree remarkably well with the GBT spectra, with their slight discrepancies likely arising from small differences in calibration or minor (undetected) extensions of high column density features. In most groups there is thus little evidence for a significant diffuse \hi \ component of the IGrM.

Having said this, there are some notable exceptions. As mentioned in Section \ref{sec:fluxdiff}, HCGs 30, 37, and 97 all show significant additional \hi \ emission in their GBT spectra that does not trace the emission from high column density features detected in the VLA data. In the cases of HCGs 30 and 37 this additional emission manifests as a spectrally broad and continuous emission feature, where no emission is detected the VLA data (Figure \ref{fig:HCG30_spec} \& \ref{fig:HCG37_spec}). 

Based on the rms noise levels in the VLA cubes, this emission should be detectable in theory (at least in the spectral range where it is brightest). However, this is only true if the emission were spatially concentrated at the scale of the VLA synthesised beam, else the effective rms noise level would be higher, decreasing the significance of the emission feature. Hence the suggestion of \citet{Borthakur+2010} that this may indicate a diffuse component. It is even possible that the emission is so spatially extended that it could be unrecoverable with the VLA. Both HCGs 30 and 37 were observed in DnC configuration, where the largest recoverable scale is expected to be $\sim$16\arcmin. Given that this is larger than the maximum separation between the galaxies, it seems unlikely that this is the reason for the non-detection of this feature in the VLA data. However, if this were to be a limiting factor then the smaller minimum baselines of MeerKAT (20~m rather than 35~m) would be a means to resolve this issue. Furthermore, the increased sensitivity of MeerKAT, relative to the VLA, offers the current best possibility for detecting this `missing' emission with an interferometer and resolving its nature.

\subsection{Evolution of \hi \ content of HCGs}
\label{sec:HIevo}

As noted above there is little difference between the distribution of \hi \ deficiencies for Phase 1 and Phase 2 groups (Figure \ref{fig:HIdefhist}). The apparent lack of a shift demonstrates that the processes that remove \hi \ from the discs of galaxies and disperse it throughout the IGrM act on a timescale significantly shorter than the timescale on which \hi \ is destroyed via evaporation or consumed through SF. As has been found by numerous works \citep[e.g.][]{Zwaan+2005,Jones+2016,Janowiecki+2017,Said+2019} the \hi \ content of galaxies is a function of their larger scale environment, which may act as a partial explanation of the large scatter in \hi \ deficiencies seen across Phase 1 \& 2. This scatter is sufficiently large that even some Phase 3 groups are not distinct (in terms of their \hi \ deficiencies) from Phase 1 \& 2.  HCG~61 has over 90\% of its \hi \ emission in extended features, but has a negligible \hi \ deficiency. Even the prototypical example of a Phase 3 group, HCG~92, only has an \hi \ deficiency of 0.46~dex, which is equalled by a handful of Phase 1 and 2 groups. Although, there is a clear trend for most Phase 3 groups to be significantly more \hi-deficient.

However, the increased \hi \ deficiency of Phase 3 is not necessarily a one-way process. As indicated in Figure \ref{fig:tidal_fraction} by the pink arrow, it appears that Phase 3c groups were likely previously Phase 3a groups, devoid of \hi, but have gained a new gas-bearing member. In some cases that new member is so gas-rich that these groups attain relatively low \hi \ deficiencies again, and would be difficult to distinguish from Phase 1 groups if it were not for the fact that they are dominated by early-type galaxies and only one galaxy contains gas. However, we reiterate here that the morphological type of the galaxies is not used to define Phase 3c, which is based solely on the \hi \ morphology of each group.

In some exceptional cases it appears these groups can undergo interactions with the newcomer galaxy, such that they effectively re-enter Phase 2 and appear to go around the cycle again. An example is HCG~79, which is classified as Phase 2 owing to 42\% of its \hi \ being in extended features. However, as can be seen in Figure \ref{fig:HCG79_split_overlay}, most of this \hi \ emission clearly originated from HCG~79d, the only late-type in the group. We note that some of the extended features in this case are close to the noise level of the data, and that there is a notable difference between the GBT spectrum and that of the VLA (Figure \ref{fig:HCG79_spec}), perhaps indicating there could be further extended features associated with HCG~79a. However, it regardless remains an evolved group that appears to have gained a new gas-rich member and now has a large fraction of its gas in extended features, analogous to a Phase 2 group.

This new step in the morphological sequence (Phase 3c) complicates the assumption that \hi \ deficiency would increase with increasing morphological phase. This hypothesis implicitly assumes that groups cannot regain neutral gas after they have become significantly evolved. However, the larger number of groups dominated by early-type galaxies, but with a single gas-rich, late-type member casts serious doubt on this assumption. Though less likely, it is even possible that some groups containing a mixture of early and late-type galaxies (e.g. HCGs 10 \& 25) represent the merger of two groups, one gas-rich and one evolved and gas-poor. In our scheme, such objects would likely be classified as either Phase 1 or Phase 2. Regardless of whether this more contrived scenario occurs frequently, the finding that the \hi \ deficiency of an individual compact group does not necessarily monotonically increase with time means that, in general, it cannot be used as a proxy for evolutionary phase. Even with the Phase 3c groups excluded, \hi \ deficiency is not a useful proxy as there appears to be more scatter than evolution in \hi \ deficiency values across the remaining phases. Only the most \hi-deficient Phase 3a groups would be distinguishable with such a proxy.

\section{Conclusions}
\label{sec:conclusions}

We have reduced and analysed archival VLA \hi \ observations of 38 HCGs and determined their \hi \ morphological phase based on an adaptation of the \citet{Verdes-Montenegro+2001} evolutionary sequence. As this sequence is thought to represent the temporal evolution of HCGs it was expected that \hi \ deficiency would act as a proxy for the evolutionary phase of a HCG, as gas is consumed or destroyed as the group evolves. However, we find that \hi \ deficiency is a very poor proxy for the \hi \ morphology of HCGs, with the exception that most Phase 3 groups are significantly deficient in \hi.

The reason for this poor correspondence appears to be two-fold. Firstly, there is very little difference between the distribution of \hi \ deficiency of Phase 1 and Phase 2 groups. This suggests that the initial \hi \ content of groups is the primary factor determining their \hi \ deficiencies in these two phases, and that the enormous morphological changes that occur between Phases 1 and 2 proceed on a timescale that is too short for a significant quantity of \hi \ to be either consumed or destroyed. Secondly, the hypothesis that \hi \ deficiency should be a proxy for \hi \ morphology relies on the assumption that gas replenishment is largely negligible. However, we find a large fraction of the HCGs, about 25\%, appear to be evolved groups dominated by early-type galaxies with a lone, late-type, gas-rich member, presumably a newcomer. In this way, once highly \hi-deficient groups can have their gas content rejuvenated, confounding the use of \hi \ deficiency as a proxy for evolutionary phase. We add a new sub-phase to the \citet{Verdes-Montenegro+2001} evolutionary sequence to classify such groups. 

We also find that there is not a clear one-to-one correspondence between the \hi \ content of HCGs and the IR activity of their galaxies. However, most groups that are dominated by active galaxies are in Phase 2 of the \hi \ evolutionary sequence, with ongoing interactions likely driving the ubiquitous IR activity. Phase 3 are likely to be dominated by passive galaxies, as is expected for these gas-poor groups. However, Phase 1 groups are a mixture of active, intermediate, and even quiescent IR classifications, as well as several groups with missing classifications.

Finally, we searched for evidence of a potential diffuse \hi \ component in HCGs, as proposed by \citet{Borthakur+2010}. While some cases appear less compelling with our revised reduction of the VLA observations, the discrepancy between the spectra of HCGs 30 and 37 in VLA versus GBT observations remain difficult to explain without the presence of a diffuse component. Deeper interferometric observations with improved surface brightness sensitivity and $uv$-coverage might be capable of revealing this component and conclusively demonstrating its reality.

\begin{acknowledgements}
We thank the anonymous referee for their helpful comments that improved this paper. MGJ thanks the NRAO helpdesk for their rapid and insightful support, in particular with some of the oldest and most problematic VLA datasets. 
MGJ was supported by a Juan de la Cierva formaci\'{o}n fellowship (FJCI-2016-29685) from the Spanish Ministerio de
Ciencia, Innovaci\'{o}n y Universidades (MCIU) during much this work. We also acknowledge support from the grants AYA2015-65973-C3-1-R (MINECO/FEDER, UE) and RTI2018-096228-B-C31 (MCIU), and from the grant IAA4SKA (Ref. R18-RT-3082) from the Economic Transformation, Industry, Knowledge and Universities Council of the Regional Government of Andalusia and the European Regional Development Fund from the European Union. This work has been supported by the State Agency for Research of the Spanish MCIU ``Centro de Excelencia Severo Ochoa'' programme under grant SEV-2017-0709.
This work used the following \texttt{Python} libraries and packages: \texttt{numpy} \citep{numpy}, \texttt{scipy} \citep{scipy}, \texttt{matplotlib} \citep{matplotlib}, \texttt{astropy} \citep{astropy}, \texttt{aplpy} \citep{aplpy}, \texttt{astroquery} \citep{astroquery}, \texttt{jupyter} \citep{jupyter}, \texttt{ruffus} \citep{ruffus}, \texttt{CGAT-core} \citep{cgatcore}, and \texttt{Anaconda} (\url{https://anaconda.com}). 
The National Radio Astronomy Observatory is a facility of the National Science Foundation operated under cooperative agreement by Associated Universities, Inc.
The Legacy Surveys consist of three individual and complementary projects: the Dark Energy Camera Legacy Survey (DECaLS; Proposal ID 2014B-0404; PIs: David Schlegel and Arjun Dey), the Beijing-Arizona Sky Survey (BASS; NOAO Prop. ID 2015A-0801; PIs: Zhou Xu and Xiaohui Fan), and the Mayall z-band Legacy Survey (MzLS; Prop. ID 2016A-0453; PI: Arjun Dey). This research has made use of the NASA/IPAC Extragalactic Database, which is funded by the National Aeronautics and Space Administration and operated by the California Institute of Technology. This research has made use of the VizieR catalogue access tool, CDS, Strasbourg, France. The original description of the VizieR service was
published in \citet{vizier}. This research has made use of the NASA/IPAC Infrared Science Archive, which is funded by the National Aeronautics and Space Administration and operated by the California Institute of Technology. Funding for the SDSS and SDSS-II has been provided by the Alfred P. Sloan Foundation, the Participating Institutions, the National Science Foundation, the U.S. Department of Energy, the National Aeronautics and Space Administration, the Japanese Monbukagakusho, the Max Planck Society, and the Higher Education Funding Council for England. The SDSS Web Site is \url{http://www.sdss.org/}. 
\end{acknowledgements}

\bibliographystyle{aa} 
\bibliography{refs.bib} 

\appendix

\section{Discrepancies with previously published VLA data}

There are some other notable differences between the VLA data presented here and in \citet{Borthakur+2010}. In particular, the spectra of HCGs 15, 23, and 31 all differ significantly. HCG~15 has already been discussed in Section \ref{sec:fluxdiff}. HCG~23 is essentially undetected in the VLA data presented in \citet{Borthakur+2010} and the bandwidth shown is too narrow to include most of the group's emission. Whereas here we find strong detections of four group members and a close match with the GBT spectrum. It is unclear why this difference exists as the observations for this group were completed in 1990 and thus the same data were presumably used in both cases. In the case of HCG~31 the difference is primarily in the flux scale. \citet{Borthakur+2010} found that this group was missing a significant amount of \hi \ emission in its VLA spectrum, however, we find a very close match. We took no special steps in the reduction of this group and used the standard flux calibration models in \texttt{CASA}. There was likely an error in the calibration of these data in their original reduction, but at this point it is difficult to know where this occurred. 

The final spectrum of HCG~58 is also quite different in the two works, with the \citet{Borthakur+2010} spectrum matching somewhat better to the GBT spectrum. In this case it is likely that it is the exact form of the correction for the GBT beam response that is most responsible for the difference as much of the flux detected in the VLA cube is near or beyond the GBT beam, resulting in enormous correction factors.

\section{Reproducibility}
\label{app:repo}

In addition to the analysis of the gas content of HCGs, we have followed best practices to support the reproducibility of the software methods used in this study. In this section we provide an overview of how we have tried to accomplish this, where barriers were met, and some lessons learnt.

\subsection{Discussion of our approach}

For our pilot study of HCG~16 \citep{Jones+2019} we constructed a full end-to-end pipeline (\url{https://github.com/AMIGA-IAA/hcg-16}), with software containers to preserve the exact software environment (in addition to the scripts themselves) in order to maximise the longevity of the pipeline. In this case, because of the larger scale of the current project, the increased volume of data, and the computation time it would take to reprocess all steps, we have instead aimed to create a flexible pipeline for processing the data of individual groups or VLA observing projects. This allowed the parameters of the data reduction to be tuned (e.g. for flagging, continuum subtraction, and imaging) for each data set and group, and the processing of each data set to be repeated and modified independently. This approach has clear advantages, but it also means that modifying the data reduction is a more manual process, as the pipeline must be executed separately for each VLA project or each HCG. In addition, the choice not to run the pipeline scripts in containers aides short term simplicity of execution, but may results in poorer long term preservation. 

Our pipeline for processing the VLA \hi \ data is available at \url{https://github.com/AMIGA-IAA/hcg_hi_pipeline}. The parameters and log file for our actual excution of the pipeline are stored in a separate \texttt{github} repository, \url{https://github.com/AMIGA-IAA/hcg_global_hi_analysis}. Instructions for the installation and use of the resources are described in the repositories themselves. In addition, the second repository includes \texttt{Python} scripts and \texttt{Jupyter} notebooks to recreate all the plots and figures in this paper. These are based on our reduction of the data, and we have therefore also included all the final \hi \ data cubes, moment zero maps, and cubelets\footnote{It show be noted that a bug in \texttt{SlicerAstro} produces a single bad pixel at the origin of each cubelet, which our scripts correct for.} of separated features in a \texttt{Zenodo} repository, \url{https://doi.org/10.5281/zenodo.6366659}. The analysis repository also contains links to the specific versions of \texttt{CASA} and \texttt{SoFiA} that were used to reduce and analyse the data.

Unfortunately, one step of the analysis process remains not fully reproducible, which is the separation of features using \texttt{SlicerAstro}. This separation is a necessarily manual and subjective process that is not straightforward to fully record in our pipeline. In the other steps of data reduction and analysis we have endeavoured to provide all the relevant parameters files such that another astronomer could modify and re-run these steps in a relatively approachable manner. However, for the separation of features step we have instead opted for preservation and exact reproduction, rather than full, independent reproduciblity. We therefore note that even if another astronomer were to repeat (and modify) all the steps of our pipeline and then run our \texttt{Jupyter} notebooks, their modifications would only be carried forward for the properties of the groups that do not rely on the separated features cubelets. In general these are the global properties, such as the total \hi \ mass and deficiency of each group, not those that depend on individual objects, such as the fraction of emission in extended features or the \hi \ masses of individual galaxies.

All of the VLA data presented in this work are publicly available in the VLA archive. However, this archive cannot be automatically queried for data downloads, and thus the data for any group must be downloaded manually. 
The logs for the reduction of each data set include the names of all files that are imported so that the exact files can be obtained from the VLA archive, thus allowing anyone to repeat and modify our reduction (using our pipeline, or otherwise) from the beginning. In general, we anticipate end users only downloading and reprocessing data for the individual groups that they are interested in. Therefore, it does not make sense to re-host the entire data set \citep[as we did for HCG~16,][]{Jones+2019} to allow automated access. In addition, creating additional, distributed copies of an archive is inefficient and impractical given the potential volume of data (for a generic VLA project). However, we point out that a more straightforward means to identify and locate individual VLA archive files, such as a digital object identifier, would streamline the process of (re-)acquiring the raw data.

For those who do not wish to repeat all the reduction, we advise downloading the reduced data products from the \texttt{Zenodo} repository. We have also included the relevant GBT \hi \ spectra of HCGs from \citet{Borthakur+2010}, as the digital versions of these spectra have not been publicly released previously. Finally, we have included the optical images from DECaLS, SDSS, and POSS that we use. Although these can all be obtained from public services, because of the small volume of these data and because some services require manual interaction, we have chosen to re-host them. These data products can all be used with the analysis \texttt{Jupyter} notebooks to reproduce or modify any of the figures or values presented in the paper. In our prior work with HCG~16 we elected to enable cloud-execution of these notebooks using the \texttt{Binder} service (\url{https://mybinder.org/}). Although, the current analysis notebooks can also be executed using \texttt{Binder}, and we do recommend attempting this for those simply wishing to reproduce a few figures or a table, we also note that given the volume of data the \texttt{Binder} container may fail to build or launch and that the success or failure may depend on the end user's location as well as the current load on the infrastructure where \texttt{Binder} attempts to allocate resources.

When generating plots and tables, where possible we have generally tried to avoid using locally stored data that another user would have to locate, download, and extract manually. Instead we have relied heavily on \texttt{astroquery} \citep{astroquery} to query data tables stored at the Centre de Donn\'{e}es astronomiques de Strasbourg (CDS, \url{https://cds.u-strasbg.fr/}). In certain cases values were not available from CDS and we instead queried the NASA/IPAC Extragalactic Database (NED, \url{https://ned.ipac.caltech.edu/}). In the extremely unusual cases where values were not available from either service they were added manually from other sources, and these additions are preserved explicitly in the \texttt{Jupyter} notebooks. 

Making use of querying services such as CDS and NED is very convenient and avoids the issue of each paper needing to replicate every data table on which they rely in order to be reproducible. However, it also raises a separate issue, which is that if any of the tables we query were to be amended or corrected then the final values we present could be altered. While this may be the desired functionality in some cases, it is also valid to want to preserve the exact form of the analysis. In practice, given the age of the data tables that we query, they are extremely unlikely to be updated. However, it would be an improvement to be able to specify whether to query from a specific version of a table (e.g. corresponding to a version on a particular date) or to simply query the latest version. As far as we are aware this option does not currently exist, but it would allow for an end user to select either functionality, depending on their aims.

To allow for either functionality, we have included the final data tables that our \texttt{Jupyter} notebooks produce (after querying CDS and NED) in the \texttt{Zenodo} repository. If the CDS/NED tables are amended, then these fixed versions of our tables can be used instead of regenerating them with the analysis notebooks. Thereby preserving the original form of the analysis.

\subsection{Outlook}

Our efforts to maximise the reproducibility of this work have taught us certain lessons regarding the storage and processing of astronomical data. We do not claim to be the first to discuss any of these issues \citep[see e.g.][]{Wilkinson+2016}, instead mostly these are lessons re-learnt or reinforced, rather than being genuinely novel.

One of the least streamlined steps in our reproducibility framework is actually the first, the retrieval of the raw data from the VLA archive. This is mostly because this archive cannot be queried automatically. As we say above, this design choice is understandable due to the considerable data volume that VLA observations can represent. However, this issue is only likely to get worse in the future as new facilities, such as Square Kilometre Array (SKA) precursors (and soon the SKA itself) are already greatly increasing the volume of this type of data. The solution to this problem, which has been discussed many times before, is to stop moving the data (via the internet or on hard disks) to the home institution of each end user, but instead to have them process it in a central computing facility that also hosts the data archive. In our case this would prevent the disconnect arising from the raw data being stored in a completely different system from where they are processed. It would likely also make sense for this same facility to act as an archive for the final data products that the end users produce when using the computing resources to reduce and analyse their data. This is essentially a (rudimentary) description of the proposed structure of SKA Regional Centres (e.g. \url{https://aussrc.org/wp-content/uploads/2021/05/SRC-White-Paper-v1.0-Final.pdf}). 

This does not, however, address the issue of additional external data that is required for certain analyses. The need for such data typically arises when performing multi-frequency analyses, as observations at different wavelengths are often stored in different archives and can follow different standards. In our case, such data is mostly in the form of optical images from large sky surveys (POSS, SDSS, and DECaLS), as well as data tables that we query from CDS and NED. The need for such multi-frequency comparisons and analyses mean that the transfer of some data is inevitable. However, typically the external data will already be reduced and therefore be much smaller in volume than in its raw form, and thus represent less of an issue. If the data portals for data at all wavelengths have automatically accessible interfaces and follow common standards (such as those described by the International Virtual Observatory Alliance) then the inclusion of external data resources would be further streamlined.

Unlike in our previous work with HCG~16 \citep{Jones+2019}, we have not constructed containers for the software. \texttt{SoFiA} is publicly available, with its version history, through \texttt{github} and we therefore consider this to be a well-preserved resource that will likely be possible to build and install for several years. Similarly, although \texttt{CASA} is not available through \texttt{github}, the National Radio Astronomy Observatory (NRAO) provide back dated versions of \texttt{CASA} going back almost a decade. Aside from these software packages (and \texttt{SlicerAstro}, which we return to below) our analysis was performed exclusively in \texttt{Python}. We make use of \texttt{conda} as a means to preserve and build the coding environment such that the results can be reproduced exactly. While these solution are all adequate and suitable for our use case, software and hardware will always continue to change over time, and at present the best means of preserving software environments is likely through containers and virtual machines (VMs). However, for each project to build their own containers and VMs for all their software is a formidable task. These are some community projects to assist with this, for example the \texttt{Kern} suite \citep{Kern} project in radio astronomy, however, ultimately once data reduction takes place predominantly in the cloud it will likely become the role of computing centres to maintain such software containers and backwards compatibility. This will make achieving reproducibility (at least of the software environment) more straightforward for the end users.

Finally, we return to the issue that the separation of features (in \texttt{SlicerAstro}) is not as fully reproducible as other steps in our pipeline because it requires manual (and subjective) input. This means that a new data product, created by modifying earlier data reduction steps, cannot automatically be propagated through the separation of features steps, and thus the results of any modifications can only be included automatically in some, but not all, of the final output of our analysis. Although in this particular scenario it is conceivable that in the future some algorithm maybe be developed to perform an equivalent separation of \hi \ features automatically, in research in general there will always be manual and subjective steps which cannot be fully included in an automated pipeline. In such cases we suggest that the best way forward is to strive for preservation and exact duplication, rather than full independent, automated reproducibility. This means storing the data products before and after such steps, along with any logs and important intermediate data products that may be manually produced during the process. Although this does not allow another researcher to interact with and modify the original reduction and analysis process, it does allow them to reliably compare to the results of each step in the process.

In summary, in our opinion the means to maximising the reproducibility of future radio astronomy projects (and in other domains) is for data, software, and computing resources to be provided in a central location which a user can access through a unified portal. As users would develop their own analysis scripts and tools in this environment, integrating version control tools and standard astronomy tools and services into the platform would help to maximise its usability. Crucially such a resource would remove much of the burden of ``recreating the wheel" that currently falls on individual teams, thus making reproducibility a much more readily achievable goal for most projects.

\section{Individual group members}
\label{app:mem_tab}

Table \ref{tab:HIdef_mem} shows the predicted and observed \hi \ masses on individual galaxies in each HCG in our sample, as well as their IR classifications from \citet{Zucker+2016}.

\begin{table*}
\centering
\caption{\hi \ deficiency of individual HCG members}
\label{tab:HIdef_mem}
\begin{tabular}{cccccccccccc}
\hline \hline
HCG & Name & RA & Dec & Type & IR class & $cz_\odot$ & $m_\mathrm{B,c}$ & $\log M_\mathrm{HI,pred}$ & $\log M_\mathrm{HI}$ & \hi-def \\
 &  & $\mathrm{deg}$ & $\mathrm{deg}$ & & & $\mathrm{km\,s^{-1}}$ & $\mathrm{mag}$ & $[\mathrm{M_\odot}]$ & $[\mathrm{M_\odot}]$ & dex \\ \hline
2 & HCG2a & 7.84955 & 8.46822 & SBd & Active & 4326 & $13.35 \pm 0.1$ & $9.59 \pm 0.2$ & 9.93 & -0.34 \\
2 & HCG2b & 7.82825 & 8.47518 & cI & Active & 4366 & $14.39 \pm 0.1$ & $9.2 \pm 0.2$ & 8.9 & 0.3 \\
2 & HCG2c & 7.87232 & 8.40068 & SBc & Active & 4235 & $14.15 \pm 0.1$ & $9.29 \pm 0.2$ & 9.36 & -0.07 \\
7 & HCG7a & 9.8065 & 0.86363 & Sb & Active & 4210 & $12.98 \pm 0.1$ & $9.73 \pm 0.2$ & 9.05 & 0.68 \\
7 & HCG7b & 9.82517 & 0.91273 & SB0 & Quiescent & 4238 & $13.74 \pm 0.2$ & $9.44 \pm 0.21$ &  &  \\
7 & HCG7c & 9.89604 & 0.85962 & SBc & Active & 4366 & $12.6 \pm 0.7$ & $9.87 \pm 0.33$ & 9.53 & 0.34 \\
7 & HCG7d & 9.82911 & 0.89148 & SBc & Active & 4116 & $14.77 \pm 0.2$ & $9.06 \pm 0.21$ & 8.9 & 0.16 \\
10 & HCG10a & 21.58943 & 34.70204 & SBb & Quiescent & 5148 & $12.62 \pm 0.1$ & $9.86 \pm 0.2$ & 9.8 & 0.06 \\
10 & HCG10b & 21.41817 & 34.71284 & E1 & Quiescent & 4862 & $12.7 \pm 0.1$ & $9.83 \pm 0.2$ &  &  \\
10 & HCG10c & 21.57849 & 34.75395 & Sc & Active & 4660 & $14.07 \pm 0.1$ & $9.32 \pm 0.2$ &  &  \\
10 & HCG10d & 21.62865 & 34.67508 & Scd & Active & 4620 & $14.69 \pm 0.2$ & $9.09 \pm 0.21$ & 8.83 & 0.26 \\
15 & HCG15a & 31.97094 & 2.16759 & Sa & Quiescent & 6967 & $14.29 \pm 0.2$ & $9.77 \pm 0.21$ &  &  \\
15 & HCG15b & 31.89217 & 2.11521 & E0 &  & 7117 & $14.74 \pm 0.1$ & $9.6 \pm 0.2$ &  &  \\
15 & HCG15c & 31.91572 & 2.14973 & E0 & Quiescent & 7222 & $14.37 \pm 0.1$ & $9.74 \pm 0.2$ &  &  \\
15 & HCG15d & 31.90634 & 2.18078 & E2 & Quiescent & 6244 & $14.65 \pm 0.2$ & $9.63 \pm 0.21$ &  &  \\
15 & HCG15e & 31.85567 & 2.11613 & Sa & Quiescent & 7197 & $15.56 \pm 0.1$ & $9.29 \pm 0.2$ &  &  \\
15 & HCG15f & 31.90778 & 2.19028 & Sbc & Active & 6242 & $15.74 \pm 0.2$ & $9.22 \pm 0.21$ & 9.09 & 0.13 \\
16 & NGC848 & 32.5735 & -10.32145 & SBab &  & 3992 & $13.37 \pm 0.1$ & $9.57 \pm 0.2$ & 9.64 & -0.07 \\
16 & HCG16a & 32.35312 & -10.13622 & SBab & Active & 4152 & $12.76 \pm 0.15$ & $9.79 \pm 0.21$ & 8.97 & 0.82 \\
16 & HCG16b & 32.3361 & -10.13309 & Sab & Quiescent & 3977 & $13.27 \pm 0.15$ & $9.6 \pm 0.21$ & 8.66 & 0.94 \\
16 & HCG16c & 32.41071 & -10.14637 & Im & Active & 3851 & $13.1 \pm 0.08$ & $9.67 \pm 0.2$ & 9.46 & 0.21 \\
16 & HCG16d & 32.42872 & -10.18394 & Im & Active & 3847 & $13.42 \pm 0.09$ & $9.55 \pm 0.2$ & 9.64 & -0.09 \\
19 & HCG19a & 40.65998 & -12.42128 & E2 & Quiescent & 4279 & $14.0 \pm 0.2$ & $9.39 \pm 0.21$ &  &  \\
19 & HCG19b & 40.67548 & -12.42777 & Scd & Active & 4210 & $15.24 \pm 0.2$ & $8.93 \pm 0.21$ & 9.17 & -0.24 \\
19 & HCG19c & 40.69507 & -12.39783 & Sdm & Canyon & 4253 & $14.46 \pm 0.1$ & $9.22 \pm 0.2$ & 8.62 & 0.6 \\
22 & NGC1188 & 45.93135 & -15.48522 & S0 &  & 2628 & $14.32 \pm 0.39$ & $8.98 \pm 0.25$ &  &  \\
22 & HCG22a & 45.91056 & -15.61369 & E2 & Quiescent & 2705 & $12.24 \pm 0.13$ & $9.76 \pm 0.21$ &  &  \\
22 & HCG22b & 45.8593 & -15.66187 & Sa &  & 2625 & $14.47 \pm 0.1$ & $8.92 \pm 0.2$ &  &  \\
22 & HCG22c & 45.85141 & -15.62332 & SBcd & Canyon & 2544 & $13.9 \pm 0.7$ & $9.14 \pm 0.33$ & 9.41 & -0.27 \\
23 & HCG23-26 & 46.78954 & -9.48784 & cI &  & 5283 & $19.26 \pm 0.5$ & $7.58 \pm 0.27$ & 9.22 & -1.64 \\
23 & HCG23a & 46.73285 & -9.54406 & Sab & Quiescent & 4798 & $14.32 \pm 0.2$ & $9.44 \pm 0.21$ & 9.28 & 0.16 \\
23 & HCG23b & 46.78976 & -9.59332 & SBc & Active & 4921 & $14.42 \pm 0.2$ & $9.4 \pm 0.21$ & 9.74 & -0.34 \\
23 & HCG23c & 46.82692 & -9.61328 & S0 & Quiescent & 5016 & $15.52 \pm 0.2$ & $8.99 \pm 0.21$ &  &  \\
23 & HCG23d & 46.73012 & -9.62953 & Sd & Active & 4562 & $16.0 \pm 0.5$ & $8.81 \pm 0.27$ & 9.31 & -0.5 \\
25 & HCG25a & 50.17978 & -1.10935 & SBc & Active & 6285 & $13.86 \pm 0.1$ & $9.8 \pm 0.2$ & 9.95 & -0.15 \\
25 & HCG25b & 50.18997 & -1.04486 & SBa & Quiescent & 6408 & $14.45 \pm 0.2$ & $9.58 \pm 0.21$ & 9.55 & 0.03 \\
25 & HCG25d & 50.16172 & -1.0354 & S0 &  & 6401 & $15.92 \pm 0.1$ & $9.03 \pm 0.2$ &  &  \\
25 & HCG25f & 50.18977 & -1.05425 & S0 &  & 6279 & $16.98 \pm 0.2$ & $8.63 \pm 0.21$ &  &  \\
26 & HCG26a & 50.4804 & -13.65085 & Scd & Active & 9678 & $16.1 \pm 0.7$ & $9.34 \pm 0.33$ & 10.17 & -0.83 \\
26 & HCG26b & 50.48811 & -13.64841 & E0 &  & 9332 & $15.61 \pm 0.2$ & $9.52 \pm 0.21$ &  &  \\
26 & HCG26c & 50.45648 & -13.64554 & S0 &  & 9618 & $17.1 \pm 0.1$ & $8.96 \pm 0.2$ &  &  \\
26 & HCG26d & 50.48558 & -13.64543 & cI &  & 9133 & $15.81 \pm 0.2$ & $9.44 \pm 0.21$ &  &  \\
26 & HCG26e & 50.46302 & -13.66473 & Im & Active & 9623 & $17.05 \pm 0.1$ & $8.98 \pm 0.2$ & 8.88 & 0.1 \\
26 & HCG26f & 50.48992 & -13.66392 & cI &  & 9626 & $18.68 \pm 0.1$ & $8.37 \pm 0.2$ &  &  \\
26 & HCG26g & 50.48012 & -13.64874 & S0 &  & 9293 & $17.4 \pm 0.7$ & $8.85 \pm 0.33$ &  &  \\
30 & HCG30a & 69.07744 & -2.83129 & SB- & Quiescent & 4697 & $12.87 \pm 0.2$ & $9.93 \pm 0.21$ &  &  \\
30 & HCG30b & 69.12619 & -2.86656 & Sa & Quiescent & 4625 & $13.65 \pm 0.1$ & $9.64 \pm 0.2$ &  &  \\
30 & HCG30c & 69.097 & -2.79985 & SBbc & Active & 4508 & $15.06 \pm 0.1$ & $9.11 \pm 0.2$ &  &  \\
30 & HCG30d & 69.15276 & -2.84302 & S0 &  & 4666 & $15.69 \pm 0.1$ & $8.87 \pm 0.2$ &  &  \\
31 & HCG31g & 75.43338 & -4.28875 & cI &  & 4011 & $15.11 \pm 0.5$ & $8.98 \pm 0.27$ & 9.0 & -0.02 \\
31 & HCG31q & 75.40974 & -4.22245 & cI &  & 4090 & $16.01 \pm 0.5$ & $8.64 \pm 0.27$ & 8.71 & -0.07 \\
31 & HCG31a & 75.41146 & -4.25946 & Sdm &  & 4042 & $14.83 \pm 0.2$ & $9.08 \pm 0.21$ & 9.22 & -0.14 \\
31 & HCG31b & 75.39756 & -4.26401 & Sm &  & 4171 & $14.31 \pm 0.2$ & $9.28 \pm 0.21$ & 9.13 & 0.15 \\
31 & HCG31c & 75.40751 & -4.2577 & Im & Active & 4068 & $12.5 \pm 0.5$ & $9.96 \pm 0.27$ & 8.96 & 1.0 \\
\hline
\end{tabular}
\tablefoot{Columns: (1) HCG ID number, (2) name of group member, (3) right ascension, (4) declination, (5) optical morphological Hubble type \citep{Hickson+1989}, (6) infrared classification \citep{Zucker+2016}, (7) redshift \citep{Hickson+1992}, (8) corrected apparent B-band magnitude \citep{Hickson+1989}, (9) predicted \hi \ mass based on B-band luminosity \citep{Jones+2018}, (10) measured \hi \ mass from VLA observations, (11) \hi \ deficiency.}
\end{table*}
\newpage

\begin{table*}[h!] \nonumber
\centering
Table C.1 continued.
\begin{tabular}{cccccccccccc}
\hline \hline
HCG & Name & RA & Dec & Type & IR class & $cz_\odot$ & $m_\mathrm{B,c}$ & $\log M_\mathrm{HI,pred}$ & $\log M_\mathrm{HI}$ & \hi-def \\
 &  & $\mathrm{deg}$ & $\mathrm{deg}$ & & & $\mathrm{km\,s^{-1}}$ & $\mathrm{mag}$ & $[\mathrm{M_\odot}]$ & $[\mathrm{M_\odot}]$ & dex \\ \hline
33 & HCG33a & 77.69998 & 18.01867 & E1 & Quiescent & 7570 & $15.35 \pm 0.2$ & $9.46 \pm 0.21$ &  &  \\
33 & HCG33b & 77.69854 & 18.02957 & E4 & Quiescent & 8006 & $15.41 \pm 0.2$ & $9.44 \pm 0.21$ &  &  \\
33 & HCG33c & 77.68831 & 18.01956 & Sd & Active & 7823 & $16.4 \pm 0.7$ & $9.06 \pm 0.33$ & 10.02 & -0.96 \\
33 & HCG33d & 77.72327 & 18.03292 & E0 &  & 7767 & $16.73 \pm 0.2$ & $8.94 \pm 0.21$ &  &  \\
37 & HCG37a & 138.41438 & 29.99243 & E7 & Quiescent & 6745 & $12.97 \pm 0.2$ & $10.27 \pm 0.21$ &  &  \\
37 & HCG37b & 138.38589 & 29.99977 & Sbc & Canyon & 6741 & $14.5 \pm 0.2$ & $9.7 \pm 0.21$ &  &  \\
37 & HCG37c & 138.40514 & 29.99954 & S0a &  & 7357 & $15.57 \pm 0.2$ & $9.3 \pm 0.21$ &  &  \\
37 & HCG37d & 138.39107 & 30.01442 & SBdm & Active & 6207 & $15.87 \pm 0.2$ & $9.18 \pm 0.21$ &  &  \\
37 & HCG37e & 138.3918 & 30.03975 & E0 & Active & 6363 & $16.21 \pm 0.1$ & $9.06 \pm 0.2$ &  &  \\
38 & HCG38a & 141.89442 & 12.26922 & Sbc & Active & 8652 & $15.25 \pm 0.1$ & $9.61 \pm 0.2$ &  &  \\
38 & HCG38b & 141.93169 & 12.28719 & SBd & Active & 8635 & $14.76 \pm 0.2$ & $9.79 \pm 0.21$ &  &  \\
38 & HCG38c & 141.93544 & 12.2879 & Im & Active & 8692 & $15.39 \pm 0.2$ & $9.56 \pm 0.21$ &  &  \\
40 & HCG40a & 144.72309 & -4.84903 & E3 & Quiescent & 6628 & $13.44 \pm 0.2$ & $10.11 \pm 0.21$ &  &  \\
40 & HCG40b & 144.72942 & -4.86608 & S0 & Quiescent & 6842 & $14.58 \pm 0.2$ & $9.68 \pm 0.21$ &  &  \\
40 & HCG40c & 144.72173 & -4.85936 & Sbc & Active & 6890 & $15.15 \pm 0.2$ & $9.47 \pm 0.21$ & 9.29 & 0.18 \\
40 & HCG40d & 144.73246 & -4.83737 & SBa & Active & 6492 & $14.53 \pm 0.2$ & $9.7 \pm 0.21$ & 9.09 & 0.61 \\
40 & HCG40e & 144.73111 & -4.85775 & Sc & Active & 6625 & $16.69 \pm 0.2$ & $8.89 \pm 0.21$ &  &  \\
47 & HCG47a & 156.4431 & 13.71686 & SBb & Active & 9581 & $14.61 \pm 0.2$ & $9.93 \pm 0.21$ &  &  \\
47 & HCG47b & 156.4527 & 13.72793 & E3 &  & 9487 & $15.67 \pm 0.2$ & $9.53 \pm 0.21$ &  &  \\
47 & HCG47c & 156.45454 & 13.75306 & Sc & Active & 9529 & $16.63 \pm 0.2$ & $9.17 \pm 0.21$ &  &  \\
47 & HCG47d & 156.4495 & 13.74866 & Sd & Active & 9471 & $16.2 \pm 0.7$ & $9.33 \pm 0.33$ &  &  \\
48 & HCG48a & 159.44682 & -27.08051 & E2 & Quiescent & 2267 & $13.21 \pm 0.2$ & $9.33 \pm 0.21$ &  &  \\
48 & HCG48b & 159.45641 & -27.12175 & Sc & Active & 2437 & $14.63 \pm 0.2$ & $8.79 \pm 0.21$ & 8.77 & 0.02 \\
49 & HCG49SDSS1 & 164.16092 & 67.15169 & cI &  & 9950 & $18.56 \pm 0.02$ & $8.47 \pm 0.2$ & 9.02 & -0.55 \\
49 & HCG49a & 164.17315 & 67.18515 & Scd & Active & 9939 & $15.87 \pm 0.2$ & $9.48 \pm 0.21$ &  &  \\
49 & HCG49b & 164.16325 & 67.18027 & Sd & Active & 9930 & $16.3 \pm 0.2$ & $9.31 \pm 0.21$ &  &  \\
49 & HCG49c & 164.15288 & 67.18123 & Im & Active & 9926 & $17.18 \pm 0.2$ & $8.98 \pm 0.21$ &  &  \\
49 & HCG49d & 164.13946 & 67.17808 & E5 & Active & 10010 & $16.99 \pm 0.2$ & $9.06 \pm 0.21$ &  &  \\
54 & A11272054 & 172.36851 & 20.63192 & cI &  & 1397 & $19.01 \pm 0.05$ & $6.96 \pm 0.2$ & 7.7 & -0.74 \\
54 & HCG54a & 172.31332 & 20.58358 & Sdm & Active & 1397 & $13.86 \pm 0.2$ & $8.89 \pm 0.21$ &  &  \\
54 & HCG54b & 172.30866 & 20.5815 & Im & Active & 1412 & $16.08 \pm 0.2$ & $8.06 \pm 0.21$ &  &  \\
54 & HCG54c & 172.31785 & 20.58646 & Im & Active & 1420 & $16.8 \pm 0.7$ & $7.79 \pm 0.33$ &  &  \\
54 & HCG54d & 172.31885 & 20.5886 & Im &  & 1670 & $18.02 \pm 0.2$ & $7.33 \pm 0.21$ &  &  \\
56 & HCG56a & 173.19443 & 52.94092 & Sc & Active & 8245 & $15.24 \pm 0.2$ & $9.57 \pm 0.21$ & 9.89 & -0.32 \\
56 & HCG56b & 173.16862 & 52.95052 & SB0 & Active & 7919 & $14.5 \pm 0.2$ & $9.84 \pm 0.21$ &  &  \\
56 & HCG56c & 173.15296 & 52.94758 & S0 & Quiescent & 8110 & $15.37 \pm 0.2$ & $9.52 \pm 0.21$ &  &  \\
56 & HCG56d & 173.14712 & 52.94725 & S0 & Active & 8346 & $16.52 \pm 0.2$ & $9.08 \pm 0.21$ &  &  \\
56 & HCG56e & 173.13647 & 52.93923 & S0 & Active & 7924 & $16.23 \pm 0.1$ & $9.19 \pm 0.2$ &  &  \\
57 & HCG57a & 174.47409 & 21.98084 & Sb & Canyon & 8727 & $13.99 \pm 0.2$ & $10.16 \pm 0.21$ &  &  \\
57 & HCG57b & 174.43198 & 22.00933 & SBb & Quiescent & 9022 & $14.32 \pm 0.2$ & $10.04 \pm 0.21$ &  &  \\
57 & HCG57c & 174.46564 & 21.97384 & E3 &  & 9081 & $14.63 \pm 0.2$ & $9.92 \pm 0.21$ &  &  \\
57 & HCG57d & 174.47966 & 21.98564 & SBc & Active & 8977 & $14.51 \pm 0.2$ & $9.96 \pm 0.21$ &  &  \\
57 & HCG57e & 174.45486 & 22.02576 & S0a & Quiescent & 8992 & $15.37 \pm 0.1$ & $9.64 \pm 0.2$ &  &  \\
57 & HCG57f & 174.47535 & 21.93609 & E4 & Quiescent & 9594 & $15.22 \pm 0.1$ & $9.7 \pm 0.2$ &  &  \\
57 & HCG57g & 174.43585 & 22.02083 & SB0 &  & 9416 & $15.84 \pm 0.1$ & $9.46 \pm 0.2$ &  &  \\
57 & HCG57h & 174.46122 & 22.01187 & SBb & Active & 9042 & $16.75 \pm 0.1$ & $9.12 \pm 0.2$ &  &  \\
58 & HCG58a & 175.54619 & 10.2777 & Sb & Active & 6138 & $13.56 \pm 0.2$ & $9.94 \pm 0.21$ & 9.67 & 0.27 \\
58 & HCG58b & 175.59827 & 10.2642 & SBab & Quiescent & 6503 & $13.4 \pm 0.2$ & $10.0 \pm 0.21$ &  &  \\
58 & HCG58c & 175.47156 & 10.30409 & SB0a & Quiescent & 6103 & $13.83 \pm 0.2$ & $9.84 \pm 0.21$ &  &  \\
58 & HCG58d & 175.52466 & 10.35087 & E1 & Quiescent & 6270 & $14.49 \pm 0.2$ & $9.59 \pm 0.21$ &  &  \\
58 & HCG58e & 175.52022 & 10.38388 & Sbc & Active & 6052 & $14.86 \pm 0.2$ & $9.45 \pm 0.21$ & 9.13 & 0.32 \\
59 & HCG59a & 177.11466 & 12.72734 & Sa & Active & 4109 & $14.52 \pm 0.1$ & $9.3 \pm 0.2$ &  &  \\
59 & HCG59b & 177.08402 & 12.71615 & E0 & Canyon & 3908 & $15.2 \pm 0.2$ & $9.04 \pm 0.21$ & 9.35 & -0.31 \\
59 & HCG59c & 177.13534 & 12.70525 & Sc &  & 4347 & $14.4 \pm 0.5$ & $9.34 \pm 0.27$ & 8.63 & 0.71 \\
59 & HCG59d & 177.12802 & 12.72981 & Im & Active & 3866 & $15.8 \pm 0.2$ & $8.82 \pm 0.21$ & 9.16 & -0.34 \\
61 & HCG61a & 183.07729 & 29.1798 & S0a & Quiescent & 3784 & $12.82 \pm 0.1$ & $9.94 \pm 0.2$ &  &  \\
61 & HCG61c & 183.12894 & 29.16854 & Sbc & Active & 3956 & $13.53 \pm 0.1$ & $9.67 \pm 0.2$ & 9.13 & 0.54 \\
61 & HCG61d & 183.11153 & 29.14912 & S0 & Quiescent & 3980 & $14.12 \pm 0.1$ & $9.45 \pm 0.2$ &  &  \\
\hline
\end{tabular}
\end{table*}
\newpage

\begin{table*}[h!] \nonumber
\centering
Table C.1 continued.
\begin{tabular}{cccccccccccc}
\hline \hline
HCG & Name & RA & Dec & Type & IR class & $cz_\odot$ & $m_\mathrm{B,c}$ & $\log M_\mathrm{HI,pred}$ & $\log M_\mathrm{HI}$ & \hi-def \\
 &  & $\mathrm{deg}$ & $\mathrm{deg}$ & & & $\mathrm{km\,s^{-1}}$ & $\mathrm{mag}$ & $[\mathrm{M_\odot}]$ & $[\mathrm{M_\odot}]$ & dex \\ \hline
62 & HCG62a & 193.27438 & -9.20458 & E3 & Quiescent & 4355 & $13.36 \pm 0.2$ & $9.73 \pm 0.21$ &  &  \\
62 & HCG62b & 193.26862 & -9.19889 & S0 & Quiescent & 3651 & $13.76 \pm 0.2$ & $9.58 \pm 0.21$ &  &  \\
62 & HCG62c & 193.29147 & -9.19809 & S0 & Quiescent & 4359 & $14.57 \pm 0.2$ & $9.28 \pm 0.21$ &  &  \\
62 & HCG62d & 193.27827 & -9.25811 & E2 & Canyon & 4123 & $15.81 \pm 0.1$ & $8.81 \pm 0.2$ &  &  \\
68 & HCG68a & 208.36072 & 40.28294 & S0 & Quiescent & 2162 & $11.84 \pm 0.08$ & $9.87 \pm 0.2$ &  &  \\
68 & HCG68b & 208.3611 & 40.30247 & E2 & Quiescent & 2635 & $12.24 \pm 0.06$ & $9.72 \pm 0.2$ &  &  \\
68 & HCG68c & 208.34055 & 40.36331 & SBbc & Active & 2313 & $11.93 \pm 0.2$ & $9.83 \pm 0.21$ & 9.95 & -0.12 \\
68 & HCG68d & 208.44021 & 40.33812 & E3 & Quiescent & 2408 & $13.73 \pm 0.1$ & $9.16 \pm 0.2$ &  &  \\
68 & HCG68e & 208.49892 & 40.27334 & S0 & Quiescent & 2401 & $14.22 \pm 0.1$ & $8.97 \pm 0.2$ &  &  \\
71 & AGC732898 & 212.76914 & 25.55967 & Sd &  & 9083 & $17.97 \pm 0.5$ & $8.64 \pm 0.27$ & 9.22 & -0.58 \\
71 & AGC242021 & 212.72563 & 25.55402 & cI &  & 9199 & $17.1 \pm 0.36$ & $8.97 \pm 0.24$ & 9.72 & -0.75 \\
71 & HCG71a & 212.73775 & 25.4967 & SBc & Canyon & 9320 & $13.75 \pm 0.2$ & $10.23 \pm 0.21$ & 10.39 & -0.16 \\
71 & HCG71b & 212.76055 & 25.51965 & Sb & Active & 9335 & $14.9 \pm 0.1$ & $9.79 \pm 0.2$ &  &  \\
71 & HCG71c & 212.77152 & 25.48257 & SBc & Active & 8827 & $15.56 \pm 0.1$ & $9.54 \pm 0.2$ & 9.85 & -0.31 \\
79 & HCG79a & 239.79777 & 20.75416 & E0 & Active & 4292 & $14.35 \pm 0.2$ & $9.35 \pm 0.21$ &  &  \\
79 & HCG79b & 239.80276 & 20.76312 & S0 & Active & 4446 & $13.78 \pm 0.2$ & $9.56 \pm 0.21$ &  &  \\
79 & HCG79c & 239.79586 & 20.76154 & S0 & Quiescent & 4146 & $14.72 \pm 0.2$ & $9.21 \pm 0.21$ &  &  \\
79 & HCG79d & 239.80026 & 20.74648 & Sdm & Active & 4503 & $15.87 \pm 0.2$ & $8.78 \pm 0.21$ & 9.26 & -0.48 \\
88 & HCG88a & 313.14714 & -5.71068 & Sb & Canyon & 6033 & $13.18 \pm 0.1$ & $9.97 \pm 0.2$ & 8.82 & 1.15 \\
88 & HCG88b & 313.12394 & -5.74655 & SBb & Quiescent & 6010 & $13.24 \pm 0.1$ & $9.95 \pm 0.2$ & 9.33 & 0.62 \\
88 & HCG88c & 313.10833 & -5.77224 & Sc & Active & 6083 & $13.87 \pm 0.1$ & $9.71 \pm 0.2$ & 9.79 & -0.08 \\
88 & HCG88d & 313.05323 & -5.79796 & Sc & Active & 6032 & $14.49 \pm 0.2$ & $9.48 \pm 0.21$ & 9.46 & 0.02 \\
90 & HCG90a & 330.50889 & -31.87014 & Sa & Active & 2575 & $12.36 \pm 0.07$ & $9.62 \pm 0.2$ & 8.94 & 0.68 \\
90 & HCG90b & 330.53631 & -31.99068 & E0 & Quiescent & 2525 & $12.57 \pm 0.13$ & $9.54 \pm 0.21$ &  &  \\
90 & HCG90c & 330.51423 & -31.97451 & E0 & Quiescent & 2696 & $12.73 \pm 0.06$ & $9.48 \pm 0.2$ &  &  \\
90 & HCG90d & 330.52602 & -31.99423 & Im & Active & 2778 & $12.81 \pm 0.15$ & $9.45 \pm 0.21$ &  &  \\
91 & HCG91a & 332.28174 & -27.80984 & SBc & Active & 7151 & $12.62 \pm 0.13$ & $10.36 \pm 0.21$ & 9.83 & 0.53 \\
91 & HCG91b & 332.3183 & -27.73134 & Sc & Active & 7196 & $14.63 \pm 0.2$ & $9.61 \pm 0.21$ & 9.67 & -0.06 \\
91 & HCG91c & 332.30881 & -27.78241 & Sc & Active & 7319 & $14.47 \pm 0.2$ & $9.67 \pm 0.21$ & 9.66 & 0.01 \\
91 & HCG91d & 332.28567 & -27.80086 & SB0 & Quiescent & 7195 & $14.99 \pm 0.2$ & $9.47 \pm 0.21$ &  &  \\
92 & HCG92b & 338.994 & 33.96595 & Sbc & Quiescent & 5774 & $13.18 \pm 0.13$ & $10.11 \pm 0.21$ &  &  \\
92 & HCG92c & 339.01604 & 33.97535 & SBc & Active & 6764 & $13.33 \pm 0.1$ & $10.05 \pm 0.2$ &  &  \\
92 & HCG92d & 338.9871 & 33.96555 & Sc & Quiescent & 6630 & $13.63 \pm 0.08$ & $9.94 \pm 0.2$ &  &  \\
92 & HCG92e & 338.96756 & 33.94459 & E1 & Quiescent & 6599 & $14.01 \pm 0.08$ & $9.79 \pm 0.2$ &  &  \\
93 & HCG93a & 348.81679 & 18.96147 & E1 & Quiescent & 5140 & $12.61 \pm 0.1$ & $10.07 \pm 0.2$ &  &  \\
93 & HCG93b & 348.82171 & 19.04158 & SBd & Active & 4672 & $13.18 \pm 0.1$ & $9.85 \pm 0.2$ & 9.53 & 0.32 \\
93 & HCG93c & 348.76515 & 18.97311 & SBa & Quiescent & 5132 & $13.94 \pm 0.1$ & $9.57 \pm 0.2$ &  &  \\
93 & HCG93d & 348.88811 & 19.0479 & SB0 &  & 5173 & $15.27 \pm 0.1$ & $9.07 \pm 0.2$ &  &  \\
95 & HCG95a & 349.87475 & 9.50793 & E3 & Quiescent & 11888 & $14.42 \pm 0.2$ & $10.1 \pm 0.21$ &  &  \\
95 & HCG95b & 349.8909 & 9.49491 & Scd & Active & 11637 & $15.34 \pm 0.1$ & $9.75 \pm 0.2$ &  &  \\
95 & HCG95c & 349.86626 & 9.49434 & Sm & Active & 11562 & $15.2 \pm 0.2$ & $9.81 \pm 0.21$ &  &  \\
95 & HCG95d & 349.87951 & 9.50274 & Sc & Active & 11593 & $16.14 \pm 0.1$ & $9.45 \pm 0.2$ &  &  \\
96 & HCG96a & 351.98711 & 8.77821 & Sc & Active & 8698 & $13.53 \pm 0.2$ & $10.21 \pm 0.21$ & 9.93 & 0.28 \\
96 & HCG96b & 352.02523 & 8.76841 & E2 & Quiescent & 8616 & $14.49 \pm 0.1$ & $9.85 \pm 0.2$ &  &  \\
96 & HCG96c & 351.99507 & 8.7828 & Sa & Active & 8753 & $15.69 \pm 0.2$ & $9.4 \pm 0.21$ &  &  \\
96 & HCG96d & 352.00084 & 8.76733 & Im & Active & 8975 & $16.56 \pm 0.1$ & $9.07 \pm 0.2$ &  &  \\
97 & HCG97a & 356.84591 & -2.30096 & E5 & Quiescent & 6910 & $14.16 \pm 0.2$ & $9.72 \pm 0.21$ &  &  \\
97 & HCG97b & 356.90753 & -2.31716 & Sc & Canyon & 6940 & $14.83 \pm 0.1$ & $9.47 \pm 0.2$ & 8.68 & 0.79 \\
97 & HCG97c & 356.84884 & -2.35143 & Sa & Quiescent & 5995 & $14.54 \pm 0.1$ & $9.58 \pm 0.2$ &  &  \\
97 & HCG97d & 356.82867 & -2.31329 & E1 &  & 6239 & $14.45 \pm 0.1$ & $9.61 \pm 0.2$ &  &  \\
97 & HCG97e & 356.83253 & -2.28102 & S0a & Canyon & 6579 & $16.31 \pm 0.2$ & $8.91 \pm 0.21$ &  &  \\
100 & MRK935 & 0.43845 & 13.10067 & cI &  & 5606 & $15.34 \pm 0.05$ & $9.08 \pm 0.2$ & 8.54 & 0.54 \\
100 & HCG100a & 0.33374 & 13.11095 & Sb &  & 5300 & $13.66 \pm 0.1$ & $9.71 \pm 0.2$ &  &  \\
100 & HCG100b & 0.35885 & 13.11278 & Sm &  & 5253 & $14.9 \pm 0.1$ & $9.25 \pm 0.2$ &  &  \\
100 & HCG100c & 0.30629 & 13.14398 & SBc &  & 5461 & $15.22 \pm 0.1$ & $9.13 \pm 0.2$ &  &  \\
100 & HCG100d & 0.31151 & 13.11264 & Scd &  & 5590 & $15.97 \pm 0.1$ & $8.84 \pm 0.2$ &  &  \\
\hline
\end{tabular}
\end{table*}
\newpage

\section{Velocity maps}
\label{app:mom1}

Maps of iso-velocity contours overlaid on the same optical images for each group in Section \ref{sec:sep_features} are shown in Figure \ref{fig:mom1_maps}. They can also be generated as described in Appendix \ref{app:repo}. The same source masks were used to produce these moment one contour maps, but only pixels with fluxes higher than $4.5 \sigma_\mathrm{rms} \sqrt{20 \mathrm{km\,s^{-1}}/\Delta v_\mathrm{chan}}$ were included. The iso-velocity contours have separations of 20~\kms \ in all cases. HCGs 30, 37, and 62 are omitted as no \hi \ was detected in their core groups.

\begin{figure*}
    \centering
    \includegraphics[width=\columnwidth]{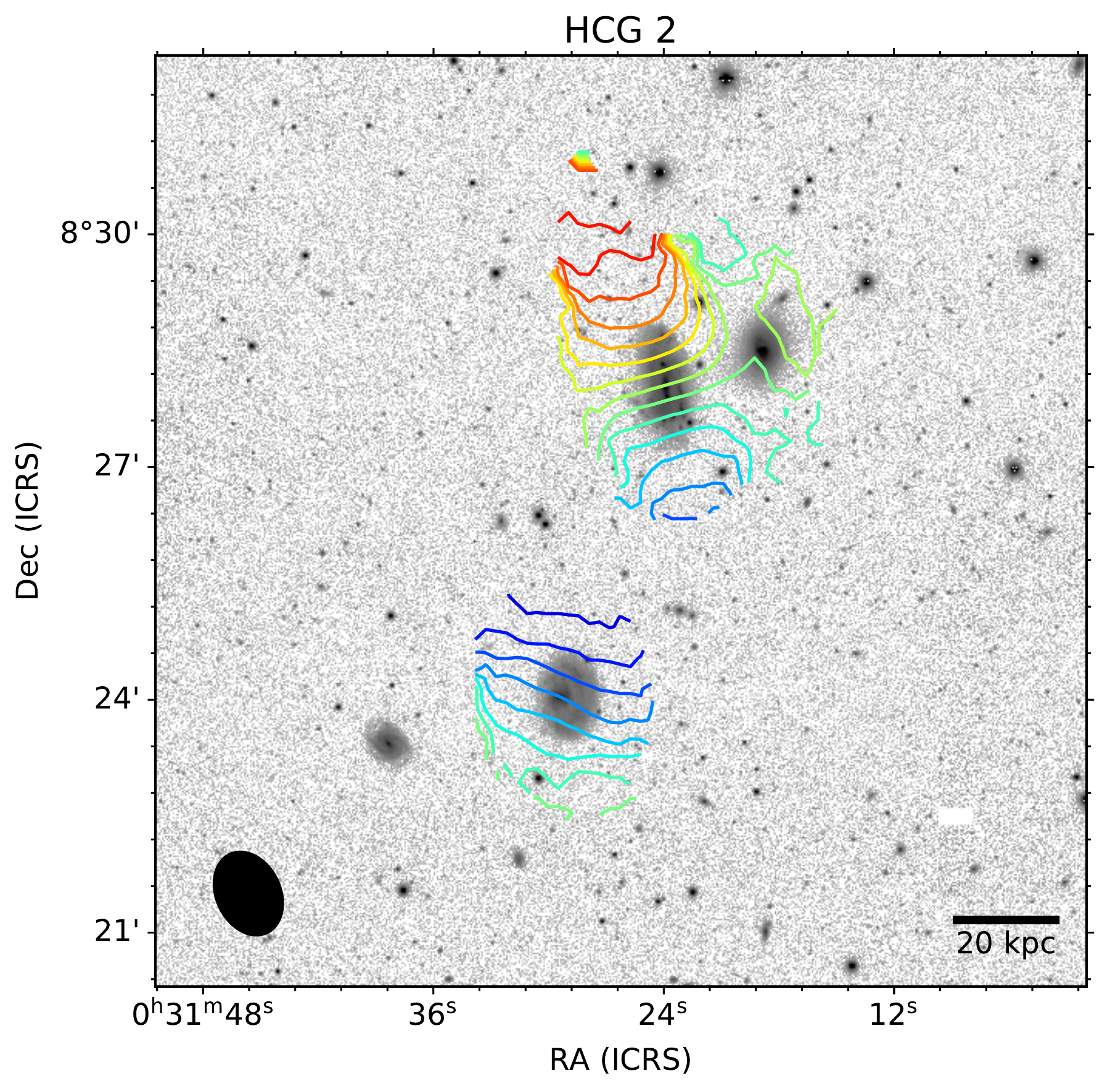}
    \includegraphics[width=\columnwidth]{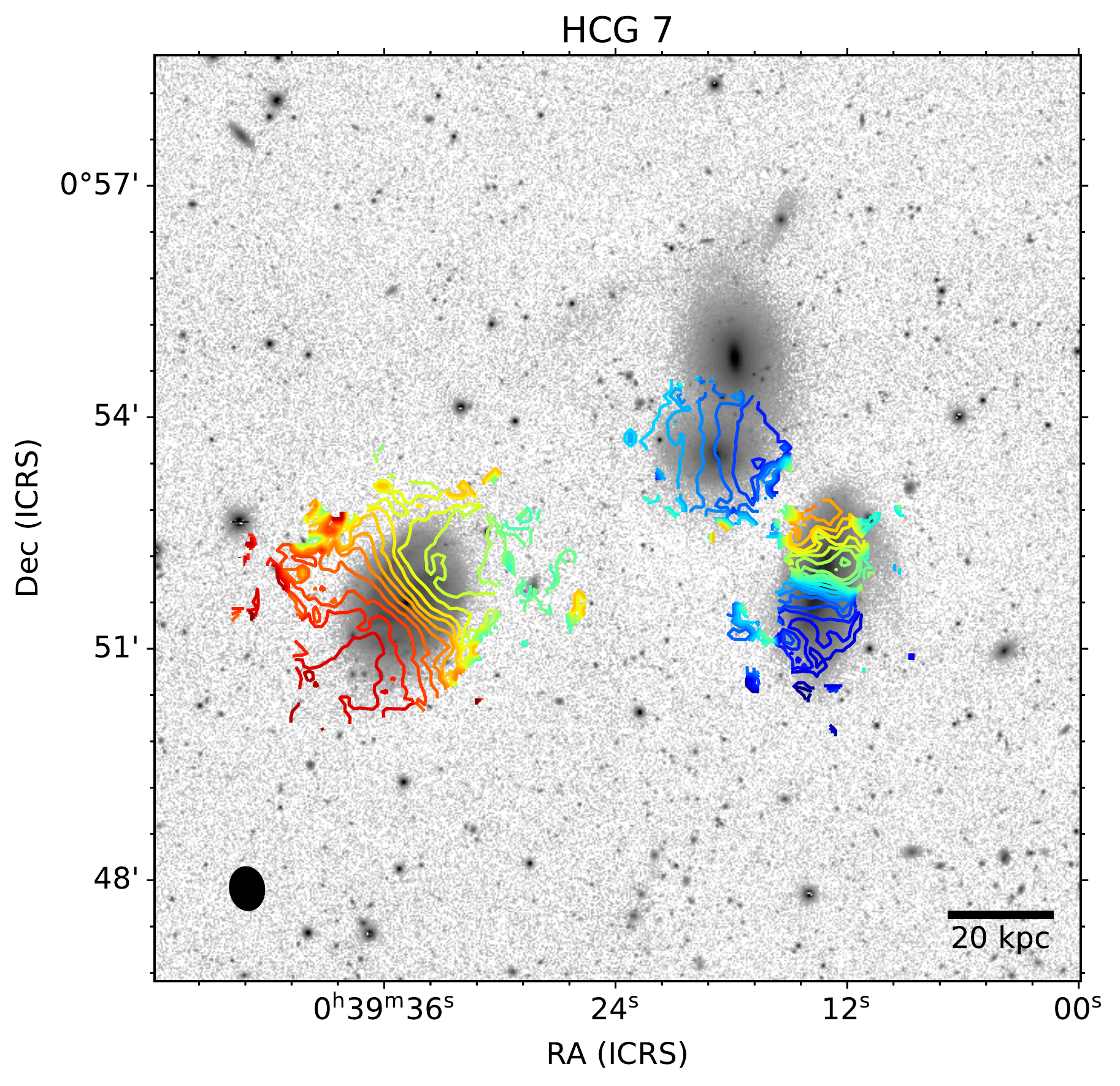}
    \includegraphics[width=\columnwidth]{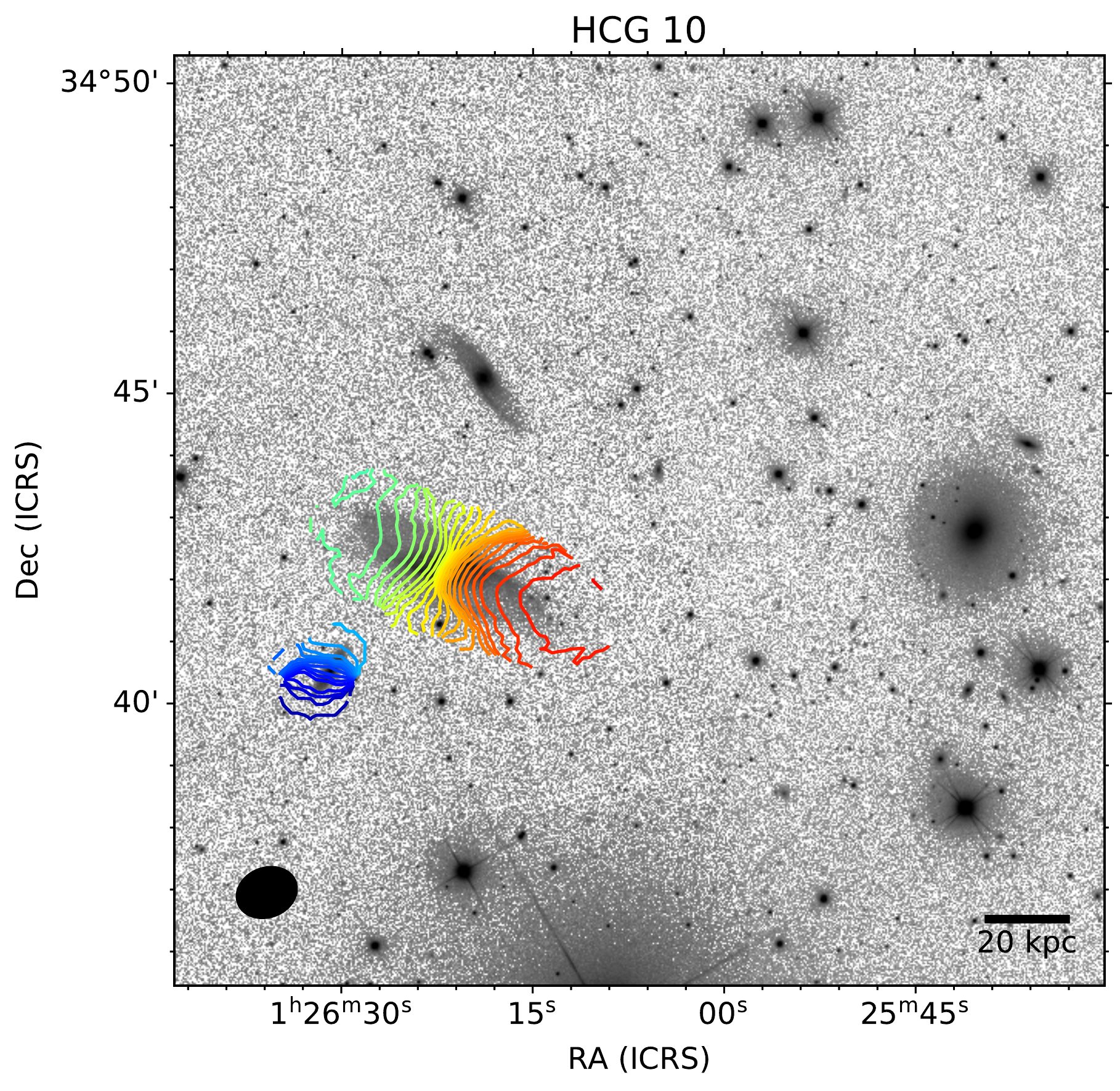}
    \includegraphics[width=\columnwidth]{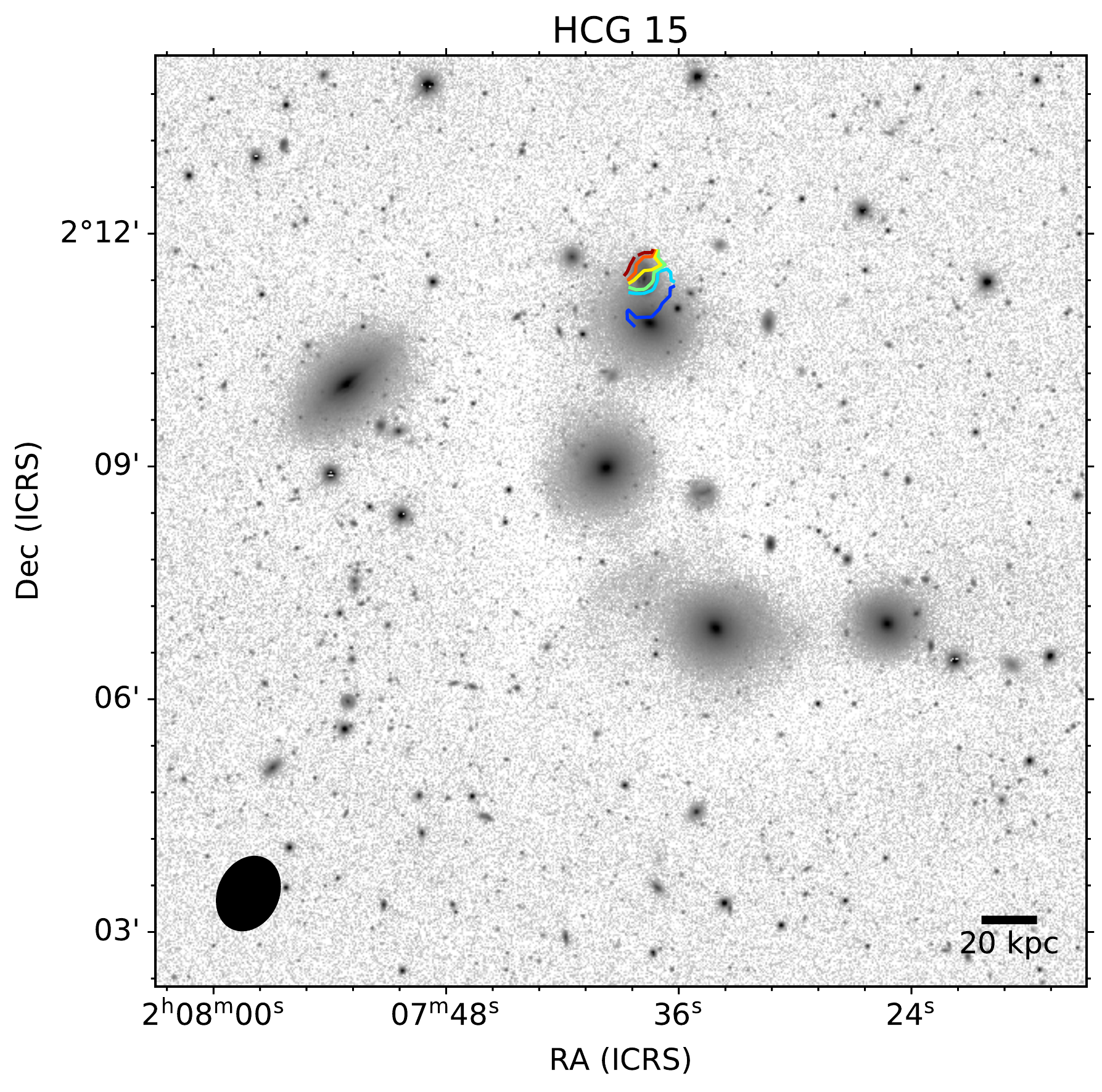}
    \caption{Iso-velocity contours (separated by 20~\kms) overlaid on DECaLS, SDSS, or POSS images depending on the group. Contours follow a rainbow colour scheme with high velocities corresponding to redder colours and lower velocities to bluer colours (note that the velocity range varies in each panel, but contours are always separated by 20~\kms).}
    \label{fig:mom1_maps}
\end{figure*}
\newpage

\begin{figure*}\nonumber
    \centering
    \includegraphics[width=\columnwidth]{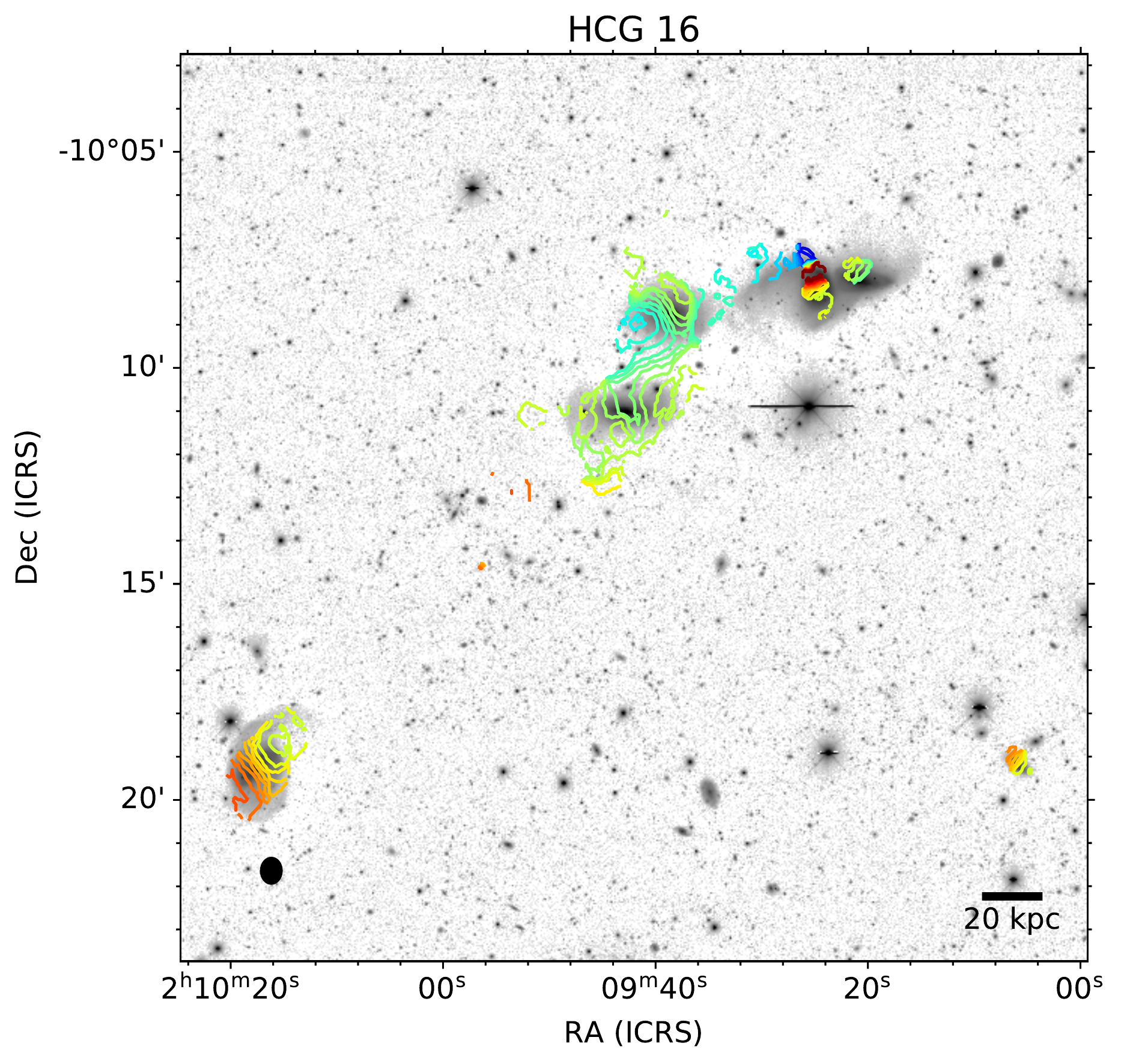}
    \includegraphics[width=\columnwidth]{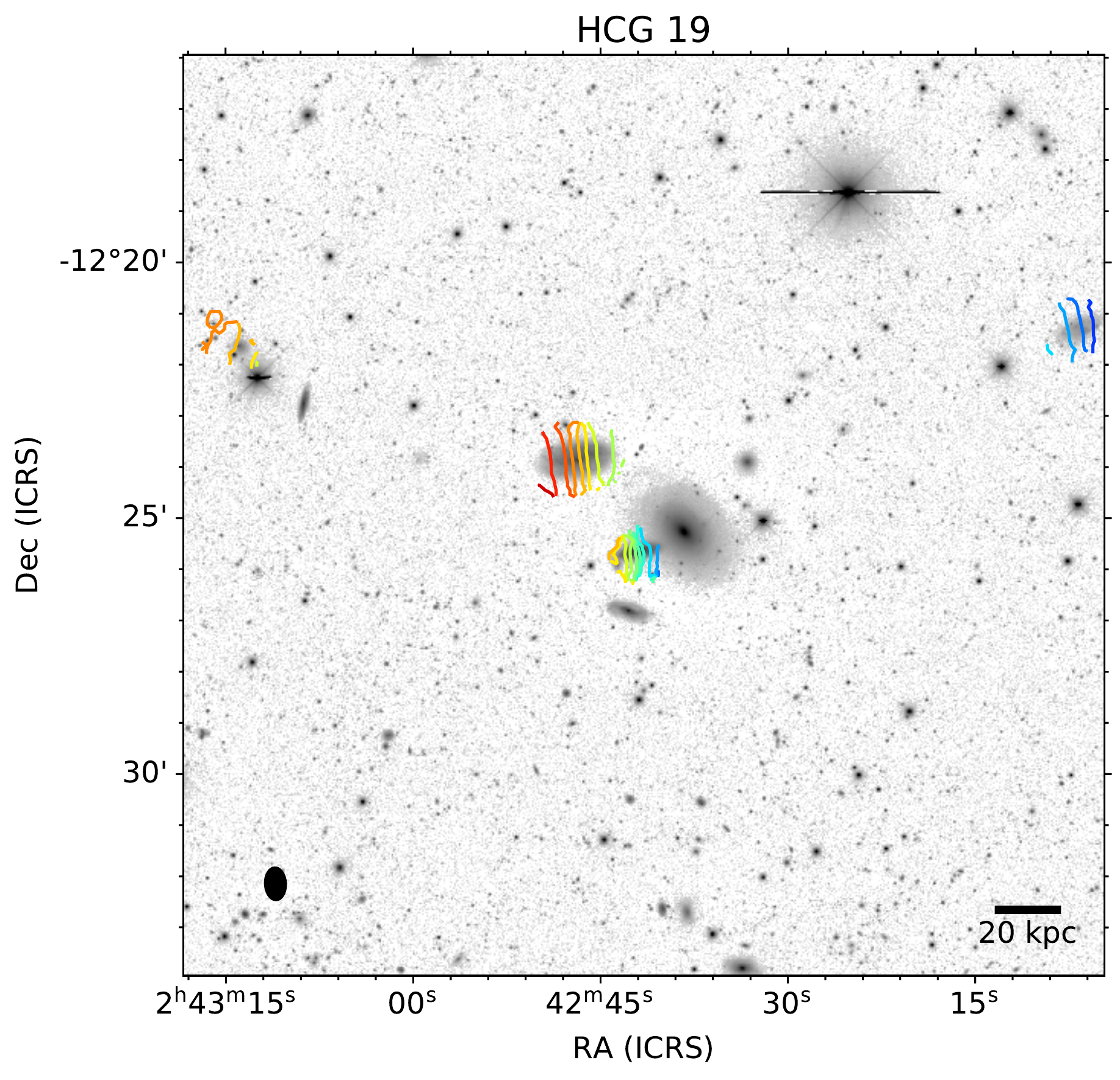}
    \includegraphics[width=\columnwidth]{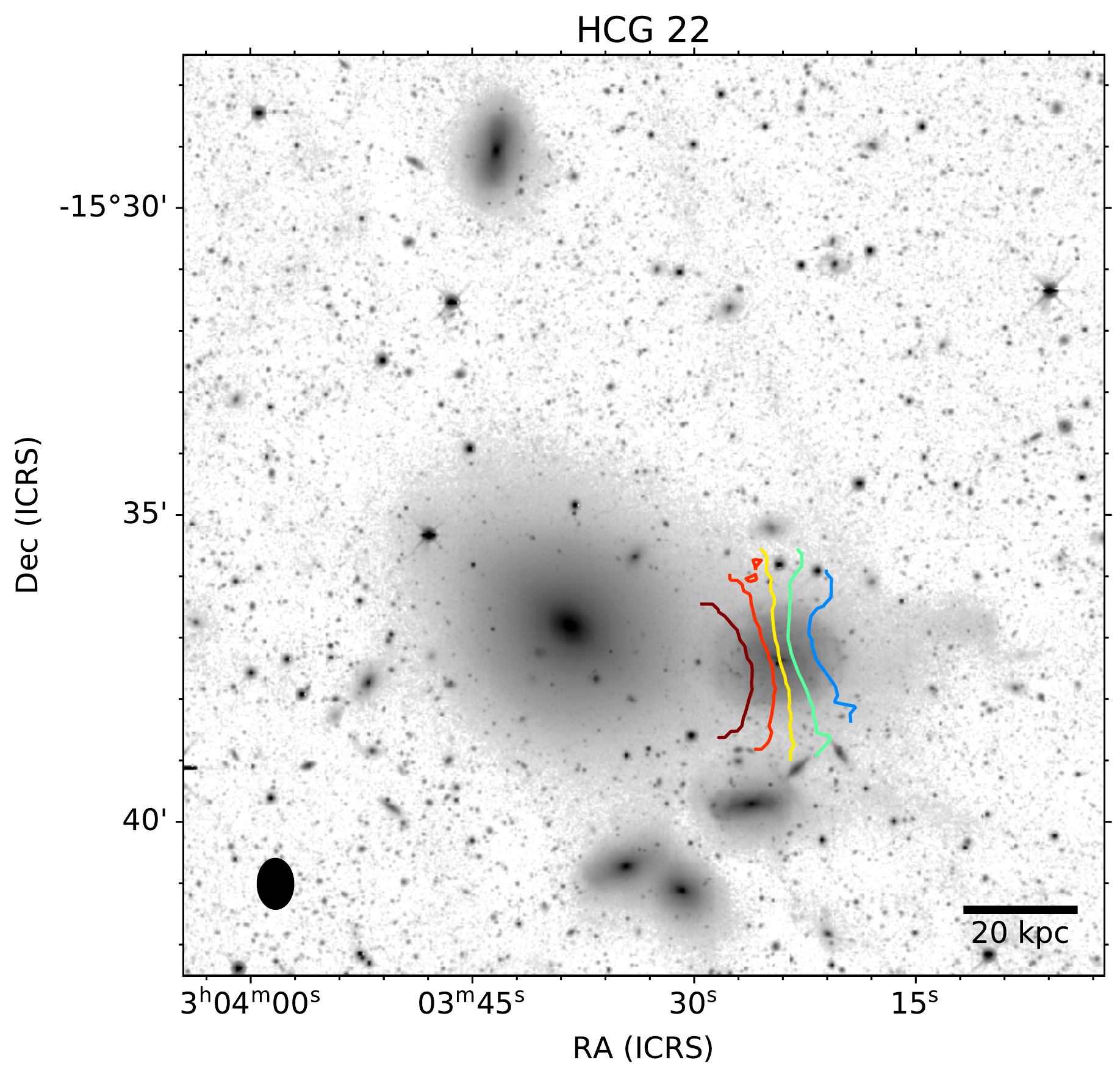}
    \includegraphics[width=\columnwidth]{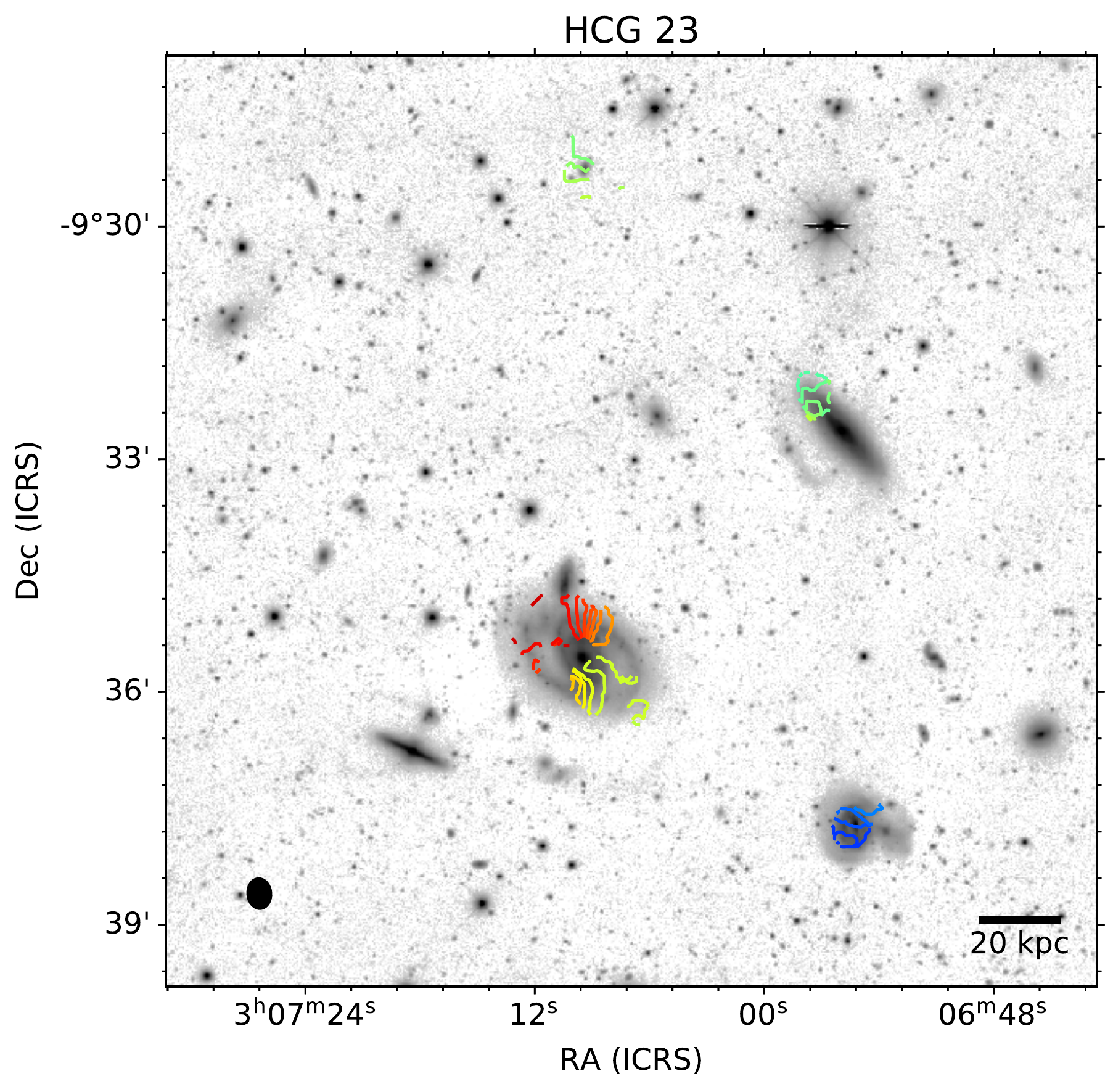}
    Figure \ref{fig:mom1_maps} continued.
\end{figure*}
\newpage

\begin{figure*}\nonumber
    \centering
    \includegraphics[width=\columnwidth]{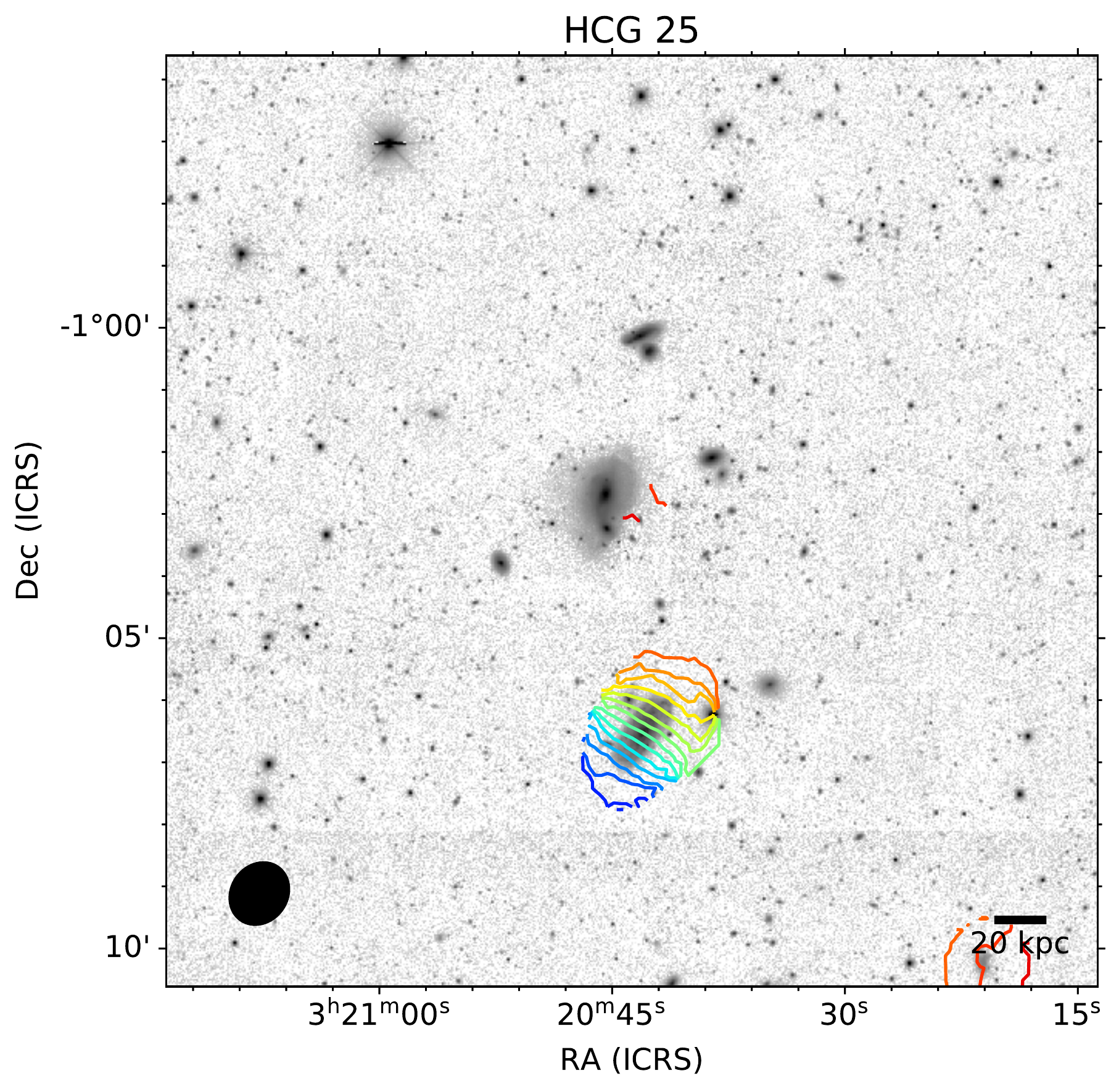}
    \includegraphics[width=\columnwidth]{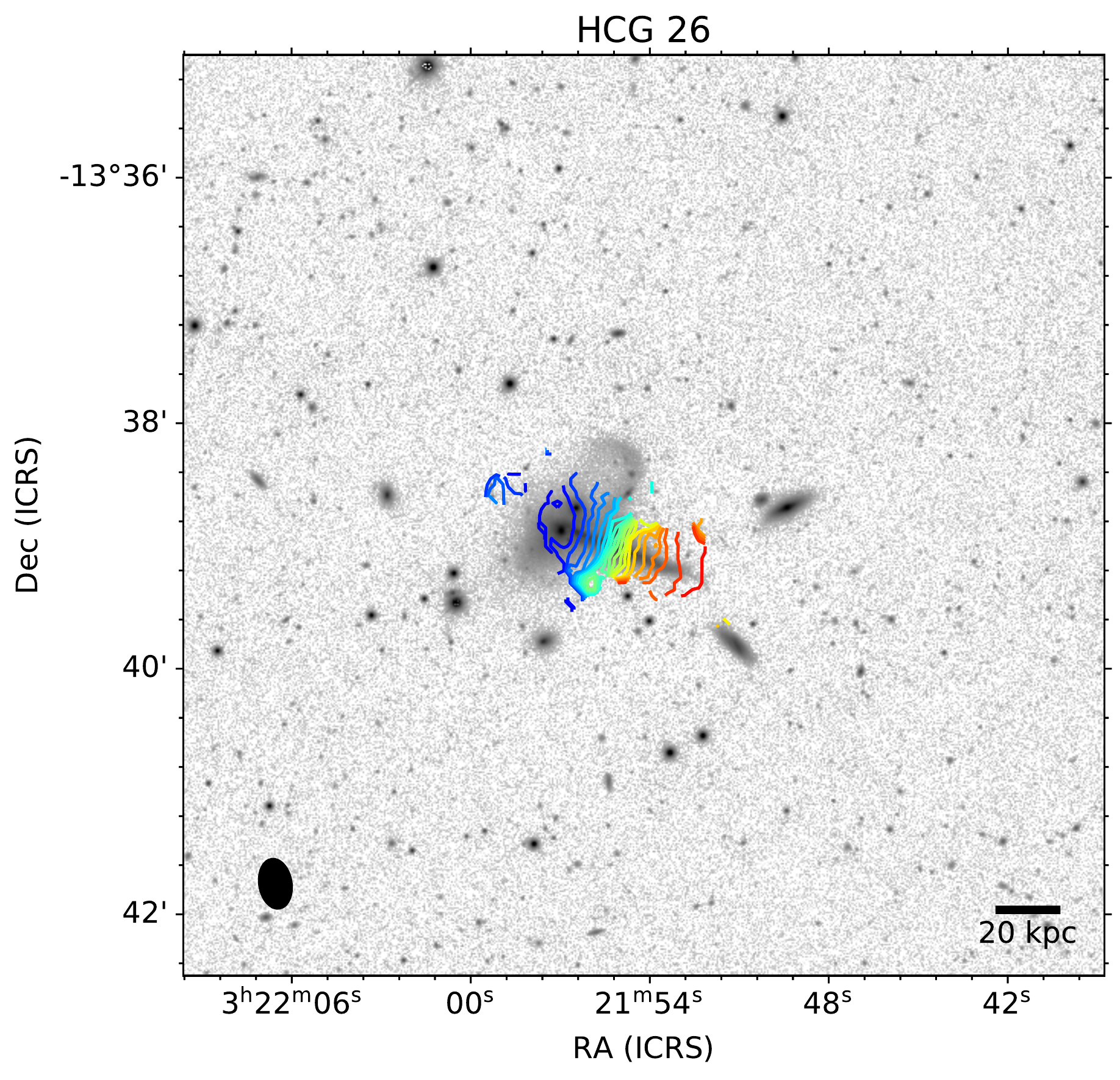}
    \includegraphics[width=\columnwidth]{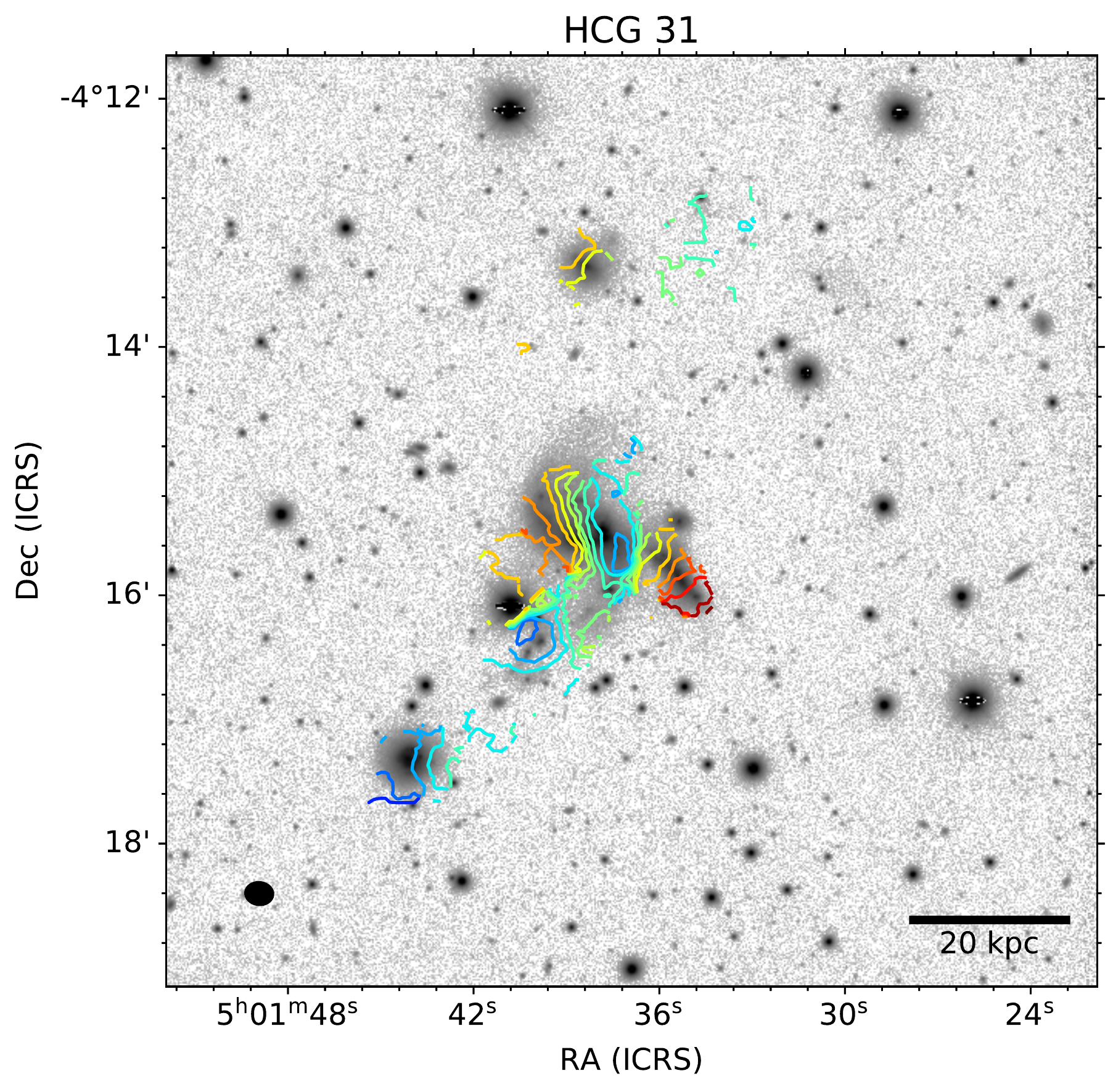}
    \includegraphics[width=\columnwidth]{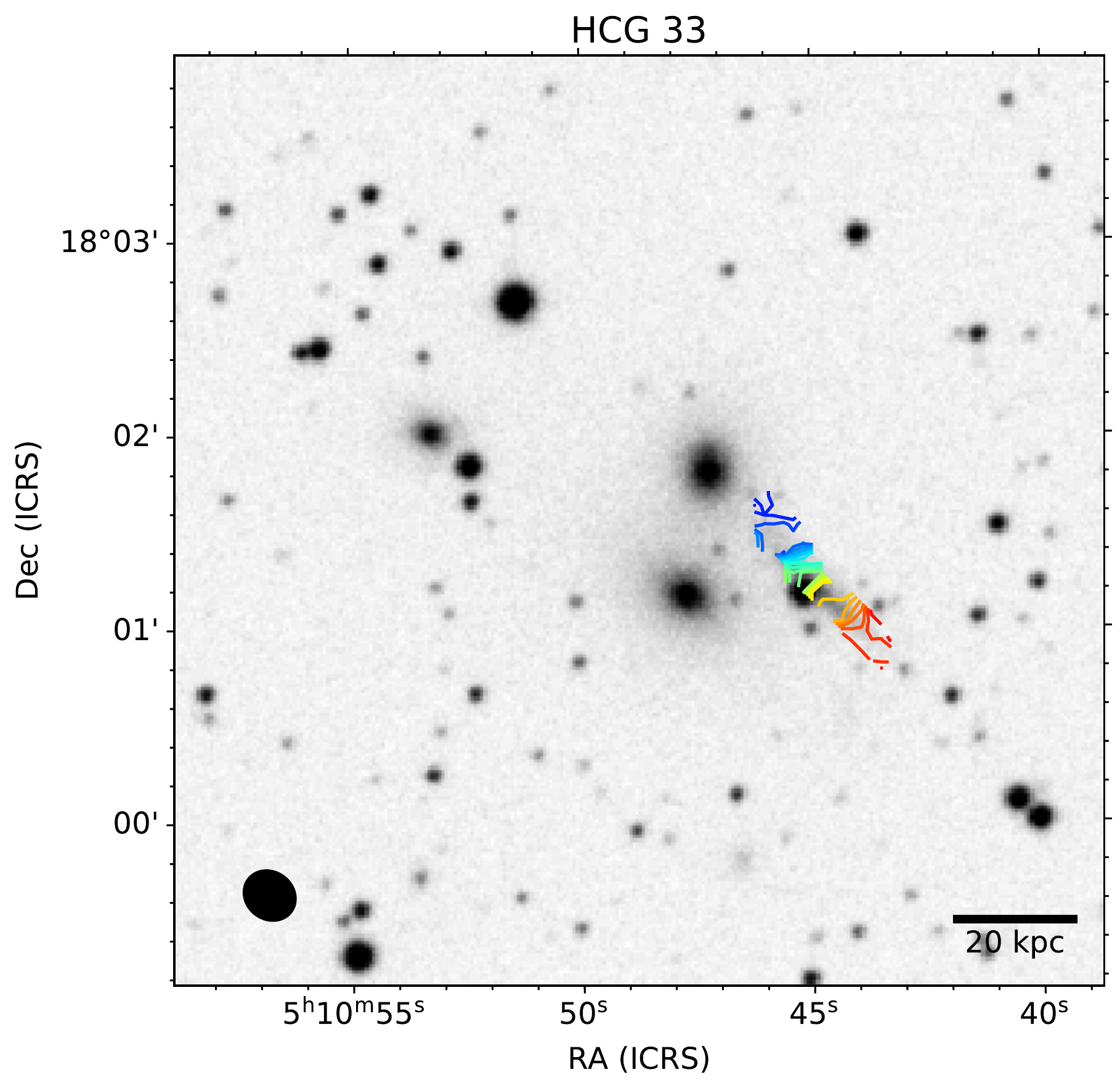}
    Figure \ref{fig:mom1_maps} continued.
\end{figure*}
\newpage

\begin{figure*}\nonumber
    \centering
    \includegraphics[width=\columnwidth]{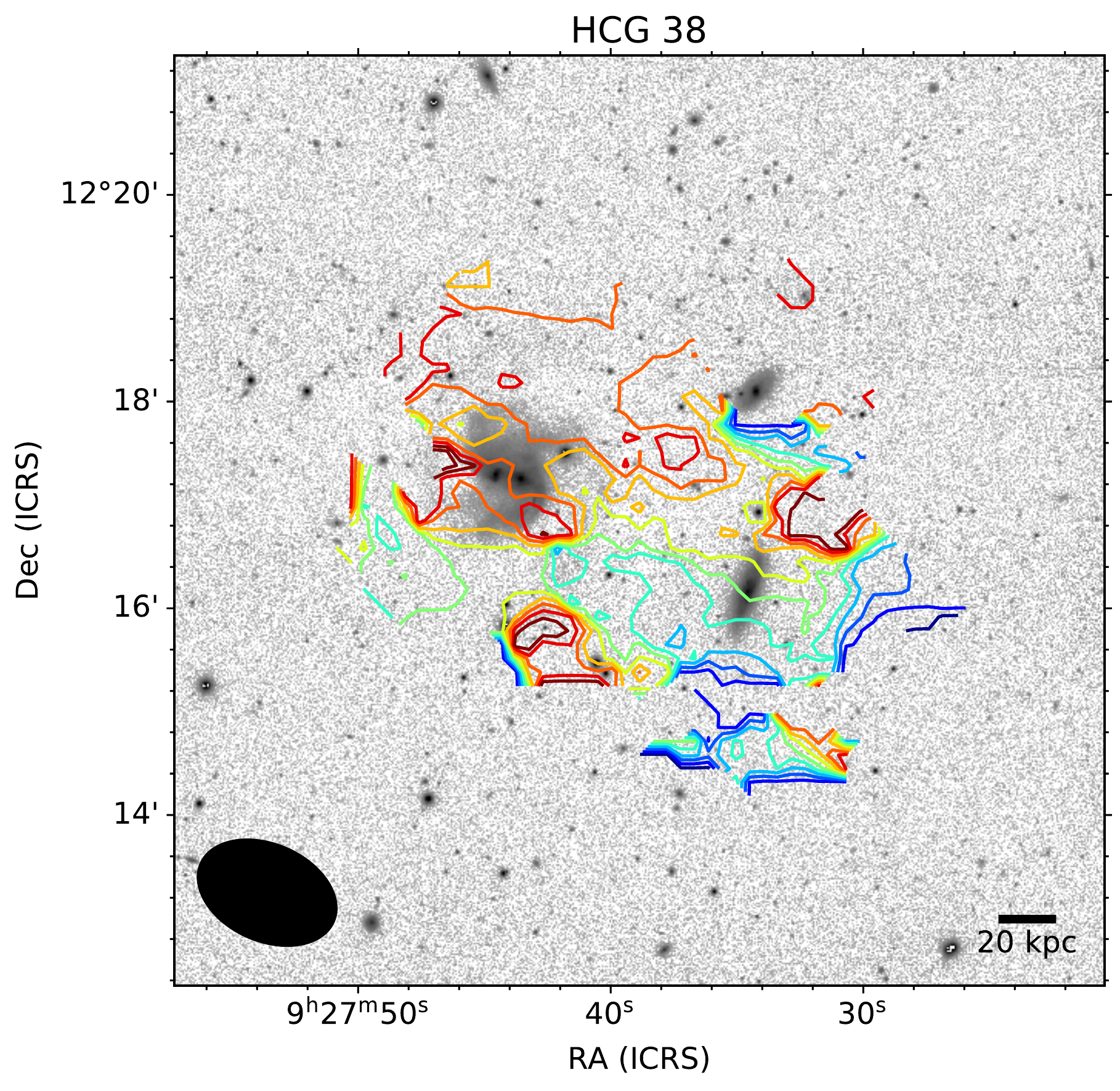}
    \includegraphics[width=\columnwidth]{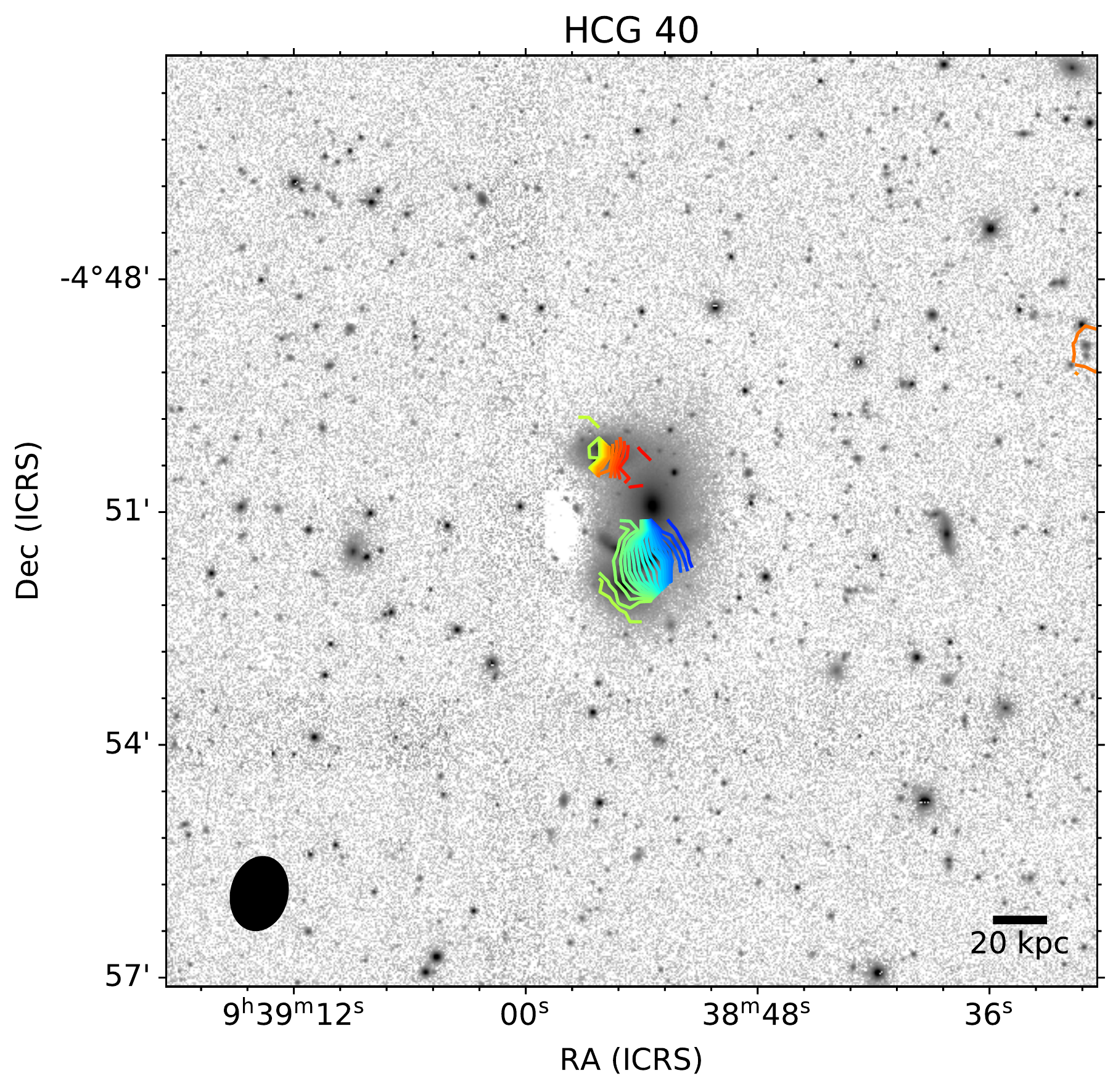}
    \includegraphics[width=\columnwidth]{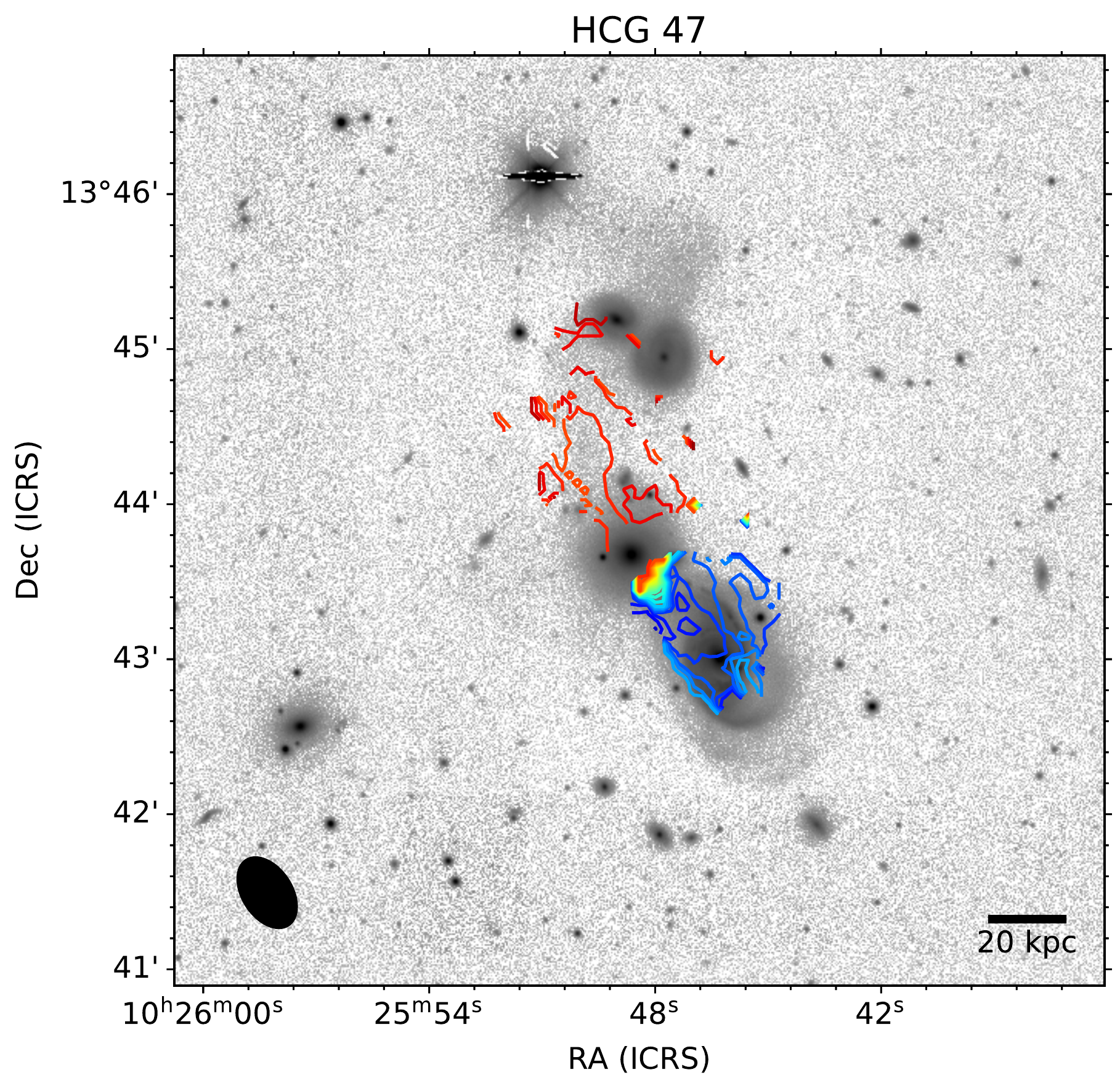}
    \includegraphics[width=\columnwidth]{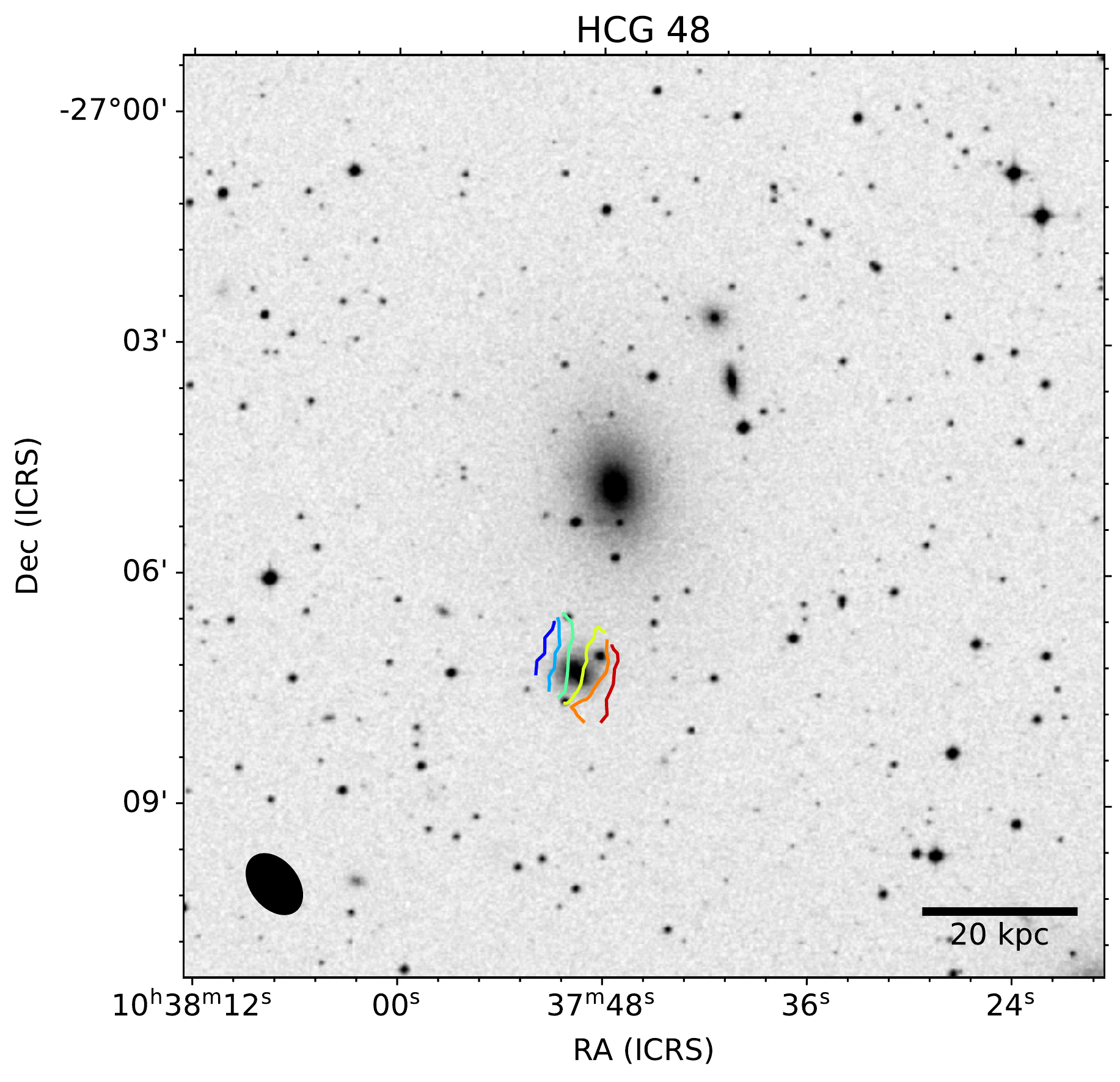}
    Figure \ref{fig:mom1_maps} continued.
\end{figure*}
\newpage

\begin{figure*}\nonumber
    \centering
    \includegraphics[width=\columnwidth]{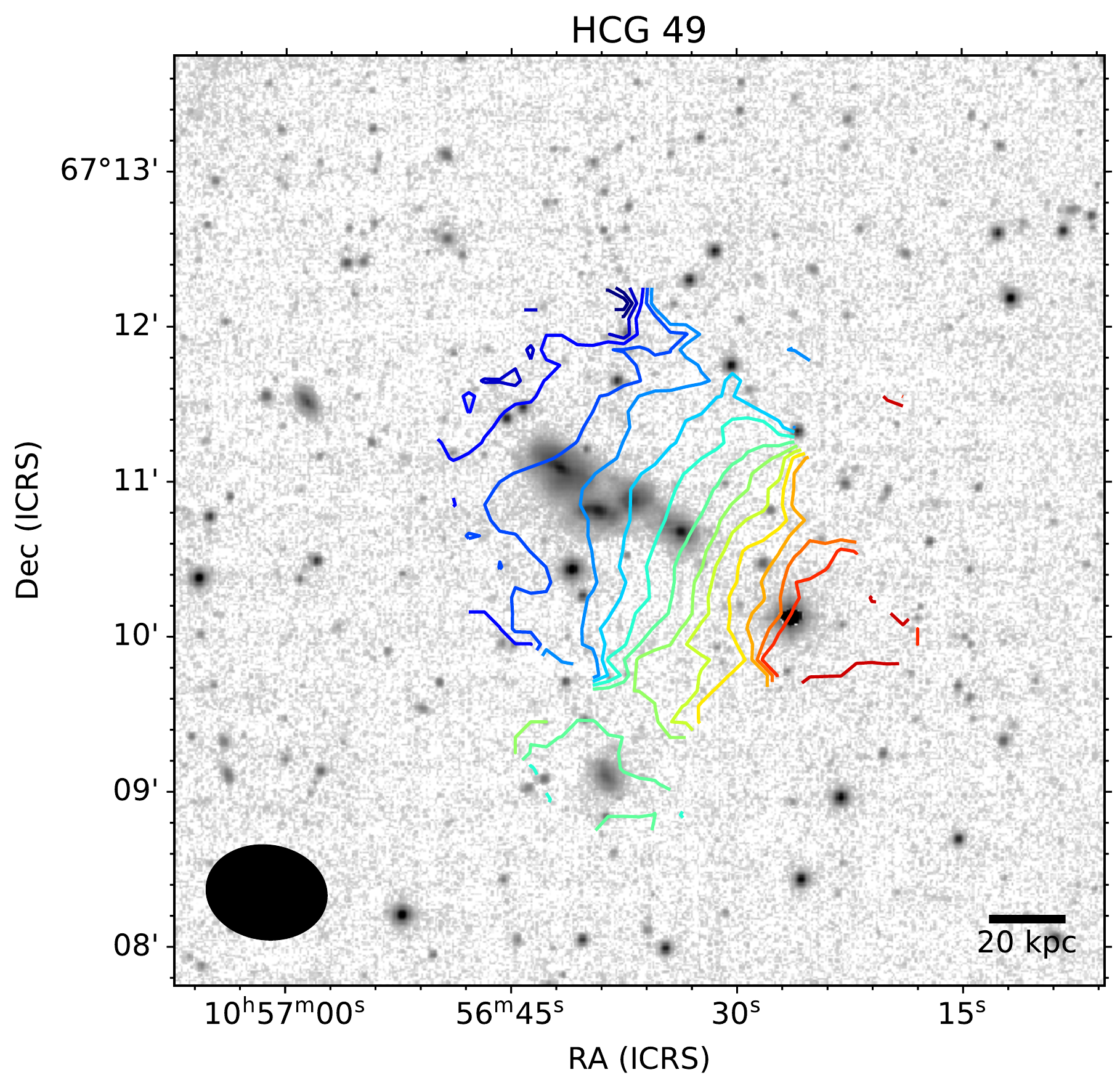}
    \includegraphics[width=\columnwidth]{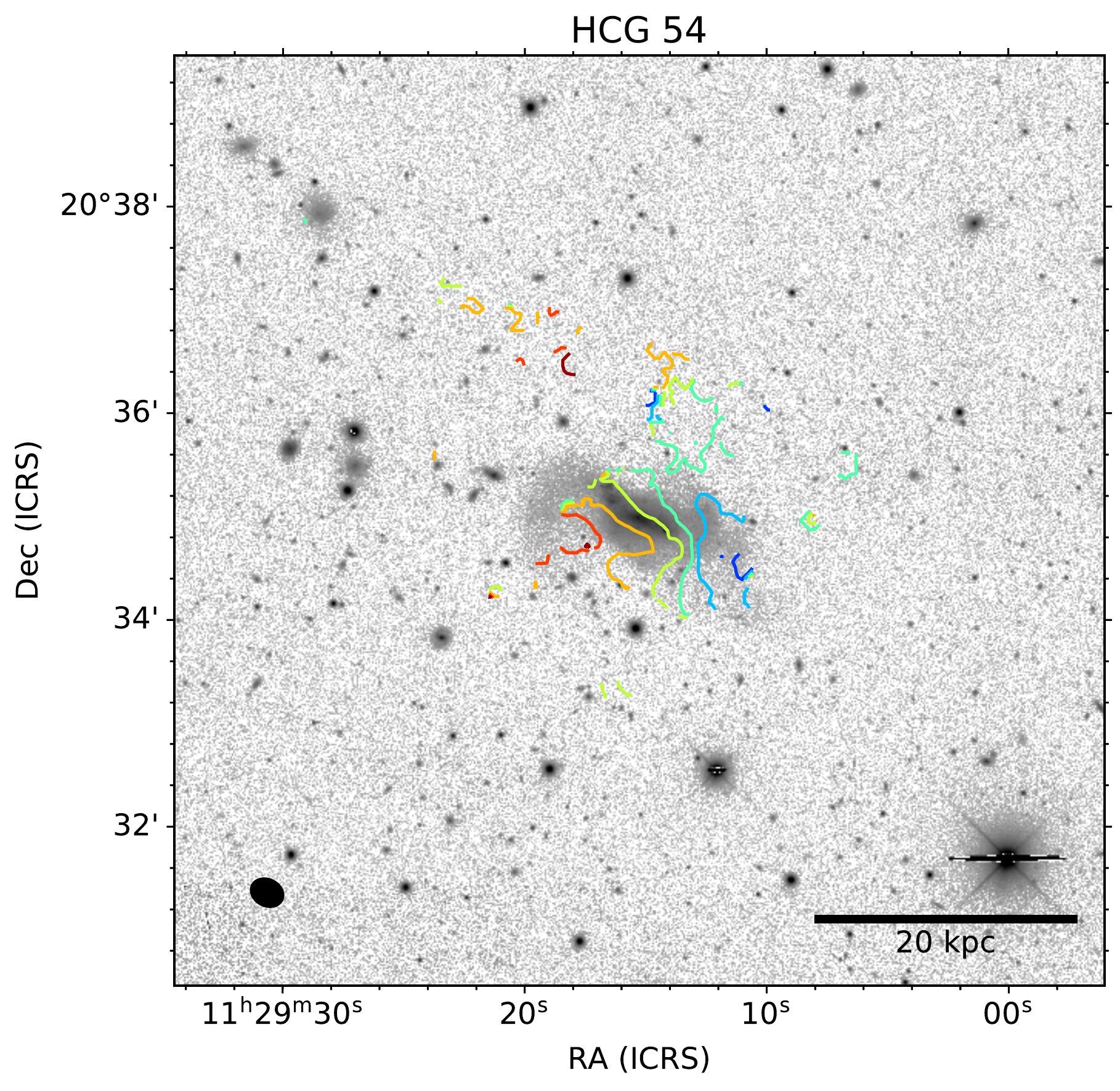}
    \includegraphics[width=\columnwidth]{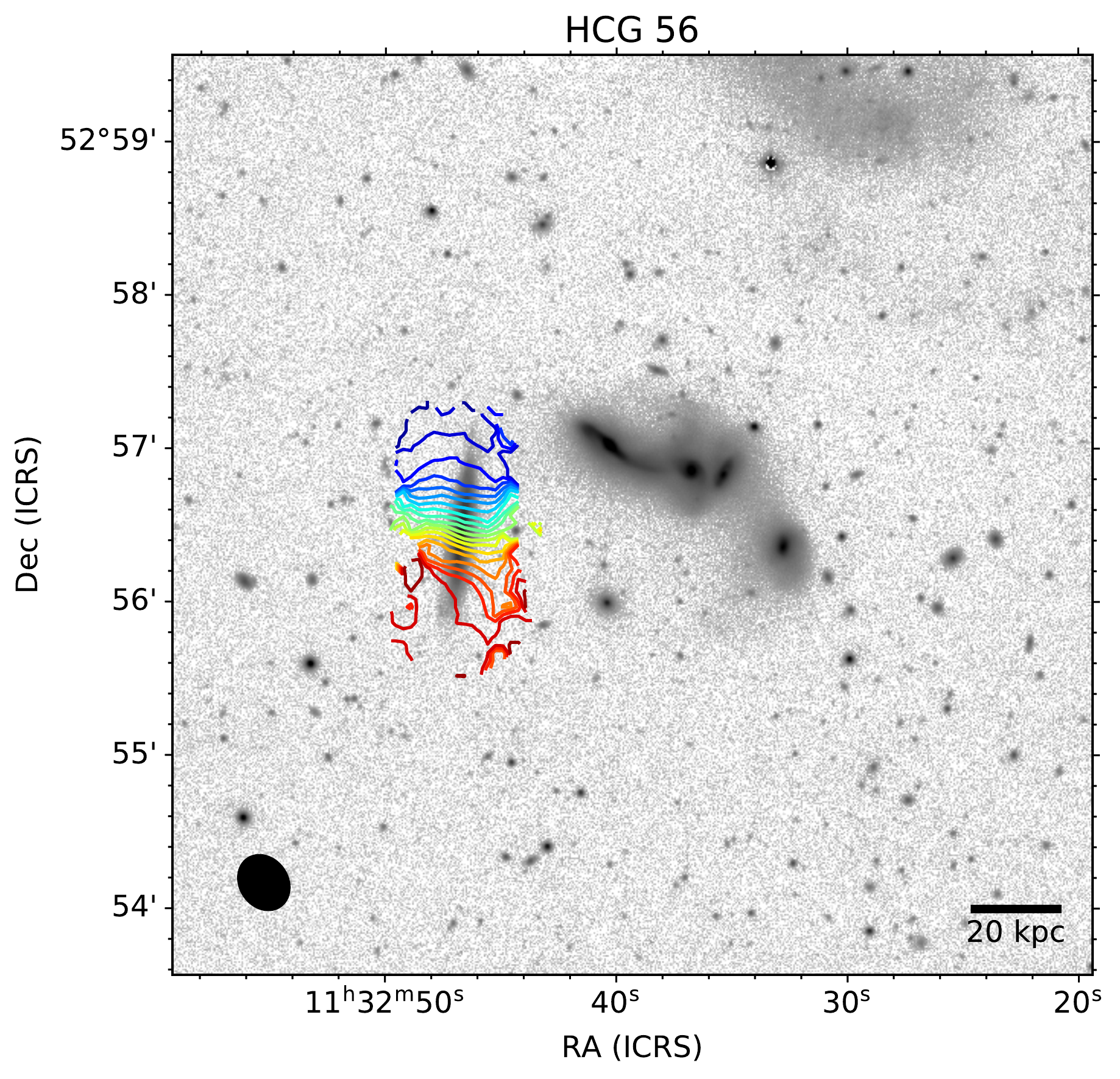}
    \includegraphics[width=\columnwidth]{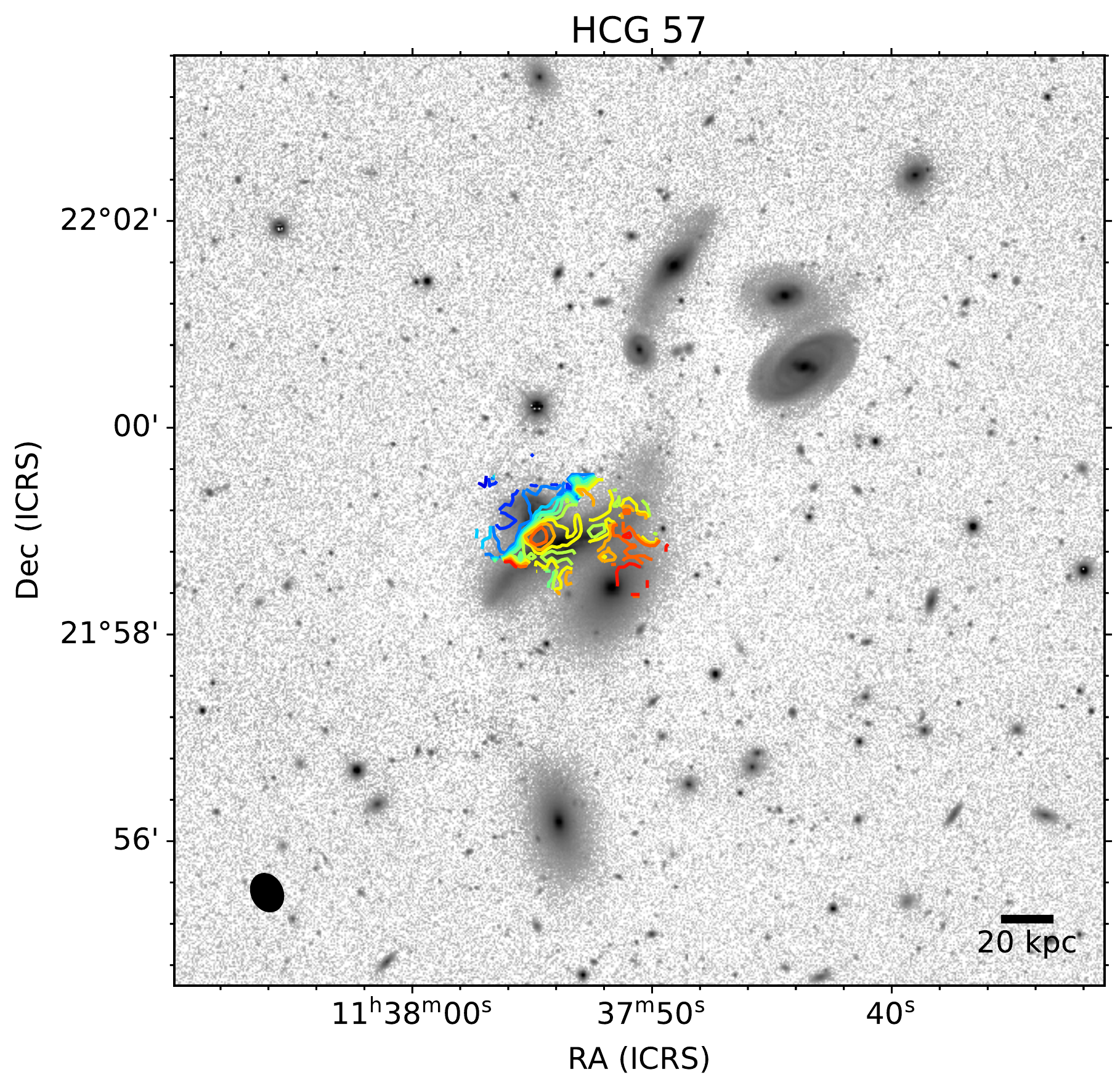}
    Figure \ref{fig:mom1_maps} continued.
\end{figure*}
\newpage

\begin{figure*}\nonumber
    \centering
    \includegraphics[width=\columnwidth]{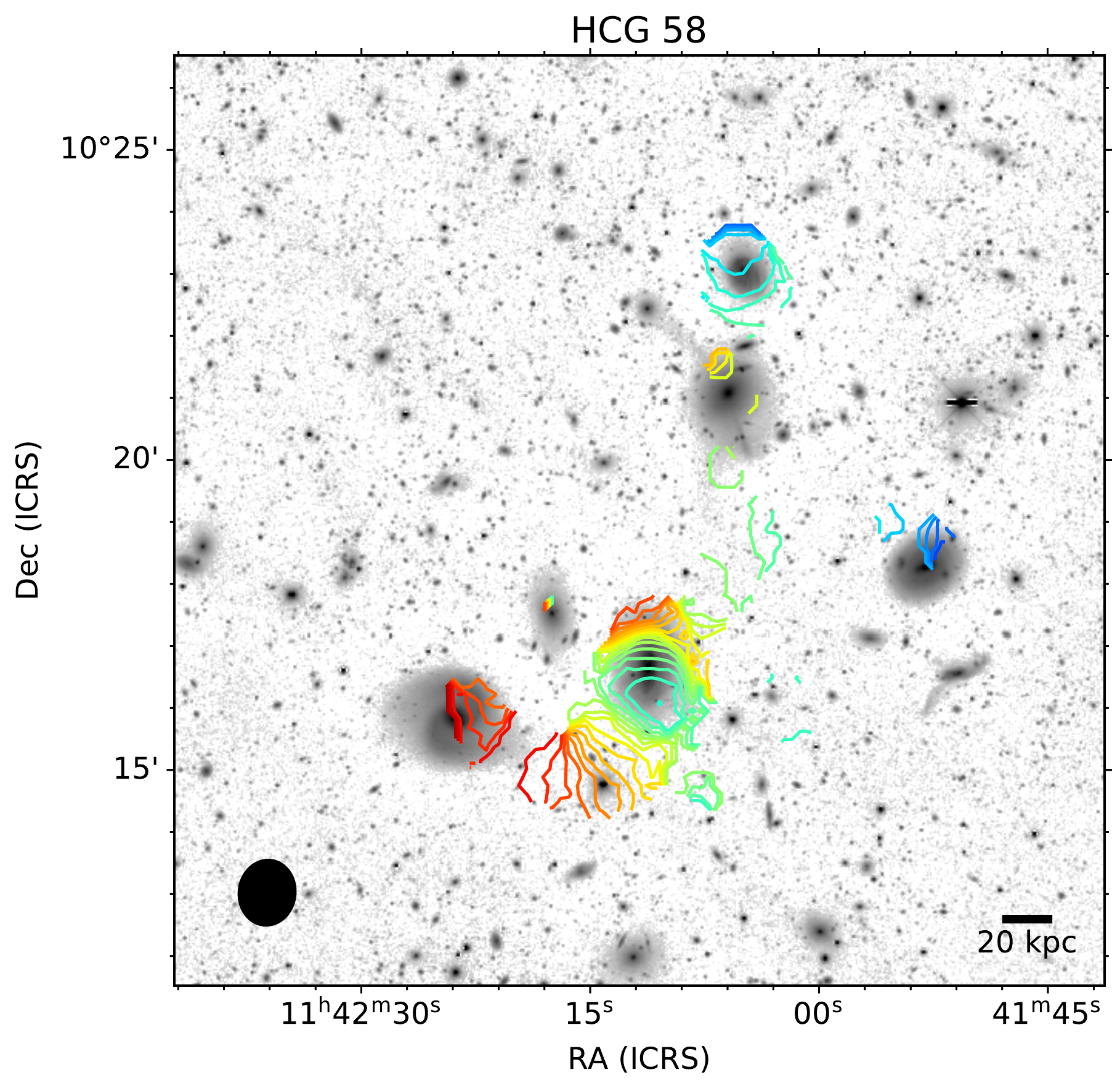}
    \includegraphics[width=\columnwidth]{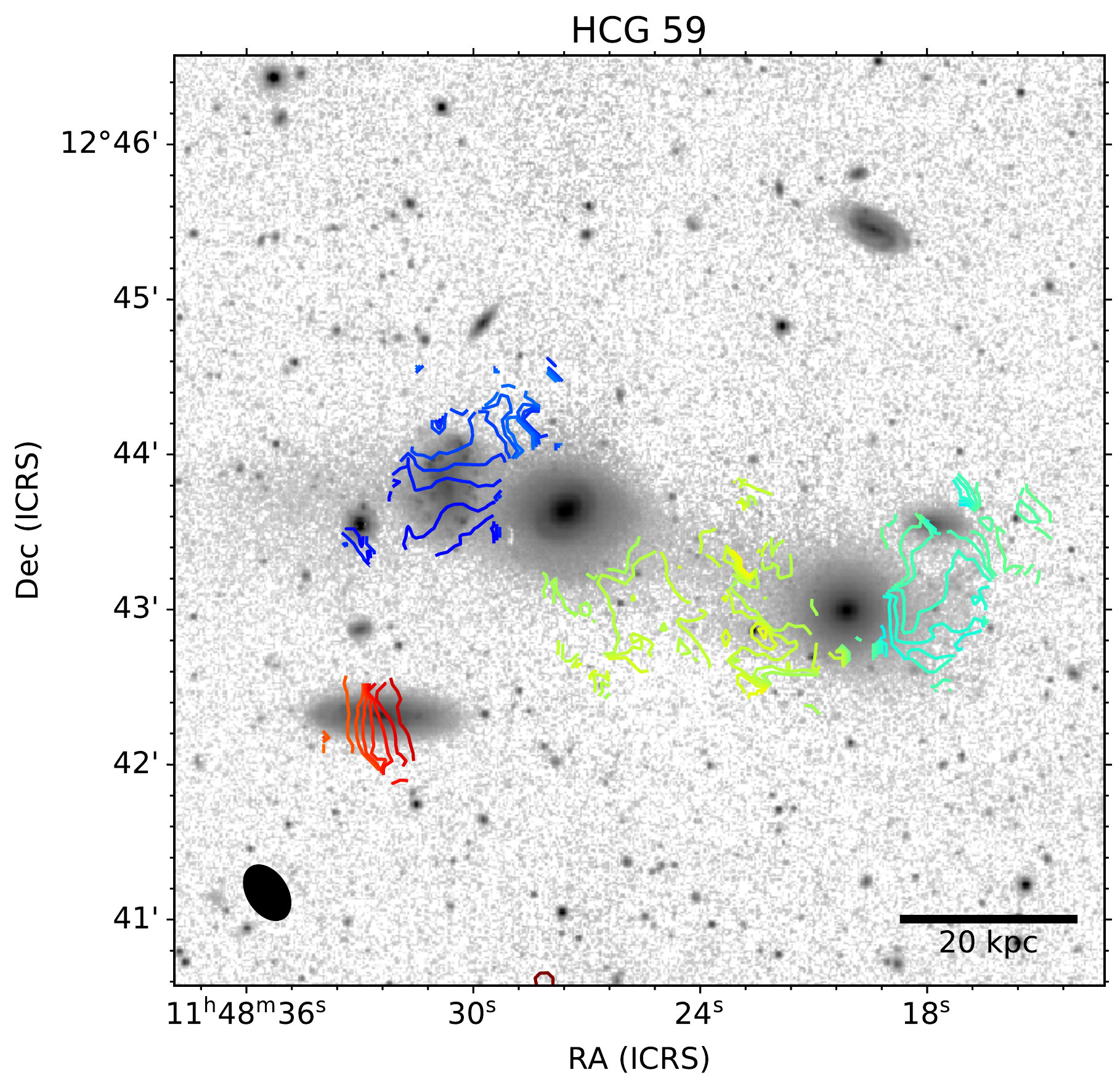}
    \includegraphics[width=\columnwidth]{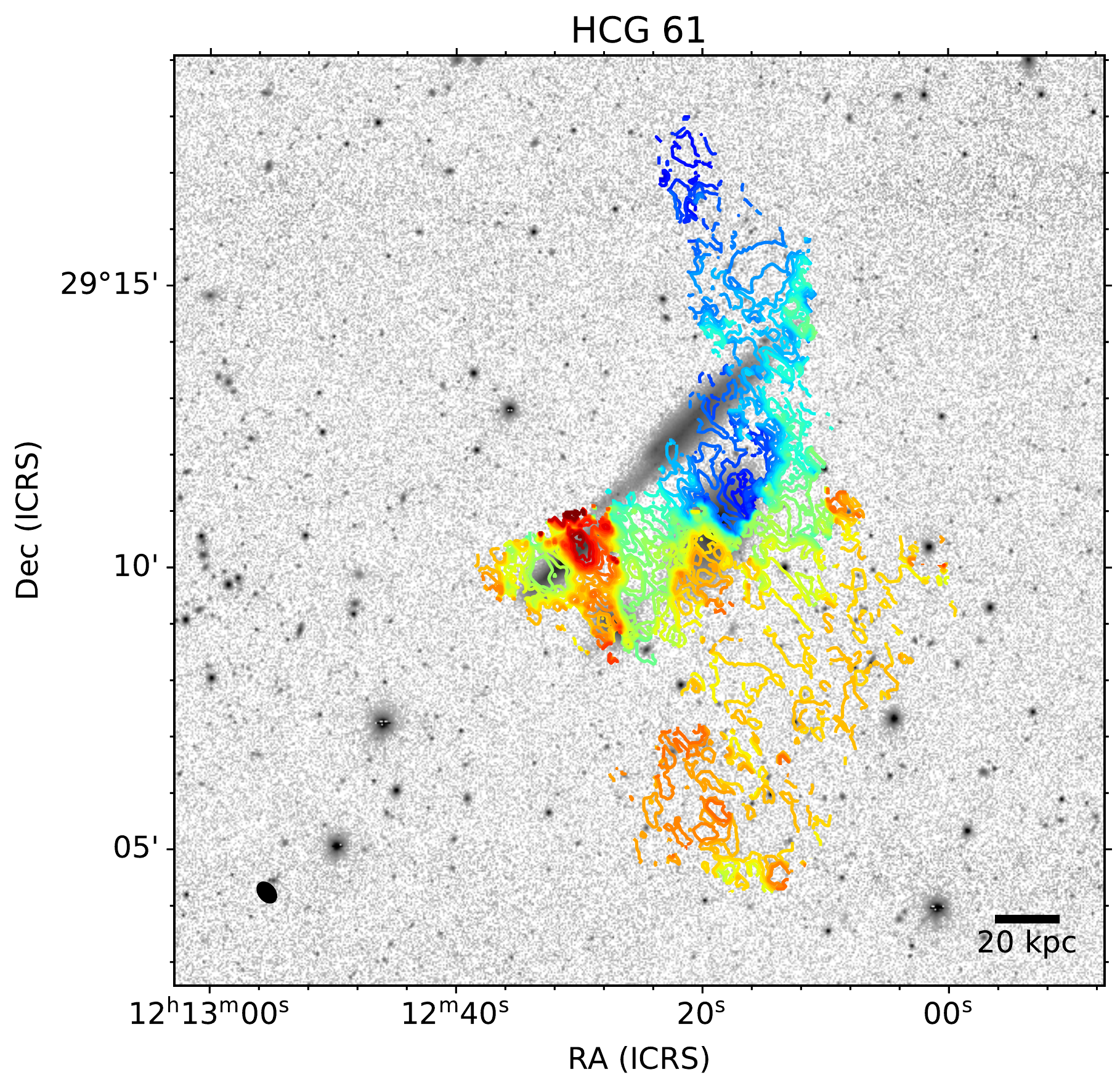}
    \includegraphics[width=\columnwidth]{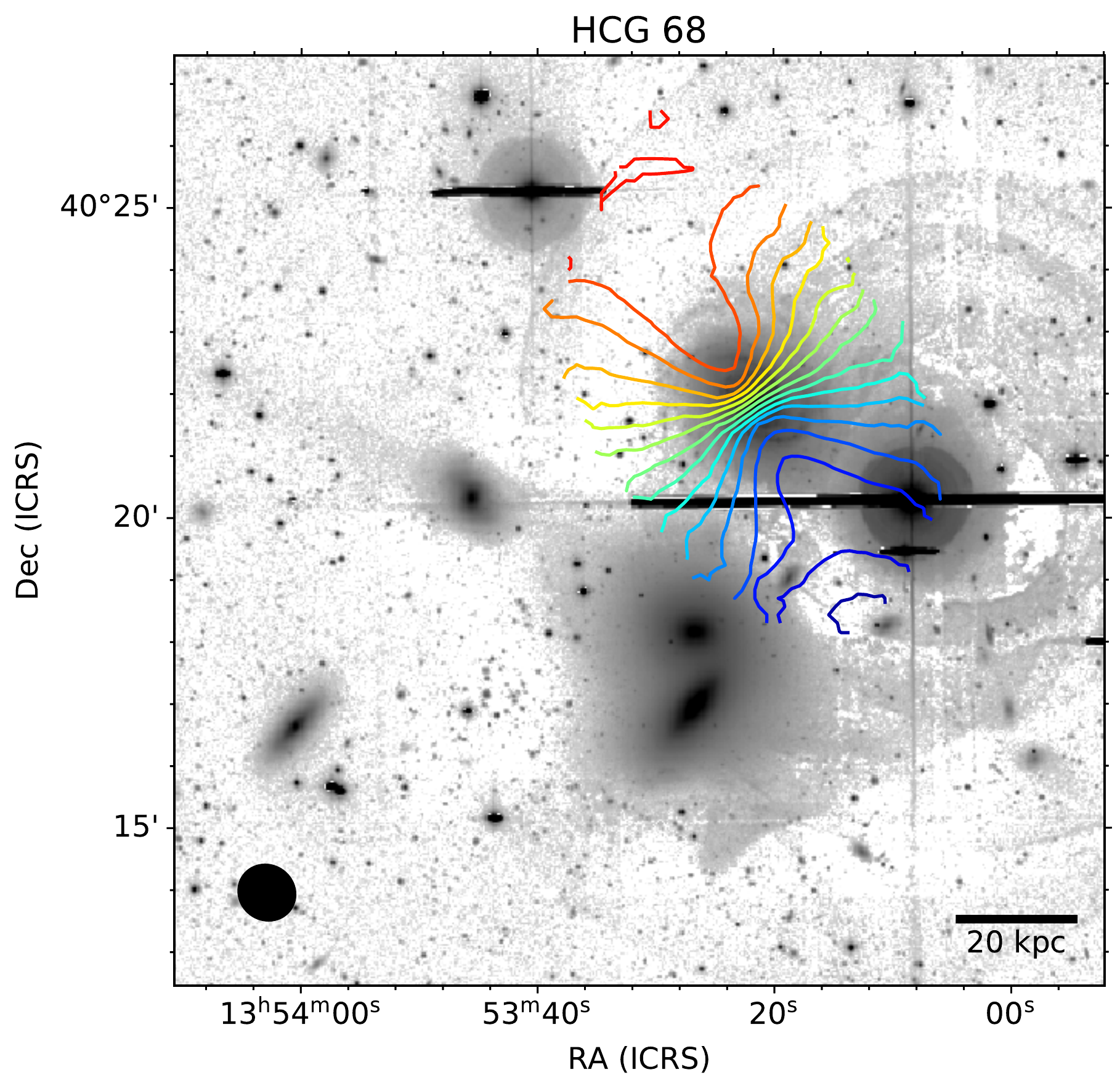}
    Figure \ref{fig:mom1_maps} continued.
\end{figure*}
\newpage

\begin{figure*}\nonumber
    \centering
    \includegraphics[width=\columnwidth]{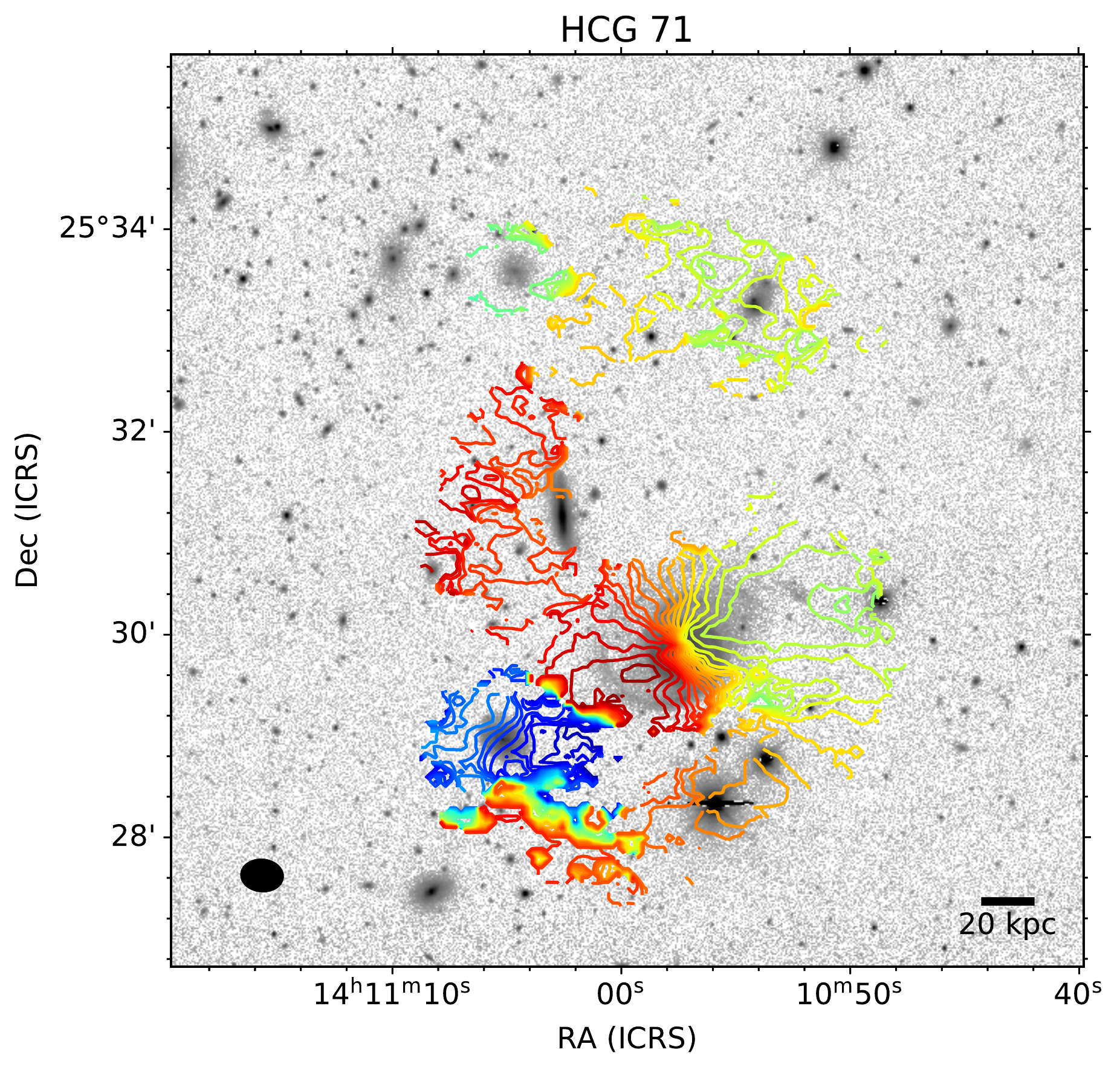}
    \includegraphics[width=\columnwidth]{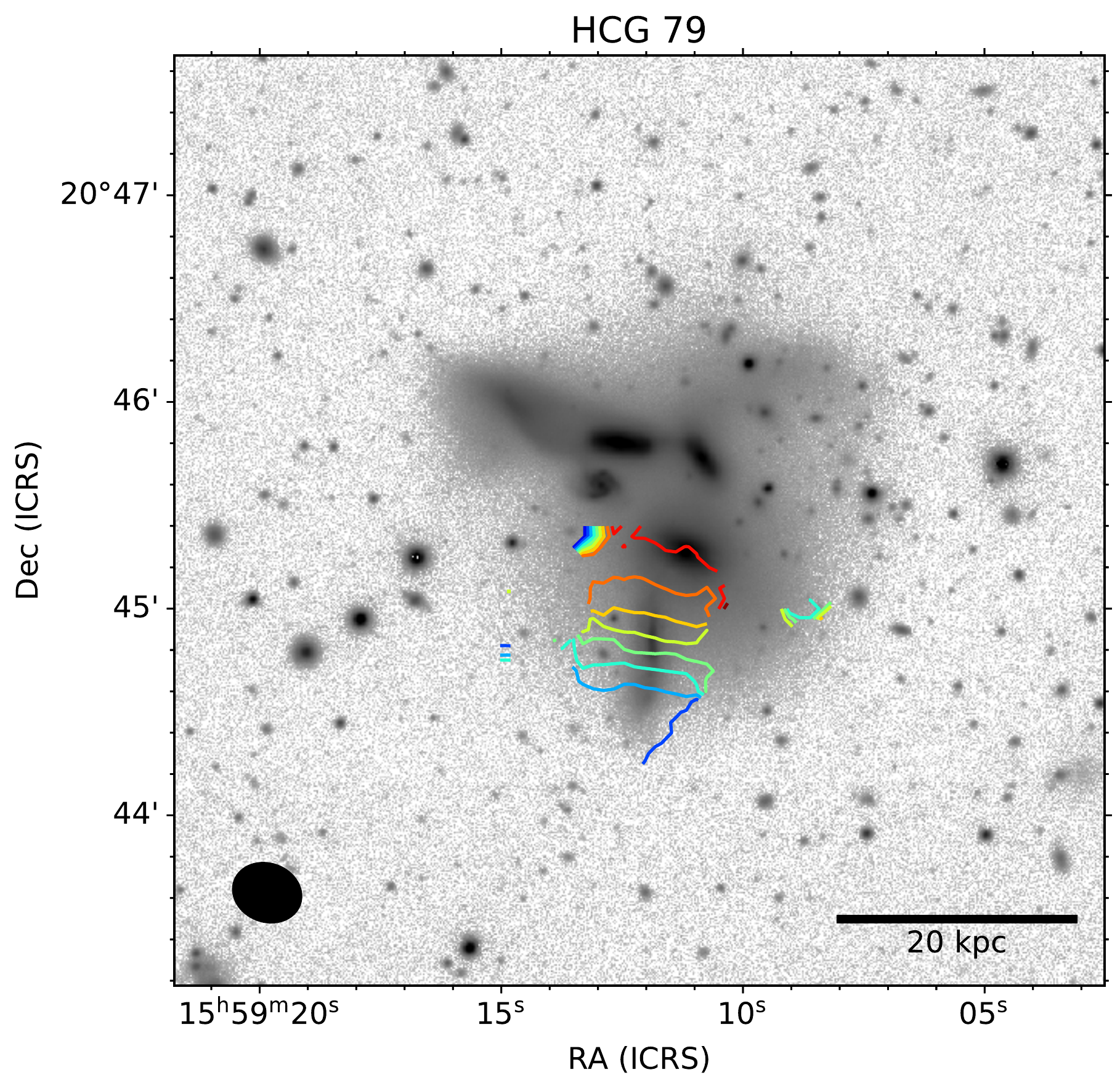}
    \includegraphics[width=\columnwidth]{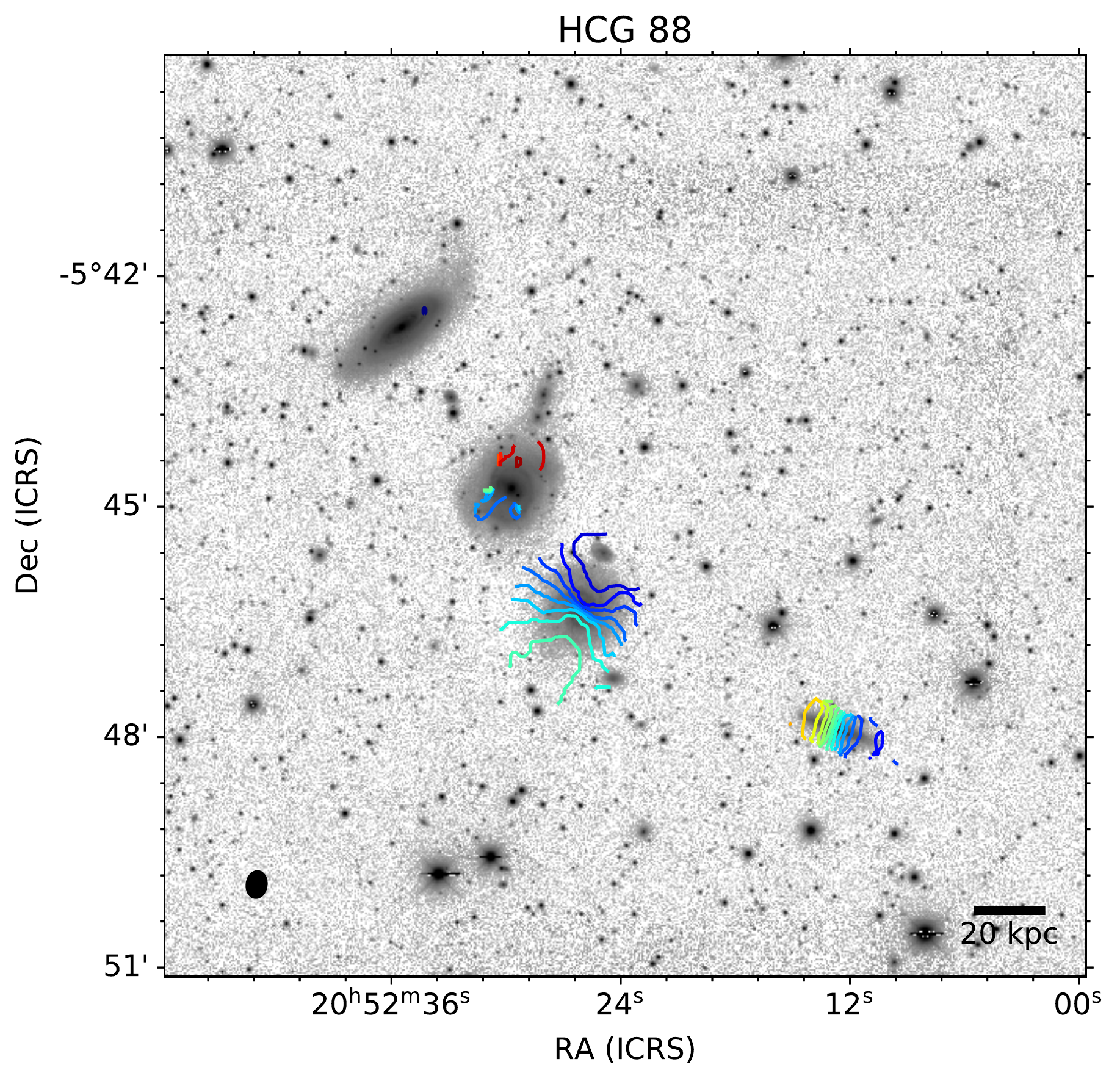}
    \includegraphics[width=\columnwidth]{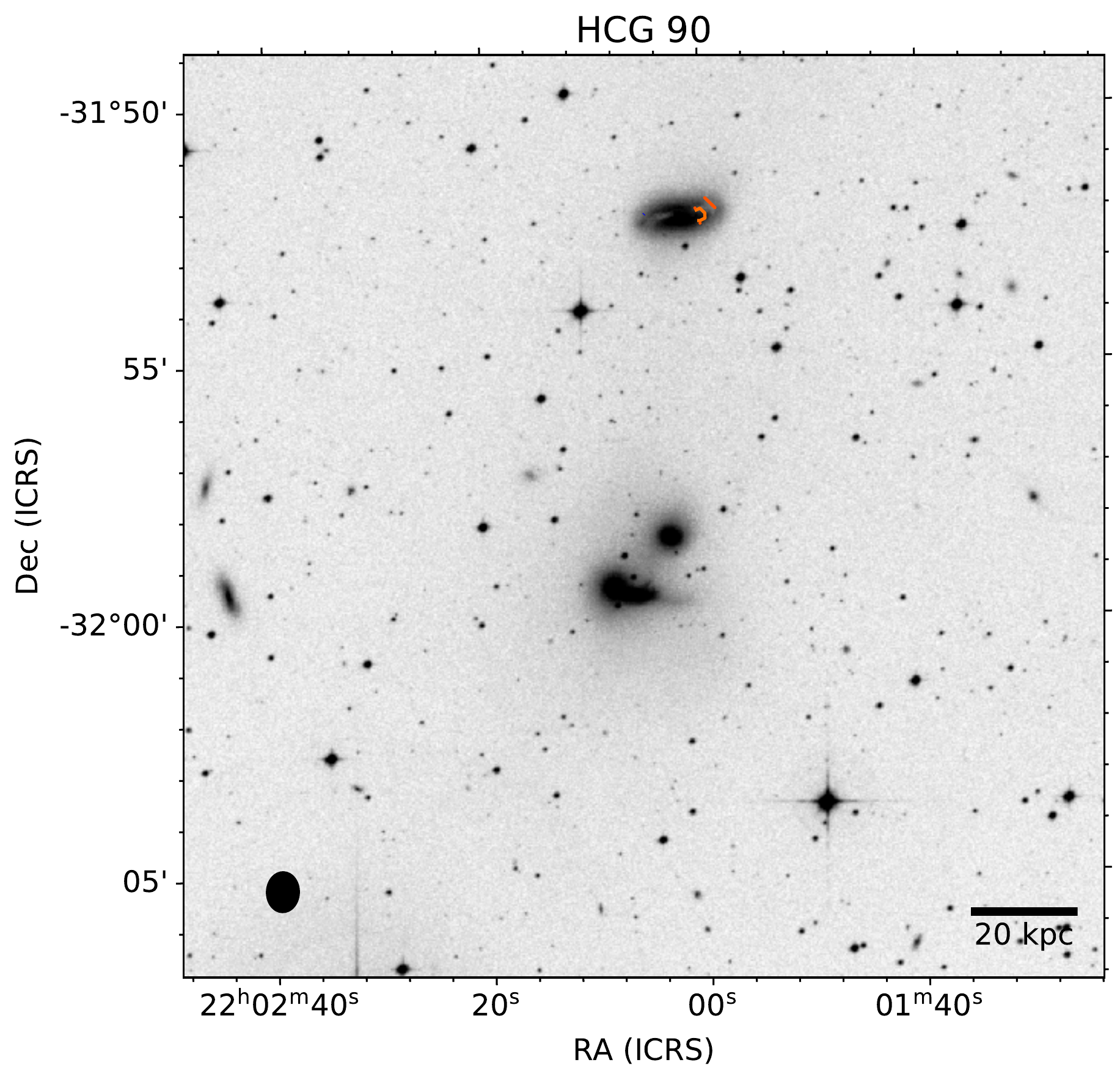}
    Figure \ref{fig:mom1_maps} continued.
\end{figure*}
\newpage

\begin{figure*}\nonumber
    \centering
    \includegraphics[width=\columnwidth]{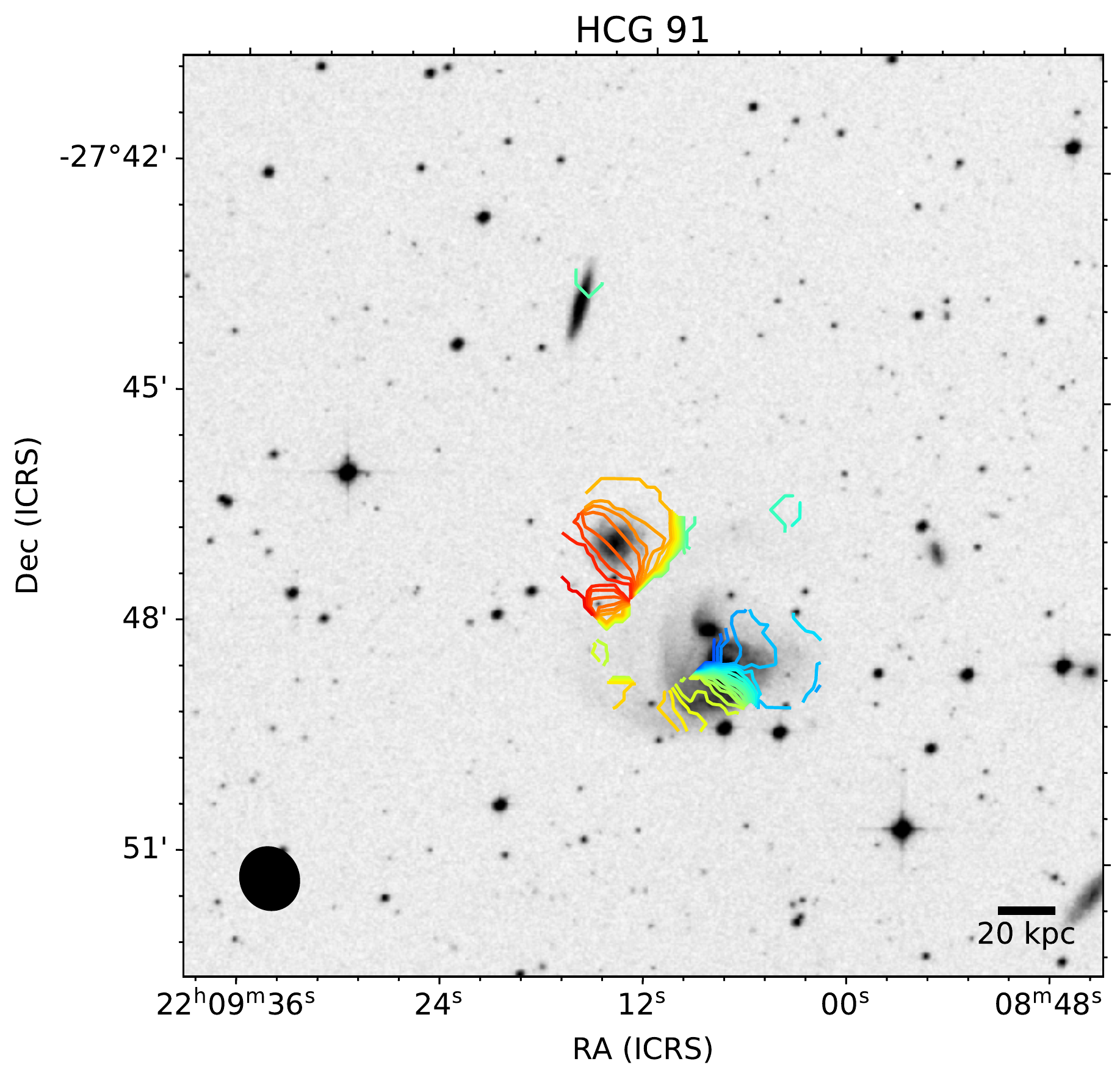}
    \includegraphics[width=\columnwidth]{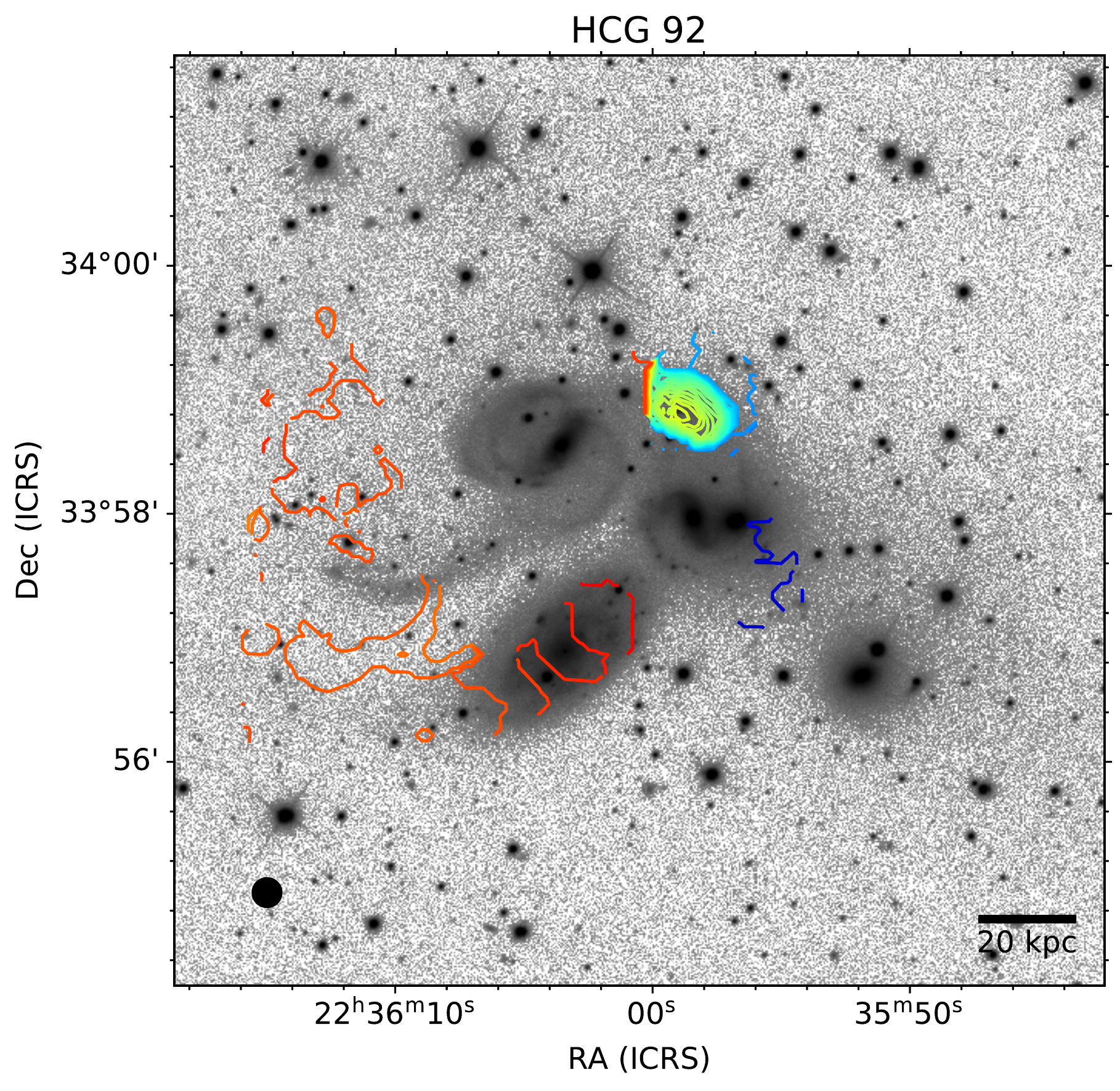}
    \includegraphics[width=\columnwidth]{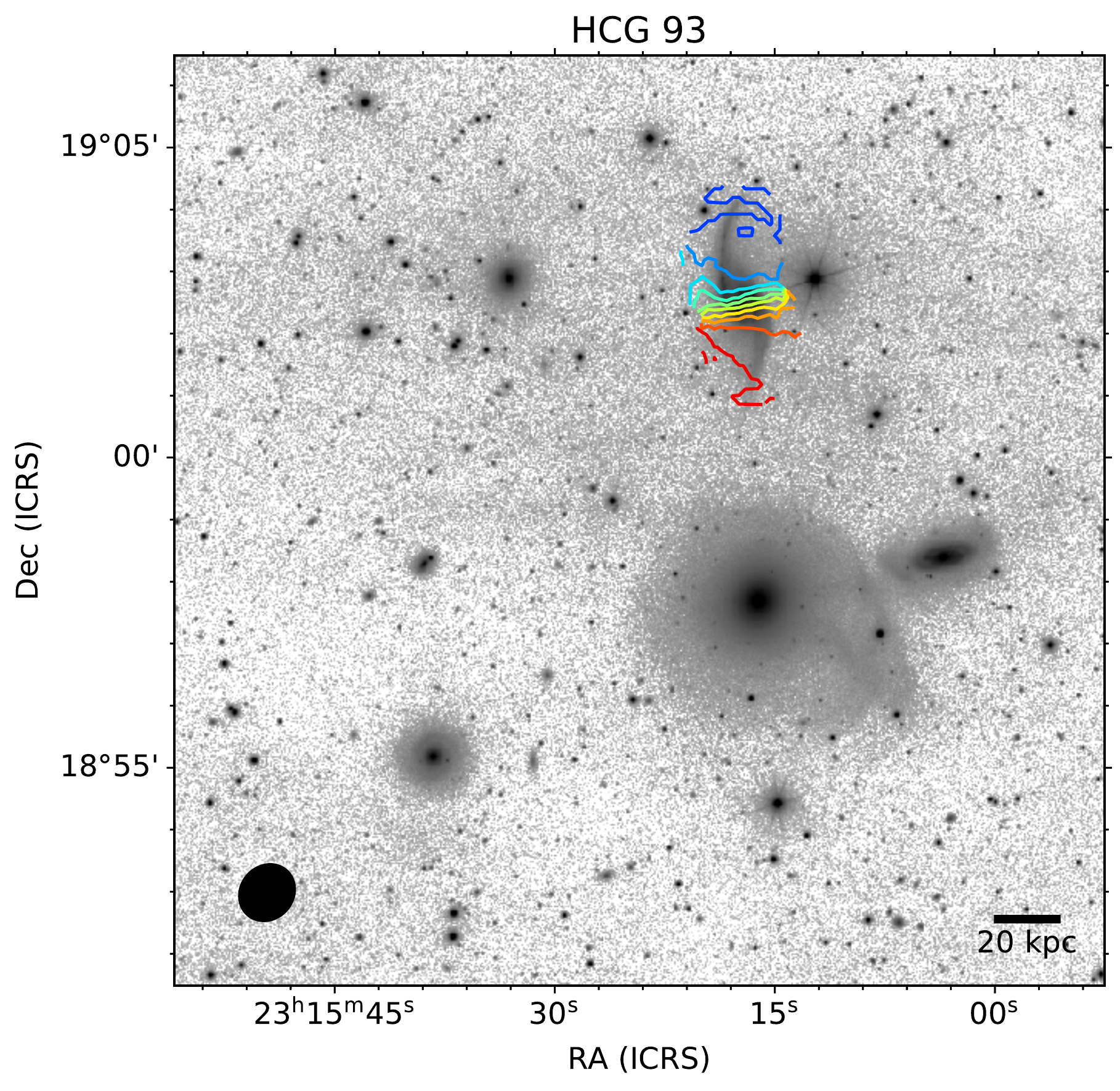}
    \includegraphics[width=\columnwidth]{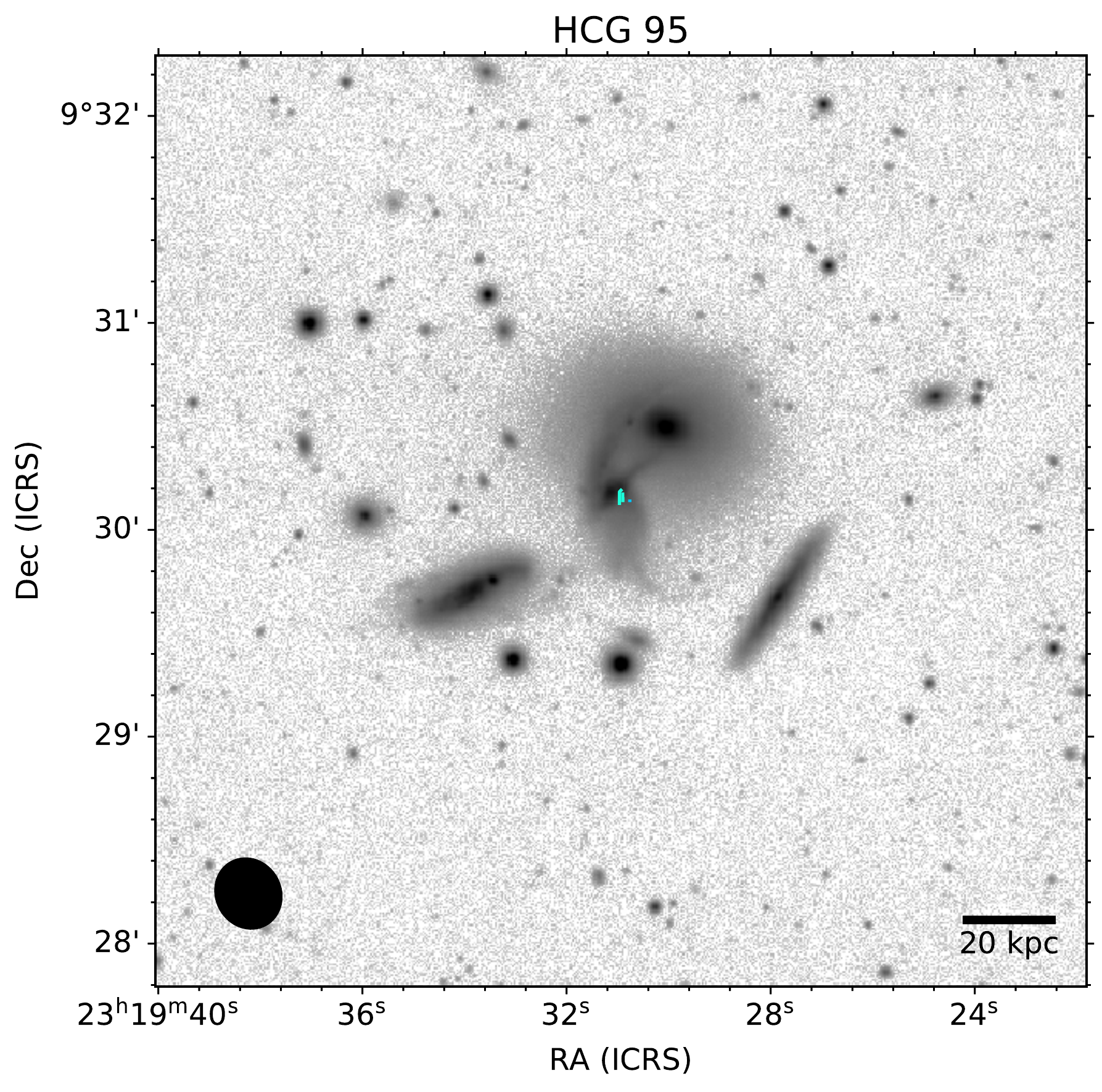}
    Figure \ref{fig:mom1_maps} continued.
\end{figure*}
\newpage

\begin{figure*}\nonumber
    \centering
    \includegraphics[width=\columnwidth]{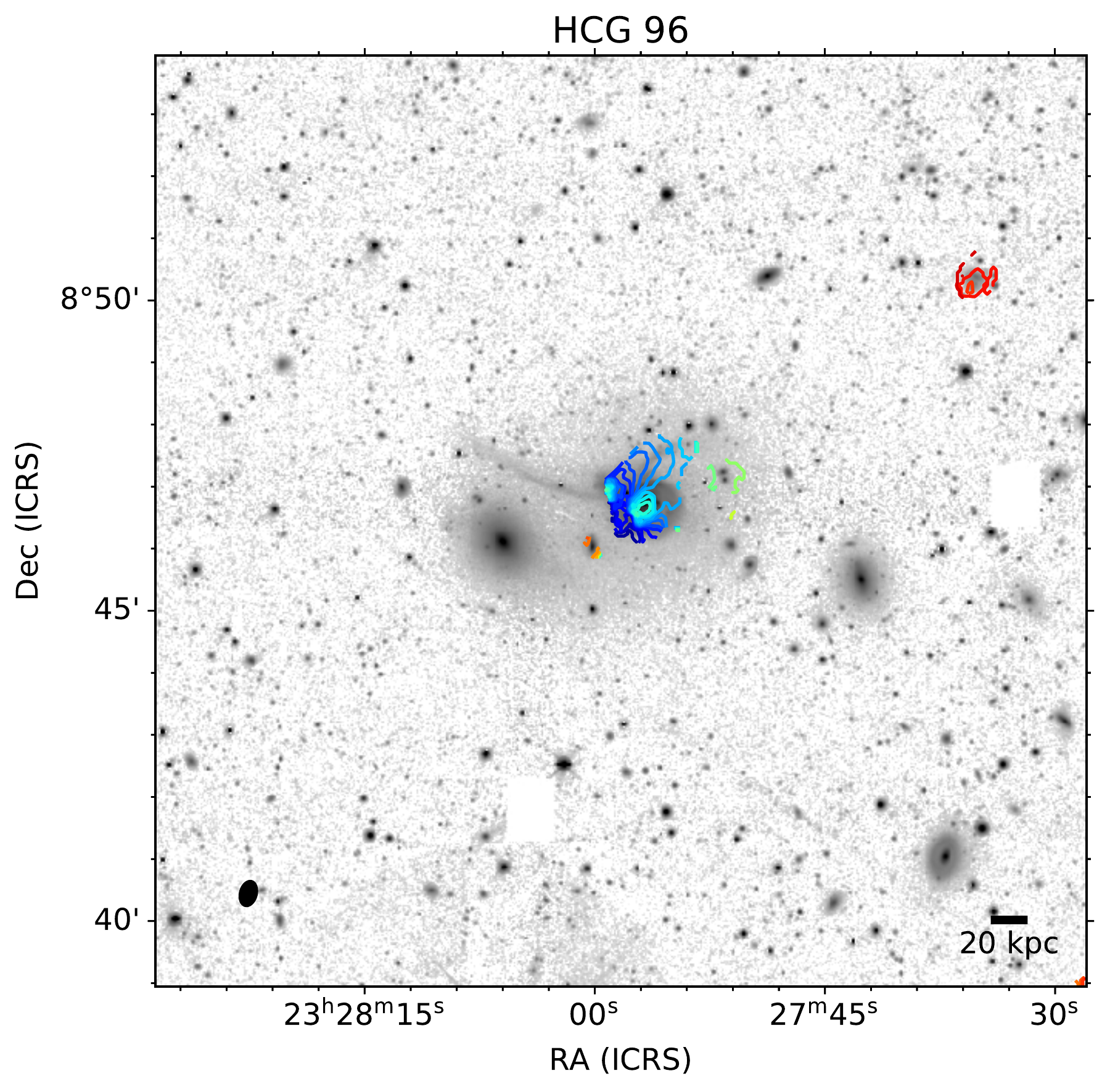}
    \includegraphics[width=\columnwidth]{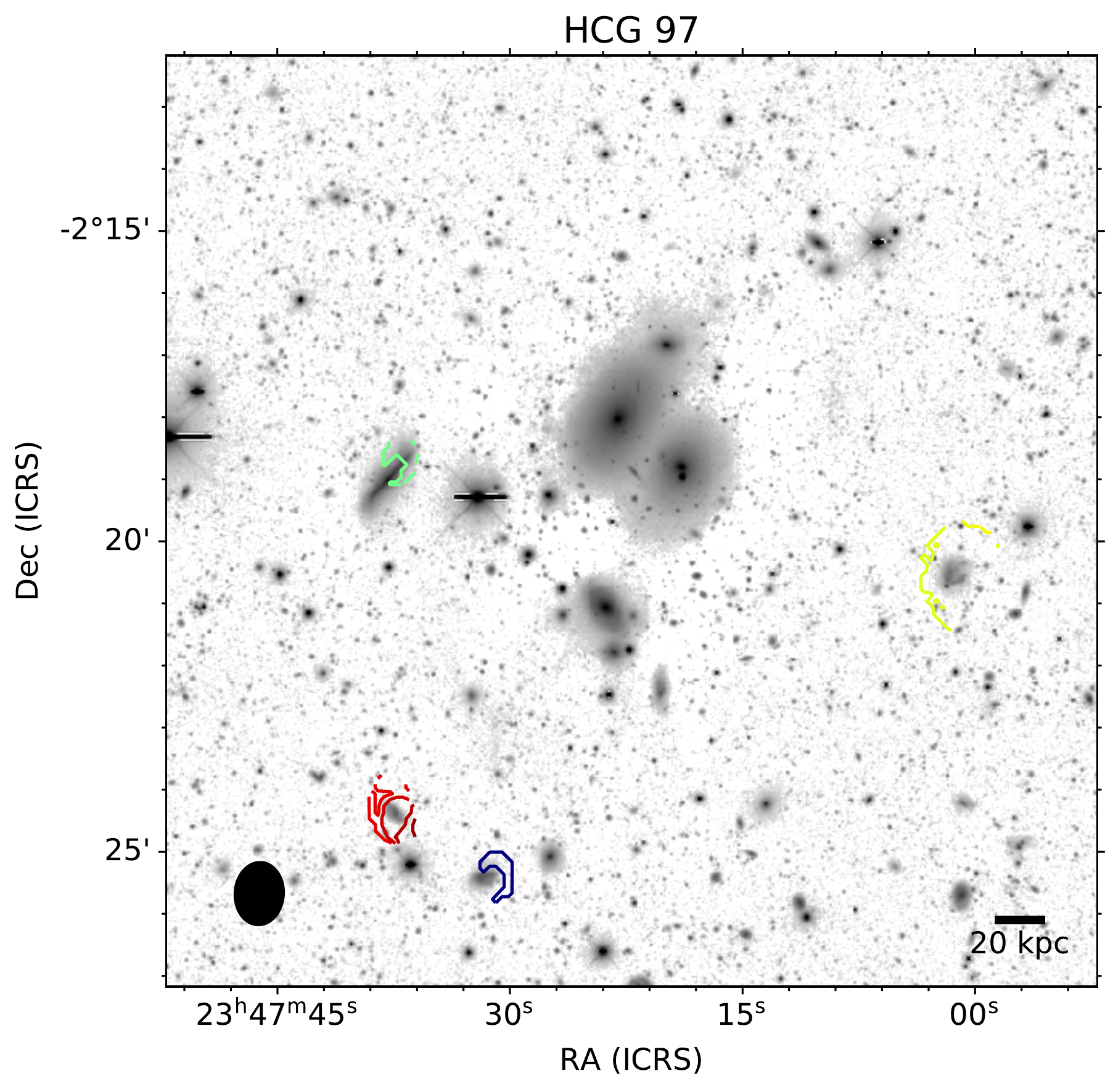}
    \includegraphics[width=\columnwidth]{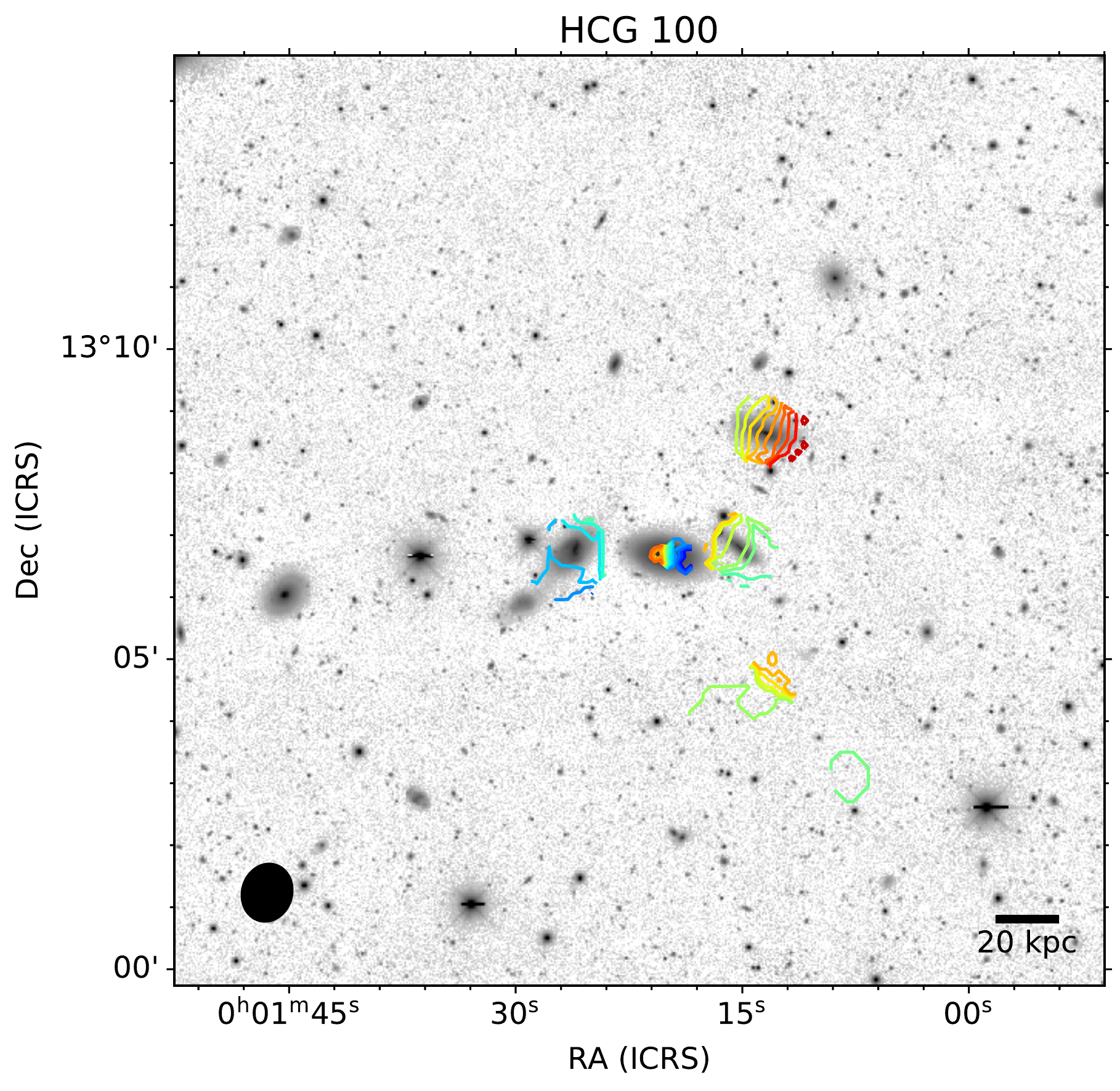}
    \\
    Figure \ref{fig:mom1_maps} continued.
\end{figure*}

\end{document}